%% file: paper.v6.tex
\documentclass[aps,nofootinbib,showpacs,preprintnumbers,amsmath,amssymb,showkeys]{revtex4-1}

\usepackage{amsmath}
\usepackage{epsf}
\usepackage{colordvi}
\usepackage{graphicx}
\usepackage{pst-all}
\usepackage{pslatex}
\usepackage{float}

\newcommand{\bsmeson}{$B_s$\ }
\newcommand{\ket}[1]{\left | #1 \right >}
\newcommand{\Ket}[1]{\Big | #1 \Big >}
\newcommand{\ten}[1]{$\ 10^{-#1}$}
\newcommand{\tten}[1]{\ 10^{-#1}}
\def\tstrut{\vrule height2.5ex depth0pt width0pt} 
\begin{document}

\title{Weak decays of $\bar B_s$ mesons}

\author{C. Albertus} 

\affiliation{Departamento de F\'\i sica At\'omica, Molecular y Nuclear.
  Universidad de Granada. \\ Avenida de Fuentenueva S/N, E-18071 Granada, Spain}


\begin{abstract}
In the present work we study the semileptonic decays of $\bar B_s$
mesons in the context of nonrelativistic constituent quark models. We
estimate the uncertainties of our calculation using different
interquark potentials to obtain the meson wave functions. We check the
results from our model against the predictions of Heavy Quark
Symmetry, in the limit of infinite heavy quark mass. We also study the
nonleptonic decays of $\bar B_s$ mesons within the factorization
approximation.
\end{abstract}

\maketitle

\section{INTRODUCTION}
Since the first claims on the existence of $B_s$ and $\bar B_s$, both their lifetimes,
decay
modes~\cite{Aaij:2013pua,Aaij:2013iua,Guadagnoli:2013mru,Blusk:2012it,
  Aaij:2013fha,Aaij:2013dda,Aaij:2013tna,Aaij:2013fpa,Dewhurst:2011zz,
Aad:2012kba,Giurgiu:2010zz,Abulencia:2005ia}
and
oscillations~\cite{GongRu:2012hh,Amhis:2012pa,vanEijk:2011mu,Kuhr:2011vm,
  Giurgiu:2010is, Batell:2010qw, Abulencia:2006mq}
have been objectives of the uttermost interest of experimental
collaborations. Being below the $B-K$ threshold, it can only decay by
means of mechanisms governed by electroweak currents, making it an
ideal system to study the physics of the weak interaction in the
presence of heavy quarks.

A considerable amount of the work devoted to the $b$-meson sector
involve the ideas of Heavy Quark Symmetry~\cite{Isgur:1989vq,
  Isgur:1989ed} (HQS). HQS is an approximate symmetry of QCD that
becomes exact in the limit in which the mass of the heavy quark
becomes infinity. This symmetry establishes that in such a limit, the
quantum numbers of the light degrees of freedom are all well defined,
and independent of the heavy quark flavor and spin. This is similar,
for instance, to what happens in atomic physics, where electron
properties are approximately independent of the mass and spin of the
nucleus for a fixed nuclear charge. Heavy Quark Symmetry can be cast
into the language of an effective theory, leading to Heavy Quark
Effective Theory~\cite{Georgi:1990um} (HQET). HQET enables a
systematic, order by order evaluation of the corrections to the
infinity mass limit in the inverse powers of the heavy quark
masses. Besides, HQET allows theoretical control of the
non-perturbative aspects of the calculation in the proximities of the
infinite quark mass limit. At leading order in an expansion on the
heavy quark mass only one form factor, the Isgur-Wise function
remains, largely simplifying the description of the decay. However,
HQS does not determine the Isgur-Wise function: one still needs to
implement some other nonperturbative method.

HQS leads to many more model independent predictions. The most
remarkable of those for the meson sector, is the fact that the masses
of pseudoscalar and vector mesons are degenerate in the heavy quark
limit. Nonrelativistic quark models fulfil this constrain: the
reduced mass of the two quarks is just the mass of the light one, and
the spin-spin terms, which can distinguish vector from pseudoscalar,
are suppressed by the mass of the heavy quark, becoming exactly zero
in the HQS limit. At this point, one important question is to what
extent do the deviations from the HQS limit, evaluated from
nonrelativistic quark models agree with the constraints predicted by
HQET. Furthermore, it is possible to make use of the HQET constrains
to improve the predictions of the quark models. In the previous work
of Ref.~\cite{Albertus:2005vd}, we studied the leptonic and
semileptonic decays of $B$ mesons, and considered the implications of
HQS. The nonleptonic and semileptonic decay of $B_c$ mesons (where
presence of two heavy quarks leads to infrared divergences that break
the flavor symmetry, and subsequently only Heavy Quark Spin Symmetry
remains), has been considered in Ref.~\cite{Hernandez:2006vv}. In
Refs.~\cite{Albertus:2006ya,Albertus:2012jt,Albertus:2011xz} we
calculated the semileptonic decay widths of baryons containing one or
two heavy quarks, and worked out the symmetry implications on the
observables.

Some of the decay modes of $B_s$ or $\bar B_s$ mesons have been
studied within the framework of relativistic constituent quark model
\cite{Faustov:2012mt, Faustov:2013ima}, perturbative QCD
\cite{Sun:2013dla, Yu:2013pua}, Bethe-Salpeter techniques
\cite{Chen:2012fi}, light front quark model \cite{Li:2009wq}, sum
rules \cite{Azizi:2008xy, Azizi:2008vt,Blasi:1993fi} or non
relativistic constituent quark model \cite{Zhao:2006at} for
instance. In this paper we study the semileptonic and nonleptonic
decay of $\bar B_s$ in the context of nonrelativistic constituent
quark model. The rest of the paper is organized as follows. In
Section~\ref{sec:mesonstates} we describe the meson states for the
different values of $J^P$ and the quark models used in this work. In
Sec.~\ref{sec:semileptonic} we give the form factor decomposition of
the weak decay matrix elements and calculate the decay width, both for
light ($e,\mu$) and heavy ($\tau$) charged lepton. We also work in the
helicity formalism \cite{Ivanov:2005fd}. Besides, we study the
implications of HQS in these decays. In Sec.~\ref{sec:nonleptonic} the
problem of nonleptonic two meson decays of \bsmeson is studied. The
meson decay constants required in Sec.~\ref{sec:nonleptonic} and the
CKM matrix elements used both in Secs.~\ref{sec:semileptonic} and
\ref{sec:nonleptonic} can be found in Tables~\ref{tab:decaycons-vckm},
respectively, while the different $D_s$ states considered in the
semileptonic and some nonleptonic decays studied in this paper and
their quantum numbers are summarized in
Table~\ref{tab:quantumnumbers}. For the $D_s^1(2460)$ and
$D_s^1(2536)$ states, we assume that they are mixing of $^3P_1$ and
$^1P_1$ $c\bar s$ states, with a mixing angle of $34.5^\circ$, as in
Ref.~\cite{Faustov:2012mt}. In Sec.~\ref{sec:summary} we present a
summary and our conclusions. The paper also includes an Appendix to
clarify some technical details of our work.
\begin{table}[hb]
\begin{tabular}{cc}
\hline\hline
                        & Mass ($\rm MeV$)  \\\hline
$\bar B_s(0^-)$         &    5366.77  \cite{PhysRevD.86.010001}            \\
$D_s^+(0^-)$             &   1968.49  \cite{PhysRevD.86.010001}            \\
$D_{s0}^{*+}(2317)(0^+)$  &    2317.8~  \cite{PhysRevD.86.010001}            \\
$D_s^{*+}(1^-)$          &    2112.3~  \cite{PhysRevD.86.010001}            \\
$D_{s1}(2460)$           &    2459.6~  \cite{PhysRevD.86.010001}            \\
$D_{s1}(2536)$           &    2535.12  \cite{PhysRevD.86.010001}            \\
$c\bar s(2^-)$          &    2806.9~                                       \\
$D_{s2}(2573)^+(2^+)$    &    2571.9~  \cite{PhysRevD.86.010001}            \\\hline
\end{tabular}
\caption{Masses of the states involved in this calculation.
\label{tab:quantumnumbers}
}
\end{table}
%
%
\begin{table}[hb]
\begin{tabular}{cccccccccc||cccccc}
\hline\hline
$f_\pi$ & $f_\rho$ & $f_K$ & $f_{K^*}$&$f_D$ & $f_{D^*}\dagger$  &$f_{D_s}$ & $f_{D_s^*}\dagger$  & $f_\Phi\dagger$ & $f_{J/\Psi}\dagger$   &$|V_{cb}|$   &  $|V_{ud}|$  & $|V_{us}|$ & $|V_{cs}|$  & $|V_{cd}|$ \\\hline
130.41 & 210     & 159.8 & 217     & 206.7  & 222       &  260    & 318        & 312.6   & 488.5       & 0.0413   & 0.9743   & 0.2240    & 0.9734 & 0.2252\\ \hline
\end{tabular}
\caption{\label{tab:decaycons-vckm} Values for the meson decay
  constants in MeV and Cabibbo-Kobayashi-Maskawa matrix elements used
  through this work. The decay constants marked with a $\dagger$ have
  been calculated using our model.  }
\end{table}

%
%
\section{MESON STATES AND INTERQUARK INTERACTIONS}
\label{sec:mesonstates}
In the context of nonrelativistic constituent quark models, the state
of a meson $M$ is written as ~\cite{Hernandez:2006gt}:
\begin{equation}
\begin{split}
\ket{M; \lambda \vec P}_{\rm NR} =& \int d^3 p \sum_{\alpha_1 \alpha_2} \hat{\phi}_{\alpha_1 \alpha_2}^{(M,\lambda)}(\vec p\,)
\frac{(-1)^{(1/2)-s_2}}{(2\pi)^{3/2}\sqrt{(2E_{f_1}(\vec{p}_1))(2E_{f_2}(\vec{p}_2))}}
\\
&\times 
\Ket{q,\alpha_1\,\vec{p}_1=\frac{m_{f_{1}}}{m_{f_{1}}+m_{f_{2}}}\vec{P}-\vec p}
\Ket{\bar{q},\alpha_2\,\vec{p}_2=\frac{m_{f_{2}}}{m_{f_{1}}+m_{f_{2}}}\vec{P}+\vec p},
\label{eq:mesonstates}
\end{split}
\end{equation}
where $\vec P$ is the meson three momentum, while $\lambda$ labels the
spin projection in the meson center of mass. The index $\alpha_i$
represent the quantum numbers of spin, flavor and color of the quark
and the antiquark, with four momentum and mass given by
$(E_{f_{i}}(\vec{p}_i),\vec{p}_i)$ and $m_{f_i}$ respectively. The
factor $(-1)^{1/2-s_2}$ ensures that the antiquark spin states have
the correct phase \footnote{Under charge conjugation {\cal C}, quark
 and antiquark states are related via ${\cal C}c_\alpha^\dagger{\cal
    C}^\dagger=(-1)^{1/2 -s}d_{\alpha}^\dagger(\vec p)$, so the antiquark states with the
  correct spin relative phase are $(-1)^{1/2 -s}d_\alpha^\dagger(\vec
  p)\ket{0} = (-1)^{1/2 -s}\ket{\bar q,\alpha\vec{p}}$}.

The normalization of quark and antiquark states is
\begin{equation}
\left\langle \alpha' \vec{p}\,'| \alpha \vec p \,\right\rangle = 2 E_f \delta_{\alpha' \alpha} \delta^3(\vec p - \vec p\,')(2\pi)^3
\end{equation}
As for the momentum wave function accounting for the relative motion
of the quark-antiquark system, the normalization is given by
\begin{equation}
\int d^3 p \sum_{\alpha_1 \alpha_2} (\hat{\phi}^{M, \lambda'}_{\alpha_1 \alpha_2}(\vec p\,))^* \hat{\phi}^{M, \lambda}_{\alpha_1 \alpha_2}(\vec p\,) = \delta_{\lambda \lambda'},
\end{equation}
and finally, the normalization of the meson states in our model is
\begin{equation}
_{\rm NR}\left<M\lambda'\vec P'|M\lambda\vec P\right>_{\rm NR} = \delta_{\lambda\lambda'}(2\pi)^3\delta(\vec P' -\vec P).
\label{eq:normnr}
\end{equation}
\\
In this calculation we will need the ground state wave function for
scalar ($0^+$), pseudoscalar ($0^-$), vector ($1^-$), axial-vector
($1^+$), tensor ($2^+$) and pseudotensor ($2^-$). Assuming always a
value for the orbital angular momentum as low as possible, we have
for a meson $M$ with scalar, pseudoscalar and vector quantum numbers
\begin{widetext}
\begin{align}
\hat{\phi}_{\alpha_1,\alpha_2}^{(M(0^+))}(\vec p\,)&=\frac{1}{\sqrt3}\delta_{c_1c_2}\hat{\phi}_{(s_1,f_1),(s_2,f_2)}^{(M(0^+))}(\vec p\,)
=\frac{i}{\sqrt3}\delta_{c_1c_2}\hat{\phi}_{f_1,f_2}^{(M(0^+))}(|\vec p\,|)\sum_m(1/2,1/2,1;s_1,s_2,-m)(1,1,0;m,-m,0)Y_{1m}(\hat p)\nonumber\\
\hat{\phi}_{\alpha_1,\alpha_2}^{(M(0^-))}(\vec p\,)&=\frac{1}{\sqrt3}\delta_{c_1c_2}\hat{\phi}_{(s_1,f_1),(s_2,f_2)}^{(M(0^-))}(\vec p\,)
=\frac{-i}{\sqrt3}\delta_{c_1c_2}\hat{\phi}_{f_1,f_2}^{(M(0^-))}(|\vec p\,|)(1/2,1/2,0;s_1,s_2,0)Y_{00}(\hat p)\nonumber\\
\hat{\phi}_{\alpha_1,\alpha_2}^{(M(1^-),\lambda)}(\vec p\,)&=\frac{1}{\sqrt3}\delta_{c_1c_2}\hat{\phi}_{(s_1,f_1),(s_2,f_2)}^{(M(1^-),\lambda)}(\vec p\,)
=\frac{-1}{\sqrt3}\delta_{c_1c_2}\hat{\phi}_{f_1,f_2}^{(M(1^-))}(|\vec p\,|)(1/2,1/2,1;s_1,s_2,0)Y_{00}(\hat p),
\end{align}
where $(j_1,j_2,j_3,m_1,m_2,m_3)$ are Clebsch-Gordan coefficients,
$Y_{lm}$ are spherical harmonics and $\hat\phi_{f_1,f_2}(|\vec p\,|)$
is the Fourier transform of the radial, coordinate space, wave
function.

Axial vector mesons require orbital angular momentum $L=1$, and in
this case the two possible values of the total quark-antiquark spin
$S_{q\bar q}=0,1$ are allowed. Thus, there are two possible states:
\begin{align}
\hat{\phi}_{\alpha_1,\alpha_2}^{(M(1^+),S_{q\bar q}=0,\lambda)}(\vec
p\,)&=\frac{1}{\sqrt3}\delta_{c_1c_2}\hat{\phi}_{(s_1,f_1),(s_2,f_2)}^{(M(1^+),S_{q\bar
    q}=0,\lambda)}(\vec p\,)
=\frac{-1}{\sqrt3}\delta_{c_1c_2}\hat{\phi}_{f_1,f_2}^{(M(1^+),S_{q\bar
    q}=0)}(|\vec p\,|)(1/2,1/2,0;s_1,s_2,0)Y_{1\lambda}(\hat
p)\nonumber\\ 
\hat{\phi}_{\alpha_1,\alpha_2}^{(M(1^+),S_{q\bar q}=1,\lambda)}(\vec p\,)&=\frac{1}{\sqrt3}\delta_{c_1c_2}\hat{\phi}_{(s_1,f_1),(s_2,f_2)}^{(M(1^+),S_{q\bar
    q}=1,\lambda)}(\vec p\,) \nonumber\\ &=\frac{-1}{\sqrt3}\delta_{c_1c_2}\hat{\phi}_{f_1,f_2}^{(M(1^+),S_{q\bar
    q}=1)}(|\vec p\,|)\sum_m(1/2,1/2,1;s_1,s_2,\lambda-m)(1,1,1;m,\lambda-m,\lambda)Y_{1m}(\hat p).
\end{align}
For tensor and pseudotensor mesons, the wave functions can be written as:
\begin{align}
\hat{\phi}_{\alpha_1,\alpha_2}^{(M(D_{s2}^*),\lambda)}(\vec p\,)&=
\frac{1}{\sqrt3}\delta_{c_1c_2}\hat{\phi}_{(s_1,f_1),(s_2,f_2)}^{(M(D_{s2}^*),\lambda)}(\vec p\,) \nonumber\\ 
&=\frac{1}{\sqrt3}\delta_{c_1c_2}\hat{\phi}_{f_1,f_2}^{(M(D_{s2}^*))}(|\vec p\,|)
\sum_m(1/2,1/2,1;s_1,s_2,\lambda-m)(1,1,2;m,\lambda-m,\lambda)Y_{1m}(\hat p)\nonumber\\
\hat{\phi}_{\alpha_1,\alpha_2}^{(M(2^-),\lambda)}(\vec p\,)&=
\frac{1}{\sqrt3}\delta_{c_1c_2}\hat{\phi}_{(s_1,f_1),(s_2,f_2)}^{(M(2^-),\lambda)}(\vec p\,) \nonumber\\ 
&=\frac{-1}{\sqrt3}\delta_{c_1c_2}\hat{\phi}_{f_1,f_2}^{(M(2^-))}(|\vec p\,|)
\sum_m(1/2,1/2,1;s_1,s_2,\lambda-m)(2,1,2;m,\lambda-m,\lambda)Y_{2m}(\hat p)
\end{align}
\end{widetext}
In the previous expressions, all phases have been introduced for later
convenience.

We consider five different interquark potentials to calculate the
coordinate space wave functions, one proposed by Bhadury
\cite{Bhaduri:1981pn} and other four proposed by Silvestre-Brac in
\cite{SilvestreBrac:1996bg}. All of them have the same structure: a
term accounting for confinement, plus Coulomb and hyperfine terms both
of them coming from one-gluon exchange. They differ from one another
in the form factors present in the hyperfine term, the power of the
confinement term, or the presence of a form factor in the Coulomb
one-gluon exchange term. All free parameters have been adjusted to
reproduce light and heavy-light meson spectra. We have successfully
used these potentials before to describe the spectra and decays of
charmed and bottom baryons.

The different results obtained with the different potentials provide
us with an estimation of the theoretical error. It has to be mentioned
that another source of theoretical uncertainty that we cannot account
for is the use of nonrelativistic kinematics in the evaluation of the
wave function. While this approximation is not, a priori, a good
choice in the presence of light quarks, one has to notice that all
nonrelativistic potentials have free parameters fitted to experimental
data. Hence, one can argue that the ignored relativistic effects are
partially included in the fitted values of the parameters.

\section{SEMILEPTONIC DECAYS}
\label{sec:semileptonic}

In this section we will consider the semileptonic decay of $\bar B_s$
mesons into different $D_s$ meson states with $0^+$, $0^-$, $1^+$, $1^-$,
$2^+$ and $2^-$ spin-parity quantum numbers. These decays correspond
to $b\to c$ transition at the quark level governed by the current
\begin{equation}
J^{cb}_\mu(0) = J^{cb}_{V\mu}(0)-J^{cb}_{A\mu}(0)=\bar\Psi_c(0)\gamma_\mu(I-\gamma_5)\Psi_b(0),
\label{eq:v-a}
\end{equation}
with $\Psi_f$ a quark field with flavor $f$.\\

\subsection{Form factor decomposition of hadronic matrix elements}

The hadronic matrix elements involved in these processes can be
parametrized in terms of form factors as:
\begin{widetext}
\begin{align}
\left<D_s^+,\vec{P}_{D_s}\left|J^{bc}_\mu(0) \right|\bar B_s,\vec{P}_{\bar B_s} \right> &=P_\mu F_+(q^2) + q_\mu F_-(q^2)\nonumber\\
\left<D_s^{*+},\lambda \vec{P}_{D_s^*}\left|J^{bc}_\mu(0) \right|\bar B_s,\vec{P}_{\bar B_s} \right> &=
\frac{-1}{m_{\bar B_s}+m_{D_s^*}}\epsilon_{\mu\nu\alpha\beta}\epsilon_{(\lambda)}^{\nu *}(\vec{P}_{c\bar s})P^\alpha q^\beta V(q^2)\nonumber\\
&-i\left\lbrace(m_{\bar B_s}-m_{D_s^*})\epsilon_{(\lambda)\mu}^*(\vec{P}_{c\bar s}) A_0(q^2) 
-\frac{P\cdot \epsilon^*_{(\lambda)}(\vec{P}_{D_s^*})}{m_{\bar B_s}+m_{D_s^*}}(P_\mu A_+(q^2) + q_\mu A_-(q^2))
\right\rbrace\nonumber\\
\left<D_{s2}^{*+},\lambda \vec{P}_{D_{s2}^*}\left|J^{bc}_\mu(0) \right|\bar B_s,\vec{P}_{\bar B_s} \right> &
=\epsilon_{\mu\nu\alpha\beta}\epsilon^{\nu\delta*}_{(\lambda)}(\vec{P}_{D_{s2}^*})P_\delta P^\alpha q^\beta T_4(q^2)\nonumber\\
&-i\left\lbrace
\epsilon_{(\lambda)\mu\delta}^*(\vec{P}_{D_{s2}^*})P^\delta T_1(q^2)+
P^\nu P^\delta \epsilon^*_{(\lambda)\nu\delta}(\vec{P}_{D_{s2}^*})(P_\mu T_2(q^2) + q_\mu T_3(q^2))
\right\rbrace,
\label{eq:matrixff}
\end{align}
\end{widetext}
where $P_{\bar B_s}$ and $P_{c\bar s}$ (with $c\bar s =D_s, D_s^*,
D_{s2}^*$) are the meson four-momenta, $m_{\bar B_s}$ and $m_{c\bar
  s}$ their masses respectively, $P=P_{\bar B_s} + P_{c\bar s}$,
$q=P_{\bar B_s} - P_{c\bar s}$. $\epsilon^{\mu\nu\alpha\beta}$ is the
fully antisymmetric tensor, for which we have taken the convention
$\epsilon^{0123}=1$. $q^2$ ranges from $q^2_{\rm min}=m_l^2$ to
$q^2_{\rm max}=(P_{\bar B_s} - P_{c\bar s})^2$. It is common to use
$\omega = (m_{\bar B_s}^2-m_{c\bar s}^2-q^2)/2m_{\bar B_s}m_{c\bar s}$
instead of $q^2$, corresponding $\omega_{\rm min}=1$ to $q^2_{\rm
  max}$.  $\epsilon_{(\lambda)\mu}(\vec P)$ and
$\epsilon_{(\lambda)\mu\nu}(\vec P)$ are the polarization vector and
tensor of vector and tensor mesons, respectively. The latter can be
evaluated as
\begin{equation}
\epsilon_{(\lambda)}^{\mu\nu}(\vec P) = \sum_m (1,1,2;m,\lambda-m,\lambda)\epsilon^{\mu}_{(\lambda)}(\vec P)\epsilon^{\nu}_{(\lambda-m)}(\vec P).
\end{equation}
The different polarization vectors used in this work can be found in
the Appendix of Ref.~\cite{Hernandez:2006gt}.

Meson states in Eq.~(\ref{eq:matrixff}) are normalized as
\begin{equation}
\left\langle M,\lambda' \vec P'| M\lambda \vec P\right\rangle = \delta_{\lambda'\lambda} 2 E_M(\vec P) (2\pi)^3\delta(\vec P - \vec P'),
\end{equation}
where $E_M(\vec P)$ is the energy of the meson $M$ with three momentum
$\vec P$. The factor $2E_M$ should be noticed, in contrast with
Eq.~(\ref{eq:normnr}).

For $0^+$, $1^+$ and $2^-$ final states the form factor decomposition
is the same as for the $0^-$, $1^-$ and $2^+$ cases above, where just
$-J_A^{cb}(0)$ is contributing instead of $J_V^{cb}(0)$ and vice versa.

\subsection{Decays into scalar and pseudoscalar states}

In this section we will consider the decay of $\bar B_s$ mesons into
pseudoscalar and scalar $c\bar s$ mesons. For $\bar B_s \to D_s^+$,
i. e. $(0^-)$ transitions, the form factors are given by
\begin{align}
F_+(q^2)&=\frac{1}{m_{\bar B_s}}\left(V^0(|\vec q\,|)+\frac{V^3(|\vec q\,|)}{|\vec q\,|}(E_{D_s}(-\vec q\,)-m_{D_s})\right)\nonumber\\
F_-(q^2)&=\frac{1}{m_{\bar B_s}}\left(V^0(|\vec q\,|)+\frac{V^3(|\vec q\,|)}{|\vec q\,|}(E_{D_s}(-\vec q\,)+m_{D_s})\right)
\label{eq:ffpseudoscalar}
\end{align}
whereas for a transition onto a $D_{s0}^{*+} (0^+)$ state we have
\begin{align}
F_+(q^2)&=\frac{-1}{m_{\bar B_s}}\left(V^0(|\vec q\,|)+\frac{V^3(|\vec q\,|)}{|\vec q\,|}(E_{D_{s0}^*}(-\vec q\,)-m_{D_{s0}^*})\right)\nonumber\\
F_-(q^2)&=\frac{-1}{m_{\bar B_s}}\left(V^0(|\vec q\,|)+\frac{V^3(|\vec q\,|)}{|\vec q\,|}(E_{D_{s0}^*}(-\vec q\,)+m_{D_{s0}^*})\right)
\end{align}
where $V^\mu(|\vec q|)$ and $A^\mu(|\vec q|)$ ($\mu = 0,3$) are calculated in our model as
\begin{align}
V^\mu(|\vec q\,|) &= \left\langle D_s^+, -|\vec q\,|\vec k\,\left|J_V^{cb\ \mu}(0)\right|\bar B_s, \vec 0 \right\rangle 
=\sqrt{4m_{\bar B_s}E_{D_s}(-\vec q\,)}_{\rm NR}\left\langle D_s^+, -|
\vec q\,|\vec k\,\left|J_V^{cb\ \mu}(0)\right|\bar B_s, \vec 0 \right\rangle_{\rm NR}
\nonumber\\
A^\mu(|\vec q\,|) &= \left\langle D_{s0}^{*+}, -|\vec q\,|\vec k\,\left|J_A^{cb\ \mu}(0)\right|\bar B_s, \vec 0 \right\rangle
=\sqrt{4m_{\bar B_s}E_{D_{s0}^*}(-\vec q\,)}_{\rm NR}\left\langle D_{s0}^{*+}, 
-|\vec q\,|\vec k\,\left|J_A^{cb\ \mu}(0)\right|\bar B_s, \vec 0 \right\rangle_{\rm NR}
\end{align}
where the expressions for the non-relativistic matrix elements are
given in the Appendix. Figure~\ref{fig:ffscalar} represents the form
factors calculated with the wave functions corresponding to the AL1
potential~\cite{SilvestreBrac:1996bg}.
\begin{widetext}
\begin{center}
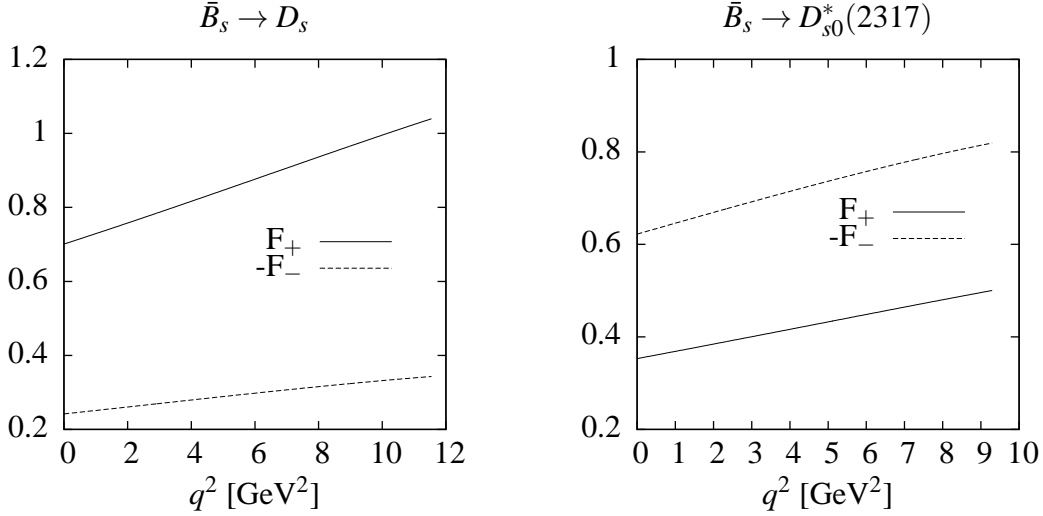
\begin{figure}[ht]{\parindent 0cm
\large
\makebox[0pt]{\input{ff.0m.0p.tex}
}
\caption{\label{fig:ffscalar}Form factors for the semileptonic decay
  of $B_s$ mesons into $0^-$ (left panel) or $0^+$ (right panel)
  $c\bar s$ states.}  }
\end{figure}
\end{center}
\end{widetext}

\subsection{Decays into vector and axial vector states}

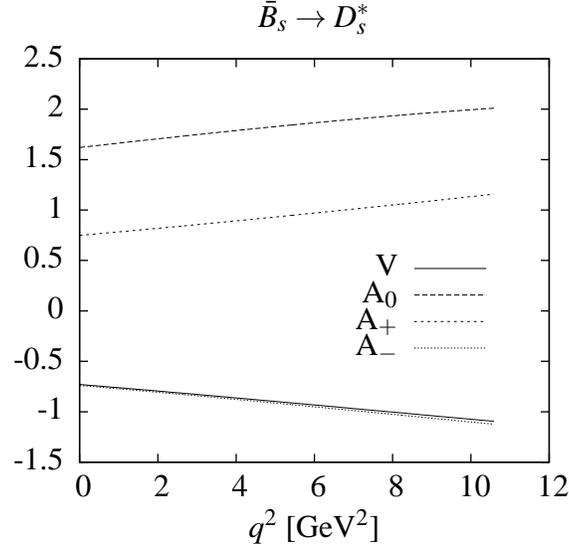
\begin{figure}{\parindent 0cm
\large
\begin{center}
\makebox[0pt]{\input{ff.1m.1.tex}
}
\caption{\label{fig:ff.vector}Form factors for the decay of $B_s$
  mesons into vector $D_s^*$ states.}
\end{center}}
\end{figure}

In the case of decays of $\bar B_s$ mesons into vector $D_s^{*+}$ the
form factors are given by:
\begin{widetext}
\begin{align}
V(q^2)&=\frac{i}{\sqrt 2}\frac{m_{\bar B_s}+m_{D_s^*}}{m_{\bar B_s}|\vec q\,|}V^1_{\lambda=-1}(|\vec q\,|)\nonumber
\\
A_+(q^2)&=i\frac{m_{\bar B_s}+m_{D_s^*}}{2m_{\bar B_s}}\frac{m_{D_s^*}}{|\vec q\,|m_{\bar B_s}}
\left\lbrace
-A_{\lambda=0}^0(|\vec q\,|)
+\frac{m_{\bar B_s}-E_{D_s^*}(-\vec q\,)}{|\vec q\,|}A^3_{\lambda=0}(|\vec q\,|)
-\sqrt 2\frac{m_{\bar B_s}E_{D_s^*}(-\vec q\,)-m_{D_s^*}^2}{|\vec q\,|m_{D_s^*}}A^1_{\lambda=-1}(|\vec q\,|)
\right\rbrace\nonumber
\end{align}
\begin{align}
A_-(q^2)&=-i\frac{m_{\bar B_s}+m_{D_s^*}}{2m_{\bar B_s}}\frac{m_{D_s^*}}{|\vec q\,|m_{\bar B_s}}
\left\lbrace
A_{\lambda=0}^0(|\vec q\,|)
+\frac{m_{\bar B_s}+E_{D_s^*}(-\vec q\,)}{|\vec q\,|}A^3_{\lambda=0}(|\vec q\,|)
-\sqrt 2\frac{m_{\bar B_s}E_{D_s^*}(-\vec q\,)+m_{D_s^*}^2}{|\vec q\,|m_{D_s^*}}A^1_{\lambda=-1}(|\vec q\,|)
\right\rbrace\nonumber\\
A_0(q^2)&=-i\sqrt 2\frac{1}{m_{\bar B_s}-m_{D_s^*}}A^1_{\lambda=-1}(|\vec q\,|)
\label{eq:vectorff}
\end{align}
\end{widetext}
with $V_{\lambda}^\mu (|\vec q\,|)$ and $A_{\lambda}^\mu (|\vec q\,|)$ calculated in our model as
\begin{align}
&V_\lambda^\mu(|\vec q\,|) = \left\langle D_s^{*+},\lambda -|\vec q\,|\vec k\,\left|J_V^{cb\ \mu}(0)\right|\bar B_s, \vec 0 \right\rangle
=\sqrt{4m_{\bar B_s}E_{D_s^*}(-\vec q\,)}_{\rm NR}\left\langle D_s^{*+}, \lambda -|\vec q\,|\vec k\,\left|J_V^{cb\ \mu}(0)\right|\bar B_s, \vec 0 \right\rangle_{\rm NR} 
\nonumber\\
&
A_\lambda^\mu(|\vec q\,|) = \left\langle D_s^{*+},\lambda -|\vec q\,|\vec k\,\left|J_A^{cb\ \mu}(0)\right|\bar B_s, \vec 0 \right\rangle 
=\sqrt{4m_{\bar B_s}E_{D_s^*}(-\vec q\,)}_{\rm NR}\left\langle D_s^{*+}, \lambda -|\vec q\,|\vec k\,\left|J_A^{cb\ \mu}(0)\right|\bar B_s, \vec 0 \right\rangle_{\rm NR}
\label{eq:ffvector}
\end{align}
for which the remaining expressions can be found in the Appendix. The
expressions for the axial vectors can be found from those in
Eq.~(\ref{eq:vectorff}), by just replacing
\begin{equation}
V^\mu_\lambda(|\vec q\,|) \leftrightarrow -A^\mu_\lambda(|\vec q\,|).
\end{equation}
Figures~\ref{fig:ff.vector} and \ref{fig:ff.pseudovector} shows the
different form factors corresponding to semileptonic decays into
vector and pseudovector states. These form factors have been
calculated with the wave functions derived from the AL1 potential. In
Fig~\ref{fig:ff.pseudovector}, the left (right) panel represents the
form factors calculated for semileptonic decays into $^1P_1$ ($^3P_1$)
states.

\begin{widetext}
\vspace{-2.8cm}
\hspace{2cm}\begin{figure}[ht]{\parindent 0cm
\large
\begin{center}
\makebox[0pt]{\input{ff.1p0.1p1.tex}}
\caption{\label{fig:ff.pseudovector}Form factors for the decay of $B_S$
  mesons into $c\bar s, J^P=1^+, S=0$ (left panel) and $c\bar s,
  J^P=1^+, S=1$ (right panel) states.}
\end{center}
}
\end{figure}
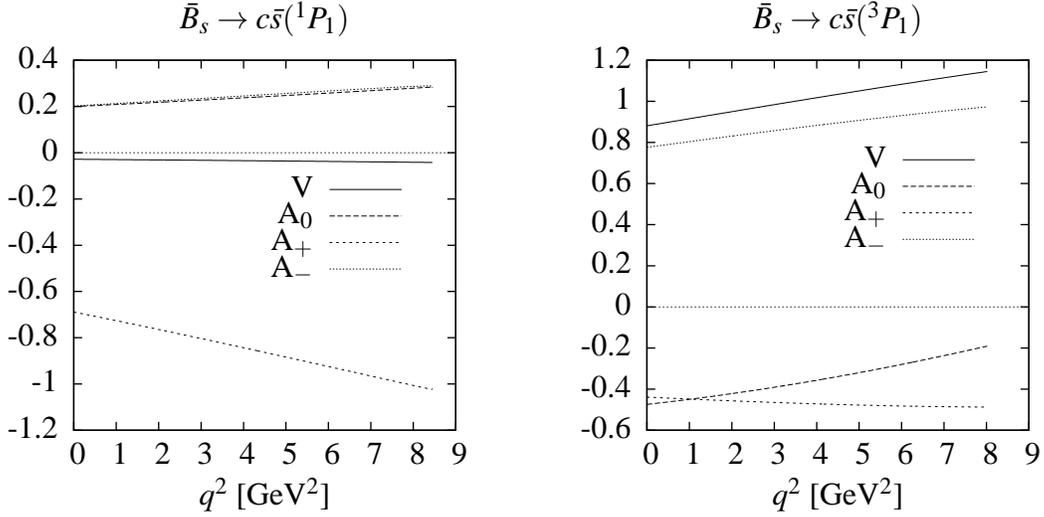
\end{widetext}
\vspace{2cm}
\subsection{Decays into tensor and pseudotensor states}
For $\bar B_s$ mesons decaying into tensor states, the form factors can be evaluated as

\begin{widetext}
\begin{align}
T_1(q^2)&=-i\frac{2m_{D_{s2}^*}}{m_{\bar B_s}|\vec q\,|}A^1_{T\lambda=+1}(|\vec q\,|)\nonumber
\\
T_2(q^2)&=i\frac{1}{2m^3_{\bar B_s}}\left\lbrace
-\sqrt{\frac32}\frac{m^2_{D_{s2}^*}}{|\vec q \,|^2}A^0_{T\lambda=0}(|\vec q\,|)
-\sqrt{\frac32}\frac{m^2_{D_{s2}^*}}{|\vec q \,|^3}(E_{D_{s2}^*}(-\vec q\,)-m_{\bar B_s})A^3_{T\lambda=0}(|\vec q\,|)
+\frac{2m_{D_{s2}^*}}{|\vec q \,|}\left(1-\frac{E_{D_{s2}^*}(-\vec q\,)}{|\vec q\,|^2}(E_{D_{s2}^*}(-\vec q\,)-m_{\bar B_s}) \right)
\right\rbrace\nonumber\\
T_3(q^2)&=i\frac{1}{2m^3_{\bar B_s}}\left\lbrace
-\sqrt{\frac32}\frac{m^2_{D_{s2}^*}}{|\vec q \,|^2}A^0_{T\lambda=0}(|\vec q\,|)
-\sqrt{\frac32}\frac{m^2_{D_{s2}^*}}{|\vec q \,|^3}(E_{D_{s2}^*}(-\vec q\,)+m_{\bar B_s})A^3_{T\lambda=0}(|\vec q\,|)
+\frac{2m_{D_{s2}^*}}{|\vec q \,|}\left(1-\frac{E_{D_{s2}^*}(-\vec q\,)}{|\vec q\,|^2}(E_{D_{s2}^*}(-\vec q\,)+m_{\bar B_s}) \right)
\right\rbrace\nonumber\\
T_4(q^2)&=i\frac{m_{D_{s2}^*}}{m^2_{\bar B_s}|\vec q\,|^2}A^1_{T\lambda=+1}(|\vec q\,|)
\end{align}
\end{widetext}
with $V_{T\lambda}^\mu(|\vec q\,|)$ and $A_{T\lambda}^\mu(|\vec q\,|)$ calculated in our model as
\begin{align*}
V_{T\lambda}^\mu(|\vec q\,|) &= \left\langle D_{s2}^{*+},\lambda -|\vec q\,|\vec k\,\left|J_V^{cb\ \mu}(0)\right|\bar B_s, \vec 0 \right\rangle 
=\sqrt{4m_{\bar B_s}E_{D_{s2}^*}(-\vec q\,)}_{\rm NR}\left\langle D_{s2}^{*+}, \lambda -|\vec q\,|\vec k\,\left|J_V^{cb\ \mu}(0)\right|\bar B_s, \vec 0 \right\rangle_{\rm NR} 
\end{align*}
\begin{align}
A_{T\lambda}^\mu(|\vec q\,|) &= \left\langle D_{s2}^{*+},\lambda -|\vec q\,|\vec k\,\left|J_A^{cb\ \mu}(0)\right|\bar B_s, \vec 0 \right\rangle 
=\sqrt{4m_{\bar B_s}E_{D_{s2}^*}(-\vec q\,)}_{\rm NR}\left\langle D_{s2}^{*+}, \lambda -|\vec q\,|\vec k\,\left|J_A^{cb\ \mu}(0)\right|\bar B_s, \vec 0 \right\rangle_{\rm NR}
\end{align}
for which the remaining expressions can be found in the
Appendix. Again, the form factor corresponding to a decay into a
pseudotensor state can be obtained from those above, just replacing
\begin{equation}
V_{T\lambda}^\mu(|\vec q\,|) \leftrightarrow -A_{T\lambda}^\mu(|\vec q\,|).
\end{equation}
In Figure~\ref{fig:ff.tensor} we have represented the form factors
corresponding to decays into tensor and pseudotensor states, with the
wave functions of the AL1 potential.

\begin{widetext}
\hspace{2cm}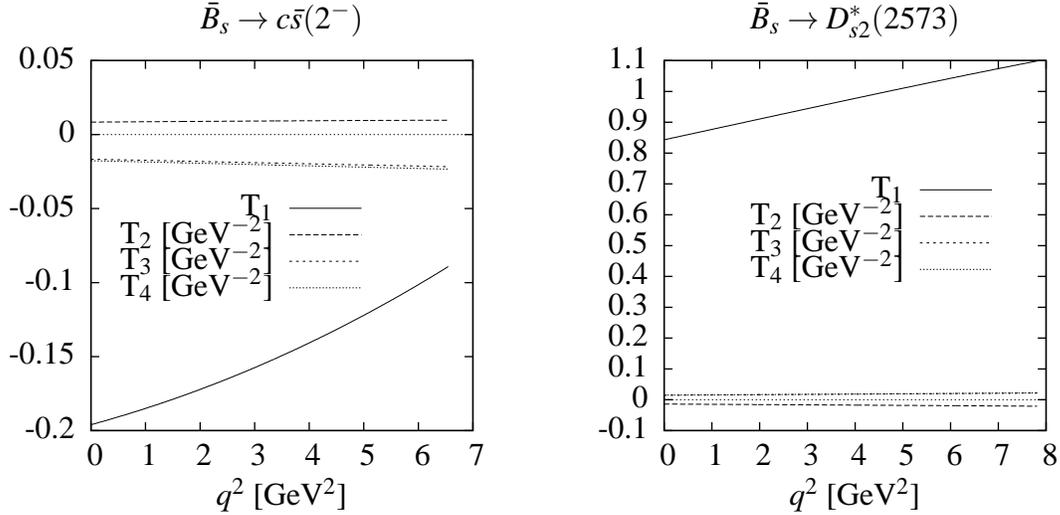
\begin{figure}[ht]{\parindent 0cm
\large
\begin{center}
\makebox[0pt]{\input{ff.2m.2p.tex}
}
\caption{\label{fig:ff.tensor}Form factors for the decay of $B_s$
  mesons into tensor (left panel) and pseudotensor (right panel)
  $c\bar s$ states.}
\end{center}}
\end{figure}
\end{widetext}

\subsection{Decay width}
Let us consider the double differential decay width with respect to
$q^2$ and the cosine, $x_l$, of the angle between the final meson
momentum and the momentum of the final charged lepton, the latter
measured in the lepton-neutrino center of mass frame (CMF). For a
$\bar B_s$ at rest, this differential decay width results to be
%
\begin{align}
\frac{d^2\Gamma}{dx_l dq^2}=&\frac{G_F^2}{64m^2_{\bar B_s}}\frac{|V_{bc}|^2}{8\pi^3}\frac{\lambda^{1/2}(q^2,m^2_{\bar B_s},m^2_{c\bar s})}{2m_{\bar B_s}}
\frac{q^2-m_l^2}{q^2}
{\cal H}_{\alpha\beta}(P_{\bar B_s},P_{c\bar s}){\cal L}^{\alpha\beta}(p_l,p_\nu),
\end{align}

\noindent where $G_F=1.16637(1)\times 10^{-5} {\rm GeV}^{-2}$
\cite{PhysRevD.86.010001} is the Fermi constant,
$\lambda(a,b,c)=(a+b-c)^2-4ab$, $m_l$ is the mass of the charged
lepton, ${\cal H}$ and ${\cal L}$ are the hadron and lepton tensors,
and $P_{\bar B_s}$, $P_{c \bar s}$, $p_l$ and $p_\nu$ are the meson and
lepton four momenta.

The lepton tensor is
\begin{equation}
{\cal L}^{\alpha\beta}(p_l,p_\nu) =8(p^\alpha_lp^\beta_\nu+p^\beta_lp^\alpha_\nu-g^{\alpha\beta}p_l\cdot p_\nu\mp i \epsilon^{\alpha\beta\sigma\rho}p_{l\sigma}p_{\nu\rho})
\end{equation}
where, in the last term, the minus (plus) sign corresponds to a decay into $l^-\bar\nu_l$ ($l^+\nu_l$).
The hadron tensor is given by
\begin{equation}
{\cal H}_{\alpha\beta} = \sum_{\lambda}h_{(\lambda)\alpha}(P_{\bar B_s},P _{c\bar s})h^{(*)}_{(\lambda)\beta}(P_{\bar B_s},P _{c\bar s})
\end{equation}
where
\begin{equation}
h_{(\lambda)\alpha}(P_{\bar B_s},P _{c\bar s})=\left<c\bar s,\lambda \vec P _{c\bar s}\left|J_{\alpha}^{cb}\right|\bar B_s \vec P_{\bar B_s}\right>.
\end{equation}
is just the corresponding matrix element of the $b\to c$ V-A weak
current given in Eq.~\ref{eq:v-a}.

To evaluate the scalar 
\begin{equation}
{\cal H}_{\alpha\beta}(P_{\bar B_s},P _{c\bar s}){\cal L}^{\alpha\beta}(p_l,p_\nu)
\label{eq:scalar}
\end{equation}
we choose $\vec P_{c\bar s}$ to be along the negative $z-$axis, which
involves that the lepton CMF moves along the positive $z-$axis.

To proceed with the calculation we shall follow \cite{Ivanov:2005fd}
and introduce the helicity components for the hadron and lepton tensor
and rewrite the scalar of the expression of Eq.~(\ref{eq:scalar}) as
\begin{equation}
{\cal H}_{\alpha\beta}(P_{\bar B_s},P _{c\bar s}){\cal L}^{\alpha\beta}(p_l,p_\nu) = {\cal H}^{\sigma\rho}(P_{\bar B_s},P _{c\bar s})g_{\sigma\alpha}g_{\beta\rho}{\cal L}^{\alpha\beta}(p_l,p_\nu)
\end{equation}
where \cite{Korner:1989qb}
\begin{align}
g_{\mu\nu}&=\sum_{r=t,\,\pm 1, 0} g_{rr} \epsilon_{(r)\mu}(q)g_{rr} \epsilon_{(r)\nu}^{(*)}(q)\nonumber\\
g_{tt} &= 1,\ \ g_{\pm 1,0}=-1
\end{align}
with $\epsilon_{(t)}^\mu(q)=q^\mu/q^2$ and
$\epsilon_{(r)}(q),\ r=\pm1,0$ are the polarization vector for an
on-shell particle with four momentum $q$ and polarization $r$.

We shall define the helicity components of the hadron and lepton tensors as
\begin{align}
{\cal H}_{rs}(P_{\bar B_s},P_{c\bar s})& = \epsilon_{(r)\sigma}^{*}(q){\cal H}^{\sigma\rho}(P_{\bar B_s},P_{c\bar s})\epsilon_{(s)\rho}(q)\nonumber\\
{\cal L}_{rs}(p_l,p_\nu)& = \epsilon_{(r)\sigma}(q){\cal L}^{\sigma\rho}(p_l,p_\nu)\epsilon^{*}_{(s)\rho}(q).
\end{align}
The contraction of lepton and hadron tensors is, using the expressions above
\begin{equation}
{\cal H}_{\alpha\beta}(P_{\bar B_s},P _{c\bar s}){\cal L}^{\alpha\beta}(p_l,p_\nu)=
\sum_{r,s=t,\pm 1,0} g_{rr}g_{ss}{\cal H}_{rs}(P_{\bar B_s},P_{c\bar s}){\cal L}_{rs}(p_l,p_\nu)
\end{equation}
We take advantage of the fact that the Wigner rotation relating the
original frame and the CMF of the final leptons is the identity. In
the latter, we have
\begin{align}
  {\cal L}_{rs}(p_l,p_\nu)=\epsilon_{(r)\alpha}(q){\cal L}^{\alpha\beta}(p_l,p_\nu)\epsilon_{(s)\beta}(q)
                        =\epsilon_{(r)\alpha}(\tilde q){\cal L}^{\alpha\beta}(\tilde p_l,\tilde p_\nu)\epsilon_{(s)\beta}(\tilde q)
\end{align}
where the tilde stands for the momentum measured in the leptons
CMF. For evaluation, we take\footnote{As we have taken the momentum of
  the final meson in the negative $z$ direction, this is in accordance
  with the definition of $x_l$.}
\begin{align}
\tilde p_l^\alpha&=(E_l(|\tilde p_l|),|\tilde p_l|\sqrt{(1-x_l^2)},0,|\tilde p_l|x_l)\nonumber\\
\tilde p_\nu^\alpha&=(|\tilde p_l|,|\tilde p_l|\sqrt{(1-x_l^2)},0,|\tilde p_l|x_l)
\end{align}
where $|\tilde p_l|$ is the modulus of the lepton three momentum in
the leptons CMF.  Now let us evaluate the lepton tensor helicity
components that we need.
\begin{align}
{\cal L}_{tt}(p_l, p_\nu)       &=4\frac{m^2_l(q^2-m^2)}{q^2}\nonumber\\
{\cal L}_{t0}(p_l, p_\nu)      &=-4x_l\frac{m^2_l(q^2-m^2)}{q^2}\nonumber\\
{\cal L}_{+1+1}(p_l, p_\nu) &=(q^2-m^2)\left(4(1\pm x_l)-2(1-x_l^2)\frac{(q^2-m^2)}{q^2}\right)\nonumber\\
{\cal L}_{-1-1}(p_l, p_\nu)   &=(q^2-m^2)\left(4(1\mp x_l)-2(1-x_l^2)\frac{(q^2-m^2)}{q^2}\right)\nonumber\\
{\cal L}_{00}(p_l, p_\nu)     &=4(q^2-m^2)\frac{1-x_l(q^2-m^2)}{q^2}
\end{align}
As for the hadron tensor, we introduce the helicity amplitudes defined as
\begin{equation}
h_{(\lambda)r}(P_{\bar B_s}, P_{c\bar s})=\epsilon^*_{(r)\alpha}h^\alpha_{(\lambda)}(P_{\bar B_s}, P_{c\bar s}),
\end{equation}
in terms of which the hadron tensor can be written as
\begin{equation}
{\cal H}_{rs}(P_{\bar B_s}, P_{c\bar s})=\sum_{\lambda}h_{(\lambda)r}(P_{\bar B_s}, P_{c\bar s})h^*_{(\lambda)s}(P_{\bar B_s}, P_{c\bar s})
\end{equation}
\begin{widetext}
The expressions for the helicity amplitudes in the original frame are given as \cite{Ivanov:2005fd, Hernandez:2006gt}:
\begin{itemize}
	\item Transitions to scalar states
	  \begin{align}
	    h_t(P_{\bar B_s}, P_{c\bar s})&=\frac{m_{\bar B_s}^2-m_{c\bar s}^2}{\sqrt{q^2}}F_+(q^2)+ \sqrt{q^2}F_-(q^2)\nonumber
\\
	    h_0(P_{\bar B_s}, P_{c\bar s})&=\frac{\lambda^{1/2}(q^2,m_{\bar B_s}^2,m_{c\bar s}^2)}{\sqrt{q^2}}F_+(q^2)\nonumber\\
	    h_{+1}(P_{\bar B_s}, P_{c\bar s}) &= h_{-1}(P_{\bar B_s}, P_{c\bar s})=0
	  \end{align}
	\item Transitions to vector states
          \begin{align}
            h_{(\lambda)t}(P_{\bar B_s}, P_{c\bar s})&=i\delta_{\lambda0}\frac{\lambda^{1/2}(q^2,m^2_{\bar B_s},m^2_{c\bar s})}{2m_{c\bar s}\sqrt{q^2}}
\left((m_{\bar B_s}-m_{c\bar s})(A_{0}(q^2)-A_+(q^2))-\frac{q^2}{m_{\bar B_s}+m_{c\bar s}}A_{-}(q^2)\right)\nonumber\\
            h_{(\lambda)+1}(P_{\bar B_s}, P_{c\bar s})&=-i\delta_{\lambda-1}\left(\frac{\lambda^{1/2}(q^2,m^2_{\bar B_s})}{m_{\bar B_s}+m_{c\bar s}}V(q^2)+(m_{\bar B_s}-m_{c\bar s})A_0(q^2)\right)\nonumber\\
            h_{(\lambda)-1}(P_{\bar B_s}, P_{c\bar s})&=-i\delta_{\lambda+1}\left(\frac{\lambda^{1/2}(q^2,m^2_{\bar B_s})}{m_{\bar B_s}+m_{c\bar s}}V(q^2)+(m_{\bar B_s}-m_{c\bar s})A_0(q^2)\right)\nonumber\\
            h_{(\lambda)0}(P_{\bar B_s}, P_{c\bar s})&=i\delta_{\lambda0}\left((m_{\bar B_s}-m_{c\bar s})\frac{m^2_{\bar B_s}-q^2-m^2_{c\bar s}}{2m_{c\bar s}\sqrt{q^2}}A_{0}(q^2)
-\frac{\lambda(q^2,m^2_{\bar B_s},m^2_{c\bar s})}{2m_{c\bar s}\sqrt{q^2}}\frac{A_+(q^2)}{m_{\bar B_s}+m_{c\bar s}}
\right)
	  \end{align}
	\item Transitions to tensor states
          \begin{align}
            h_{(\lambda)t}(P_{\bar B_s}, P_{c\bar s})&=-i\delta_{\lambda0}\sqrt{\frac23}\frac{\lambda(q^2,m^2_{\bar B_s},m^2_{c\bar s})}{4m^2_{c\bar s}\sqrt{q^2}}
\left(T_1(q^2)+(m^2_{\bar B_s}-m^2_{c \bar s})T_2(q^2)+q^2T_{3}(q^2)\right)\nonumber\\
            h_{(\lambda)+1}(P_{\bar B_s}, P_{c\bar s})&=i\delta_{\lambda-1}\frac{1}{\sqrt2}\frac{\lambda^{1/2}(q^2,m^2_{\bar B_s},m^2_{c\bar s})}{2m_{c\bar s}}
\left(T_1(q^2)-\lambda^{1/2}(q^2,m^2_{\bar B_s}m^2_{c\bar s})T_4(q^2)\right)\nonumber\\
            h_{(\lambda)-1}(P_{\bar B_s}, P_{c\bar s})&=i\delta_{\lambda+1}\frac{1}{\sqrt2}\frac{\lambda^{1/2}(q^2,m^2_{\bar B_s},m^2_{c\bar s})}{2m_{c\bar s}}
\left(T_1(q^2)+\lambda^{1/2}(q^2,m^2_{\bar B_s}m^2_{c\bar s})T_4(q^2)\right)\nonumber\\
            h_{(\lambda)0}(P_{\bar B_s}, P_{c\bar s})&=-i\delta_{\lambda0}\sqrt{\frac23}\frac{\lambda^{1/2}(q^2,m^2_{\bar B_s},m^2_{c\bar s})}{4m^2_{c\bar s}\sqrt{q^2}}
\left((m^2_{\bar B_s}-q^2-m_{c\bar s})T_1(q^2) + \lambda(q^2,m^2_{\bar B_s},m^2_{c\bar s})T_2(q^2)\right)
          \end{align}
\end{itemize}
\end{widetext}
Where we shall remark that the helicity amplitudes, and thus the
components of the hadron tensor depend only on $q^2$. We define the
following combinations for further convenience:
\begin{align}
H_U&={\cal H}_{+1+1}+{\cal H}_{-1-1}\nonumber\\
H_P&={\cal H}_{+1+1}-{\cal H}_{-1-1}\nonumber\\
H_L&={\cal H}_{00};\ H_S=3{\cal H}_{tt};\ H_{SL}={\cal H}_{t0}\nonumber\\
\tilde H_J&=\frac{m^2_l}{2q^2}{\cal H};\ \ J=U,L,S,SL
\end{align}
with $U$, $L$, $P$, $S$ and $SL$ representing, respectively,
unpolarized-transverse, longitudinal, parity-odd, scalar and
scalar-longitudinal interference.

The double differential decay width can be written in terms of the
combination above as
\begin{align}
\frac{d^2\Gamma}{dq^2 dx_l}&=\frac{G_F^2}{8\pi^3}|V_{bc}|^2\frac{(q^2-m_l^2)^2}{12m^2_{\bar B_s}q^2}\frac{\lambda^{1/2}(q^2,m_{\bar B_s}^2,m_{c\bar s}^2)}{2m_{\bar B_s}}\nonumber\\
&\times\left\lbrace
\frac38(1+x_l^2)H_U + \frac34(1-x_l^2)H_L\pm\frac34H_P
+\frac34(1-x_l^2)\tilde H_U + \frac32 x_l^2 \tilde H_l+\frac12\tilde H_S + 3x_l\tilde H_{sl}
\right\rbrace
\end{align}
The term $H_P$ changes sign for antiparticle decay, in contrast to the
rest of the helicity components. This extra sign compensates the $\mp$
sign in the lepton tensor, leading to an expression for the double
differential decay which is the same for particle or antiparticle
decay.

Finally, we obtain the differential decay width integrating over $x_l$.
\begin{align}
\frac{d\Gamma}{dq^2}&=\frac{G_F^2}{8\pi^3}|V_{bc}|^2\frac{(q^2-m_l^2)^2}{12m^2_{\bar B_s}q^2}\frac{\lambda^{1/2}(q^2,m_{\bar B_s}^2,m_{c\bar s}^2)}{2m_{\bar B_s}}
\left\lbrace H_U + H_L +\tilde H_U + \tilde H_L+ \tilde H_S \right\rbrace,
\end{align}
from where we obtain the total decay width integrating over $q^2$,
that can be written as
\begin{equation}
\Gamma=\Gamma_U+\Gamma_L+\tilde\Gamma_U+\tilde\Gamma_L+\tilde\Gamma_S,
\end{equation}
with $\Gamma$ and $\Gamma_J$ partial helicity widths defined as
\begin{equation}
\Gamma_J=\int dq^2 \frac{G_F^2}{8\pi^3}|V_{bc}|^2\frac{(q^2-m_l^2)^2}{12m^2_{\bar B_s}q^2}\frac{\lambda^{1/2}(q^2,m_{\bar B_s}^2,m_{c\bar s}^2)}{2m_{\bar B_s}} H_J
\end{equation}
and similarly for $\tilde \Gamma_J$ in terms of $\tilde H_J$.

The forward-backward asymmetry of the charged leptons, measured in the
leptons CMF, which in terms of partial helicity widths, can be written
as
\begin{equation}
A_{FB}=\frac{\Gamma_{x_l>0}-\Gamma_{x_l<0}}{\Gamma_{x_l>0}+\Gamma_{x_l<0}}=\frac34 \frac{\pm \Gamma_P + 4\tilde\Gamma_{SL}}{\Gamma_U+\Gamma_L+\tilde\Gamma_U+\tilde\Gamma_L+\tilde\Gamma_S}
\end{equation}
As $\Gamma_P$ changes sign for antiparticle decay, $A_{FB}$ is the
same for a negative charged lepton as for a positive.

\subsection {Results}

Table~\ref{tab:semilept.tw} summarizes the values for the total decay
widths calculated with our model. We give the semileptonic decay
widths for the different leptons in the final state, in units of
$\tten{15} {\rm GeV}$. The central values have been calculated using the AL1
potential of \cite{SilvestreBrac:1996bg}, while the theoretical
uncertainties have been estimated by considering other potential
models (see Ref.~\cite{SilvestreBrac:1996bg}). Table~\ref{tab:bra1}
shows the corresponding values for branching fractions.

In Table~\ref{tab:comp1} we compare with previous results. In
Refs.~\cite{Faustov:2012mt} the authors adopt the relativistic quark
model. Chen {\it et al.} solve the instantaneous Bethe-Salpeter
equation in \cite{Chen:2012fi} to estimate the weak transition form
factors. In \cite{Li:2009wq} the authors work out the form factors
within the covariant light front quark model.  Azizi {\it et al.} in
\cite{Azizi:2008xy, Azizi:2008vt} and Blasi {\it et al.} in
\cite{Blasi:1993fi} apply the sum rules technique to obtain the form
factors and branching fractions. In \cite{Zhao:2006at}, work within
the Constituent Quark Model, as in this work. In
Ref.~\cite{segovia:2011dg} we studied some of the decays into
orbitally excited final $D_s$ states, using the potential
model of~\cite{Vijande:2004he}.

The results of this work are in a systematic good agreement with those
from the relativistic quark model of \cite{Faustov:2012mt}. The
agreement is also good with the quark model calculation of
\cite{Zhao:2006at}. It is worth to mention that our results for decays
into orbitally excited final $D_s$ mesons are in rather good agreement
with our previous results from \cite{segovia:2011dg}, though in that
work the potential model that have been used is much more
sophisticated, even enabling the posibility to consider non-$q\bar q$
components for these orbitally excited states. Our results also
compare well to the sum-rules calculation of \cite{Azizi:2008xy,
  Azizi:2008vt}, while the result of \cite{Blasi:1993fi} is lower by
about one half. The same happens if we compare with the results of
\cite{Chen:2012fi} or \cite{Li:2009wq}.

In Tables~\ref{tab:helicitypartial} and \ref{tab:fb} we give our
results for partial helicity widths corresponding to $\bar B_s^0$, and
the values we obtain for the forward-backward asymetry,
respectively. In Table~\ref{tab:helicitypartial} the "P" column
changes sign for $B_s^0$ decay. As before, the central values have
been evaluated with the AL1 potential.

In the different panels of Figures~\ref{fig:dg.scalar} to
\ref{fig:dg.tensor} we plot the differential decay widths that we
obtain for the different $J^P$ $c\bar s$ final states, with $e^+$ or
$\tau^+$, accounting for the leptons.
\begin{widetext}
\begin{table}
\begin{tabular}{c|ccc}
\hline\hline
$\bar B_S\to M' l^-\bar \nu_l$            &             \multicolumn{3}{c} {$\Gamma [10^{-15} {\rm GeV}]$}      \\\hline
    $M'$        &      $l=e$                  &  $l= \mu$                &  $l=\tau$                \\\hline
$D_s^+$          &  $10.37_{-0.2}^{+0.15}$       &  $10.32_{-0.10}^{+0.16}$     &  $2.99_{-0.03}^{+0.01}$     \\
$D_{s0}^{*+} $    &  $1.75_{-0.07}^{+0.03}$        &  $1.74_{-0.08}^{+0.03}$      &  $0.20_{-0.003}^{+0.003}$    \\
$D_s^{*+} $      &  $28.02_{-0.48}^{+0.24}$       &  $27.90_{-0.48}^{+0.86}$     &  $6.86_{-0.09}^{+0.12}$     \\
$D_{s1}^+(2460)$ &  $2.07_{-0.09}$              &  $2.05_{-0.08}$             &  $0.17_{-0.008}$            \\
$D_{s1}^+(2536)$ &  $1.40_{-0.07}$              &  $1.39_{-0.07}$             &  $0.12_{-0.006}$     \\
$c\bar s(2^-) $ &  $4.11_{-0.56}\tten{2}$      &  $4.06_{-0.64}\tten{2}$     &  $9.02_{-2.39}\tten{4}$     \\
$D_{s2}^{*+} $    &  $1.97_{-0.15}$              &  $1.95_{-0.14}$             &  $0.12_{-0.02}$             \tstrut\\
\hline
\end{tabular}
\caption{Decay widths in units of $10^{-15}\ {\rm GeV}$ for semileptonic
  $\bar B_s \to c\bar s $ decays. The central value has been obtained with
  the AL1 potential.}
\label{tab:semilept.tw}
\end{table}
\begin{table}
\begin{center}
\begin{tabular}{c|cc}
\hline\hline
    $M'$        &   $l=e,\mu$        &  $l=\tau$                \\\hline
$D_s^+ $         &   2.32           &   0.67                     \\
$D_{s0}^{*+}  $   &   0.39           &   0.04                     \\
$D_s^{*+} $      &   6.26           &   1.53                     \\
$D_{s1}^+(2460)$ &   0.47           &   0.04                     \\
$D_{s1}^+(2536)$ &   0.32           &   0.03                     \\
$c\bar s(2^-) $ &   9.2\ten{3}     &   2.0\ten{4}               \\
$D_{s2}^{*+} $    &   0.44           &   0.03                     \\\hline
\end{tabular} 
\caption{Branching fractions for the indicated decay channels, in percentage.\label{tab:bra1}}
\end{center}
\end{table}

\begin{table}
\begin{center}
\begin{tabular}{c|cccccccc}
\hline\hline
                                      & This work    &   \cite{Faustov:2012mt} & \cite{Chen:2012fi} & \cite{Li:2009wq} & \cite{Blasi:1993fi} & \cite{Zhao:2006at}  & \cite{Azizi:2008xy},\cite{Azizi:2008vt} &\cite{segovia:2011dg} \\\hline
$\bar B_s \to D_s^+ e^- \bar \nu_e$  &  2.32   &  $2.1\pm0.2$   &  1.4-1.7    & $1.0_{-0.3}^{+0.4}$    & $1.35\pm0.21$ & 2.73-3.00  & 2.8-3.8 &\\
$\bar B_s \to D_s^{*+} e^-\bar \nu_e$ &  6.26   &  $5.3\pm0.5$   &  5.1-5.8    &                     & $2.5\pm0.1$   & 7.49-7.66  & 1.89-6.61 &\\
$\bar B_s \to D_s^+ \tau^- \bar \nu_\tau$   &  0.67   &  $0.62\pm0.05$ &  0.47-0.55  & $0.33_{-0.11}^{+0.14}$ &               &            &  &\\
$\bar B_s \to D_s^{*+} \tau^-\bar \nu_\tau$  &  1.53   &  $1.3\pm0.1$   &  1.2-1.3    &                     &               &            &  &\\\hline
$\bar B_s \to D_{s0}^{*+} \mu^-\bar \nu_\mu$       &0.39  & &&&&& & 0.44\\
$\bar B_s \to D_{s1}^{*+}(2460) \mu^-\bar \nu_\mu$ &0.47  & &&&&& & 0.17-0.5\\
$\bar B_s \to D_{s1}^{*+}(2536) \mu^-\bar \nu_\mu$ &0.32  & &&&&& & 0.4\\
$\bar B_s \to D_{s2}^{*+} \mu^-\bar \nu_\mu$       &0.44  & &&&&& & 0.37\\
\hline
\end{tabular}
\caption{\label{tab:comp1} Branching fractions for the indicated decay channels, in percentage.\label{tab:bracomp}}
\end{center}
\end{table}

\begin{table}[ht]
\begin{center}
\begin{tabular}{l|ccccccc}
\hline\hline
                                     & $\Gamma_U$ & $\tilde\Gamma_U$ & $\Gamma_L$ & $\tilde\Gamma_L$ & $\Gamma_P$ & $\tilde\Gamma_S$ & $\tilde\Gamma_{SL}$
\\\hline
$\bar B_s \to D_s^+ e^-\bar \nu_e$         &  0          &  0           & 10.37       & 2.29\ten{6}  &    0        & 7.43\ten{6}  & 2.36\ten{6}  \\
$\bar B_s \to D_{s0}^{*+} e^-\bar \nu_e$    &  0          &  0           & 1.75        & 4.82\ten{7}  &    0        & 1.44\ten{6}  & 4.81\ten{7}  \\
$\bar B_s \to D_s^{*+} e^-\bar \nu_e$      & 13.87       &  4.15\ten{7} & 14.16       & 2.13\ten{6}  & -7.32       & 5.90\ten{6}  & 2.03\ten{6}  \\
$\bar B_s \to D_{s1}^+(2460) e^-\bar \nu_e$& 0.32        &  1.6\ten{8} & 1.75        & 6.41\ten{7}  &-0.22        & 1.98\ten{6}  & 6.51\ten{7}  \\
$\bar B_s \to D_{s1}^+(2536)e^-\bar \nu_e$ & 0.56        & 2.97\ten{8}  & 0.84        & 3.04\ten{7}  &-0.44        & 9.40\ten{7} & 3.08\ten{7}   \\
$\bar B_s \to c\bar s(2^-) e^-\bar \nu_e$ & 3.95\ten{2} & 3.37\ten{9}  & 1.58\ten{3} & 3.55\ten{10} &-3.24\ten{2} & 8.76\ten{10} & 3.15\ten{10} \\
$\bar B_s \to D_{s2}^{*+} e^-\bar \nu_e$    & 0.67        & 3.76\ten{8}  & 1.30        & 4.35\ten{7}  &-0.35        & 1.25\ten{6}  & 4.24\ten{7}  \\\hline 
$\bar B_s \to D_s^+ \mu^-\bar\nu_\mu$        &   0        &  0          &  10.11      & 4.72\ten{2}  & 0           & 0.16        & 5.05\ten{2} \\
$\bar B_s \to D_{s0}^{*+} \mu^-\bar\nu_\mu$   &   0        &  0          &   1.70      & 9.47\ten{3}  & 0           & 2.80\ten{2} & 9.40\ten{3} \\  
$\bar B_s \to D_s^{*+} \mu^-\bar\nu_\mu$     & 13.80      & 1.74\ten{2} & 13.91       & 4.68\ten{2}  & -7.28       & 0.12        & 4.28\ten{2} \\ 
$\bar B_s \to D_{s1}^+(2460)\mu^-\bar\nu_\mu$& 0.32       & 6.6\ten{4} & 1.68        & 1.7\ten{2}  &-0.21        & 3.79\ten{2} & 1.22\ten{2} \\
$\bar B_s \to D_{s1}^+(2536)\mu^-\bar \nu_\mu$& 0.55       & 1.23\ten{3} & 0.81        & 5.58\ten{3}  &-0.44        & 1.79\ten{3} & 5.76\ten{3} \\
$\bar B_s \to c\bar s(2^-) \mu^-\bar \nu_\mu$& 3.89\ten{2}& 1.37\ten{4} & 1.55\ten{3} & 7.49\ten{6}  &-3.20\ten{2} & 1.46\ten{5} & 5.82\ten{6} \\
$\bar B_s \to D_{s2}^{*+} \mu^-\bar \nu_\mu$  & 0.66       & 1.55\ten{3} & 1.26        & 8.15\ten{3}  &-0.35        & 2.20\ten{2} & 7.69\ten{3} \\\hline
$\bar B_s \to D_s^+ \tau^-\bar \nu_\tau$        & 0           & 0           & 0.94        & 0.22        & 0           & 1.82        & 0.36       \\
$\bar B_s \to D_{s0}^{*+} \tau^-\bar \nu_\tau$   & 0           & 0           & 9.28\ten{2} & 2.51\ten{2} & 0           & 8.26\ten{2} & 2.61\ten{2}\\
$\bar B_s \to D_s^{*+} \tau^-\bar \nu_\tau$     & 3.18        & 0.68        & 2.06        & 0.46        &-1.39        & 0.49        & 0.26       \\  
$\bar B_s \to D_{s1}^+(2460)\tau^-\bar \nu_\tau$& 0.03        & 8.22\ten{3} & 5.19\ten{2} & 1.50\ten{2} &-1.71\ten{2} & 0.07        & 1.88\ten{2}\\ 
$\bar B_s \to D_{s1}^+(2536)\tau^-\bar \nu_\tau$& 4.48\ten{2} & 1.25\ten{2} & 2.60\ten{2} & 7.49\ten{3} &-3.39\ten{2} & 3.53\ten{2} & 8.96\ten{3}\\
$\bar B_s \to c\bar s(2^-) \tau^-\bar \nu_\tau$& 6.26\ten{4} & 2.14\ten{4} & 4.42\ten{5} & 1.44\ten{5} &-5.34\ten{4} & 3.05\ten{8} & 3.68\ten{6}\\
$\bar B_s \to D_{s2}^{*+} \tau^-\bar \nu_\tau$  & 4.26\ten{2} & 1.25\ten{2} & 3.85\ten{2} & 1.15\ten{2} &-1.75\ten{2} & 1.22\ten{2} & 6.73\ten{3}\\\hline
\end{tabular}
\end{center}
\caption{\label{tab:helicitypartial}
Partial helicity widths in units of $10^{-15}$ GeV. These results have been calculated using the AL1 potential.}
\end{table}
\begin{table}
\begin{tabular}{l|ccc}
\hline\hline
                                                   &\multicolumn{3}{c} {$A_{\rm FB}$}\\
                                                   & $l=e$ & $l=\mu$ & $l=\tau$  \\\hline
$\bar B_s \to  D_s^+      l^-\bar \nu_l $           & 6.86\ten{7}    & 1.47\ten{2}     & 0.36              \\
$\bar B_s \to  D_{s0}^{*+} l^- \bar \nu_l $          & 8.22\ten{7}    & 1.62\ten{2}     & 0.39              \\
$\bar B_s \to  D_s^{*+}    l^- \bar \nu_l $         & -0.20          & -0.19           & -3.71\ten{2}      \\
$\bar B_s \to  D_{s1}^+(2460) l^- \bar \nu_l$       & -0.19          & -0.18           & 0.10              \\
$\bar B_s \to  D_{s1}^+(2536) l^- \bar \nu_l$       & -0.41          & -0.40           & -0.20             \\
$\bar B_s \to  c \bar s(2^-)      l^- \bar \nu_l$  & -0.59          & -0.59           & -0.43             \\
$\bar B_s \to  D_{s2}^{*+}  l^- \bar \nu_l$         & -0.14          & -0.12           & 6.03\ten{2}       \\\hline
\end{tabular}
\caption{\label{tab:fb}Forward-Backward asymmetry parameters for the semileptonic $B_s$ decays, obtained for the AL1 potential.}
\end{table} 

\vspace{-2cm}
\hspace{2cm}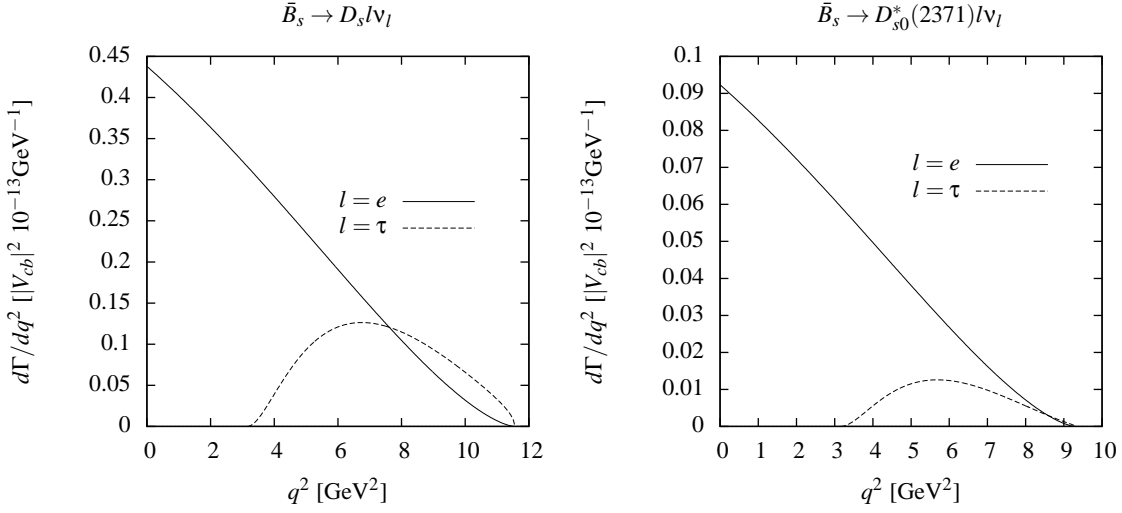
\begin{figure}[ht]{\parindent 0cm
\begin{center}
\makebox[0pt]{\input{dg.0m.0p.tex}
}
\caption{\label{fig:dg.scalar}Differential decay width for the $\bar B_s$ into $0^-$ (left panel) and $0^+$ (right panel) states.}
\end{center}
}
\end{figure}
%
\hspace{2cm}\begin{figure}[ht]{\parindent 0cm
\large
\begin{center}
\makebox[0pt]{\input{dg.1m1m.tex}
}
\caption{\label{fig:dg.vector}Differential decay width for the semileptonic $\bar B_s\to D^*$ process.}
\end{center}
}
\end{figure}
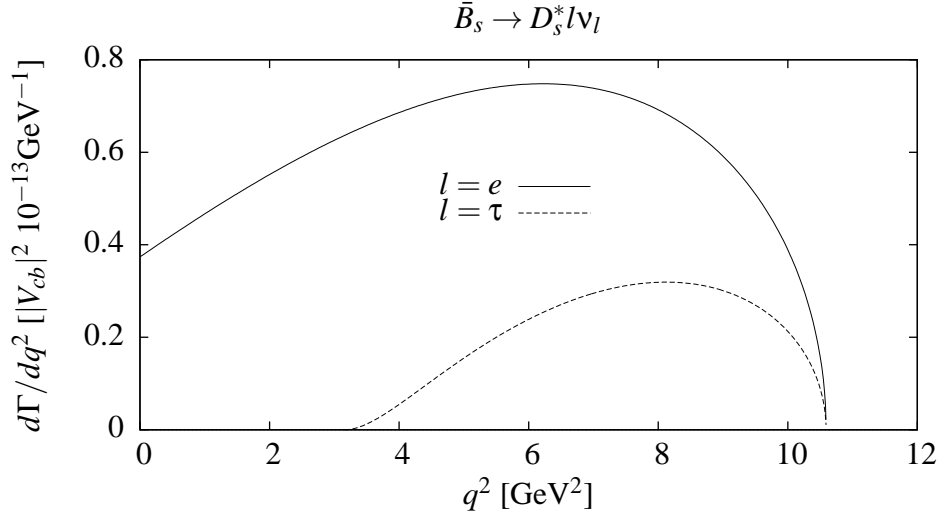
%
\hspace{2cm}\begin{figure}[ht]{\parindent 0cm
\large
\begin{center}
\makebox[0pt]{\input{dg.1ps0s1.tex}
}
\caption{\label{fig:dg.pseudovector}Differential decay widths for the semileptonic decays of
  $\bar B_s$ into $J^P=1^+, S=0$ (left panel) and $J^P=1^+, S=1$ (right
  panel) states.}
\end{center}
}
\end{figure}
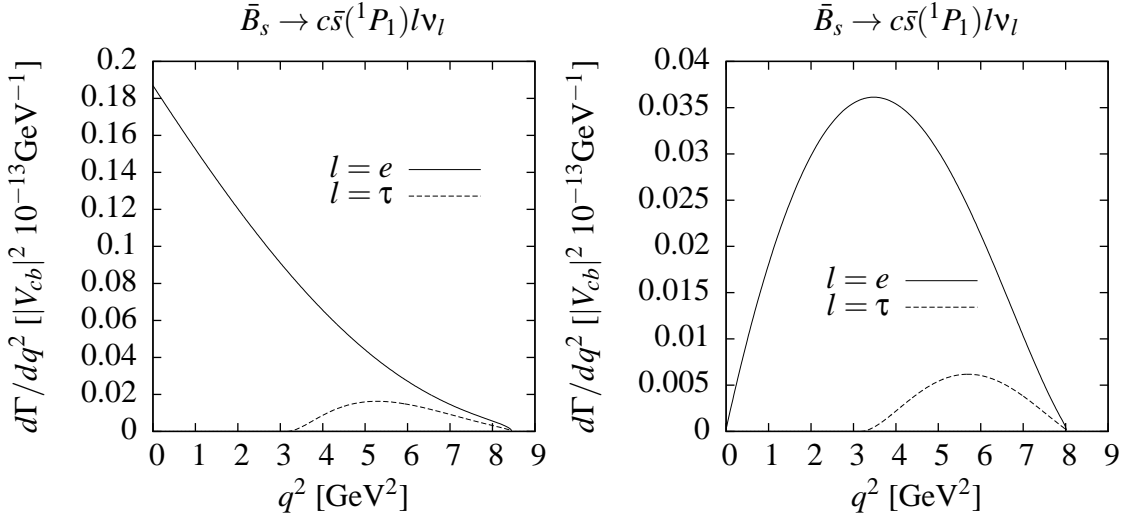
\hspace{3cm}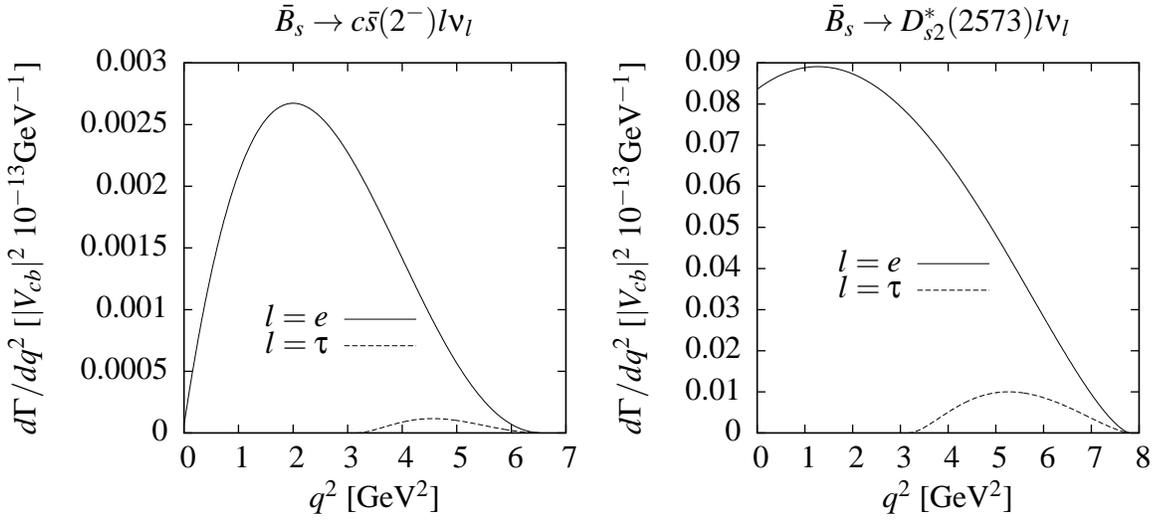
\begin{figure}[ht]{\parindent 0cm
\large
\begin{center}
\makebox[0pt]{\input{dg.2m.2p.tex}
}
\caption{\label{fig:dg.tensor}Differential decay widths for the semileptonic decays of
  $\bar B_s$ into $J^P=2^-$ (left panel) and $J^P=2^+$ (right panel)
  states.}
\end{center}
}
\end{figure}
\end{widetext}
\vspace{2cm}
\subsection{Heavy Quark Symmetry}
In systems with a quark with mass much larger than the QCD scale
($\Lambda_{\rm QCD}$), the dynamics of the light degrees of freedom
becomes independent of the heavy quark flavor and spin.

The six form factors involved in the $\bar B_s$ decays into pseudoscalar and
vector mesons are related by HQS, which reduces their evaluation to
that of a single function, $\xi$. In particular, HQS predicts~\cite{Isgur:1989vq, Isgur:1989ed}:
\begin{align}
h_+(\omega)&=h_{V}(\omega)=h_{A_1}(\omega)=h_{A_3}(\omega)=\xi(\omega)\nonumber\\
h_-(\omega)&=h_{A_2}(\omega)=0.
\label{eq:hqs1}
\end{align}
The $h$ form factors are just a redefinition of the those above, given by
\begin{equation}
h_\pm(\omega)=\frac{2m_{\bar B_s}}{\sqrt{2m_{\bar B_s}m_{c\bar s}}}f_{\pm}(\omega)
\end{equation}
for decays into pseudoscalar states, and 
\begin{align}
h_{V} (\omega) &= \sqrt2 \sqrt{\frac{M_{D_s^*}}{M_{\bar B_s}}}\frac{V^2_{\lambda=-1}(|\vec q|)}{|\vec q|}\nonumber\\
h_{A_1}(\omega) &= i\frac{\sqrt2}{w+1}\frac{1}{\sqrt{M_{\bar B_s}M_{D_s^*}}}A^1_{\lambda=-1}(|\vec q|)\nonumber\\
h_{A_2}(\omega) &= i\sqrt{\frac{M_{D_s^*}}{M_{\bar B_s}}}\left(-\frac{A^0_{\lambda=0}(|\vec q|)}{|\vec q|}+\frac{E_{D_s^*}(|\vec q|)A_{\lambda=0}^3(|\vec q|)}{|\vec q|^2}
-\sqrt2M_{D^*}\frac{A_{\lambda = -1}^1(|\vec q|)}{|\vec q|^2}\right)\nonumber\\
h_{A_3}(\omega) &= i\frac{M^2_{D_s^*}}{\sqrt{M_{D_s^*}M_{\bar B_s}}}\left(-\frac{A^3_{\lambda=0}(|\vec q|)}{|\vec q|^2}+\frac{\sqrt2}{M_{D^*}}\frac{A_{\lambda = -1}^1(|\vec q|)}{|\vec q|^2}\right)
\end{align}
for decays into vector states \cite{Albertus:2005vd}.

Conservation of vector current in the equal mass case, provides
another constrain, in the form of a normalization condition:
\begin{equation}
\xi(\omega=1) = 1.
\label{eq:hqs2}
\end{equation}

The purpose of this section is test the form factors we have obtained
previously against the HQS predictions. In the left panel of
Fig.~\ref{fig:hqs1} we plot our values for the $h$ form factors. These
values have been obtained with the wave functions of the AL1
potential.  In the right panel of Fig.~\ref{fig:hqs1}, we also
evaluate the ratios
\begin{align}
R_1(\omega)&=\frac{h_V(\omega)}{h_{A_1}(\omega)}\nonumber\\
R_2(\omega)&=\frac{h_{A_3}(\omega)+rh_{A_2}(\omega)}{h_{A_1}(\omega)}
\end{align}
where $r=m_{c\bar s}/m_{\bar B_s}$. These ratios are expected to vary
smoothly with $\omega$.

\begin{widetext}
\hspace{2cm}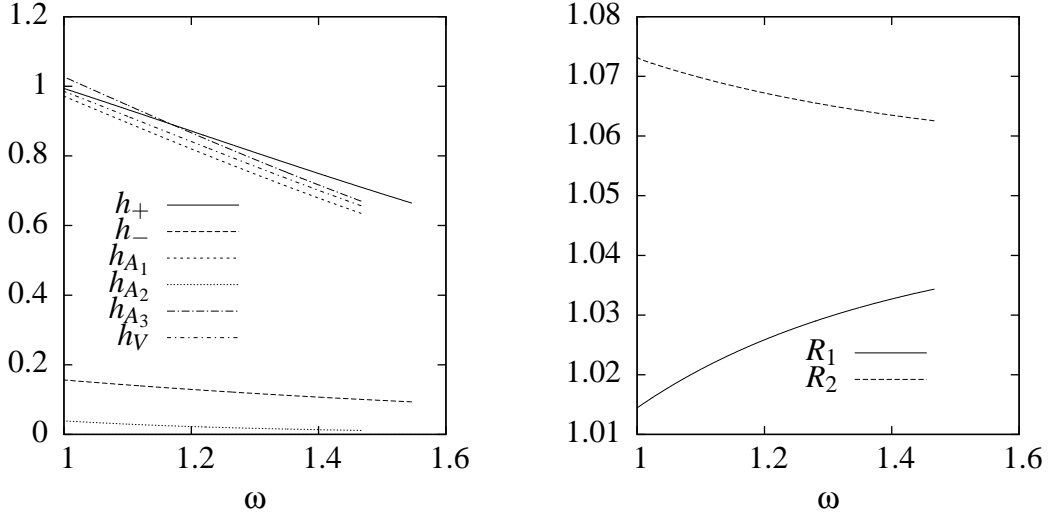
\begin{figure}[ht]{\parindent 0cm
\large
\begin{center}
\makebox[0pt]{\input{hqs1.tex}
}
\caption{\label{fig:hqs1} $h$ form factors for the decay of $\bar B_s$
  mesons into pseudoscalar and vector $c\bar s$ states (left panel)
  and ratios $R_1$ and $R_2$ (right panel).}
\end{center}
}
\end{figure}
\end{widetext}

In the case of the semileptonic decays of $B$ mesons, one expects
discrepancies of the order of $10-15 \%$ from the predictions of HQET
at most.  For $B_s$ mesons, one should expect, in principle the same
kind of unaccuracies than in the $B$ case. Figure~\ref{fig:hqs1} shows
that this is indeed the case.

At most, $h_-$ or $h_{a_2}$ differ from 0 at the level of the 15\%
approximatelly.  On the other hand the ratio $R_1$ gives an estimate
of the discrepancies from the Isgur-Wise function, being these much
smaller.  These negligible corrections to the heavy quark limit
predictions in the case of $h_+$ at $\omega=1$ were also found in the
similar calculation carried out in Ref.~\cite{Albertus:2005vd} for the
semileptonic $B\to D$ decay, and this quite small violations might be
related to the NRCQM approach.  Sum rule and lattice calculation
predicts somehow larger corrections, though much smaller than those
that affect to $h_-$. This is because Luke's theorem, that guaranties
that corrections to $h_+$(1) are order ${\cal O}(1/M_Q^2)$.

\section{SEMILEPTONIC $\bar B_s$ TO $B^-$ AND $B^{*-}$ DECAYS}

In principle, one could also consider those weak processes of \bsmeson
driven by the $\bar s\to \bar u$ decays at the quark level. In this case, due
to the similar value of the masses of the $\bar B_s$, $B$ and $B^*$ mesons
($m_{\bar B_s} - m_{B^-} = 87$ MeV, $m_{\bar B_s} - m_{B^{*-}} = 41$ MeV), the only
decay modes allowed are the semileptonic $\bar B_s \to B^- e^+ \nu_e$
and $\bar B_s \to B^{*-} e^+ \nu_e$, as the muon, for
instance, lay beyond the scope of the available phase space, so that
other semileptonic or nonleptonic processes are forbidden.

Let us consider first the $\bar B_s \to B^-$ transition. This process involve
a pseudoscalar to pseudoscalar transition, so we take the following
form factor decomposition
\begin{equation}
\left<B^-\vec{P}_{B^-}\left|J^{su}_\mu(0) \right|\bar B_s,\vec{P}_{\bar B_s} \right> =P_\mu F_+(q^2) + q_\mu F_-(q^2).
\end{equation}
The expressions of the form factors are exactly the same of
those of Eq.~\ref{eq:ffpseudoscalar}.

The total decay width of this process results to be
\begin{equation}
\Gamma_{B_s \to B^- e^+ \nu_{e}} = 1.7 \tten{20}\ \rm{GeV}
\end{equation}
For the process  $\bar B_s \to B^{*-} e^+
\nu_e$ process. Again, the form factor decomposition is the same
as that of the $B_s \to D_s^*$ decay,
\begin{widetext}
\begin{align}
\left<B^*\lambda \vec{P}_{c\bar s}\left|J^{su}_\mu(0) \right|B_s,\vec{P}_{B_s} \right> &=
\frac{-1}{m_{B_s}+m_{c\bar s}}\epsilon_{\mu\nu\alpha\beta}\epsilon_{(\lambda)}^{\nu *}(\vec{P}_{c\bar s})P^\alpha q^\beta V(q^2)\nonumber\\
&-i\left\lbrace(m_{B_s}-m_{c\bar s})\epsilon_{(\lambda)\mu}^*(\vec{P}_{c\bar s}) A_0(q^2) 
-\frac{P\cdot \epsilon^*_{(\lambda)}(\vec{P}_{c\bar s})}{m_{B_s}+m_{c\bar s}}(P_\mu A_+(q^2) + q_\mu A_-(q^2))
\right\rbrace,
\end{align}
\end{widetext}
and the expression of the form factor is that of
Eq.~\ref{eq:ffvector}. Now we obtain
\begin{equation}
\Gamma_{B_s \to B^* e^- \bar\nu_{e}} = 7.6\tten{22}\ \rm{GeV}
\end{equation}

The decay widths of these transitions are several orders of magnitude
smaller than other corresponding to reactions involving a $b \to c$
transition. One could expect this fact due to the reduced phase space
available for reactions driven by a $s \to u$ transition at the quark
level.

\section{NONLEPTONIC $B_s \to c\bar s M_F$ TWO MESON DECAYS}
\label{sec:nonleptonic}
\begin{figure}[ht]
\begin{center}
\includegraphics[width=0.4\columnwidth, angle=0]{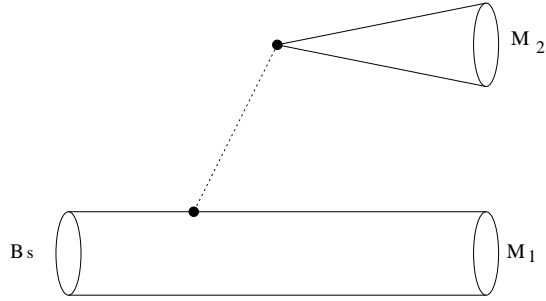}
\caption{\label{fig:nonleptonic}
Diagrammatic representation of $\bar B_s$ two meson decay in the factorization approximation.}
\end{center}
\end{figure}

In this section we evaluate decay widths for nonleptonic $\bar B_s \to
c\bar s M_F$ two-meson decays where $M_F$ is a pseudoscalar or vector
meson. These decays correspond to a $b \to c$ transition at the quark
level. These transitions are governed, neglecting penguin operators,
by the effective Hamiltonian \cite{Ebert:2006nz,Beneke:1999br}
\begin{equation}
H_{\rm
  eff}=\frac{G_F}{\sqrt2}\left(V_{cb}\left[c_1(\mu)Q_1^{cb}+c_2(\mu)Q_2^{cb}\right]+H.c.\right),
\end{equation}
where $c_{1,2}$ are scale-dependent Wilson coefficients, and $Q_{1,2}$
are local four-quark operators given by
\begin{align}
Q_1^{cb}=&\bar{\Psi}_c(0)\gamma_\mu(I-\gamma_5)\Psi_b(0)
\left[V_{ud}^*\bar{\Psi}_d(0)\gamma^\mu(I-\gamma_5)\Psi_u(0)
+V_{us}^*\bar{\Psi}_s(0)\gamma^\mu(I-\gamma_5)\Psi_u(0)
\right.
\nonumber\\&
\hspace{3.2cm}+V_{cd}^*\bar{\Psi}_d(0)\gamma^\mu(I-\gamma_5)\Psi_c(0)
\left.
+V_{cs}^*\bar{\Psi}_s(0)\gamma^\mu(I-\gamma_5)\Psi_c(0)\right]
\nonumber\\
Q_2^{cb}=&\bar{\Psi}_d(0)\gamma_\mu(I-\gamma_5)\Psi_b(0)\left[
V_{ud}^*\bar{\Psi}_c(0)\gamma^\mu(I-\gamma_5)\Psi_u(0)
+V^*_{cd}\bar{\Psi}_c(0)\gamma^\mu(I-\gamma_5)\Psi_c(0)\right]\nonumber\\
+&\bar{\Psi}_s(0)\gamma_\mu(I-\gamma_5)\Psi_b(0)\left[
V_{us}^*\bar{\Psi}_c(0)\gamma^\mu(I-\gamma_5)\Psi_u(0)
+V^*_{cs}\bar{\Psi}_c(0)\gamma^\mu(I-\gamma_5)\Psi_c(0)\right],
\end{align}
where $V_{ij}$ are CKM matrix elements. We shall work in the
factorization approximation, i. e., the hadron matrix elements of the
effective Hamiltonian are evaluated as a product of quark-current
matrix elements. One of these is the matrix element of the $B_s$
transition to one of the final mesons, while the other corresponds to
the transition to the vacuum to the other final mesons, which is given
by the corresponding meson decay constant. This is depicted in
Fig.~\ref{fig:nonleptonic}.

When writing the factorization amplitude one has to take into account
the Fierz reordered contribution so that the relevant coefficients are
not $c_1$ and $c_2$, but the combinations
\begin{align}
a_1(\mu)=c_1(\mu)+\frac{1}{N_C}c_2(\mu)\nonumber\\
a_2(\mu)=c_2(\mu)+\frac{1}{N_C}c_1(\mu)
\end{align}
with $N_C=3$ the number of colors. The appropriate energy scale
($\mu$) in our case is $\mu \approx m_b$, providing the following
values for $a_{1,2}$ \cite{Ivanov:2006ni}:
\begin{equation}
a_{1}=1.14\hspace{1cm}a_2=-0.20.
\end{equation}

\subsection{$M_F=\pi,\rho,K,K^*$}
For final states containing one of these mesons, the decay width is given by
\begin{align}
\Gamma&=\frac{G_F^2}{16\pi m_{B_s}^2}|V_{bc}|^2|V_{F}|^2
\frac{\lambda^{1/2}(m_{B_s}^2,m_{c\bar s}^2,m_{M_F})^2}{2m_{B_s}}a_1^2
{\cal H}_{\alpha\beta}(P_{B_s},P_{c\bar s})\hat{\cal H}^{\alpha\beta}(P_{F}),
\end{align}
where $m_f$ is the mass of the $M_F$ meson, $V_F$ is $V_{ud}$ for
$M_F=\pi,\rho$ and $V_{us}$ for $M_F=K,K^*$. ${\cal
  H}_{\alpha\beta}(P_{B_s},P_{c\bar s})$ is the hadron tensor
accounting for the $B_s \to c\bar s$ transition, while the other,
${\cal H}^{\alpha\beta}(P_{F})$ corresponds to a vacuum $\to M_F$
transition. This is equal to
\begin{equation}
\hat{\cal H}^{\alpha\beta}(P_F)=p_F^\alpha p_F^\beta f_F^2
\end{equation}
for a pseudoscalar $M_F$, and
\begin{equation}
\hat{\cal H}^{\alpha\beta}(P_F)=(p_F^\alpha p_F^\beta -m_F^2g^{\alpha\beta}) f_F^2
\end{equation}
for a vector $M_F$. All the necessary meson decay constants can be found in Table~\ref{tab:decaycons-vckm}.

As we did in the case of semileptonic decays, the contraction of the
two hadron tensors can be written in terms of helicity amplitudes.
For a pseudoscalar $M_F$, this is
\begin{equation}
{\cal H}_{\alpha\beta} \tilde{\cal H}^{\alpha\beta} = {\cal H}_{tt}(m_F^2)m_F^2f_F^2,
\label{eq:nlc1}
\end{equation}
and for a vector $M_F$,
\begin{equation}
{\cal H}_{\alpha\beta} \tilde{\cal H}^{\alpha\beta} = m_F^2f_F^2\left(
{\cal H}_{+1+1}(m_F^2)
+{\cal H}_{-1-1}(m_F^2)
+{\cal H}_{00}(m_F^2)
\right)
\label{eq:nlc2}
\end{equation}
In Table~\ref{tab:nl.res1} we show the values for the decay widths we
obtain for the nonleptonic decay widths of the different channels
considered in units of $\tten{15}{\rm GeV}$. In
Table~\ref{tab:brcomp2} we express our results as branching fractions
and compare with other calculations. As shown there, our results agree
with those from Refs.~\cite{Faustov:2012mt} and \cite{Blasi:1993fi} in
which relativistic CQM and QCD sum rules techniques were used,
respectively.  Our results for decays with a vector $D_s^*$ in the
final state also agree finely with those from \cite{Chen:2012fi},
although our values for final states with a pseudoscalar $D_s$ meson
in the final state are about a factor 2 larger than those from
\cite{Chen:2012fi}, who also works in the context of nonrelativistic
constituent quark models. The values calculated in this work are
larger by a factor 2 or more than the results from
Refs.~\cite{Li:2009wq}, \cite{Azizi:2008ty} and \cite{Li:2008ts}, in
which a light cone sum rules, QCD sum rules and covariant light front
quark model approaches have been used. Finally we compare our results
with the experimental measurements enclosed in
Ref.~\cite{PhysRevD.86.010001}.
\begin{table}
\begin{center}
\begin{tabular}{lc|lc|lc}
\hline\hline
                         & $\Gamma$ [$10^{-15}$ GeV] &  & $\Gamma$ [$10^{-15}$ GeV]&    & $\Gamma$ [$10^{-15}$ GeV] \\\hline
$\bar B_s \to D_s^+ \pi^-$       & $1.84_{-0.03}^{+0.04}\,a_1^2$      & $\bar B_s \to D_s^{*+} \pi^-$       & $1.56_{-0.04}^{+0.1}\,a_1^2$     &$\bar B_s \to (2^-)^+ \pi^-$       & $2.46_{-0.27}^{+0.09}\tten{4}\,a_1^2$ \\
$\bar B_s \to D_s^+ \rho^-$      & $4.53_{-0.09}^{+0.1}\,a_1^2$       & $\bar B_s \to D_s^{*+} \rho^-$      & $4.67_{-0.11}^{+0.3}\,a_1^2$     &$\bar B_s \to (2^-)^+ \rho^-$      & $1.62_{-0.26}\tten{2}\,a_1^2$     \\
$\bar B_s \to D_s^+ K^-$         & $0.14^{+0.01}\,a_1^2$             & $\bar B_s \to D_s^{*+} K^-$         & $0.12_{-0.01}^{+0.01}\,a_1^2$    &$\bar B_s \to (2^-)^+ K^-$         & $1.82^{+0.06}_{-0.2}\tten{5}\,a_1^2$      \\
$\bar B_s \to D_s^+ K^{*-}$       & $0.25_{-0.01}^{0.01}\,a_1^2$       & $\bar B_s \to D_s^{*+} K^{*-}$       & $0.27_{-0.01}^{+0.02}\,a_1^2$    &$\bar B_s \to (2^-)^+ K^{*-}$       & $1.11_{-0.12}\tten{3}\,a_1^2$   \\
\hline
$\bar B_s \to D_{s0}^{*+} \pi^-$       & $0.39_{-0.01}^{+0.01}\,a_1^2$ & $\bar B_s \to D_{s1}^+(2460) \pi^-$   & $0.53_{-0.02}\,a_1^2$            &$\bar B_s \to D_{s2}^{*+} \pi^-$       & $0.35_{-0.03}\,a_1^2$ \\
$\bar B_s \to D_{s0}^{*+} \rho^-$      & $0.94_{-0.04}^{+0.04}\,a_1^2$  & $\bar B_s \to D_{s1}^+(2460) \rho^-$  & $1.26_{-0.06}\,a_1^2$            & $\bar B_s \to D_{s2}^{*+} \rho^-$      & $0.95_{-0.07}\,a_1^2$            \\
$\bar B_s \to D_{s0}^{*+} K^-$         & $2.98_{-0.15}^{+0.11}\tten{2}\,a_1^2$ & $\bar B_s \to D_{s1}^+(2460) K^-$     & $4.09_{-0.1}\tten{2}\,a_1^2$     & $\bar B_s \to D_{s2}^{*+} K^-$         & $2.61_{-0.17}\tten{2}\,a_1^2$     \\
$\bar B_s \to D_{s0}^{*+} K^{*-}$       & $5.19_{-0.24}^{+0.18}\tten{2}\,a_1^2$ & $\bar B_s \to D_{s1}^+(2460) K^{*-}$   & $6.93_{-0.3}\tten{2}\,a_1^2$     & $\bar B_s \to D_{s2}^{*+} K^{*-}$       & $5.41_{-0.37}\tten{2}\,a_1^2$     \\
\hline
&&$\bar B_s \to D_{s1}^+(2536) \pi^-$   & $0.25_{-0.01}\,a_1^2$&&\\
&&$\bar B_s \to D_{s1}^+(2536) \rho^-$  & $0.66_{-0.03}\,a_1^2$     &&\\
&&$\bar B_s \to D_{s1}^+(2536) K^-$     & $1.94_{-0.08}\tten{2}\,a_1^2$     &&\\
&&$\bar B_s \to D_{s1}^+(2536) K^{*-}$   & $3.75_{-0.2}\tten{2}\,a_1^2$     &&\\\hline
\end{tabular}
\end{center}
\caption{\label{tab:nl.res1} Total nonleptonic decay widths of $B_s$
  mesons for generic values of the Wilson parameter $a_1$. The central values
  have been calculated using the AL1 potential.}
\end{table}

\begin{table}
\begin{tabular}{lccccccc|c}
\hline\hline
                         &  This work & \cite{Faustov:2012mt} & \cite{Blasi:1993fi} & \cite{Chen:2012fi} &\cite{Li:2009wq} 
&\cite{Azizi:2008ty}& \cite{Li:2008ts} & Experiment \cite{PhysRevD.86.010001}\\\hline
$\bar B_s \to D_s^+ \pi^-$       & 0.53 & 0.35  & 0.5  & $0.27_{-0.03}^{+0.07}$   & $0.17_{-0.06}^{+0.07}$     & $0.142\pm0.57$    & $0.196_{-0.097}^{+0.123}$ & $0.32\pm0.4$ \\
$\bar B_s \to D_s^+ \rho^-$      & 1.26 & 0.94  & 1.3  & $0.64_{-0.11}^{+0.17}$   & $0.42_{-1.4}^{+1.7}$      &                   & $0.47_{-2.3}^{+2.9}$ & $0.74\pm0.17$\\
$\bar B_s \to D_s^+ K^-$         & 0.04 & 0.028 & 0.04 & $0.021_{-0.002}^{+0.002}$ & $0.013_{-0.004}^{+0.005}$ & $0.0103\pm0.0051$ & $0.017_{-0.0066}^{+0.0087}$& \\
$\bar B_s \to D_s^+ K^{*-}$       & 0.08 & 0.047 & 0.06 & $0.038_{-0.005}^{+0.005}$ & $0.028_{-0.08}^{+0.01}$   & $0.005\pm0.0022$  & \\
$\bar B_s \to D_{s0}^{*+} \pi^-$       & 0.10     &0.09                        &  $0.052_{-0.021}^{+0.25}$   &&&&&\\
$\bar B_s \to D_{s0}^{*+} \rho^-$      & 0.27     &0.22                            &  $0.013_{-0.05}^{+0.06}$   &&&&&\\
$\bar B_s \to D_{s0}^{*+} K^-$         & 0.009    &0.007                           &  $0.004_{-0.002}^{+0.002}$ &&&&&\\
$\bar B_s \to D_{s0}^{*+} K^{*-}$       & 0.16     &0.012                           &  $0.008_{-0.003}^{+0.004}$  &&&&&\\
$\bar B_s \to D_s^{*+} \pi^-$       & 0.45 & 0.27   & 0.2  & $0.31_{-0.02}^{+0.03}$    &                       & $0.211\pm0.073$  & $0.189_{-0.093}^{+0.120}$ & $0.21\pm0.06$\\
$\bar B_s \to D_s^{*+} \rho^-$      & 1.35 & 0.87   & 1.3  & $0.9_{-1.5}^{+1.5}$      &                        &                  & $0.523_{-0.256}^{+0.334}$ & $1.03\pm2.6$ \\
$\bar B_s \to D_s^{*+} K^-$         & 0.04 & 0.021  & 0.02 & $0.024_{-0.002}^{+0.002}$ &                        &                  & \\
$\bar B_s \to D_s^{*+} K^{*-}$       & 0.08 & 0.048  & 0.06 & $0.056_{-0.007}^{+0.006}$ &                        &                  & \\
$\bar B_s \to D_{s1}^+(2460) \pi^-$   &0.15           &0.19  &                       &                        &&&&\\
$\bar B_s \to D_{s1}^+(2460) \rho^-$  &0.36           &0.49  &                       &                        &&&&\\
$\bar B_s \to D_{s1}^+(2460) K^-$     &0.012          &0.014 &                       &                        &&&&\\
$\bar B_s \to D_{s1}^+(2460) K^{*-}$   &0.020          &0.026 &                       &                        &&&&\\
$\bar B_s \to D_{s1}^+(2536) \pi^-$   &0.07          &0.029  &                       &                        &&&&\\
$\bar B_s \to D_{s1}^+(2536) \rho^-$  &0.19          &0.083  &                       &                        &&&&\\
$\bar B_s \to D_{s1}^+(2536) K^-$     &0.0054        &0.0021 &                       &                        &&&&\\
$\bar B_s \to D_{s1}^+(2536) K^{*-}$   &0.01          &0.0044 &                       &                        &&&&\\
$\bar B_s \to (2^-)^+ \pi^-$       &7.1\ten{5}    &&&&&&&\\
$\bar B_s \to (2^-)^+ \rho^-$      &0.0047        &&&&&&&\\
$\bar B_s \to (2^-)^+ K^-$         &5.2\ten{6}    &&&&&&&\\
$\bar B_s \to (2^-)^+ K^{*-}$       &2.2\ten{8}    &&&&&&&\\
$\bar B_s \to D_{s2}^{*+} \pi^-$       &0.1          &0.16   &                       &                        &&&&\\
$\bar B_s \to D_{s2}^{*+} \rho^-$      &0.27         &0.42   &                       &                        &&&&\\
$\bar B_s \to D_{s2}^{*+} K^-$         &0.008        &0.012  &                       &                        &&&&\\
$\bar B_s \to D_{s2}^{*+} K^{*-}$       &0.016        &0.022  &                       &                        &&&&\\
\hline
\end{tabular}
\caption{\label{tab:brcomp2} Branching ratios for the decays above.}
\end{table}

\subsection{$M_F=D,D_s,D^*,D_s^*$}

In the same way, we can calculate the nonleptonic decay width of the
processes $\bar B_s \to D_s D$, $\bar B_s\to D_s D^*$, $\bar B_s\to
D_s^*D_s$, $\bar B_s\to D_s^* D_s^*$, $\bar B_s\to D_s D_s^*$, and
$\bar B_s\to D_s^*D_s^*$ decays. As in the previous case, there is
only one contribution proportional to the coefficient $a_1$, with
momentum transfer ranges between $m_D^2$ and $m_{D^*}^2$. These
momentum transfers are neither too high (so there is no need to
involve a $B_s^*$ resonance) nor too low (with a high trimomentum
transfer). For $M_F = D, D^*,D_S, D_s^*$, the relevant contractions
for the hadron tensors can be obtained from Eqs.~\ref{eq:nlc1} and
\ref{eq:nlc2}, performing straigthforward substitutions.
%
%
As in the previous case, the decay constants relevant for these
calculations can be found in Table~\ref{tab:decaycons-vckm}.

Results are enclosed in Table~\ref{tab:nl2}. In Table~\ref{tab:br3},
we present our values as branching fractions, and compare with other
results, in the case of decays with two $D_s^{(*)}$ mesons in the
final state. We have found a fair agreement with the branching
fractions calculated in \cite{Faustov:2012mt} and \cite{Blasi:1993fi},
and larger differences with the values given in \cite{Chen:2012fi},
\cite{Li:2009wq} and \cite{Ivanov:2011aa}. Most of the values
calculated here and those found in the literature differ from the
experimental results by a factor around 2. Appart from the
inaccuracies of the factorization approximation, the results are
sensitive, not only two the Wilson parameter $a_1$, also on the value
that have been used for the mesons decay constant and on the overlap
among the wave functions used to calculate the matrix elements.

\begin{table}
\begin{tabular}{lc}
\hline\hline
                         & $\Gamma$ [$10^{-15}$ GeV] \\\hline
$\bar B_s \to D_s^+   D_s^-$        &  $7.35_{-0.14}^{+0.04}\,a_1^2$      \\
$\bar B_s \to D_s^+   D_s^{*-}$      &  $6.89_{-0.13}^{+0.02}\,a_1^2$      \\
$\bar B_s \to D_s^{*+}  D_s^-$       &  $4.23_{-0.38}\,a_1^2$            \\
$\bar B_s \to D_s^{*+} D_s^{*-}$     &  $18.79_{-1.6}\,a_1^2$            \\\hline
$\bar B_s \to D_s^+   D^-$         &  $0.25_{-0.01}\,a_1^2$            \\
$\bar B_s \to D_s^+   D^{*-}$       &  $0.19_{-0.01}\,a_1^2$            \\
$\bar B_s \to D_s^{*+}  D^-$        &  $0.15_{-0.01}^{+0.01}\,a_1^2$      \\
$\bar B_s \to D_s^{*+} D^{*-}$       &  $0.46_{-0.01}^{+0.02}\,a_1^2$      \\
\hline
\end{tabular}
\caption{\label{tab:nl2} Nonleptonic decay widths for the indicated processes
  indicated, using the factorization approximation. We give our results for generic values of the 
  parameter $a_1$.}
\end{table}

\begin{table}
\begin{tabular}{l|cccccc|c}
\hline\hline
                         & This work & \cite{Faustov:2012mt} & \cite{Blasi:1993fi} &  \cite{Chen:2012fi} & \cite{Li:2009wq} & 
\cite{Ivanov:2011aa} & Experiment \cite{PhysRevD.86.010001} \\\hline
$\bar B_s \to D_s^+   D_s^-$         &    2.1   & 1.1   & 1.0 & $0.83_{-0.1}^{+0.1}$  & 1.65     &  $0.217\pm0.082$ & $0.53\pm0.09$ \\
$\bar B_s \to D_s^+   D_s^{*-}$       &    2.0   & 1.0   & 0.8 & $0.84_{-0.12}^{+0.12}$ &          &  $0.262\pm0.93$  &               \\
$\bar B_s \to D_s^{*+}  D_s^-$        &    1.24   & 0.61  & 0.4 & $0.7_{-0.15}^{+1.6}$   &          &  $0.254\pm0.57$  &               \\
$\bar B_s \to D_s^+   D_s^{*-} +D_s^{*+}  D_s^-$& 3.24 & 1.61  & 1.2 & $1.54_{-0.19}^{+0.2}$   & 2.4      &  $5.16\pm0.11$   & $1.24\pm0.21$ \\
$\bar B_s \to D_s^{*+} D_s^{*-}$       &    5.45   & 2.5   & 1.6 & $2.4_{-0.4}^{+0.4}$    & 3.18     &  $2.77\pm0.76$   & $1.88\pm0.34$ \\
$\bar B_s \to D_s^{(*)+} D_s^{(*)-}$    &    10.8  &5.21  & 3.8 & $4.77_{-0.46}^{+0.46}$ & 7.23     & $3.5\pm0.78$     & $4.5\pm1.4$   \\
\hline
$\bar B_s \to D_s^+   D^-$         &    0.08   &      &     & & & &\\
$\bar B_s \to D_s^+   D^{*-}$       &    0.05   &      &     & & & &\\
$\bar B_s \to D_s^{*+}  D^-$        &    0.04   &      &     & & & &\\
$\bar B_s \to D_s^{*+} D^{*-}$       &    0.13   &      &     & & & &\\
\hline
\end{tabular}
\caption{\label{tab:br3} Branching ratios in \% for the decays indicated above. We also compare with other calculations.}
\end{table}

\section{Other nonleptonic decays}
\label{sec:othernl}
The calculation of decay channels
\begin{align}
\bar B_s &\to \phi J/\Psi\nonumber\\
\bar B_s &\to K^0 J/\Psi\nonumber\\
\bar B_s &\to K^{*0} J/\Psi\nonumber\\
\end{align}
in the factorization approximation can be easily performed. Their
decay width are summarized in Table~\ref{tab:resnl3}.
\begin{table}
\begin{tabular}{lcc|c}
\hline\hline
                                 & $\Gamma$ [$10^{-15}$ GeV]         &  BR in \%             &  Experiment \cite{PhysRevD.86.010001}   \\\hline
$\bar B_s \to \phi J/\Psi $      & $11.80_{-0.8}^{+1.9}\,a_2^2$        & $0.11$                &  $(0.109_{-0.23}^{+0.28})$         \\
$\bar B_s \to K^0 J/\Psi $       & $8.1_{-1.3}^{+0.5}\tten{2}\,a_2^2$  &  $7.25\tten{4}$       &  $(3.6\pm0.8)\tten{3}$          \\
$\bar B_s \to K^{*0} J/\Psi $     & $0.51_{-0.3}\,a_2^2$              &  $4.6\tten{3}$        &  $(9\pm 4)\tten{3}$             \\
\hline
\end{tabular}
\caption{\label{tab:resnl3} Branching ratios in \% for the reactions
  indicated above. We give our results for generic values of $a_2$.}
\end{table}

In Table~\ref{tab:resnl3} we also give the branching fractions of
these channels in \%, for generic values of the Wilson parameter
$a_2$, and compare with the experimental measurements. Our result for
the branching ratio corresponding to the decay into $K^0 J/\Psi$
states reproduces roughly the order of magnitude of the corresponding
experimental value ($\approx10^{-3}$). In contrast, our results for the
branching fractions for the $\bar B_s \to \phi J/\Psi $ and $\bar B_s
\to K^{*0} J/\Psi $ decays agree with the experimental data of
Ref.~\cite{PhysRevD.86.010001}.

\section{Summary and conclusions}
\label{sec:summary}
In this paper we have studied the semileptonic decays of the $\bar
B_s$ meson into $c\bar s$ states with $J^P=0^-,0^+,1^-, 1^+, 2^-$ and
$2^+$. We have worked in the context of nonrelativistic constituent
quark models. We compare with the experimental results enclosed in
Ref.~\cite{PhysRevD.86.010001} when possible.  We have also computed
several nonleptonic decay modes of $\bar B_s$ mesons. We work in the
factorization approximation, as the momenta involved does not involve
resonances or high trimomentum transfer.  We give results for general
values of the Wilson coefficients. We give an estimate of our
theoretical uncertainties by considering different sets of wave
functions derived from the quark--antiquark potentials of
Ref.~\cite{SilvestreBrac:1996bg}.  The results that we obtain for the
semileptonic decay width are in general in good agreement with
previous calculations and with the available experimental
measurements. In the case of the nonleptonic decay channels that we
have studied, we have found reasonable agreement with previous
calculations. The nonleptonic decays of $\bar B_s$ mesons into $\phi
J/\Psi$, $K^0 J/\Psi$ and $K^{*0}(892) J/\Psi$ have been considered in
this work, finding a good agreement with the experimental results.

\acknowledgments 

The author thanks J. Nieves, E. Hern\'andez, M. \'A. P\'erez-Garc\'\i
a, I. Vida\~na and S. Chiacchiera for useful discussions and kind
hospitality.  The author also thanks the Physics Department at the
University of Coimbra and Departamento de F\'\i sica Fundamental at
University of Salamanca for their hospitality.  The author
acknowledges a contract from the CPAN project and support from Junta
de Andalucía under contract FQM-225.

\begin{widetext}
\newpage
\appendix

\section{Expressions for the matrix elements}

\begin{itemize}
\item {\large Case $J^\pi=0^-$}
\end{itemize}
\begin{align*}
V^0(|\vec{q}\,|)&=\sqrt{2m_I2E_F(-\vec{q}\,)}\ \int\,d^3p\ \frac{1}{4\pi}
\left(\hat{\phi}^{(M_F(0^-))}_{f'_1,\,f_2}(|\vec{p}\,|)\right)^*\,
\hat{\phi}^{(M_I(0^-))}_{f_1,\,f_2}\left(\bigg|\,\vec{p}-\frac{m_{f_2}}{m_{f'_1}+m_{f_2}}
|\vec{q}\,|
\vec{k} \bigg|\right)\nonumber\\
&\hspace{3cm}\sqrt{\frac{\widehat{E}_{f'_1}\widehat{E}_{f1}}{4E_{f'_1}
E_{f_1}}}
\left(
1+\frac{(-\frac{m_{f'_1}}{m_{f'_1}+m_{f_2}}\,|\vec{q}\,|\vec{k}-\vec{p}\,)
\cdot(\frac{m_{f_2}}{m_{f'_1}+m_{f_2}}\,|\vec{q}\,|\vec{k}-\vec{p}\,)}{\widehat{E}_{f'_1}\widehat{E}_{f_1}}
\right) \nonumber
\end{align*}
\begin{align}
V^3(|\vec{q}\,|)&=\sqrt{2m_I2E_F(-\vec{q}\,)}\ \int\,d^3p\ \frac{1}{4\pi}
\left(\hat{\phi}^{(M_F(0^-))}_{f'_1,\,f_2}(|\vec{p}\,|)\right)^*\,
 \hat{\phi}^{(M_I
 (0^-))}_{f_1,\,f_2}\left(\bigg|\,\vec{p}-\frac{m_{f_2}}{m_{f'_1}+m_{f_2}}
|\vec{q}\,|
\vec{k} \bigg|\right)
 \nonumber\\
&\hspace{3cm}\sqrt{\frac{\widehat{E}_{f'_1}\widehat{E}_{f1}}{4E_{f'_1}
E_{f_1}}}
\left(\frac{\frac{m_{f_2}}{m_{f'_1}+m_{f_2}}\,|\vec{q}\,|-p_z}{\widehat{E}_{f_1}}+
\frac{-\frac{m_{f'_1}}{m_{f'_1}+m_{f_2}}\,|\vec{q}\,|-p_z}{\widehat{E}_{f'_1}}
\right) \nonumber\\
\end{align}
\begin{itemize}
\item {\large Case $J^\pi= 0^+$}
\end{itemize}
\begin{align*}
A^0(|\vec{q}\,|)&=\sqrt{2m_I2E_F(-\vec{q}\,)}\ \int\,d^3p\ \frac{1}{4\pi
|\vec{p}\,|}
\left(\hat{\phi}^{(M_F(0^+))}_{f'_1,\,f_2}(|\vec{p}\,|)\right)^*\,
\hat{\phi}^{(M_I
(0^-))}_{f_1,\,f_2}\left(\bigg|\,\vec{p}-\frac{m_{f_2}}{m_{f'_1}+m_{f_2}}
|\vec{q}\,|
\vec{k} \bigg|\right)\nonumber\\
&\hspace{1cm}\sqrt{\frac{\widehat{E}_{f'_1}\widehat{E}_{f1}}{4E_{f'_1}
E_{f_1}}}
\left(\frac{\vec{p}\cdot(\frac{m_{f_2}}{m_{f'_1}+m_{f_2}}\,|\vec{q}\,|\vec{k}-\vec{p}\,)}{\widehat{E}_{f_1}}
+\frac{\vec{p}\cdot(-\frac{m_{f'_1}}{m_{f'_1}+m_{f_2}}\,|\vec{q}\,|\vec{k}-\vec{p}\,)}
{\widehat{E}_{f'_1}}
\right) \nonumber
\end{align*}
\begin{align}
A^3(|\vec{q}\,|)&=\sqrt{2m_I2E_F(-\vec{q}\,)}\ \int\,d^3p\
\frac{1}{4\pi|\vec{p}\,|}
\left(\hat{\phi}^{(M_F (0^+))}_{f'_1,\,f_2}(|\vec{p}\,|)\right)^*\,
\hat{\phi}^{(M_I(0^-))}_{f_1,\,f_2}\left(\bigg|\,\vec{p}-\frac{m_{f_2}}{m_{f'_1}+m_{f_2}}
|\vec{q}\,|
\vec{k} \bigg|\right)
 \nonumber\\
&\hspace{1cm}\sqrt{\frac{\widehat{E}_{f'_1}\widehat{E}_{f1}}{4E_{f'_1}
E_{f_1}}}\bigg\{\
p_z\bigg(  
1-\frac{(-\frac{m_{f'_1}}{m_{f'_1}+m_{f_2}}\,|\vec{q}\,|\vec{k}-\vec{p}\,)
\cdot(\frac{m_{f_2}}{m_{f'_1}+m_{f_2}}\,|\vec{q}\,|\vec{k}-\vec{p}\,)}{\widehat{E}_{f'_1}\widehat{E}_{f_1}}
\bigg)\nonumber\\ 
&\hspace{3.cm}+\frac{1}{\widehat{E}_{f'_1}\widehat{E}_{f1}}
\bigg[\hspace{.35cm} (-\frac{m_{f'_1}}{m_{f'_1}+m_{f_2}}|\vec{q}\,|-p_z)\ \ \
\vec{p}\cdot\bigg(\hspace{.4cm}\frac{m_{f_2}}{m_{f'_1}+m_{f_2}}|\vec{q}\,|\vec{k}-\vec{p}
\bigg)\nonumber\\
&\hspace{4.75cm}+(\hspace{.4cm} \frac{m_{f_2}}{m_{f'_1}+m_{f_2}}|\vec{q}\,|-p_z
)\ \ \
\vec{p}\cdot\bigg(-\frac{m_{f'_1}}{m_{f'_1}+m_{f_2}}|\vec{q}\,|\vec{k}-\vec{p}
\bigg)
\bigg]\bigg\}\nonumber\\
\end{align}

\begin{itemize}
\item {\large Case $J^\pi=1^-$}
\end{itemize}

\begin{align}
V^{(1^-)\,1}_{\lambda=-1}(|\vec{q}\,|)&=\frac{-i}{\sqrt2}
\sqrt{2m_I2E_F(-\vec{q}\,)}\ \int\,d^3p\ \frac{1}{4\pi}
\left(\hat{\phi}^{(M_F(1^-))}_{f'_1,\,f_2}(|\vec{p}\,|)\right)^*\,
\hat{\phi}^{(M_I(0^-))}_{f_1,\,f_2}\left(\bigg|\,\vec{p}-\frac{m_{f_2}}{m_{f'_1}+m_{f_2}}
|\vec{q}\,|
\vec{k} \bigg|\right)\nonumber\\
&\hspace{3cm}\sqrt{\frac{\widehat{E}_{f'_1}\widehat{E}_{f1}}{4E_{f'_1}
E_{f_1}}}
\left(-\frac{\frac{m_{f_2}}{m_{f'_1}+m_{f_2}}\,|\vec{q}\,|-p_z}{\widehat{E}_{f_1}}+
\frac{-\frac{m_{f'_1}}{m_{f'_1}+m_{f_2}}\,|\vec{q}\,|-p_z}{\widehat{E}_{f'_1}}
\right)\nonumber\\
\end{align}
%
%
\begin{align}
A^{(1^-)\,0}_{\lambda=0}(|\vec{q}\,|)&=i\
\sqrt{2m_I2E_F(-\vec{q}\,)}\ \int\,d^3p\ \frac{1}{4\pi}
\left(\hat{\phi}^{(M_F(1^-))}_{f'_1,\,f_2}(|\vec{p}\,|)\right)^*\,
\hat{\phi}^{(M_I(0^-))}_{f_1,\,f_2}\left(\bigg|\,\vec{p}-\frac{m_{f_2}}{m_{f'_1}+m_{f_2}}
|\vec{q}\,|
\vec{k} \bigg|\right)\nonumber\\
&\hspace{3cm}\sqrt{\frac{\widehat{E}_{f'_1}\widehat{E}_{f1}}{4E_{f'_1}
E_{f_1}}}
\left(\frac{\frac{m_{f_2}}{m_{f'_1}+m_{f_2}}\,|\vec{q}\,|-p_z}{\widehat{E}_{f_1}}+
\frac{-\frac{m_{f'_1}}{m_{f'_1}+m_{f_2}}\,|\vec{q}\,|-p_z}{\widehat{E}_{f'_1}}
\right)\nonumber
\\
A^{(1^-)\,1}_{\lambda=-1}(|\vec{q}\,|)&=\frac{i}{\sqrt2}
\sqrt{2m_I2E_F(-\vec{q}\,)}\ \int\,d^3p\ \frac{1}{4\pi}
\left(\hat{\phi}^{(M_F(1^-))}_{f'_1,\,f_2}(|\vec{p}\,|)\right)^*\,
\hat{\phi}^{(M_I(0^-))}_{f_1,\,f_2}\left(\bigg|\,\vec{p}-\frac{m_{f_2}}{m_{f'_1}+m_{f_2}}
|\vec{q}\,|
\vec{k} \bigg|\right)\nonumber\\
&\hspace{3cm}\sqrt{\frac{\widehat{E}_{f'_1}\widehat{E}_{f1}}{4E_{f'_1}
E_{f_1}}}
\left(1+\frac{2p_x^2-(-\frac{m_{f'_1}}{m_{f'_1}+m_{f_2}}\,|\vec{q}\,|\vec{k}-\vec{p}\,)
\cdot(\frac{m_{f_2}}{m_{f'_1}+m_{f_2}}\,|\vec{q}\,|\vec{k}-\vec{p}\,)}
{\widehat{E}_{f'_1}\widehat{E}_{f_1}}
\right)
\nonumber\\
A^{(1^-)\,3}_{\lambda=0}(|\vec{q}\,|)&=i
\sqrt{2m_I2E_F(-\vec{q}\,)}\ \int\,d^3p\ \frac{1}{4\pi}
\left(\hat{\phi}^{(M_F(1^-))}_{f'_1,\,f_2}(|\vec{p}\,|)\right)^*\,
\hat{\phi}^{(M_I(0^-))}_{f_1,\,f_2}\left(\bigg|\,\vec{p}-\frac{m_{f_2}}{m_{f'_1}+m_{f_2}}
|\vec{q}\,|
\vec{k} \bigg|\right)\nonumber\\
&\hspace{3cm}\sqrt{\frac{\widehat{E}_{f'_1}\widehat{E}_{f1}}{4E_{f'_1}
E_{f_1}}}
\ \Bigg(
1+\frac{2(-\frac{m_{f'_1}}{m_{f'_1}+m_{f_2}}\,|\vec{q}\,|-p_z\,)
\cdot(\frac{m_{f_2}}{m_{f'_1}+m_{f_2}}\,|\vec{q}\,|-p_z\,)}
{\widehat{E}_{f'_1}\widehat{E}_{f_1}}\nonumber\\
&\hspace{5.45cm}-\frac{(-\frac{m_{f'_1}}{m_{f'_1}+m_{f_2}}\,|\vec{q}\,|\vec{k}-\vec{p}\,)
\cdot(\frac{m_{f_2}}{m_{f'_1}+m_{f_2}}\,|\vec{q}\,|\vec k-\vec{p}\,)}
{\widehat{E}_{f'_1}\widehat{E}_{f_1}}
\Bigg)\nonumber\\
\end{align}

\begin{itemize}
\item {\large Case $J^\pi=1^+$}
\end{itemize}

\begin{align}
V^{(1^+,S_{q\bar{q}}=0)\,0}_{\lambda=0}(|\vec{q}\,|)&=i\sqrt3
\sqrt{2m_I2E_F(-\vec{q}\,)}\ \int\,d^3p\ \frac{1}{4\pi|\vec p\,|}
\left(\hat{\phi}^{(M_F(1^+,S_{q\bar{q}}=0))}_{f'_1,\,f_2}(|\vec{p}\,|)\right)^*\,
\hat{\phi}^{(M_I(0^-))}_{f_1,\,f_2}\left(\bigg|\,\vec{p}-\frac{m_{f_2}}{m_{f'_1}+m_{f_2}}
|\vec{q}\,|
\vec{k} \bigg|\right)\nonumber\\
&\hspace{3cm}\sqrt{\frac{\widehat{E}_{f'_1}\widehat{E}_{f1}}{4E_{f'_1}
E_{f_1}}}\,p_z\,
\left(
1+\frac{(-\frac{m_{f'_1}}{m_{f'_1}+m_{f_2}}\,|\vec{q}\,|\vec{k}-\vec{p}\,)
\cdot(\frac{m_{f_2}}{m_{f'_1}+m_{f_2}}\,|\vec{q}\,|\vec{k}-\vec{p}\,)}{\widehat{E}_{f'_1}\widehat{E}_{f_1}}
\right)
\nonumber\\
V^{(1^+,S_{q\bar{q}}=1)\,0}_{\lambda=0}(|\vec{q}\,|)&=-i\sqrt{\frac{3}{2}}
\sqrt{2m_I2E_F(-\vec{q}\,)}\ \int\,d^3p\ \frac{1}{4\pi|\vec p\,|}
\left(\hat{\phi}^{(M_F(1^+,S_{q\bar{q}}=1))}_{f'_1,\,f_2}(|\vec{p}\,|)\right)^*\,
\hat{\phi}^{(M_I(0^-))}_{f_1,\,f_2}\left(\bigg|\,\vec{p}-\frac{m_{f_2}}{m_{f'_1}+m_{f_2}}
|\vec{q}\,|
\vec{k} \bigg|\right)\nonumber\\
&\hspace{3cm}\sqrt{\frac{\widehat{E}_{f'_1}\widehat{E}_{f1}}{4E_{f'_1}
E_{f_1}}}
\ \frac{|\vec{q}\,|
(p_z^2-\vec{p}^{\,2})}{\widehat{E}_{f'_1}\widehat{E}_{f1}}\nonumber\\
V^{(1^+,S_{q\bar{q}}=0)\,1}_{\lambda=-1}(|\vec{q}\,|)&=-i\sqrt{\frac{3}{2}}
\sqrt{2m_I2E_F(-\vec{q}\,)}\ \int\,d^3p\ \frac{1}{4\pi|\vec{p}\,|}
\left(\hat{\phi}^{(M_F(1^+,S_{q\bar{q}}=0))}_{f'_1,\,f_2}(|\vec{p}\,|)\right)^*\,
\hat{\phi}^{(M_I(0^-))}_{f_1,\,f_2}\left(\bigg|\,\vec{p}-\frac{m_{f_2}}{m_{f'_1}+m_{f_2}}
|\vec{q}\,|
\vec{k} \bigg|\right)\nonumber\\
&\hspace{3cm}\sqrt{\frac{\widehat{E}_{f'_1}\widehat{E}_{f1}}{4E_{f'_1}
E_{f_1}}}
\ p_x^2\,\left(\frac{1}{\widehat{E}_{f_1}}+\frac{1}{\widehat{E}_{f'_1}}
\right)\nonumber\\%
V^{(1^+,S_{q\bar{q}}=1)\,1}_{\lambda=-1}(|\vec{q}\,|)&=i\frac{\sqrt{3}}{2}
\sqrt{2m_I2E_F(-\vec{q}\,)}\ \int\,d^3p\ \frac{1}{4\pi|\vec{p}\,|}
\left(\hat{\phi}^{(M_F(1^+,S_{q\bar{q}}=1))}_{f'_1,\,f_2}(|\vec{p}\,|)\right)^*\,
\hat{\phi}^{(M_I(0^-))}_{f_1,\,f_2}\left(\bigg|\,\vec{p}-\frac{m_{f_2}}
{m_{f'_1}+m_{f_2}}
|\vec{q}\,|
\vec{k} \bigg|\right)\nonumber\\
&\hspace{3cm}\sqrt{\frac{\widehat{E}_{f'_1}\widehat{E}_{f1}}{4E_{f'_1}
E_{f_1}}}
\ \left(\frac{p_y^2+p_z^2+p_z|\vec{q}\,|
\frac{m_{f'_1}}{m_{f'_1}+m_{f_2}}
}{\widehat{E}_{f'_1}}-\frac{p_y^2+p_z^2-p_z|\vec{q}\,|
\frac{m_{f_2}}{m_{f'_1}+m_{f_2}}}{\widehat{E}_{f_1}}
\right)\nonumber\\%
\nonumber
\end{align}
\begin{align}
V^{(1^+,S_{q\bar{q}}=0)\,3}_{\lambda=0}(|\vec{q}\,|)&=i\sqrt3
\sqrt{2m_I2E_F(-\vec{q}\,)}\ \int\,d^3p\ \frac{1}{4\pi|\vec{p}\,|}
\left(\hat{\phi}^{(M_F(1^+,S_{q\bar{q}}=0))}_{f'_1,\,f_2}(|\vec{p}\,|)\right)^*\,
\hat{\phi}^{(M_I(0^-))}_{f_1,\,f_2}\left(\bigg|\,\vec{p}-\frac{m_{f_2}}{m_{f'_1}+m_{f_2}}
|\vec{q}\,|
\vec{k} \bigg|\right)\nonumber\\
&\hspace{3cm}\sqrt{\frac{\widehat{E}_{f'_1}\widehat{E}_{f1}}{4E_{f'_1}
E_{f_1}}}\
p_z\,\Bigg(\frac{\frac{m_{f_2}}{m_{f'_1}+m_{f_2}}\,|\vec{q}\,|-p_z}{\widehat{E}_{f_1}}+
\frac{-\frac{m_{f'_1}}{m_{f'_1}+m_{f_2}}\,|\vec{q}\,|-p_z}{\widehat{E}_{f'_1}}
\Bigg)\nonumber\\%
V^{(1^+,S_{q\bar{q}}=1)\,3}_{\lambda=0}(|\vec{q}\,|)&=-i\sqrt{\frac{3}{2}}
\sqrt{2m_I2E_F(-\vec{q}\,)}\ \int\,d^3p\ \frac{1}{4\pi|\vec{p}\,|}
\left(\hat{\phi}^{(M_F(1^+,S_{q\bar{q}}=1))}_{f'_1,\,f_2}(|\vec{p}\,|)\right)^*\,
\hat{\phi}^{(M_I(0^-))}_{f_1,\,f_2}\left(\bigg|\,\vec{p}-\frac{m_{f_2}}{m_{f'_1}+m_{f_2}}
|\vec{q}\,|
\vec{k} \bigg|\right)\nonumber\\
&\hspace{3cm}\sqrt{\frac{\widehat{E}_{f'_1}\widehat{E}_{f1}}{4E_{f'_1}
E_{f_1}}}
\ (p_x^2+p_y^2)\,\left(\frac{1}{\widehat{E}_{f_1}}-\frac{1}{\widehat{E}_{f'_1}}
\right)\nonumber\\
\nonumber
\end{align}
\begin{eqnarray}
A^{(1^+,S_{q\bar{q}}=0)\,1}_{\lambda=-1}(|\vec{q}\,|)&=&-i\sqrt{\frac{3}{2}}
\sqrt{2m_I2E_F(-\vec{q}\,)}\ \int\,d^3p\ \frac{1}{4\pi|\vec{p}\,|}
\left(\hat{\phi}^{(M_F(1^+,S_{q\bar{q}}=0))}_{f'_1,\,f_2}(|\vec{p}\,|)\right)^*\,
\hat{\phi}^{(M_I(0^-))}_{f_1,\,f_2}\left(\bigg|\,\vec{p}-\frac{m_{f_2}}{m_{f'_1}+m_{f_2}}
|\vec{q}\,|
\vec{k} \bigg|\right)\nonumber\\
&&\hspace{3cm}\sqrt{\frac{\widehat{E}_{f'_1}\widehat{E}_{f1}}{4E_{f'_1}
E_{f_1}}}
\ \frac{p_y^2|\vec{q}\,|}{\widehat{E}_{f_1}\widehat{E}_{f'_1}}
\nonumber\\%
A^{(1^+,S_{q\bar{q}}=1)1}_{\lambda=-1}(|\vec{q}\,|)&=&i\frac{\sqrt3}{2}\sqrt{2m_I2E_F(-\vec{q}\,)}\
\int\,d^3p\
\frac{1}{4\pi|\vec{p}\,|}
\left(\hat{\phi}^{(M_F (1^+,S_{q\bar{q}}=1))}_{f'_1,\,f_2}(|\vec{p}\,|)
\right)^*\,
\hat{\phi}^{(M_I(0^-))}_{f_1,\,f_2}\left(\bigg|\,\vec{p}-\frac{m_{f_2}}{m_{f'_1}+m_{f_2}}
|\vec{q}\,|
\vec{k} \bigg|\right)
 \nonumber\\
&&\hspace{3cm}\sqrt{\frac{\widehat{E}_{f'_1}\widehat{E}_{f1}}{4E_{f'_1}
E_{f_1}}}\bigg\{\
p_z\bigg(  
1-\frac{(-\frac{m_{f'_1}}{m_{f'_1}+m_{f_2}}\,|\vec{q}\,|\vec{k}-\vec{p}\,)
\cdot(\frac{m_{f_2}}{m_{f'_1}+m_{f_2}}\,|\vec{q}\,|\vec{k}-\vec{p}\,)}{\widehat{E}_{f'_1}\widehat{E}_{f_1}}
\bigg)\nonumber\\ 
&&\hspace{5.cm}+\frac{m_{f_2}-m_{f'_1}}{m_{f'_1}+m_{f_2}}
\frac{p_x^2|\vec{q}\,|}{\widehat{E}_{f'_1}\widehat{E}_{f1}}
\bigg\}\nonumber\\
\end{eqnarray}

\begin{itemize}
\item {\large Case $J^\pi= 2^-$}
\end{itemize}

\begin{eqnarray}
V^{(2^-)\,0}_{T\lambda=0}(|\vec{q}\,|)&=&i\,\sqrt{\frac{15}{2}}
\sqrt{2m_I2E_F(-\vec{q}\,)}\ \int\,d^3p\ \frac{1}{4\pi|\vec{p}\,|^2}
\left(\hat{\phi}^{(M_F(2^-))}_{f'_1,\,f_2}(|\vec{p}\,|)\right)^*\,
\hat{\phi}^{(M_I(0^-))}_{f_1,\,f_2}\left(\bigg|\,\vec{p}-\frac{m_{f_2}}{m_{f'_1}+m_{f_2}}
|\vec{q}\,|
\vec{k} \bigg|\right)\nonumber\\
&&\hspace{3cm}\sqrt{\frac{\widehat{E}_{f'_1}\widehat{E}_{f1}}{4E_{f'_1}
E_{f_1}}}\
\frac{p_z\left(p_x^2+p_y^2\right)|\vec{q}\,|}
{\widehat{E}_{f'_1}\widehat{E}_{f1}}\nonumber\\
V^{(2^-)\,1}_{T\lambda=+1}(|\vec{q}\,|)&=&i\,\frac{\sqrt5}{2}
\sqrt{2m_I2E_F(-\vec{q}\,)}\ \int\,d^3p\ \frac{1}{4\pi|\vec{p}\,|^2}
\left(\hat{\phi}^{(M_F(2^-))}_{f'_1,\,f_2}(|\vec{p}\,|)\right)^*\,
\hat{\phi}^{(M_I(0^-))}_{f_1,\,f_2}\left(\bigg|\,\vec{p}-\frac{m_{f_2}}{m_{f'_1}+m_{f_2}}
|\vec{q}\,|
\vec{k} \bigg|\right)\nonumber\\
&&\hspace{3cm}\sqrt{\frac{\widehat{E}_{f'_1}\widehat{E}_{f1}}{4E_{f'_1}
E_{f_1}}}
\Bigg\{\left(p_z^2-p_x^2\right)
\left(\ \frac{-p_z-\frac{m_{f'_1}}{m_{f'_1}+m_{f_2}}|\vec{q}\,|}
{\widehat{E}_{f'_1}}
-\frac{-p_z+\frac{m_{f_2}}{m_{f'_1}+m_{f_2}}|\vec{q}\,|}
{\widehat{E}_{f_1}}
\right)\nonumber\\
&&\hspace{5cm}-p_zp_y^2\left(\frac{1}{\widehat{E}_{f'_1}}-\frac{1}{\widehat{E}_{f_1}}
\right)\Bigg\}\nonumber\\
V^{(2^-)\,3}_{T\lambda=0}(|\vec{q}\,|)&=&i\,\sqrt{\frac{15}{2}}
\sqrt{2m_I2E_F(-\vec{q}\,)}\ \int\,d^3p\ \frac{1}{4\pi|\vec{p}\,|^2}
\left(\hat{\phi}^{(M_F(2^-))}_{f'_1,\,f_2}(|\vec{p}\,|)\right)^*\,
\hat{\phi}^{(M_I(0^-))}_{f_1,\,f_2}\left(\bigg|\,\vec{p}-\frac{m_{f_2}}{m_{f'_1}+m_{f_2}}
|\vec{q}\,|
\vec{k} \bigg|\right)\nonumber\\
&&\hspace{3cm}\sqrt{\frac{\widehat{E}_{f'_1}\widehat{E}_{f1}}{4E_{f'_1}
E_{f_1}}}\ p_z\left(p_x^2+p_y^2\right)
\left(
\frac{1}
{\widehat{E}_{f'_1}}-\frac{1}
{\widehat{E}_{f_1}}
\right)\nonumber\\
\nonumber
\end{eqnarray}
\begin{eqnarray}
A^{(2^-)\,1}_{T\lambda=+1}(|\vec{q}\,|)&=&i\,\frac{\sqrt5}{2}
\sqrt{2m_I2E_F(-\vec{q}\,)}\ \int\,d^3p\ \frac{1}{4\pi|\vec{p}\,|^2}
\left(\hat{\phi}^{(M_F(2^-))}_{f'_1,\,f_2}(|\vec{p}\,|)\right)^*\,
\hat{\phi}^{(M_I(0^-))}_{f_1,\,f_2}\left(\bigg|\,\vec{p}-\frac{m_{f_2}}{m_{f'_1}+m_{f_2}}
|\vec{q}\,|
\vec{k} \bigg|\right)\nonumber\\
&&\hspace{3cm}\sqrt{\frac{\widehat{E}_{f'_1}\widehat{E}_{f1}}{4E_{f'_1}
E_{f_1}}}
\Bigg\{\left(p_z^2-p_y^2\right)
\left(1-
\frac{(-\frac{m_{f'_1}}{m_{f'_1}+m_{f_2}}\,|\vec{q}\,|\vec{k}-\vec{p}\,)
\cdot(\frac{m_{f_2}}{m_{f'_1}+m_{f_2}}\,|\vec{q}\,|\vec
k-\vec{p}\,)}{\widehat{E}_{f'_1}\widehat{E}_{f1}}
\right)
\nonumber\\
&&\hspace{5cm}-p_zp_x^2|\vec{q}\,|\,\frac{m_{f'_1}-m_{f_2}}{m_{f'_1}+m_{f_2}}
\,\frac{1}
{\widehat{E}_{f'_1}\widehat{E}_{f_1}}\ 
\Bigg\}\nonumber\\
\end{eqnarray}

\begin{itemize}
\item  {\large  Case $J^\pi= 2^+$}
\end{itemize}

\begin{eqnarray}
V^{(D_{s2}^*)\,1}_{T\lambda=+1}(|\vec{q}\,|)&=&i\,\frac{\sqrt3}{2}
\sqrt{2m_I2E_F(-\vec{q}\,)}\ \int\,d^3p\ \frac{1}{4\pi|\vec{p}\,|}
\left(\hat{\phi}^{(M_F(D_{s2}^*))}_{f'_1,\,f_2}(|\vec{p}\,|)\right)^*\,
\hat{\phi}^{(M_I(0^-))}_{f_1,\,f_2}\left(\bigg|\,\vec{p}-\frac{m_{f_2}}{m_{f'_1}+m_{f_2}}
|\vec{q}\,|
\vec{k} \bigg|\right)\nonumber\\
&&\hspace{3cm}\sqrt{\frac{\widehat{E}_{f'_1}\widehat{E}_{f1}}{4E_{f'_1}
E_{f_1}}}
\left(\frac{p_y^2-p_z^2-p_z|\vec{q}\,|\frac{m_{f'_1}}{m_{f'_1}+m_{f_2}}}
{\widehat{E}_{f'_1}}
-\frac{p_y^2-p_z^2+p_z|\vec{q}\,|\frac{m_{f_2}}{m_{f'_1}+m_{f_2}}}
{\widehat{E}_{f_1}}
\right)\nonumber\\
\nonumber
\end{eqnarray}
%
\begin{eqnarray}
A^{(D_{s2}^*)\,0}_{T\lambda=0}(|\vec{q}\,|)&=&\frac{-i}{\sqrt2}
\sqrt{2m_I2E_F(-\vec{q}\,)}\ \int\,d^3p\ \frac{1}{4\pi|\vec{p}\,|}
\left(\hat{\phi}^{(M_F(D_{s2}^*))}_{f'_1,\,f_2}(|\vec{p}\,|)\right)^*\,
\hat{\phi}^{(M_I(0^-))}_{f_1,\,f_2}\left(\bigg|\,\vec{p}-\frac{m_{f_2}}{m_{f'_1}+m_{f_2}}
|\vec{q}\,|
\vec{k} \bigg|\right)\nonumber\\
&&\hspace{2cm}\sqrt{\frac{\widehat{E}_{f'_1}\widehat{E}_{f1}}{4E_{f'_1}
E_{f_1}}}
\left(\frac{p_x^2+p_y^2-2p_z^2-2p_z|\vec{q}\,|\frac{m_{f'_1}}{m_{f'_1}+m_{f_2}}}
{\widehat{E}_{f'_1}}
+\frac{p_x^2+p_y^2-2p_z^2+2p_z|\vec{q}\,|\frac{m_{f_2}}{m_{f'_1}+m_{f_2}}}
{\widehat{E}_{f_1}}
\right)\nonumber\\
A^{(D_{s2}^*)\,1}_{T\lambda=+1}(|\vec{q}\,|)&=&i\,\frac{\sqrt3}{2}
\sqrt{2m_I2E_F(-\vec{q}\,)}\ \int\,d^3p\ \frac{1}{4\pi|\vec{p}\,|}
\left(\hat{\phi}^{(M_F(D_{s2}^*))}_{f'_1,\,f_2}(|\vec{p}\,|)\right)^*\,
\hat{\phi}^{(M_I(0^-))}_{f_1,\,f_2}\left(\bigg|\,\vec{p}-\frac{m_{f_2}}{m_{f'_1}+m_{f_2}}
|\vec{q}\,|
\vec{k} \bigg|\right)\nonumber\\
&&\hspace{3cm}\sqrt{\frac{\widehat{E}_{f'_1}\widehat{E}_{f1}}{4E_{f'_1}
E_{f_1}}}\Bigg\{\ \,p_z\,\left(1-
\frac{(-\frac{m_{f'_1}}{m_{f'_1}+m_{f_2}}\,|\vec{q}\,|\vec{k}-\vec{p}\,)
\cdot(\frac{m_{f_2}}{m_{f'_1}+m_{f_2}}\,|\vec{q}\,|\vec
k-\vec{p}\,)}{\widehat{E}_{f'_1}\widehat{E}_{f1}}
\right)\nonumber\\
&&\hspace{5.cm}
+\frac{4p_zp_x^2-p_x^2|\vec{q}\,|
\frac{m_{f_2}-m_{f'_1}}{m_{f'_1}+m_{f_2}}}
{\widehat{E}_{f'_1}\widehat{E}_{f1}}\Bigg\}\nonumber\\
A^{(D_{s2}^*)\,3}_{T\lambda=0}(|\vec{q}\,|)&=&-i\,\sqrt2
\sqrt{2m_I2E_F(-\vec{q}\,)}\ \int\,d^3p\ \frac{1}{4\pi|\vec{p}\,|}
\left(\hat{\phi}^{(M_F(D_{s2}^*))}_{f'_1,\,f_2}(|\vec{p}\,|)\right)^*\,
\hat{\phi}^{(M_I(0^-))}_{f_1,\,f_2}\left(\bigg|\,\vec{p}-\frac{m_{f_2}}{m_{f'_1}+m_{f_2}}
|\vec{q}\,|
\vec{k} \bigg|\right)\nonumber\\
&&\hspace{2cm}\sqrt{\frac{\widehat{E}_{f'_1}\widehat{E}_{f1}}{4E_{f'_1}
E_{f_1}}}\Bigg\{\ \,p_z\,\left(1-
\frac{(-\frac{m_{f'_1}}{m_{f'_1}+m_{f_2}}\,|\vec{q}\,|\vec{k}-\vec{p}\,)
\cdot(\frac{m_{f_2}}{m_{f'_1}+m_{f_2}}\,|\vec{q}\,|\vec
k-\vec{p}\,)}{\widehat{E}_{f'_1}\widehat{E}_{f1}}
\right)\nonumber\\
&&\hspace{4.cm}
+\frac{1}
{\widehat{E}_{f'_1}\widehat{E}_{f1}}
\Bigg[
 \ \,2p_z(-\frac{m_{f'_1}}{m_{f'_1}+m_{f_2}}\,|\vec{q}\,|-p_z)
\cdot (\frac{m_{f_2}}{m_{f'_1}+m_{f_2}}\,|\vec{q}\,|-p_z)\nonumber\\
&&\hspace{5.75cm}+\left(p_x^2+p_y^2\right)\bigg(
-p_z+\frac{m_{f_2}-m_{f'_1}}{2(m_{f'_1}+m_{f_2})}|\vec{q}\,|
\bigg)
\Bigg]
\ \Bigg\}\nonumber\\
\end{eqnarray}

\end{widetext}
\bibliography{biblist1}{}

\end{document}

%% file: ff.0m.0p.tex
\begingroup%
\makeatletter%
\newcommand{\GNUPLOTspecial}{%
  \@sanitize\catcode`\%=14\relax\special}%
\setlength{\unitlength}{0.0500bp}%
\begin{picture}(7200,5040)(0,0)%
  {\GNUPLOTspecial{"
/gnudict 256 dict def
gnudict begin
%
%
/Color false def
/Blacktext true def
/Solid false def
/Dashlength 1 def
/Landscape false def
/Level1 false def
/Rounded false def
/ClipToBoundingBox false def
/SuppressPDFMark false def
/TransparentPatterns false def
/gnulinewidth 5.000 def
/userlinewidth gnulinewidth def
/Gamma 1.0 def
/BackgroundColor {-1.000 -1.000 -1.000} def
/vshift -66 def
/dl1 {
  10.0 Dashlength mul mul
  Rounded { currentlinewidth 0.75 mul sub dup 0 le { pop 0.01 } if } if
} def
/dl2 {
  10.0 Dashlength mul mul
  Rounded { currentlinewidth 0.75 mul add } if
} def
/hpt_ 31.5 def
/vpt_ 31.5 def
/hpt hpt_ def
/vpt vpt_ def
/doclip {
  ClipToBoundingBox {
    newpath 0 0 moveto 360 0 lineto 360 252 lineto 0 252 lineto closepath
    clip
  } if
} def
%
%
%
/M {moveto} bind def
/L {lineto} bind def
/R {rmoveto} bind def
/V {rlineto} bind def
/N {newpath moveto} bind def
/Z {closepath} bind def
/C {setrgbcolor} bind def
/f {rlineto fill} bind def
/g {setgray} bind def
/Gshow {show} def   
/vpt2 vpt 2 mul def
/hpt2 hpt 2 mul def
/Lshow {currentpoint stroke M 0 vshift R 
	Blacktext {gsave 0 setgray show grestore} {show} ifelse} def
/Rshow {currentpoint stroke M dup stringwidth pop neg vshift R
	Blacktext {gsave 0 setgray show grestore} {show} ifelse} def
/Cshow {currentpoint stroke M dup stringwidth pop -2 div vshift R 
	Blacktext {gsave 0 setgray show grestore} {show} ifelse} def
/UP {dup vpt_ mul /vpt exch def hpt_ mul /hpt exch def
  /hpt2 hpt 2 mul def /vpt2 vpt 2 mul def} def
/DL {Color {setrgbcolor Solid {pop []} if 0 setdash}
 {pop pop pop 0 setgray Solid {pop []} if 0 setdash} ifelse} def
/BL {stroke userlinewidth 2 mul setlinewidth
	Rounded {1 setlinejoin 1 setlinecap} if} def
/AL {stroke userlinewidth 2 div setlinewidth
	Rounded {1 setlinejoin 1 setlinecap} if} def
/UL {dup gnulinewidth mul /userlinewidth exch def
	dup 1 lt {pop 1} if 10 mul /udl exch def} def
/PL {stroke userlinewidth setlinewidth
	Rounded {1 setlinejoin 1 setlinecap} if} def
3.8 setmiterlimit
/LCw {1 1 1} def
/LCb {0 0 0} def
/LCa {0 0 0} def
/LC0 {1 0 0} def
/LC1 {0 1 0} def
/LC2 {0 0 1} def
/LC3 {1 0 1} def
/LC4 {0 1 1} def
/LC5 {1 1 0} def
/LC6 {0 0 0} def
/LC7 {1 0.3 0} def
/LC8 {0.5 0.5 0.5} def
/LTw {PL [] 1 setgray} def
/LTb {BL [] LCb DL} def
/LTa {AL [1 udl mul 2 udl mul] 0 setdash LCa setrgbcolor} def
/LT0 {PL [] LC0 DL} def
/LT1 {PL [4 dl1 2 dl2] LC1 DL} def
/LT2 {PL [2 dl1 3 dl2] LC2 DL} def
/LT3 {PL [1 dl1 1.5 dl2] LC3 DL} def
/LT4 {PL [6 dl1 2 dl2 1 dl1 2 dl2] LC4 DL} def
/LT5 {PL [3 dl1 3 dl2 1 dl1 3 dl2] LC5 DL} def
/LT6 {PL [2 dl1 2 dl2 2 dl1 6 dl2] LC6 DL} def
/LT7 {PL [1 dl1 2 dl2 6 dl1 2 dl2 1 dl1 2 dl2] LC7 DL} def
/LT8 {PL [2 dl1 2 dl2 2 dl1 2 dl2 2 dl1 2 dl2 2 dl1 4 dl2] LC8 DL} def
/Pnt {stroke [] 0 setdash gsave 1 setlinecap M 0 0 V stroke grestore} def
/Dia {stroke [] 0 setdash 2 copy vpt add M
  hpt neg vpt neg V hpt vpt neg V
  hpt vpt V hpt neg vpt V closepath stroke
  Pnt} def
/Pls {stroke [] 0 setdash vpt sub M 0 vpt2 V
  currentpoint stroke M
  hpt neg vpt neg R hpt2 0 V stroke
 } def
/Box {stroke [] 0 setdash 2 copy exch hpt sub exch vpt add M
  0 vpt2 neg V hpt2 0 V 0 vpt2 V
  hpt2 neg 0 V closepath stroke
  Pnt} def
/Crs {stroke [] 0 setdash exch hpt sub exch vpt add M
  hpt2 vpt2 neg V currentpoint stroke M
  hpt2 neg 0 R hpt2 vpt2 V stroke} def
/TriU {stroke [] 0 setdash 2 copy vpt 1.12 mul add M
  hpt neg vpt -1.62 mul V
  hpt 2 mul 0 V
  hpt neg vpt 1.62 mul V closepath stroke
  Pnt} def
/Star {2 copy Pls Crs} def
/BoxF {stroke [] 0 setdash exch hpt sub exch vpt add M
  0 vpt2 neg V hpt2 0 V 0 vpt2 V
  hpt2 neg 0 V closepath fill} def
/TriUF {stroke [] 0 setdash vpt 1.12 mul add M
  hpt neg vpt -1.62 mul V
  hpt 2 mul 0 V
  hpt neg vpt 1.62 mul V closepath fill} def
/TriD {stroke [] 0 setdash 2 copy vpt 1.12 mul sub M
  hpt neg vpt 1.62 mul V
  hpt 2 mul 0 V
  hpt neg vpt -1.62 mul V closepath stroke
  Pnt} def
/TriDF {stroke [] 0 setdash vpt 1.12 mul sub M
  hpt neg vpt 1.62 mul V
  hpt 2 mul 0 V
  hpt neg vpt -1.62 mul V closepath fill} def
/DiaF {stroke [] 0 setdash vpt add M
  hpt neg vpt neg V hpt vpt neg V
  hpt vpt V hpt neg vpt V closepath fill} def
/Pent {stroke [] 0 setdash 2 copy gsave
  translate 0 hpt M 4 {72 rotate 0 hpt L} repeat
  closepath stroke grestore Pnt} def
/PentF {stroke [] 0 setdash gsave
  translate 0 hpt M 4 {72 rotate 0 hpt L} repeat
  closepath fill grestore} def
/Circle {stroke [] 0 setdash 2 copy
  hpt 0 360 arc stroke Pnt} def
/CircleF {stroke [] 0 setdash hpt 0 360 arc fill} def
/C0 {BL [] 0 setdash 2 copy moveto vpt 90 450 arc} bind def
/C1 {BL [] 0 setdash 2 copy moveto
	2 copy vpt 0 90 arc closepath fill
	vpt 0 360 arc closepath} bind def
/C2 {BL [] 0 setdash 2 copy moveto
	2 copy vpt 90 180 arc closepath fill
	vpt 0 360 arc closepath} bind def
/C3 {BL [] 0 setdash 2 copy moveto
	2 copy vpt 0 180 arc closepath fill
	vpt 0 360 arc closepath} bind def
/C4 {BL [] 0 setdash 2 copy moveto
	2 copy vpt 180 270 arc closepath fill
	vpt 0 360 arc closepath} bind def
/C5 {BL [] 0 setdash 2 copy moveto
	2 copy vpt 0 90 arc
	2 copy moveto
	2 copy vpt 180 270 arc closepath fill
	vpt 0 360 arc} bind def
/C6 {BL [] 0 setdash 2 copy moveto
	2 copy vpt 90 270 arc closepath fill
	vpt 0 360 arc closepath} bind def
/C7 {BL [] 0 setdash 2 copy moveto
	2 copy vpt 0 270 arc closepath fill
	vpt 0 360 arc closepath} bind def
/C8 {BL [] 0 setdash 2 copy moveto
	2 copy vpt 270 360 arc closepath fill
	vpt 0 360 arc closepath} bind def
/C9 {BL [] 0 setdash 2 copy moveto
	2 copy vpt 270 450 arc closepath fill
	vpt 0 360 arc closepath} bind def
/C10 {BL [] 0 setdash 2 copy 2 copy moveto vpt 270 360 arc closepath fill
	2 copy moveto
	2 copy vpt 90 180 arc closepath fill
	vpt 0 360 arc closepath} bind def
/C11 {BL [] 0 setdash 2 copy moveto
	2 copy vpt 0 180 arc closepath fill
	2 copy moveto
	2 copy vpt 270 360 arc closepath fill
	vpt 0 360 arc closepath} bind def
/C12 {BL [] 0 setdash 2 copy moveto
	2 copy vpt 180 360 arc closepath fill
	vpt 0 360 arc closepath} bind def
/C13 {BL [] 0 setdash 2 copy moveto
	2 copy vpt 0 90 arc closepath fill
	2 copy moveto
	2 copy vpt 180 360 arc closepath fill
	vpt 0 360 arc closepath} bind def
/C14 {BL [] 0 setdash 2 copy moveto
	2 copy vpt 90 360 arc closepath fill
	vpt 0 360 arc} bind def
/C15 {BL [] 0 setdash 2 copy vpt 0 360 arc closepath fill
	vpt 0 360 arc closepath} bind def
/Rec {newpath 4 2 roll moveto 1 index 0 rlineto 0 exch rlineto
	neg 0 rlineto closepath} bind def
/Square {dup Rec} bind def
/Bsquare {vpt sub exch vpt sub exch vpt2 Square} bind def
/S0 {BL [] 0 setdash 2 copy moveto 0 vpt rlineto BL Bsquare} bind def
/S1 {BL [] 0 setdash 2 copy vpt Square fill Bsquare} bind def
/S2 {BL [] 0 setdash 2 copy exch vpt sub exch vpt Square fill Bsquare} bind def
/S3 {BL [] 0 setdash 2 copy exch vpt sub exch vpt2 vpt Rec fill Bsquare} bind def
/S4 {BL [] 0 setdash 2 copy exch vpt sub exch vpt sub vpt Square fill Bsquare} bind def
/S5 {BL [] 0 setdash 2 copy 2 copy vpt Square fill
	exch vpt sub exch vpt sub vpt Square fill Bsquare} bind def
/S6 {BL [] 0 setdash 2 copy exch vpt sub exch vpt sub vpt vpt2 Rec fill Bsquare} bind def
/S7 {BL [] 0 setdash 2 copy exch vpt sub exch vpt sub vpt vpt2 Rec fill
	2 copy vpt Square fill Bsquare} bind def
/S8 {BL [] 0 setdash 2 copy vpt sub vpt Square fill Bsquare} bind def
/S9 {BL [] 0 setdash 2 copy vpt sub vpt vpt2 Rec fill Bsquare} bind def
/S10 {BL [] 0 setdash 2 copy vpt sub vpt Square fill 2 copy exch vpt sub exch vpt Square fill
	Bsquare} bind def
/S11 {BL [] 0 setdash 2 copy vpt sub vpt Square fill 2 copy exch vpt sub exch vpt2 vpt Rec fill
	Bsquare} bind def
/S12 {BL [] 0 setdash 2 copy exch vpt sub exch vpt sub vpt2 vpt Rec fill Bsquare} bind def
/S13 {BL [] 0 setdash 2 copy exch vpt sub exch vpt sub vpt2 vpt Rec fill
	2 copy vpt Square fill Bsquare} bind def
/S14 {BL [] 0 setdash 2 copy exch vpt sub exch vpt sub vpt2 vpt Rec fill
	2 copy exch vpt sub exch vpt Square fill Bsquare} bind def
/S15 {BL [] 0 setdash 2 copy Bsquare fill Bsquare} bind def
/D0 {gsave translate 45 rotate 0 0 S0 stroke grestore} bind def
/D1 {gsave translate 45 rotate 0 0 S1 stroke grestore} bind def
/D2 {gsave translate 45 rotate 0 0 S2 stroke grestore} bind def
/D3 {gsave translate 45 rotate 0 0 S3 stroke grestore} bind def
/D4 {gsave translate 45 rotate 0 0 S4 stroke grestore} bind def
/D5 {gsave translate 45 rotate 0 0 S5 stroke grestore} bind def
/D6 {gsave translate 45 rotate 0 0 S6 stroke grestore} bind def
/D7 {gsave translate 45 rotate 0 0 S7 stroke grestore} bind def
/D8 {gsave translate 45 rotate 0 0 S8 stroke grestore} bind def
/D9 {gsave translate 45 rotate 0 0 S9 stroke grestore} bind def
/D10 {gsave translate 45 rotate 0 0 S10 stroke grestore} bind def
/D11 {gsave translate 45 rotate 0 0 S11 stroke grestore} bind def
/D12 {gsave translate 45 rotate 0 0 S12 stroke grestore} bind def
/D13 {gsave translate 45 rotate 0 0 S13 stroke grestore} bind def
/D14 {gsave translate 45 rotate 0 0 S14 stroke grestore} bind def
/D15 {gsave translate 45 rotate 0 0 S15 stroke grestore} bind def
/DiaE {stroke [] 0 setdash vpt add M
  hpt neg vpt neg V hpt vpt neg V
  hpt vpt V hpt neg vpt V closepath stroke} def
/BoxE {stroke [] 0 setdash exch hpt sub exch vpt add M
  0 vpt2 neg V hpt2 0 V 0 vpt2 V
  hpt2 neg 0 V closepath stroke} def
/TriUE {stroke [] 0 setdash vpt 1.12 mul add M
  hpt neg vpt -1.62 mul V
  hpt 2 mul 0 V
  hpt neg vpt 1.62 mul V closepath stroke} def
/TriDE {stroke [] 0 setdash vpt 1.12 mul sub M
  hpt neg vpt 1.62 mul V
  hpt 2 mul 0 V
  hpt neg vpt -1.62 mul V closepath stroke} def
/PentE {stroke [] 0 setdash gsave
  translate 0 hpt M 4 {72 rotate 0 hpt L} repeat
  closepath stroke grestore} def
/CircE {stroke [] 0 setdash 
  hpt 0 360 arc stroke} def
/Opaque {gsave closepath 1 setgray fill grestore 0 setgray closepath} def
/DiaW {stroke [] 0 setdash vpt add M
  hpt neg vpt neg V hpt vpt neg V
  hpt vpt V hpt neg vpt V Opaque stroke} def
/BoxW {stroke [] 0 setdash exch hpt sub exch vpt add M
  0 vpt2 neg V hpt2 0 V 0 vpt2 V
  hpt2 neg 0 V Opaque stroke} def
/TriUW {stroke [] 0 setdash vpt 1.12 mul add M
  hpt neg vpt -1.62 mul V
  hpt 2 mul 0 V
  hpt neg vpt 1.62 mul V Opaque stroke} def
/TriDW {stroke [] 0 setdash vpt 1.12 mul sub M
  hpt neg vpt 1.62 mul V
  hpt 2 mul 0 V
  hpt neg vpt -1.62 mul V Opaque stroke} def
/PentW {stroke [] 0 setdash gsave
  translate 0 hpt M 4 {72 rotate 0 hpt L} repeat
  Opaque stroke grestore} def
/CircW {stroke [] 0 setdash 
  hpt 0 360 arc Opaque stroke} def
/BoxFill {gsave Rec 1 setgray fill grestore} def
/Density {
  /Fillden exch def
  currentrgbcolor
  /ColB exch def /ColG exch def /ColR exch def
  /ColR ColR Fillden mul Fillden sub 1 add def
  /ColG ColG Fillden mul Fillden sub 1 add def
  /ColB ColB Fillden mul Fillden sub 1 add def
  ColR ColG ColB setrgbcolor} def
/BoxColFill {gsave Rec PolyFill} def
/PolyFill {gsave Density fill grestore grestore} def
/h {rlineto rlineto rlineto gsave closepath fill grestore} bind def
%
%
/PatternFill {gsave /PFa [ 9 2 roll ] def
  PFa 0 get PFa 2 get 2 div add PFa 1 get PFa 3 get 2 div add translate
  PFa 2 get -2 div PFa 3 get -2 div PFa 2 get PFa 3 get Rec
  gsave 1 setgray fill grestore clip
  currentlinewidth 0.5 mul setlinewidth
  /PFs PFa 2 get dup mul PFa 3 get dup mul add sqrt def
  0 0 M PFa 5 get rotate PFs -2 div dup translate
  0 1 PFs PFa 4 get div 1 add floor cvi
	{PFa 4 get mul 0 M 0 PFs V} for
  0 PFa 6 get ne {
	0 1 PFs PFa 4 get div 1 add floor cvi
	{PFa 4 get mul 0 2 1 roll M PFs 0 V} for
 } if
  stroke grestore} def
/languagelevel where
 {pop languagelevel} {1} ifelse
 2 lt
	{/InterpretLevel1 true def}
	{/InterpretLevel1 Level1 def}
 ifelse
%
%
/Level2PatternFill {
/Tile8x8 {/PaintType 2 /PatternType 1 /TilingType 1 /BBox [0 0 8 8] /XStep 8 /YStep 8}
	bind def
/KeepColor {currentrgbcolor [/Pattern /DeviceRGB] setcolorspace} bind def
<< Tile8x8
 /PaintProc {0.5 setlinewidth pop 0 0 M 8 8 L 0 8 M 8 0 L stroke} 
>> matrix makepattern
/Pat1 exch def
<< Tile8x8
 /PaintProc {0.5 setlinewidth pop 0 0 M 8 8 L 0 8 M 8 0 L stroke
	0 4 M 4 8 L 8 4 L 4 0 L 0 4 L stroke}
>> matrix makepattern
/Pat2 exch def
<< Tile8x8
 /PaintProc {0.5 setlinewidth pop 0 0 M 0 8 L
	8 8 L 8 0 L 0 0 L fill}
>> matrix makepattern
/Pat3 exch def
<< Tile8x8
 /PaintProc {0.5 setlinewidth pop -4 8 M 8 -4 L
	0 12 M 12 0 L stroke}
>> matrix makepattern
/Pat4 exch def
<< Tile8x8
 /PaintProc {0.5 setlinewidth pop -4 0 M 8 12 L
	0 -4 M 12 8 L stroke}
>> matrix makepattern
/Pat5 exch def
<< Tile8x8
 /PaintProc {0.5 setlinewidth pop -2 8 M 4 -4 L
	0 12 M 8 -4 L 4 12 M 10 0 L stroke}
>> matrix makepattern
/Pat6 exch def
<< Tile8x8
 /PaintProc {0.5 setlinewidth pop -2 0 M 4 12 L
	0 -4 M 8 12 L 4 -4 M 10 8 L stroke}
>> matrix makepattern
/Pat7 exch def
<< Tile8x8
 /PaintProc {0.5 setlinewidth pop 8 -2 M -4 4 L
	12 0 M -4 8 L 12 4 M 0 10 L stroke}
>> matrix makepattern
/Pat8 exch def
<< Tile8x8
 /PaintProc {0.5 setlinewidth pop 0 -2 M 12 4 L
	-4 0 M 12 8 L -4 4 M 8 10 L stroke}
>> matrix makepattern
/Pat9 exch def
/Pattern1 {PatternBgnd KeepColor Pat1 setpattern} bind def
/Pattern2 {PatternBgnd KeepColor Pat2 setpattern} bind def
/Pattern3 {PatternBgnd KeepColor Pat3 setpattern} bind def
/Pattern4 {PatternBgnd KeepColor Landscape {Pat5} {Pat4} ifelse setpattern} bind def
/Pattern5 {PatternBgnd KeepColor Landscape {Pat4} {Pat5} ifelse setpattern} bind def
/Pattern6 {PatternBgnd KeepColor Landscape {Pat9} {Pat6} ifelse setpattern} bind def
/Pattern7 {PatternBgnd KeepColor Landscape {Pat8} {Pat7} ifelse setpattern} bind def
} def
%
%
%
/PatternBgnd {
  TransparentPatterns {} {gsave 1 setgray fill grestore} ifelse
} def
%
%
/Level1PatternFill {
/Pattern1 {0.250 Density} bind def
/Pattern2 {0.500 Density} bind def
/Pattern3 {0.750 Density} bind def
/Pattern4 {0.125 Density} bind def
/Pattern5 {0.375 Density} bind def
/Pattern6 {0.625 Density} bind def
/Pattern7 {0.875 Density} bind def
} def
%
%
Level1 {Level1PatternFill} {Level2PatternFill} ifelse
/Symbol-Oblique /Symbol findfont [1 0 .167 1 0 0] makefont
dup length dict begin {1 index /FID eq {pop pop} {def} ifelse} forall
currentdict end definefont pop
Level1 SuppressPDFMark or 
{} {
/SDict 10 dict def
systemdict /pdfmark known not {
  userdict /pdfmark systemdict /cleartomark get put
} if
SDict begin [
  /Title (paper/ff.0m.0p.tex)
  /Subject (gnuplot plot)
  /Creator (gnuplot 4.6 patchlevel 0)
  /Author (conrado)
  /CreationDate (Sat Oct 12 12:34:16 2013)
  /DOCINFO pdfmark
end
} ifelse
end
gnudict begin
gsave
doclip
0 0 translate
0.050 0.050 scale
0 setgray
newpath
BackgroundColor 0 lt 3 1 roll 0 lt exch 0 lt or or not {BackgroundColor C 1.000 0 0 7200.00 5040.00 BoxColFill} if
1.000 UL
LTb
0 640 M
63 0 V
2816 0 R
-63 0 V
0 1198 M
63 0 V
2816 0 R
-63 0 V
0 1756 M
63 0 V
2816 0 R
-63 0 V
0 2315 M
63 0 V
2816 0 R
-63 0 V
0 2873 M
63 0 V
2816 0 R
-63 0 V
0 3431 M
63 0 V
2816 0 R
-63 0 V
0 640 M
0 63 V
0 3431 M
0 -63 V
480 640 M
0 63 V
0 2728 R
0 -63 V
960 640 M
0 63 V
0 2728 R
0 -63 V
1440 640 M
0 63 V
0 2728 R
0 -63 V
1919 640 M
0 63 V
0 2728 R
0 -63 V
2399 640 M
0 63 V
0 2728 R
0 -63 V
2879 640 M
0 63 V
0 2728 R
0 -63 V
stroke
0 3431 N
0 640 L
2879 0 V
0 2791 V
0 3431 L
Z stroke
LCb setrgbcolor
LTb
1.000 UP
1.000 UL
LTb
1.000 UL
LT0
LCb setrgbcolor
LT0
1928 2056 M
543 0 V
0 2038 M
2 1 V
4 1 V
5 2 V
7 2 V
7 2 V
9 3 V
9 3 V
10 3 V
11 4 V
11 3 V
10 4 V
10 3 V
10 3 V
8 3 V
8 2 V
6 3 V
5 1 V
4 1 V
2 1 V
1 0 V
2 1 V
4 1 V
5 2 V
6 2 V
8 3 V
8 2 V
10 4 V
10 3 V
10 3 V
11 4 V
11 3 V
10 4 V
9 3 V
9 3 V
7 2 V
7 2 V
5 2 V
4 1 V
2 1 V
1 0 V
2 1 V
3 1 V
5 2 V
7 2 V
7 2 V
9 3 V
9 3 V
11 4 V
10 3 V
11 4 V
10 3 V
10 4 V
10 3 V
8 3 V
8 2 V
6 2 V
6 2 V
3 1 V
2 1 V
1 0 V
2 1 V
4 1 V
5 2 V
6 2 V
8 3 V
8 3 V
10 3 V
10 3 V
11 4 V
10 3 V
11 4 V
10 3 V
9 3 V
9 3 V
8 3 V
6 2 V
5 2 V
4 1 V
2 1 V
1 0 V
2 1 V
3 1 V
5 2 V
7 2 V
7 2 V
9 3 V
10 4 V
10 3 V
10 4 V
11 3 V
10 4 V
10 3 V
10 3 V
8 3 V
8 3 V
7 2 V
5 2 V
3 1 V
2 1 V
1 0 V
2 1 V
4 1 V
stroke 699 2271 M
5 2 V
6 2 V
8 2 V
9 3 V
9 4 V
10 3 V
11 4 V
10 3 V
11 4 V
10 3 V
9 4 V
9 3 V
8 2 V
6 2 V
5 2 V
4 1 V
2 1 V
1 0 V
2 1 V
3 1 V
5 2 V
7 2 V
7 3 V
9 3 V
10 3 V
10 4 V
10 3 V
11 4 V
10 4 V
10 3 V
10 3 V
9 3 V
7 3 V
7 2 V
5 2 V
3 1 V
2 1 V
1 0 V
2 1 V
4 1 V
5 2 V
6 2 V
8 3 V
9 3 V
9 3 V
10 3 V
11 4 V
10 4 V
11 3 V
10 4 V
9 3 V
9 3 V
8 3 V
6 2 V
5 2 V
4 1 V
2 1 V
1 0 V
2 1 V
3 1 V
5 2 V
7 2 V
7 3 V
9 3 V
10 3 V
10 3 V
10 4 V
11 4 V
10 3 V
10 4 V
10 3 V
9 3 V
7 3 V
7 2 V
5 2 V
3 1 V
2 1 V
1 0 V
2 1 V
4 1 V
5 2 V
6 2 V
8 3 V
9 3 V
9 3 V
10 4 V
11 3 V
10 4 V
11 4 V
10 3 V
9 4 V
9 3 V
8 2 V
6 2 V
5 2 V
4 1 V
2 1 V
1 1 V
2 0 V
3 1 V
5 2 V
7 2 V
8 3 V
8 3 V
stroke 1419 2520 M
10 3 V
10 4 V
10 4 V
11 3 V
10 4 V
10 4 V
10 3 V
9 3 V
7 3 V
7 2 V
5 2 V
3 1 V
2 1 V
1 0 V
2 1 V
4 1 V
5 2 V
6 2 V
8 3 V
9 3 V
9 3 V
10 3 V
11 4 V
10 4 V
11 3 V
10 4 V
9 3 V
9 3 V
8 3 V
6 2 V
5 2 V
4 1 V
2 1 V
1 0 V
2 1 V
3 1 V
6 2 V
6 2 V
8 3 V
8 3 V
10 3 V
10 4 V
10 3 V
11 4 V
10 4 V
11 3 V
9 4 V
9 3 V
7 2 V
7 3 V
5 1 V
3 2 V
2 0 V
1 1 V
2 0 V
4 2 V
5 1 V
6 3 V
8 2 V
9 3 V
9 4 V
10 3 V
11 4 V
10 3 V
11 4 V
10 4 V
10 3 V
8 3 V
8 3 V
6 2 V
5 2 V
4 1 V
2 1 V
1 0 V
2 1 V
4 1 V
5 2 V
6 2 V
8 3 V
8 3 V
10 3 V
10 3 V
10 4 V
11 4 V
10 3 V
11 4 V
9 3 V
9 3 V
7 3 V
7 2 V
5 2 V
3 1 V
2 1 V
1 0 V
2 1 V
4 1 V
5 2 V
7 2 V
7 3 V
9 2 V
9 4 V
10 3 V
11 4 V
11 4 V
stroke 2153 2776 M
10 3 V
10 4 V
10 3 V
8 3 V
8 2 V
6 3 V
5 1 V
4 2 V
2 0 V
1 1 V
2 0 V
4 2 V
5 1 V
6 3 V
8 2 V
8 3 V
10 3 V
10 4 V
10 3 V
11 4 V
11 4 V
10 3 V
9 3 V
9 3 V
7 3 V
7 2 V
5 2 V
4 1 V
2 1 V
1 0 V
2 1 V
3 1 V
5 2 V
7 2 V
7 2 V
9 3 V
9 4 V
11 3 V
10 4 V
11 3 V
10 4 V
10 3 V
10 3 V
8 3 V
8 3 V
6 2 V
5 2 V
4 1 V
2 1 V
1 0 V
2 1 V
4 1 V
5 1 V
6 3 V
8 2 V
8 3 V
10 3 V
10 4 V
11 3 V
10 4 V
11 3 V
10 4 V
9 3 V
9 3 V
8 2 V
6 2 V
5 2 V
4 1 V
2 1 V
1 0 V
2 1 V
3 1 V
5 2 V
7 2 V
7 2 V
9 3 V
9 3 V
11 4 V
10 3 V
11 4 V
10 3 V
10 3 V
10 3 V
8 3 V
8 3 V
6 2 V
6 1 V
3 2 V
2 0 V
stroke
LT1
LCb setrgbcolor
LT1
1928 1856 M
543 0 V
0 757 M
2 0 V
4 1 V
5 0 V
7 1 V
7 1 V
9 1 V
9 1 V
10 1 V
11 1 V
11 1 V
10 1 V
10 1 V
10 1 V
8 1 V
8 1 V
6 1 V
5 0 V
4 1 V
2 0 V
1 0 V
2 0 V
4 1 V
5 0 V
6 1 V
8 1 V
8 1 V
10 1 V
10 1 V
10 1 V
11 1 V
11 1 V
10 1 V
9 1 V
9 1 V
7 1 V
7 1 V
5 0 V
4 1 V
2 0 V
1 0 V
2 0 V
3 1 V
5 0 V
7 1 V
7 1 V
9 1 V
9 1 V
11 1 V
10 1 V
11 1 V
10 1 V
10 1 V
10 1 V
8 1 V
8 1 V
6 1 V
6 0 V
3 1 V
2 0 V
1 0 V
2 0 V
4 1 V
5 0 V
6 1 V
8 1 V
8 1 V
10 1 V
10 1 V
11 1 V
10 1 V
11 1 V
10 2 V
9 1 V
9 1 V
8 0 V
6 1 V
5 1 V
4 0 V
2 0 V
1 0 V
2 1 V
3 0 V
5 1 V
7 0 V
7 1 V
9 1 V
10 1 V
10 1 V
10 1 V
11 2 V
10 1 V
10 1 V
10 1 V
8 1 V
8 1 V
7 0 V
5 1 V
3 0 V
2 0 V
1 1 V
2 0 V
4 0 V
stroke 699 833 M
5 1 V
6 0 V
8 1 V
9 1 V
9 1 V
10 1 V
11 2 V
10 1 V
11 1 V
10 1 V
9 1 V
9 1 V
8 1 V
6 0 V
5 1 V
4 0 V
2 1 V
1 0 V
2 0 V
3 0 V
5 1 V
7 1 V
7 0 V
9 1 V
10 1 V
10 2 V
10 1 V
11 1 V
10 1 V
10 1 V
10 1 V
9 1 V
7 1 V
7 1 V
5 0 V
3 1 V
2 0 V
1 0 V
2 0 V
4 0 V
5 1 V
6 1 V
8 1 V
9 1 V
9 1 V
10 1 V
11 1 V
10 1 V
11 1 V
10 1 V
9 1 V
9 1 V
8 1 V
6 1 V
5 0 V
4 1 V
2 0 V
1 0 V
2 0 V
3 1 V
5 0 V
7 1 V
7 1 V
9 1 V
10 1 V
10 1 V
10 1 V
11 1 V
10 1 V
10 1 V
10 1 V
9 1 V
7 1 V
7 1 V
5 0 V
3 1 V
2 0 V
1 0 V
2 0 V
4 0 V
5 1 V
6 1 V
8 1 V
9 0 V
9 1 V
10 2 V
11 1 V
10 1 V
11 1 V
10 1 V
9 1 V
9 1 V
8 1 V
6 0 V
5 1 V
4 0 V
2 1 V
1 0 V
2 0 V
3 0 V
5 1 V
7 1 V
8 0 V
8 1 V
stroke 1419 911 M
10 1 V
10 1 V
10 1 V
11 2 V
10 1 V
10 1 V
10 1 V
9 1 V
7 0 V
7 1 V
5 1 V
3 0 V
2 0 V
1 0 V
2 1 V
4 0 V
5 0 V
6 1 V
8 1 V
9 1 V
9 1 V
10 1 V
11 1 V
10 1 V
11 1 V
10 1 V
9 1 V
9 1 V
8 1 V
6 1 V
5 0 V
4 0 V
2 1 V
1 0 V
2 0 V
3 0 V
6 1 V
6 1 V
8 0 V
8 1 V
10 1 V
10 1 V
10 1 V
11 1 V
10 1 V
11 1 V
9 1 V
9 1 V
7 1 V
7 1 V
5 0 V
3 1 V
2 0 V
1 0 V
2 0 V
4 1 V
5 0 V
6 1 V
8 0 V
9 1 V
9 1 V
10 1 V
11 1 V
10 1 V
11 1 V
10 1 V
10 1 V
8 1 V
8 1 V
6 1 V
5 0 V
4 1 V
2 0 V
1 0 V
2 0 V
4 0 V
5 1 V
6 1 V
8 0 V
8 1 V
10 1 V
10 1 V
10 1 V
11 1 V
10 1 V
11 1 V
9 1 V
9 1 V
7 1 V
7 0 V
5 1 V
3 0 V
2 0 V
1 0 V
2 1 V
4 0 V
5 0 V
7 1 V
7 1 V
9 1 V
9 1 V
10 0 V
11 1 V
11 1 V
stroke 2153 985 M
10 1 V
10 1 V
10 1 V
8 1 V
8 1 V
6 0 V
5 1 V
4 0 V
2 0 V
1 1 V
2 0 V
4 0 V
5 1 V
6 0 V
8 1 V
8 1 V
10 1 V
10 0 V
10 1 V
11 1 V
11 1 V
10 1 V
9 1 V
9 1 V
7 1 V
7 0 V
5 1 V
4 0 V
2 0 V
1 0 V
2 0 V
3 1 V
5 0 V
7 1 V
7 0 V
9 1 V
9 1 V
11 1 V
10 1 V
11 1 V
10 1 V
10 1 V
10 1 V
8 0 V
8 1 V
6 0 V
5 1 V
4 0 V
2 0 V
1 0 V
2 1 V
4 0 V
5 0 V
6 1 V
8 1 V
8 0 V
10 1 V
10 1 V
11 1 V
10 1 V
11 1 V
10 0 V
9 1 V
9 1 V
8 1 V
6 0 V
5 1 V
4 0 V
2 0 V
1 0 V
2 0 V
3 1 V
5 0 V
7 1 V
7 0 V
9 1 V
9 1 V
11 0 V
10 1 V
11 1 V
10 1 V
10 1 V
10 0 V
8 1 V
8 1 V
6 0 V
6 1 V
3 0 V
2 0 V
stroke
LTb
0 3431 N
0 640 L
2879 0 V
0 2791 V
0 3431 L
Z stroke
1.000 UP
1.000 UL
LTb
1.000 UL
LTb
4320 640 M
63 0 V
2816 0 R
-63 0 V
4320 1338 M
63 0 V
2816 0 R
-63 0 V
4320 2036 M
63 0 V
2816 0 R
-63 0 V
4320 2733 M
63 0 V
2816 0 R
-63 0 V
4320 3431 M
63 0 V
2816 0 R
-63 0 V
4320 640 M
0 63 V
0 2728 R
0 -63 V
4608 640 M
0 63 V
0 2728 R
0 -63 V
4896 640 M
0 63 V
0 2728 R
0 -63 V
5184 640 M
0 63 V
0 2728 R
0 -63 V
5472 640 M
0 63 V
0 2728 R
0 -63 V
5760 640 M
0 63 V
0 2728 R
0 -63 V
6047 640 M
0 63 V
0 2728 R
0 -63 V
6335 640 M
0 63 V
0 2728 R
0 -63 V
6623 640 M
0 63 V
0 2728 R
0 -63 V
6911 640 M
0 63 V
0 2728 R
0 -63 V
7199 640 M
0 63 V
0 2728 R
0 -63 V
stroke
4320 3431 N
0 -2791 V
2879 0 V
0 2791 V
-2879 0 V
Z stroke
LCb setrgbcolor
LTb
1.000 UP
1.000 UL
LTb
1.000 UL
LT0
LCb setrgbcolor
LT0
6248 2280 M
543 0 V
4320 1174 M
2 1 V
4 0 V
5 1 V
6 1 V
7 2 V
9 1 V
9 2 V
10 2 V
10 2 V
10 2 V
10 2 V
10 1 V
9 2 V
8 2 V
8 1 V
6 1 V
5 1 V
3 1 V
2 0 V
1 0 V
2 1 V
4 1 V
5 0 V
6 2 V
7 1 V
9 2 V
9 1 V
9 2 V
11 2 V
10 2 V
10 2 V
10 2 V
9 2 V
8 1 V
8 2 V
6 1 V
5 1 V
3 0 V
2 1 V
1 0 V
2 0 V
4 1 V
4 1 V
7 1 V
7 1 V
8 2 V
10 2 V
9 2 V
10 2 V
11 1 V
10 2 V
10 2 V
9 2 V
8 2 V
7 1 V
7 1 V
5 1 V
3 1 V
2 0 V
1 0 V
2 1 V
3 0 V
5 1 V
6 1 V
8 2 V
8 1 V
9 2 V
10 2 V
10 2 V
10 2 V
11 2 V
9 2 V
10 2 V
8 1 V
7 2 V
7 1 V
4 1 V
4 0 V
2 1 V
1 0 V
2 0 V
3 1 V
5 1 V
6 1 V
8 2 V
8 1 V
9 2 V
10 2 V
10 2 V
10 2 V
10 2 V
10 2 V
9 1 V
9 2 V
7 1 V
6 2 V
5 0 V
4 1 V
2 1 V
1 0 V
2 0 V
3 1 V
stroke 4995 1303 M
5 1 V
6 1 V
7 1 V
9 2 V
9 2 V
10 2 V
10 2 V
10 2 V
10 1 V
10 2 V
9 2 V
9 2 V
7 1 V
6 1 V
5 1 V
4 1 V
1 0 V
1 1 V
2 0 V
4 1 V
5 1 V
6 1 V
7 1 V
9 2 V
9 2 V
10 1 V
10 2 V
10 2 V
10 2 V
10 2 V
9 2 V
8 2 V
8 1 V
6 1 V
5 1 V
3 1 V
2 0 V
1 0 V
2 1 V
4 1 V
4 0 V
7 2 V
7 1 V
8 2 V
10 2 V
9 1 V
11 2 V
10 2 V
10 2 V
10 2 V
9 2 V
8 2 V
8 1 V
6 1 V
5 1 V
3 1 V
2 0 V
1 0 V
2 1 V
3 0 V
5 1 V
7 2 V
7 1 V
8 2 V
9 1 V
10 2 V
10 2 V
11 2 V
10 2 V
9 2 V
10 2 V
8 2 V
7 1 V
7 1 V
4 1 V
4 1 V
2 0 V
1 0 V
2 1 V
3 1 V
5 0 V
6 2 V
8 1 V
8 2 V
9 2 V
10 1 V
10 2 V
10 2 V
11 2 V
9 2 V
9 2 V
9 2 V
7 1 V
6 1 V
5 1 V
4 1 V
2 0 V
1 1 V
2 0 V
3 1 V
5 1 V
6 1 V
8 1 V
8 2 V
stroke 5691 1438 M
9 2 V
10 1 V
10 2 V
10 2 V
10 2 V
10 2 V
9 2 V
9 2 V
7 1 V
6 1 V
5 1 V
4 1 V
2 0 V
0 1 V
2 0 V
4 1 V
5 1 V
6 1 V
7 1 V
9 2 V
9 2 V
10 2 V
10 1 V
10 2 V
10 2 V
10 2 V
9 2 V
8 2 V
8 1 V
6 1 V
5 1 V
3 1 V
2 0 V
1 1 V
2 0 V
4 1 V
5 1 V
6 1 V
7 1 V
9 2 V
9 2 V
9 2 V
11 2 V
10 2 V
10 1 V
10 2 V
9 2 V
8 2 V
8 1 V
6 1 V
5 1 V
3 1 V
2 0 V
1 1 V
2 0 V
4 1 V
4 1 V
7 1 V
7 1 V
8 2 V
10 2 V
9 2 V
10 1 V
11 2 V
10 2 V
10 2 V
9 2 V
8 2 V
7 1 V
7 1 V
5 1 V
3 1 V
2 0 V
1 1 V
2 0 V
3 1 V
5 1 V
6 1 V
8 1 V
8 2 V
9 2 V
10 1 V
10 2 V
10 2 V
11 2 V
9 2 V
10 2 V
8 2 V
7 1 V
7 1 V
4 1 V
4 1 V
2 0 V
1 0 V
2 1 V
3 0 V
5 1 V
6 2 V
8 1 V
8 2 V
9 1 V
10 2 V
10 2 V
10 2 V
stroke 6399 1575 M
10 2 V
10 2 V
9 2 V
9 1 V
7 2 V
6 1 V
5 1 V
4 1 V
2 0 V
1 0 V
2 0 V
3 1 V
5 1 V
6 1 V
7 2 V
9 1 V
9 2 V
10 2 V
10 2 V
10 2 V
10 2 V
10 2 V
9 1 V
9 2 V
7 1 V
6 1 V
5 1 V
4 1 V
1 0 V
1 1 V
2 0 V
4 1 V
5 1 V
6 1 V
7 1 V
9 2 V
9 2 V
10 1 V
10 2 V
10 2 V
10 2 V
10 2 V
9 2 V
8 1 V
8 2 V
6 1 V
5 1 V
3 0 V
2 1 V
1 0 V
2 0 V
4 1 V
5 1 V
6 1 V
7 2 V
8 1 V
10 2 V
9 2 V
11 2 V
10 2 V
10 1 V
10 2 V
9 2 V
8 2 V
8 1 V
6 1 V
5 1 V
3 1 V
2 0 V
1 0 V
2 1 V
3 0 V
5 1 V
7 1 V
7 2 V
8 1 V
9 2 V
10 2 V
10 2 V
11 2 V
10 2 V
9 1 V
10 2 V
8 2 V
7 1 V
7 1 V
5 1 V
3 1 V
2 0 V
stroke
LT1
LCb setrgbcolor
LT1
6248 2080 M
543 0 V
-2471 33 R
2 1 V
4 1 V
5 1 V
6 2 V
7 2 V
9 3 V
9 2 V
10 3 V
10 3 V
10 3 V
10 3 V
10 3 V
9 2 V
8 3 V
8 2 V
6 1 V
5 2 V
3 1 V
2 0 V
1 1 V
2 0 V
4 1 V
5 2 V
6 1 V
7 2 V
9 3 V
9 2 V
9 3 V
11 3 V
10 3 V
10 3 V
10 3 V
9 2 V
8 3 V
8 2 V
6 2 V
5 1 V
3 1 V
2 1 V
1 0 V
2 0 V
4 1 V
4 2 V
7 1 V
7 3 V
8 2 V
10 3 V
9 2 V
10 3 V
11 3 V
10 3 V
10 3 V
9 2 V
8 3 V
7 2 V
7 2 V
5 1 V
3 1 V
2 0 V
1 1 V
2 0 V
3 1 V
5 2 V
6 1 V
8 2 V
8 3 V
9 2 V
10 3 V
10 3 V
10 3 V
11 3 V
9 3 V
10 2 V
8 3 V
7 2 V
7 1 V
4 2 V
4 1 V
2 0 V
1 1 V
2 0 V
3 1 V
5 1 V
6 2 V
8 2 V
8 3 V
9 2 V
10 3 V
10 3 V
10 3 V
10 2 V
10 3 V
9 3 V
9 2 V
7 2 V
6 2 V
5 1 V
4 1 V
2 1 V
1 0 V
2 1 V
3 1 V
stroke 4995 2305 M
5 1 V
6 2 V
7 2 V
9 2 V
9 3 V
10 2 V
10 3 V
10 3 V
10 3 V
10 3 V
9 2 V
9 2 V
7 2 V
6 2 V
5 2 V
4 0 V
1 1 V
1 0 V
2 1 V
4 1 V
5 1 V
6 2 V
7 2 V
9 2 V
9 3 V
10 2 V
10 3 V
10 3 V
10 3 V
10 3 V
9 2 V
8 2 V
8 2 V
6 2 V
5 1 V
3 1 V
2 1 V
1 0 V
2 1 V
4 1 V
4 1 V
7 2 V
7 2 V
8 2 V
10 3 V
9 2 V
11 3 V
10 3 V
10 3 V
10 2 V
9 3 V
8 2 V
8 2 V
6 2 V
5 1 V
3 1 V
2 0 V
1 1 V
2 0 V
3 1 V
5 1 V
7 2 V
7 2 V
8 2 V
9 3 V
10 3 V
10 2 V
11 3 V
10 3 V
9 2 V
10 3 V
8 2 V
7 2 V
7 2 V
4 1 V
4 1 V
2 1 V
1 0 V
2 0 V
3 1 V
5 2 V
6 1 V
8 2 V
8 2 V
9 3 V
10 2 V
10 3 V
10 3 V
11 3 V
9 2 V
9 3 V
9 2 V
7 2 V
6 1 V
5 2 V
4 1 V
2 0 V
1 0 V
2 1 V
3 1 V
5 1 V
6 2 V
8 2 V
8 2 V
stroke 5691 2494 M
9 2 V
10 3 V
10 2 V
10 3 V
10 3 V
10 2 V
9 3 V
9 2 V
7 2 V
6 1 V
5 2 V
4 1 V
2 0 V
2 1 V
4 1 V
5 1 V
6 2 V
7 2 V
9 2 V
9 2 V
10 3 V
10 2 V
10 3 V
10 2 V
10 3 V
9 2 V
8 2 V
8 2 V
6 2 V
5 1 V
3 1 V
2 1 V
1 0 V
2 0 V
4 1 V
5 1 V
6 2 V
7 2 V
9 2 V
9 2 V
9 3 V
11 2 V
10 3 V
10 2 V
10 3 V
9 2 V
8 2 V
8 2 V
6 2 V
5 1 V
3 1 V
2 0 V
1 1 V
2 0 V
4 1 V
4 1 V
7 2 V
7 1 V
8 3 V
10 2 V
9 2 V
10 3 V
11 2 V
10 3 V
10 2 V
9 2 V
8 2 V
7 2 V
7 2 V
5 1 V
3 1 V
2 0 V
1 1 V
2 0 V
3 1 V
5 1 V
6 2 V
8 1 V
8 2 V
9 3 V
10 2 V
10 2 V
10 3 V
11 2 V
9 3 V
10 2 V
8 2 V
7 2 V
7 1 V
4 1 V
4 1 V
2 1 V
1 0 V
2 0 V
3 1 V
5 1 V
6 2 V
8 1 V
8 2 V
9 3 V
10 2 V
10 2 V
10 3 V
10 2 V
stroke 6409 2673 M
10 2 V
9 2 V
9 2 V
7 2 V
6 1 V
5 2 V
4 0 V
2 1 V
1 0 V
2 1 V
3 0 V
5 1 V
6 2 V
7 2 V
9 2 V
9 2 V
10 2 V
10 2 V
10 2 V
10 3 V
10 2 V
9 2 V
9 2 V
7 2 V
6 1 V
5 1 V
4 1 V
1 0 V
1 1 V
2 0 V
4 1 V
5 1 V
6 1 V
7 2 V
9 2 V
9 2 V
10 2 V
10 2 V
10 2 V
10 3 V
10 2 V
9 2 V
8 2 V
8 1 V
6 1 V
5 2 V
3 0 V
2 1 V
1 0 V
2 0 V
4 1 V
5 1 V
6 1 V
7 2 V
8 2 V
10 2 V
9 2 V
11 2 V
10 2 V
10 2 V
10 2 V
9 2 V
8 2 V
8 2 V
6 1 V
5 1 V
3 1 V
2 0 V
1 0 V
2 1 V
3 0 V
5 1 V
7 2 V
7 1 V
8 2 V
9 2 V
10 2 V
10 2 V
11 2 V
10 2 V
9 2 V
10 2 V
8 1 V
7 2 V
7 1 V
5 1 V
3 1 V
2 0 V
stroke
LTb
4320 3431 N
0 -2791 V
2879 0 V
0 2791 V
-2879 0 V
Z stroke
1.000 UP
1.000 UL
LTb
stroke
grestore
end
showpage
  }}%
  \put(6128,2080){\makebox(0,0)[r]{\strut{}-F$_-$}}%
  \put(6128,2280){\makebox(0,0)[r]{\strut{}F$_+$}}%
  \put(5759,3731){\makebox(0,0){\strut{}$\bar B_s\to D_{s0}^*(2317)$}}%
  \put(5759,140){\makebox(0,0){\strut{}$q^2$ [$\rm GeV^2$]}}%
  \put(7199,440){\makebox(0,0){\strut{} 10}}%
  \put(6911,440){\makebox(0,0){\strut{} 9}}%
  \put(6623,440){\makebox(0,0){\strut{} 8}}%
  \put(6335,440){\makebox(0,0){\strut{} 7}}%
  \put(6047,440){\makebox(0,0){\strut{} 6}}%
  \put(5760,440){\makebox(0,0){\strut{} 5}}%
  \put(5472,440){\makebox(0,0){\strut{} 4}}%
  \put(5184,440){\makebox(0,0){\strut{} 3}}%
  \put(4896,440){\makebox(0,0){\strut{} 2}}%
  \put(4608,440){\makebox(0,0){\strut{} 1}}%
  \put(4320,440){\makebox(0,0){\strut{} 0}}%
  \put(4200,3431){\makebox(0,0)[r]{\strut{} 1}}%
  \put(4200,2733){\makebox(0,0)[r]{\strut{} 0.8}}%
  \put(4200,2036){\makebox(0,0)[r]{\strut{} 0.6}}%
  \put(4200,1338){\makebox(0,0)[r]{\strut{} 0.4}}%
  \put(4200,640){\makebox(0,0)[r]{\strut{} 0.2}}%
  \put(1808,1856){\makebox(0,0)[r]{\strut{}-F$_-$}}%
  \put(1808,2056){\makebox(0,0)[r]{\strut{}F$_+$}}%
  \put(1439,3731){\makebox(0,0){\strut{}$\bar B_s\to D_s$}}%
  \put(1439,140){\makebox(0,0){\strut{}$q^2$ [$\rm GeV^2$]}}%
  \put(2879,440){\makebox(0,0){\strut{} 12}}%
  \put(2399,440){\makebox(0,0){\strut{} 10}}%
  \put(1919,440){\makebox(0,0){\strut{} 8}}%
  \put(1440,440){\makebox(0,0){\strut{} 6}}%
  \put(960,440){\makebox(0,0){\strut{} 4}}%
  \put(480,440){\makebox(0,0){\strut{} 2}}%
  \put(0,440){\makebox(0,0){\strut{} 0}}%
  \put(-120,3431){\makebox(0,0)[r]{\strut{} 1.2}}%
  \put(-120,2873){\makebox(0,0)[r]{\strut{} 1}}%
  \put(-120,2315){\makebox(0,0)[r]{\strut{} 0.8}}%
  \put(-120,1756){\makebox(0,0)[r]{\strut{} 0.6}}%
  \put(-120,1198){\makebox(0,0)[r]{\strut{} 0.4}}%
  \put(-120,640){\makebox(0,0)[r]{\strut{} 0.2}}%
\end{picture}%
\endgroup
 

%% file: ff.1m.1.tex
\begingroup%
\makeatletter%
\newcommand{\GNUPLOTspecial}{%
  \@sanitize\catcode`\%=14\relax\special}%
\setlength{\unitlength}{0.0500bp}%
\begin{picture}(4680,4284)(0,0)%
  {\GNUPLOTspecial{"
/gnudict 256 dict def
gnudict begin
%
%
/Color false def
/Blacktext true def
/Solid false def
/Dashlength 1 def
/Landscape false def
/Level1 false def
/Rounded false def
/ClipToBoundingBox false def
/SuppressPDFMark false def
/TransparentPatterns false def
/gnulinewidth 5.000 def
/userlinewidth gnulinewidth def
/Gamma 1.0 def
/BackgroundColor {-1.000 -1.000 -1.000} def
/vshift -66 def
/dl1 {
  10.0 Dashlength mul mul
  Rounded { currentlinewidth 0.75 mul sub dup 0 le { pop 0.01 } if } if
} def
/dl2 {
  10.0 Dashlength mul mul
  Rounded { currentlinewidth 0.75 mul add } if
} def
/hpt_ 31.5 def
/vpt_ 31.5 def
/hpt hpt_ def
/vpt vpt_ def
/doclip {
  ClipToBoundingBox {
    newpath 0 0 moveto 234 0 lineto 234 214 lineto 0 214 lineto closepath
    clip
  } if
} def
%
%
%
/M {moveto} bind def
/L {lineto} bind def
/R {rmoveto} bind def
/V {rlineto} bind def
/N {newpath moveto} bind def
/Z {closepath} bind def
/C {setrgbcolor} bind def
/f {rlineto fill} bind def
/g {setgray} bind def
/Gshow {show} def   
/vpt2 vpt 2 mul def
/hpt2 hpt 2 mul def
/Lshow {currentpoint stroke M 0 vshift R 
	Blacktext {gsave 0 setgray show grestore} {show} ifelse} def
/Rshow {currentpoint stroke M dup stringwidth pop neg vshift R
	Blacktext {gsave 0 setgray show grestore} {show} ifelse} def
/Cshow {currentpoint stroke M dup stringwidth pop -2 div vshift R 
	Blacktext {gsave 0 setgray show grestore} {show} ifelse} def
/UP {dup vpt_ mul /vpt exch def hpt_ mul /hpt exch def
  /hpt2 hpt 2 mul def /vpt2 vpt 2 mul def} def
/DL {Color {setrgbcolor Solid {pop []} if 0 setdash}
 {pop pop pop 0 setgray Solid {pop []} if 0 setdash} ifelse} def
/BL {stroke userlinewidth 2 mul setlinewidth
	Rounded {1 setlinejoin 1 setlinecap} if} def
/AL {stroke userlinewidth 2 div setlinewidth
	Rounded {1 setlinejoin 1 setlinecap} if} def
/UL {dup gnulinewidth mul /userlinewidth exch def
	dup 1 lt {pop 1} if 10 mul /udl exch def} def
/PL {stroke userlinewidth setlinewidth
	Rounded {1 setlinejoin 1 setlinecap} if} def
3.8 setmiterlimit
/LCw {1 1 1} def
/LCb {0 0 0} def
/LCa {0 0 0} def
/LC0 {1 0 0} def
/LC1 {0 1 0} def
/LC2 {0 0 1} def
/LC3 {1 0 1} def
/LC4 {0 1 1} def
/LC5 {1 1 0} def
/LC6 {0 0 0} def
/LC7 {1 0.3 0} def
/LC8 {0.5 0.5 0.5} def
/LTw {PL [] 1 setgray} def
/LTb {BL [] LCb DL} def
/LTa {AL [1 udl mul 2 udl mul] 0 setdash LCa setrgbcolor} def
/LT0 {PL [] LC0 DL} def
/LT1 {PL [4 dl1 2 dl2] LC1 DL} def
/LT2 {PL [2 dl1 3 dl2] LC2 DL} def
/LT3 {PL [1 dl1 1.5 dl2] LC3 DL} def
/LT4 {PL [6 dl1 2 dl2 1 dl1 2 dl2] LC4 DL} def
/LT5 {PL [3 dl1 3 dl2 1 dl1 3 dl2] LC5 DL} def
/LT6 {PL [2 dl1 2 dl2 2 dl1 6 dl2] LC6 DL} def
/LT7 {PL [1 dl1 2 dl2 6 dl1 2 dl2 1 dl1 2 dl2] LC7 DL} def
/LT8 {PL [2 dl1 2 dl2 2 dl1 2 dl2 2 dl1 2 dl2 2 dl1 4 dl2] LC8 DL} def
/Pnt {stroke [] 0 setdash gsave 1 setlinecap M 0 0 V stroke grestore} def
/Dia {stroke [] 0 setdash 2 copy vpt add M
  hpt neg vpt neg V hpt vpt neg V
  hpt vpt V hpt neg vpt V closepath stroke
  Pnt} def
/Pls {stroke [] 0 setdash vpt sub M 0 vpt2 V
  currentpoint stroke M
  hpt neg vpt neg R hpt2 0 V stroke
 } def
/Box {stroke [] 0 setdash 2 copy exch hpt sub exch vpt add M
  0 vpt2 neg V hpt2 0 V 0 vpt2 V
  hpt2 neg 0 V closepath stroke
  Pnt} def
/Crs {stroke [] 0 setdash exch hpt sub exch vpt add M
  hpt2 vpt2 neg V currentpoint stroke M
  hpt2 neg 0 R hpt2 vpt2 V stroke} def
/TriU {stroke [] 0 setdash 2 copy vpt 1.12 mul add M
  hpt neg vpt -1.62 mul V
  hpt 2 mul 0 V
  hpt neg vpt 1.62 mul V closepath stroke
  Pnt} def
/Star {2 copy Pls Crs} def
/BoxF {stroke [] 0 setdash exch hpt sub exch vpt add M
  0 vpt2 neg V hpt2 0 V 0 vpt2 V
  hpt2 neg 0 V closepath fill} def
/TriUF {stroke [] 0 setdash vpt 1.12 mul add M
  hpt neg vpt -1.62 mul V
  hpt 2 mul 0 V
  hpt neg vpt 1.62 mul V closepath fill} def
/TriD {stroke [] 0 setdash 2 copy vpt 1.12 mul sub M
  hpt neg vpt 1.62 mul V
  hpt 2 mul 0 V
  hpt neg vpt -1.62 mul V closepath stroke
  Pnt} def
/TriDF {stroke [] 0 setdash vpt 1.12 mul sub M
  hpt neg vpt 1.62 mul V
  hpt 2 mul 0 V
  hpt neg vpt -1.62 mul V closepath fill} def
/DiaF {stroke [] 0 setdash vpt add M
  hpt neg vpt neg V hpt vpt neg V
  hpt vpt V hpt neg vpt V closepath fill} def
/Pent {stroke [] 0 setdash 2 copy gsave
  translate 0 hpt M 4 {72 rotate 0 hpt L} repeat
  closepath stroke grestore Pnt} def
/PentF {stroke [] 0 setdash gsave
  translate 0 hpt M 4 {72 rotate 0 hpt L} repeat
  closepath fill grestore} def
/Circle {stroke [] 0 setdash 2 copy
  hpt 0 360 arc stroke Pnt} def
/CircleF {stroke [] 0 setdash hpt 0 360 arc fill} def
/C0 {BL [] 0 setdash 2 copy moveto vpt 90 450 arc} bind def
/C1 {BL [] 0 setdash 2 copy moveto
	2 copy vpt 0 90 arc closepath fill
	vpt 0 360 arc closepath} bind def
/C2 {BL [] 0 setdash 2 copy moveto
	2 copy vpt 90 180 arc closepath fill
	vpt 0 360 arc closepath} bind def
/C3 {BL [] 0 setdash 2 copy moveto
	2 copy vpt 0 180 arc closepath fill
	vpt 0 360 arc closepath} bind def
/C4 {BL [] 0 setdash 2 copy moveto
	2 copy vpt 180 270 arc closepath fill
	vpt 0 360 arc closepath} bind def
/C5 {BL [] 0 setdash 2 copy moveto
	2 copy vpt 0 90 arc
	2 copy moveto
	2 copy vpt 180 270 arc closepath fill
	vpt 0 360 arc} bind def
/C6 {BL [] 0 setdash 2 copy moveto
	2 copy vpt 90 270 arc closepath fill
	vpt 0 360 arc closepath} bind def
/C7 {BL [] 0 setdash 2 copy moveto
	2 copy vpt 0 270 arc closepath fill
	vpt 0 360 arc closepath} bind def
/C8 {BL [] 0 setdash 2 copy moveto
	2 copy vpt 270 360 arc closepath fill
	vpt 0 360 arc closepath} bind def
/C9 {BL [] 0 setdash 2 copy moveto
	2 copy vpt 270 450 arc closepath fill
	vpt 0 360 arc closepath} bind def
/C10 {BL [] 0 setdash 2 copy 2 copy moveto vpt 270 360 arc closepath fill
	2 copy moveto
	2 copy vpt 90 180 arc closepath fill
	vpt 0 360 arc closepath} bind def
/C11 {BL [] 0 setdash 2 copy moveto
	2 copy vpt 0 180 arc closepath fill
	2 copy moveto
	2 copy vpt 270 360 arc closepath fill
	vpt 0 360 arc closepath} bind def
/C12 {BL [] 0 setdash 2 copy moveto
	2 copy vpt 180 360 arc closepath fill
	vpt 0 360 arc closepath} bind def
/C13 {BL [] 0 setdash 2 copy moveto
	2 copy vpt 0 90 arc closepath fill
	2 copy moveto
	2 copy vpt 180 360 arc closepath fill
	vpt 0 360 arc closepath} bind def
/C14 {BL [] 0 setdash 2 copy moveto
	2 copy vpt 90 360 arc closepath fill
	vpt 0 360 arc} bind def
/C15 {BL [] 0 setdash 2 copy vpt 0 360 arc closepath fill
	vpt 0 360 arc closepath} bind def
/Rec {newpath 4 2 roll moveto 1 index 0 rlineto 0 exch rlineto
	neg 0 rlineto closepath} bind def
/Square {dup Rec} bind def
/Bsquare {vpt sub exch vpt sub exch vpt2 Square} bind def
/S0 {BL [] 0 setdash 2 copy moveto 0 vpt rlineto BL Bsquare} bind def
/S1 {BL [] 0 setdash 2 copy vpt Square fill Bsquare} bind def
/S2 {BL [] 0 setdash 2 copy exch vpt sub exch vpt Square fill Bsquare} bind def
/S3 {BL [] 0 setdash 2 copy exch vpt sub exch vpt2 vpt Rec fill Bsquare} bind def
/S4 {BL [] 0 setdash 2 copy exch vpt sub exch vpt sub vpt Square fill Bsquare} bind def
/S5 {BL [] 0 setdash 2 copy 2 copy vpt Square fill
	exch vpt sub exch vpt sub vpt Square fill Bsquare} bind def
/S6 {BL [] 0 setdash 2 copy exch vpt sub exch vpt sub vpt vpt2 Rec fill Bsquare} bind def
/S7 {BL [] 0 setdash 2 copy exch vpt sub exch vpt sub vpt vpt2 Rec fill
	2 copy vpt Square fill Bsquare} bind def
/S8 {BL [] 0 setdash 2 copy vpt sub vpt Square fill Bsquare} bind def
/S9 {BL [] 0 setdash 2 copy vpt sub vpt vpt2 Rec fill Bsquare} bind def
/S10 {BL [] 0 setdash 2 copy vpt sub vpt Square fill 2 copy exch vpt sub exch vpt Square fill
	Bsquare} bind def
/S11 {BL [] 0 setdash 2 copy vpt sub vpt Square fill 2 copy exch vpt sub exch vpt2 vpt Rec fill
	Bsquare} bind def
/S12 {BL [] 0 setdash 2 copy exch vpt sub exch vpt sub vpt2 vpt Rec fill Bsquare} bind def
/S13 {BL [] 0 setdash 2 copy exch vpt sub exch vpt sub vpt2 vpt Rec fill
	2 copy vpt Square fill Bsquare} bind def
/S14 {BL [] 0 setdash 2 copy exch vpt sub exch vpt sub vpt2 vpt Rec fill
	2 copy exch vpt sub exch vpt Square fill Bsquare} bind def
/S15 {BL [] 0 setdash 2 copy Bsquare fill Bsquare} bind def
/D0 {gsave translate 45 rotate 0 0 S0 stroke grestore} bind def
/D1 {gsave translate 45 rotate 0 0 S1 stroke grestore} bind def
/D2 {gsave translate 45 rotate 0 0 S2 stroke grestore} bind def
/D3 {gsave translate 45 rotate 0 0 S3 stroke grestore} bind def
/D4 {gsave translate 45 rotate 0 0 S4 stroke grestore} bind def
/D5 {gsave translate 45 rotate 0 0 S5 stroke grestore} bind def
/D6 {gsave translate 45 rotate 0 0 S6 stroke grestore} bind def
/D7 {gsave translate 45 rotate 0 0 S7 stroke grestore} bind def
/D8 {gsave translate 45 rotate 0 0 S8 stroke grestore} bind def
/D9 {gsave translate 45 rotate 0 0 S9 stroke grestore} bind def
/D10 {gsave translate 45 rotate 0 0 S10 stroke grestore} bind def
/D11 {gsave translate 45 rotate 0 0 S11 stroke grestore} bind def
/D12 {gsave translate 45 rotate 0 0 S12 stroke grestore} bind def
/D13 {gsave translate 45 rotate 0 0 S13 stroke grestore} bind def
/D14 {gsave translate 45 rotate 0 0 S14 stroke grestore} bind def
/D15 {gsave translate 45 rotate 0 0 S15 stroke grestore} bind def
/DiaE {stroke [] 0 setdash vpt add M
  hpt neg vpt neg V hpt vpt neg V
  hpt vpt V hpt neg vpt V closepath stroke} def
/BoxE {stroke [] 0 setdash exch hpt sub exch vpt add M
  0 vpt2 neg V hpt2 0 V 0 vpt2 V
  hpt2 neg 0 V closepath stroke} def
/TriUE {stroke [] 0 setdash vpt 1.12 mul add M
  hpt neg vpt -1.62 mul V
  hpt 2 mul 0 V
  hpt neg vpt 1.62 mul V closepath stroke} def
/TriDE {stroke [] 0 setdash vpt 1.12 mul sub M
  hpt neg vpt 1.62 mul V
  hpt 2 mul 0 V
  hpt neg vpt -1.62 mul V closepath stroke} def
/PentE {stroke [] 0 setdash gsave
  translate 0 hpt M 4 {72 rotate 0 hpt L} repeat
  closepath stroke grestore} def
/CircE {stroke [] 0 setdash 
  hpt 0 360 arc stroke} def
/Opaque {gsave closepath 1 setgray fill grestore 0 setgray closepath} def
/DiaW {stroke [] 0 setdash vpt add M
  hpt neg vpt neg V hpt vpt neg V
  hpt vpt V hpt neg vpt V Opaque stroke} def
/BoxW {stroke [] 0 setdash exch hpt sub exch vpt add M
  0 vpt2 neg V hpt2 0 V 0 vpt2 V
  hpt2 neg 0 V Opaque stroke} def
/TriUW {stroke [] 0 setdash vpt 1.12 mul add M
  hpt neg vpt -1.62 mul V
  hpt 2 mul 0 V
  hpt neg vpt 1.62 mul V Opaque stroke} def
/TriDW {stroke [] 0 setdash vpt 1.12 mul sub M
  hpt neg vpt 1.62 mul V
  hpt 2 mul 0 V
  hpt neg vpt -1.62 mul V Opaque stroke} def
/PentW {stroke [] 0 setdash gsave
  translate 0 hpt M 4 {72 rotate 0 hpt L} repeat
  Opaque stroke grestore} def
/CircW {stroke [] 0 setdash 
  hpt 0 360 arc Opaque stroke} def
/BoxFill {gsave Rec 1 setgray fill grestore} def
/Density {
  /Fillden exch def
  currentrgbcolor
  /ColB exch def /ColG exch def /ColR exch def
  /ColR ColR Fillden mul Fillden sub 1 add def
  /ColG ColG Fillden mul Fillden sub 1 add def
  /ColB ColB Fillden mul Fillden sub 1 add def
  ColR ColG ColB setrgbcolor} def
/BoxColFill {gsave Rec PolyFill} def
/PolyFill {gsave Density fill grestore grestore} def
/h {rlineto rlineto rlineto gsave closepath fill grestore} bind def
%
%
/PatternFill {gsave /PFa [ 9 2 roll ] def
  PFa 0 get PFa 2 get 2 div add PFa 1 get PFa 3 get 2 div add translate
  PFa 2 get -2 div PFa 3 get -2 div PFa 2 get PFa 3 get Rec
  gsave 1 setgray fill grestore clip
  currentlinewidth 0.5 mul setlinewidth
  /PFs PFa 2 get dup mul PFa 3 get dup mul add sqrt def
  0 0 M PFa 5 get rotate PFs -2 div dup translate
  0 1 PFs PFa 4 get div 1 add floor cvi
	{PFa 4 get mul 0 M 0 PFs V} for
  0 PFa 6 get ne {
	0 1 PFs PFa 4 get div 1 add floor cvi
	{PFa 4 get mul 0 2 1 roll M PFs 0 V} for
 } if
  stroke grestore} def
/languagelevel where
 {pop languagelevel} {1} ifelse
 2 lt
	{/InterpretLevel1 true def}
	{/InterpretLevel1 Level1 def}
 ifelse
%
%
/Level2PatternFill {
/Tile8x8 {/PaintType 2 /PatternType 1 /TilingType 1 /BBox [0 0 8 8] /XStep 8 /YStep 8}
	bind def
/KeepColor {currentrgbcolor [/Pattern /DeviceRGB] setcolorspace} bind def
<< Tile8x8
 /PaintProc {0.5 setlinewidth pop 0 0 M 8 8 L 0 8 M 8 0 L stroke} 
>> matrix makepattern
/Pat1 exch def
<< Tile8x8
 /PaintProc {0.5 setlinewidth pop 0 0 M 8 8 L 0 8 M 8 0 L stroke
	0 4 M 4 8 L 8 4 L 4 0 L 0 4 L stroke}
>> matrix makepattern
/Pat2 exch def
<< Tile8x8
 /PaintProc {0.5 setlinewidth pop 0 0 M 0 8 L
	8 8 L 8 0 L 0 0 L fill}
>> matrix makepattern
/Pat3 exch def
<< Tile8x8
 /PaintProc {0.5 setlinewidth pop -4 8 M 8 -4 L
	0 12 M 12 0 L stroke}
>> matrix makepattern
/Pat4 exch def
<< Tile8x8
 /PaintProc {0.5 setlinewidth pop -4 0 M 8 12 L
	0 -4 M 12 8 L stroke}
>> matrix makepattern
/Pat5 exch def
<< Tile8x8
 /PaintProc {0.5 setlinewidth pop -2 8 M 4 -4 L
	0 12 M 8 -4 L 4 12 M 10 0 L stroke}
>> matrix makepattern
/Pat6 exch def
<< Tile8x8
 /PaintProc {0.5 setlinewidth pop -2 0 M 4 12 L
	0 -4 M 8 12 L 4 -4 M 10 8 L stroke}
>> matrix makepattern
/Pat7 exch def
<< Tile8x8
 /PaintProc {0.5 setlinewidth pop 8 -2 M -4 4 L
	12 0 M -4 8 L 12 4 M 0 10 L stroke}
>> matrix makepattern
/Pat8 exch def
<< Tile8x8
 /PaintProc {0.5 setlinewidth pop 0 -2 M 12 4 L
	-4 0 M 12 8 L -4 4 M 8 10 L stroke}
>> matrix makepattern
/Pat9 exch def
/Pattern1 {PatternBgnd KeepColor Pat1 setpattern} bind def
/Pattern2 {PatternBgnd KeepColor Pat2 setpattern} bind def
/Pattern3 {PatternBgnd KeepColor Pat3 setpattern} bind def
/Pattern4 {PatternBgnd KeepColor Landscape {Pat5} {Pat4} ifelse setpattern} bind def
/Pattern5 {PatternBgnd KeepColor Landscape {Pat4} {Pat5} ifelse setpattern} bind def
/Pattern6 {PatternBgnd KeepColor Landscape {Pat9} {Pat6} ifelse setpattern} bind def
/Pattern7 {PatternBgnd KeepColor Landscape {Pat8} {Pat7} ifelse setpattern} bind def
} def
%
%
%
/PatternBgnd {
  TransparentPatterns {} {gsave 1 setgray fill grestore} ifelse
} def
%
%
/Level1PatternFill {
/Pattern1 {0.250 Density} bind def
/Pattern2 {0.500 Density} bind def
/Pattern3 {0.750 Density} bind def
/Pattern4 {0.125 Density} bind def
/Pattern5 {0.375 Density} bind def
/Pattern6 {0.625 Density} bind def
/Pattern7 {0.875 Density} bind def
} def
%
%
Level1 {Level1PatternFill} {Level2PatternFill} ifelse
/Symbol-Oblique /Symbol findfont [1 0 .167 1 0 0] makefont
dup length dict begin {1 index /FID eq {pop pop} {def} ifelse} forall
currentdict end definefont pop
Level1 SuppressPDFMark or 
{} {
/SDict 10 dict def
systemdict /pdfmark known not {
  userdict /pdfmark systemdict /cleartomark get put
} if
SDict begin [
  /Title (paper/ff.1m.1.tex)
  /Subject (gnuplot plot)
  /Creator (gnuplot 4.6 patchlevel 0)
  /Author (conrado)
  /CreationDate (Mon Nov  4 12:00:14 2013)
  /DOCINFO pdfmark
end
} ifelse
end
gnudict begin
gsave
doclip
0 0 translate
0.050 0.050 scale
0 setgray
newpath
BackgroundColor 0 lt 3 1 roll 0 lt exch 0 lt or or not {BackgroundColor C 1.000 0 0 4680.00 4284.00 BoxColFill} if
1.000 UL
LTb
780 640 M
63 0 V
3476 0 R
-63 0 V
780 1020 M
63 0 V
3476 0 R
-63 0 V
780 1401 M
63 0 V
3476 0 R
-63 0 V
780 1781 M
63 0 V
3476 0 R
-63 0 V
780 2162 M
63 0 V
3476 0 R
-63 0 V
780 2542 M
63 0 V
3476 0 R
-63 0 V
780 2922 M
63 0 V
3476 0 R
-63 0 V
780 3303 M
63 0 V
3476 0 R
-63 0 V
780 3683 M
63 0 V
3476 0 R
-63 0 V
780 640 M
0 63 V
0 2980 R
0 -63 V
1370 640 M
0 63 V
0 2980 R
0 -63 V
1960 640 M
0 63 V
0 2980 R
0 -63 V
2550 640 M
0 63 V
0 2980 R
0 -63 V
3139 640 M
0 63 V
0 2980 R
0 -63 V
3729 640 M
0 63 V
0 2980 R
0 -63 V
4319 640 M
0 63 V
0 2980 R
0 -63 V
stroke
780 3683 N
780 640 L
3539 0 V
0 3043 V
-3539 0 V
Z stroke
LCb setrgbcolor
LTb
1.000 UP
1.000 UL
LTb
1.000 UL
LT0
LCb setrgbcolor
LT0
3302 2101 M
543 0 V
781 1226 M
2 0 V
4 0 V
6 -1 V
7 0 V
8 -1 V
10 -1 V
11 -1 V
11 -1 V
12 -1 V
12 -1 V
12 -1 V
11 -1 V
11 0 V
10 -1 V
8 -1 V
8 -1 V
5 0 V
4 0 V
3 -1 V
1 0 V
2 0 V
4 0 V
6 -1 V
7 0 V
9 -1 V
9 -1 V
11 -1 V
11 -1 V
12 -1 V
12 -1 V
12 -1 V
11 -1 V
11 0 V
10 -1 V
9 -1 V
7 -1 V
6 0 V
4 0 V
2 -1 V
1 0 V
2 0 V
4 0 V
6 -1 V
7 0 V
9 -1 V
10 -1 V
10 -1 V
12 -1 V
11 -1 V
12 -1 V
12 -1 V
12 -1 V
10 -1 V
10 -1 V
9 0 V
7 -1 V
6 0 V
4 -1 V
2 0 V
1 0 V
2 0 V
4 -1 V
6 0 V
7 -1 V
9 0 V
10 -1 V
10 -1 V
12 -1 V
12 -1 V
12 -1 V
11 -1 V
12 -1 V
11 -1 V
9 -1 V
9 -1 V
7 0 V
6 -1 V
4 0 V
2 0 V
1 0 V
3 -1 V
4 0 V
5 0 V
8 -1 V
8 -1 V
10 -1 V
11 -1 V
11 -1 V
12 -1 V
12 -1 V
12 -1 V
11 -1 V
11 -1 V
9 0 V
9 -1 V
7 -1 V
6 0 V
4 -1 V
2 0 V
1 0 V
3 0 V
4 -1 V
stroke 1568 1159 M
5 0 V
8 -1 V
8 0 V
10 -1 V
11 -1 V
11 -1 V
12 -1 V
12 -1 V
12 -1 V
11 -1 V
11 -1 V
10 -1 V
8 -1 V
8 -1 V
5 0 V
4 0 V
3 -1 V
1 0 V
2 0 V
4 0 V
6 -1 V
7 0 V
9 -1 V
9 -1 V
11 -1 V
11 -1 V
12 -1 V
12 -1 V
12 -1 V
11 -1 V
11 -1 V
10 -1 V
8 -1 V
8 0 V
5 -1 V
4 0 V
3 0 V
1 0 V
2 -1 V
4 0 V
6 -1 V
7 0 V
9 -1 V
9 -1 V
11 -1 V
12 -1 V
11 -1 V
12 -1 V
12 -1 V
12 -1 V
10 -1 V
10 -1 V
9 -1 V
7 0 V
6 -1 V
4 0 V
2 0 V
1 0 V
2 -1 V
4 0 V
6 0 V
7 -1 V
9 -1 V
10 -1 V
10 -1 V
12 -1 V
12 -1 V
12 -1 V
11 -1 V
12 -1 V
10 -1 V
10 -1 V
9 -1 V
7 0 V
6 -1 V
4 0 V
2 0 V
1 0 V
2 -1 V
4 0 V
6 0 V
7 -1 V
9 -1 V
10 -1 V
11 -1 V
11 -1 V
12 -1 V
12 -1 V
12 -1 V
11 -1 V
11 -1 V
9 -1 V
9 -1 V
7 0 V
6 -1 V
4 0 V
2 0 V
1 0 V
3 -1 V
4 0 V
5 -1 V
8 0 V
8 -1 V
10 -1 V
stroke 2380 1087 M
11 -1 V
11 -1 V
12 -1 V
12 -1 V
12 -1 V
11 -1 V
11 -1 V
10 -1 V
8 -1 V
7 0 V
6 -1 V
4 0 V
2 -1 V
2 0 V
2 0 V
4 0 V
6 -1 V
7 0 V
8 -1 V
10 -1 V
11 -1 V
11 -1 V
12 -1 V
12 -1 V
12 -1 V
11 -1 V
11 -1 V
10 -1 V
8 -1 V
8 -1 V
5 0 V
4 -1 V
3 0 V
1 0 V
2 0 V
4 0 V
6 -1 V
7 -1 V
9 0 V
9 -1 V
11 -1 V
11 -1 V
12 -1 V
12 -1 V
12 -2 V
11 -1 V
11 -1 V
10 0 V
9 -1 V
7 -1 V
6 0 V
4 -1 V
2 0 V
1 0 V
2 0 V
4 -1 V
6 0 V
7 -1 V
9 -1 V
10 -1 V
10 -1 V
12 -1 V
11 -1 V
12 -1 V
12 -1 V
12 -1 V
10 -1 V
10 -1 V
9 -1 V
7 0 V
6 -1 V
4 0 V
2 0 V
1 0 V
2 -1 V
4 0 V
6 -1 V
7 0 V
9 -1 V
10 -1 V
10 -1 V
12 -1 V
12 -1 V
12 -1 V
11 -1 V
12 -1 V
11 -1 V
9 -1 V
9 -1 V
7 -1 V
6 0 V
4 0 V
2 -1 V
1 0 V
3 0 V
4 0 V
5 -1 V
8 0 V
8 -1 V
10 -1 V
11 -1 V
11 -1 V
12 -1 V
12 -1 V
stroke 3207 1012 M
12 -1 V
11 -1 V
11 -1 V
9 -1 V
9 -1 V
7 -1 V
6 0 V
4 -1 V
2 0 V
1 0 V
3 0 V
4 -1 V
5 0 V
8 -1 V
8 -1 V
10 0 V
11 -1 V
11 -1 V
12 -2 V
12 -1 V
12 -1 V
11 -1 V
11 -1 V
10 -1 V
8 0 V
8 -1 V
5 -1 V
4 0 V
3 0 V
1 0 V
2 0 V
4 -1 V
6 0 V
7 -1 V
9 -1 V
9 -1 V
11 -1 V
11 -1 V
12 -1 V
12 -1 V
12 -1 V
11 -1 V
11 -1 V
10 -1 V
8 -1 V
8 0 V
5 -1 V
4 0 V
3 0 V
1 0 V
2 -1 V
4 0 V
6 -1 V
7 0 V
9 -1 V
9 -1 V
11 -1 V
12 -1 V
11 -1 V
12 -1 V
12 -1 V
12 -1 V
10 -1 V
10 -1 V
9 -1 V
7 0 V
6 -1 V
4 0 V
2 -1 V
1 0 V
2 0 V
4 0 V
6 -1 V
7 0 V
9 -1 V
10 -1 V
10 -1 V
12 -1 V
12 -1 V
12 -1 V
11 -1 V
12 -1 V
10 -1 V
10 -1 V
9 -1 V
7 0 V
6 -1 V
4 0 V
2 -1 V
stroke
LT1
LCb setrgbcolor
LT1
3302 1901 M
543 0 V
781 3014 M
2 0 V
4 0 V
6 1 V
7 1 V
8 1 V
10 1 V
11 1 V
11 1 V
12 2 V
12 1 V
12 1 V
11 2 V
11 1 V
10 1 V
8 1 V
8 1 V
5 0 V
4 1 V
3 0 V
1 0 V
2 0 V
4 1 V
6 1 V
7 0 V
9 1 V
9 1 V
11 2 V
11 1 V
12 1 V
12 2 V
12 1 V
11 1 V
11 1 V
10 1 V
9 1 V
7 1 V
6 1 V
4 0 V
2 1 V
1 0 V
2 0 V
4 0 V
6 1 V
7 1 V
9 1 V
10 1 V
10 1 V
12 1 V
11 2 V
12 1 V
12 1 V
12 1 V
10 2 V
10 1 V
9 1 V
7 0 V
6 1 V
4 0 V
2 1 V
1 0 V
2 0 V
4 1 V
6 0 V
7 1 V
9 1 V
10 1 V
10 1 V
12 1 V
12 2 V
12 1 V
11 1 V
12 1 V
11 2 V
9 1 V
9 1 V
7 0 V
6 1 V
4 1 V
2 0 V
1 0 V
3 0 V
4 1 V
5 0 V
8 1 V
8 1 V
10 1 V
11 1 V
11 1 V
12 2 V
12 1 V
12 1 V
11 1 V
11 2 V
9 1 V
9 0 V
7 1 V
6 1 V
4 0 V
2 1 V
1 0 V
3 0 V
4 0 V
stroke 1568 3100 M
5 1 V
8 1 V
8 1 V
10 1 V
11 1 V
11 1 V
12 1 V
12 2 V
12 1 V
11 1 V
11 1 V
10 1 V
8 1 V
8 1 V
5 0 V
4 1 V
3 0 V
1 0 V
2 0 V
4 1 V
6 1 V
7 0 V
9 1 V
9 1 V
11 1 V
11 2 V
12 1 V
12 1 V
12 1 V
11 1 V
11 2 V
10 1 V
8 0 V
8 1 V
5 1 V
4 0 V
3 0 V
1 1 V
2 0 V
4 0 V
6 1 V
7 1 V
9 0 V
9 1 V
11 2 V
12 1 V
11 1 V
12 1 V
12 1 V
12 2 V
10 1 V
10 1 V
9 1 V
7 0 V
6 1 V
4 0 V
2 1 V
1 0 V
2 0 V
4 0 V
6 1 V
7 1 V
9 0 V
10 1 V
10 2 V
12 1 V
12 1 V
12 1 V
11 1 V
12 1 V
10 1 V
10 1 V
9 1 V
7 1 V
6 1 V
4 0 V
2 0 V
1 0 V
2 1 V
4 0 V
6 1 V
7 0 V
9 1 V
10 1 V
11 1 V
11 1 V
12 1 V
12 2 V
12 1 V
11 1 V
11 1 V
9 1 V
9 1 V
7 1 V
6 0 V
4 0 V
2 1 V
1 0 V
3 0 V
4 0 V
5 1 V
8 1 V
8 1 V
10 0 V
stroke 2380 3183 M
11 1 V
11 2 V
12 1 V
12 1 V
12 1 V
11 1 V
11 1 V
10 1 V
8 1 V
7 1 V
6 0 V
4 1 V
2 0 V
2 0 V
2 0 V
4 0 V
6 1 V
7 1 V
8 1 V
10 0 V
11 1 V
11 2 V
12 1 V
12 1 V
12 1 V
11 1 V
11 1 V
10 1 V
8 1 V
8 0 V
5 1 V
4 0 V
3 0 V
1 1 V
2 0 V
4 0 V
6 1 V
7 0 V
9 1 V
9 1 V
11 1 V
11 1 V
12 1 V
12 1 V
12 1 V
11 1 V
11 1 V
10 1 V
9 1 V
7 1 V
6 0 V
4 1 V
2 0 V
1 0 V
2 0 V
4 0 V
6 1 V
7 1 V
9 0 V
10 1 V
10 1 V
12 1 V
11 1 V
12 1 V
12 1 V
12 1 V
10 1 V
10 1 V
9 1 V
7 1 V
6 0 V
4 0 V
2 1 V
1 0 V
2 0 V
4 0 V
6 1 V
7 0 V
9 1 V
10 1 V
10 1 V
12 1 V
12 1 V
12 1 V
11 1 V
12 1 V
11 1 V
9 1 V
9 0 V
7 1 V
6 1 V
4 0 V
2 0 V
1 0 V
3 0 V
4 1 V
5 0 V
8 1 V
8 0 V
10 1 V
11 1 V
11 1 V
12 1 V
12 1 V
stroke 3207 3258 M
12 1 V
11 1 V
11 1 V
9 1 V
9 0 V
7 1 V
6 0 V
4 1 V
2 0 V
1 0 V
3 0 V
4 1 V
5 0 V
8 1 V
8 0 V
10 1 V
11 1 V
11 1 V
12 1 V
12 1 V
12 1 V
11 1 V
11 0 V
10 1 V
8 1 V
8 0 V
5 1 V
4 0 V
3 0 V
1 0 V
2 1 V
4 0 V
6 0 V
7 1 V
9 1 V
9 0 V
11 1 V
11 1 V
12 1 V
12 1 V
12 1 V
11 1 V
11 0 V
10 1 V
8 1 V
8 0 V
5 1 V
4 0 V
3 0 V
1 0 V
2 1 V
4 0 V
6 0 V
7 1 V
9 0 V
9 1 V
11 1 V
12 1 V
11 1 V
12 0 V
12 1 V
12 1 V
10 1 V
10 1 V
9 0 V
7 1 V
6 0 V
4 0 V
2 1 V
1 0 V
2 0 V
4 0 V
6 1 V
7 0 V
9 1 V
10 0 V
10 1 V
12 1 V
12 1 V
12 1 V
11 0 V
12 1 V
10 1 V
10 0 V
9 1 V
7 1 V
6 0 V
4 0 V
2 0 V
stroke
LT2
LCb setrgbcolor
LT2
3302 1701 M
543 0 V
781 2351 M
2 0 V
4 1 V
6 0 V
7 1 V
8 0 V
10 1 V
11 1 V
11 1 V
12 1 V
12 1 V
12 1 V
11 1 V
11 1 V
10 1 V
8 1 V
8 1 V
5 0 V
4 1 V
3 0 V
1 0 V
2 0 V
4 0 V
6 1 V
7 1 V
9 0 V
9 1 V
11 1 V
11 1 V
12 1 V
12 1 V
12 1 V
11 1 V
11 1 V
10 1 V
9 1 V
7 1 V
6 0 V
4 1 V
2 0 V
1 0 V
2 0 V
4 0 V
6 1 V
7 1 V
9 0 V
10 1 V
10 1 V
12 1 V
11 1 V
12 1 V
12 2 V
12 1 V
10 1 V
10 0 V
9 1 V
7 1 V
6 0 V
4 1 V
2 0 V
1 0 V
2 0 V
4 1 V
6 0 V
7 1 V
9 1 V
10 1 V
10 1 V
12 1 V
12 1 V
12 1 V
11 1 V
12 1 V
11 1 V
9 1 V
9 1 V
7 0 V
6 1 V
4 0 V
2 0 V
1 1 V
3 0 V
4 0 V
5 1 V
8 0 V
8 1 V
10 1 V
11 1 V
11 1 V
12 1 V
12 1 V
12 1 V
11 1 V
11 1 V
9 1 V
9 1 V
7 1 V
6 0 V
4 1 V
2 0 V
1 0 V
3 0 V
4 1 V
stroke 1568 2423 M
5 0 V
8 1 V
8 1 V
10 1 V
11 1 V
11 1 V
12 1 V
12 1 V
12 1 V
11 1 V
11 1 V
10 1 V
8 1 V
8 1 V
5 0 V
4 1 V
3 0 V
1 0 V
2 0 V
4 1 V
6 0 V
7 1 V
9 1 V
9 1 V
11 1 V
11 1 V
12 1 V
12 1 V
12 1 V
11 1 V
11 1 V
10 1 V
8 1 V
8 1 V
5 0 V
4 1 V
3 0 V
1 0 V
2 0 V
4 1 V
6 0 V
7 1 V
9 1 V
9 1 V
11 1 V
12 1 V
11 1 V
12 1 V
12 1 V
12 1 V
10 1 V
10 1 V
9 1 V
7 1 V
6 0 V
4 1 V
2 0 V
1 0 V
2 0 V
4 1 V
6 0 V
7 1 V
9 1 V
10 1 V
10 1 V
12 1 V
12 1 V
12 1 V
11 2 V
12 1 V
10 1 V
10 1 V
9 1 V
7 0 V
6 1 V
4 0 V
2 0 V
1 1 V
2 0 V
4 0 V
6 1 V
7 0 V
9 1 V
10 1 V
11 1 V
11 1 V
12 2 V
12 1 V
12 1 V
11 1 V
11 1 V
9 1 V
9 1 V
7 1 V
6 0 V
4 1 V
2 0 V
1 0 V
3 0 V
4 1 V
5 0 V
8 1 V
8 1 V
10 1 V
stroke 2380 2502 M
11 1 V
11 1 V
12 1 V
12 1 V
12 2 V
11 1 V
11 1 V
10 1 V
8 1 V
7 1 V
6 0 V
4 1 V
2 0 V
2 0 V
2 0 V
4 0 V
6 1 V
7 1 V
8 1 V
10 1 V
11 1 V
11 1 V
12 1 V
12 1 V
12 2 V
11 1 V
11 1 V
10 1 V
8 1 V
8 0 V
5 1 V
4 0 V
3 1 V
1 0 V
2 0 V
4 0 V
6 1 V
7 1 V
9 1 V
9 1 V
11 1 V
11 1 V
12 1 V
12 1 V
12 2 V
11 1 V
11 1 V
10 1 V
9 1 V
7 0 V
6 1 V
4 0 V
2 1 V
1 0 V
2 0 V
4 0 V
6 1 V
7 1 V
9 1 V
10 1 V
10 1 V
12 1 V
11 1 V
12 1 V
12 2 V
12 1 V
10 1 V
10 1 V
9 1 V
7 1 V
6 0 V
4 1 V
2 0 V
1 0 V
2 0 V
4 1 V
6 0 V
7 1 V
9 1 V
10 1 V
10 1 V
12 1 V
12 1 V
12 2 V
11 1 V
12 1 V
11 1 V
9 1 V
9 1 V
7 1 V
6 1 V
4 0 V
2 0 V
1 0 V
3 1 V
4 0 V
5 1 V
8 0 V
8 1 V
10 1 V
11 1 V
11 2 V
12 1 V
12 1 V
stroke 3207 2587 M
12 1 V
11 2 V
11 1 V
9 1 V
9 1 V
7 0 V
6 1 V
4 1 V
2 0 V
1 0 V
3 0 V
4 1 V
5 0 V
8 1 V
8 1 V
10 1 V
11 1 V
11 1 V
12 1 V
12 2 V
12 1 V
11 1 V
11 1 V
10 1 V
8 1 V
8 1 V
5 1 V
4 0 V
3 0 V
1 0 V
2 1 V
4 0 V
6 1 V
7 1 V
9 0 V
9 1 V
11 2 V
11 1 V
12 1 V
12 1 V
12 2 V
11 1 V
11 1 V
10 1 V
8 1 V
8 1 V
5 0 V
4 1 V
3 0 V
1 0 V
2 0 V
4 1 V
6 1 V
7 0 V
9 1 V
9 1 V
11 1 V
12 2 V
11 1 V
12 1 V
12 1 V
12 2 V
10 1 V
10 1 V
9 1 V
7 1 V
6 0 V
4 1 V
2 0 V
1 0 V
2 0 V
4 1 V
6 0 V
7 1 V
9 1 V
10 1 V
10 1 V
12 2 V
12 1 V
12 1 V
11 1 V
12 2 V
10 1 V
10 1 V
9 1 V
7 1 V
6 0 V
4 1 V
2 0 V
stroke
LT3
LCb setrgbcolor
LT3
3302 1501 M
543 0 V
781 1219 M
2 0 V
4 -1 V
6 0 V
7 -1 V
8 -1 V
10 0 V
11 -1 V
11 -1 V
12 -1 V
12 -1 V
12 -1 V
11 -1 V
11 -1 V
10 -1 V
8 -1 V
8 0 V
5 -1 V
4 0 V
3 0 V
1 0 V
2 -1 V
4 0 V
6 -1 V
7 0 V
9 -1 V
9 -1 V
11 -1 V
11 -1 V
12 -1 V
12 -1 V
12 -1 V
11 -1 V
11 -1 V
10 -1 V
9 0 V
7 -1 V
6 0 V
4 -1 V
2 0 V
1 0 V
2 0 V
4 -1 V
6 0 V
7 -1 V
9 0 V
10 -1 V
10 -1 V
12 -1 V
11 -1 V
12 -1 V
12 -1 V
12 -1 V
10 -1 V
10 -1 V
9 -1 V
7 -1 V
6 0 V
4 0 V
2 -1 V
1 0 V
2 0 V
4 0 V
6 -1 V
7 0 V
9 -1 V
10 -1 V
10 -1 V
12 -1 V
12 -1 V
12 -1 V
11 -1 V
12 -1 V
11 -1 V
9 -1 V
9 -1 V
7 0 V
6 -1 V
4 0 V
2 -1 V
1 0 V
3 0 V
4 0 V
5 -1 V
8 0 V
8 -1 V
10 -1 V
11 -1 V
11 -1 V
12 -1 V
12 -1 V
12 -1 V
11 -1 V
11 -1 V
9 -1 V
9 -1 V
7 0 V
6 -1 V
4 0 V
2 -1 V
1 0 V
3 0 V
4 0 V
stroke 1568 1150 M
5 -1 V
8 0 V
8 -1 V
10 -1 V
11 -1 V
11 -1 V
12 -1 V
12 -1 V
12 -1 V
11 -1 V
11 -1 V
10 -1 V
8 -1 V
8 -1 V
5 0 V
4 -1 V
3 0 V
1 0 V
2 0 V
4 0 V
6 -1 V
7 -1 V
9 0 V
9 -1 V
11 -1 V
11 -1 V
12 -1 V
12 -2 V
12 -1 V
11 -1 V
11 -1 V
10 -1 V
8 0 V
8 -1 V
5 -1 V
4 0 V
3 0 V
1 0 V
2 0 V
4 -1 V
6 0 V
7 -1 V
9 -1 V
9 -1 V
11 -1 V
12 -1 V
11 -1 V
12 -1 V
12 -1 V
12 -1 V
10 -1 V
10 -1 V
9 -1 V
7 0 V
6 -1 V
4 0 V
2 -1 V
1 0 V
2 0 V
4 0 V
6 -1 V
7 0 V
9 -1 V
10 -1 V
10 -1 V
12 -1 V
12 -1 V
12 -2 V
11 -1 V
12 -1 V
10 -1 V
10 -1 V
9 0 V
7 -1 V
6 -1 V
4 0 V
2 0 V
1 0 V
2 -1 V
4 0 V
6 0 V
7 -1 V
9 -1 V
10 -1 V
11 -1 V
11 -1 V
12 -1 V
12 -1 V
12 -1 V
11 -1 V
11 -1 V
9 -1 V
9 -1 V
7 -1 V
6 0 V
4 -1 V
2 0 V
1 0 V
3 0 V
4 -1 V
5 0 V
8 -1 V
8 -1 V
10 -1 V
stroke 2380 1074 M
11 -1 V
11 -1 V
12 -1 V
12 -1 V
12 -1 V
11 -1 V
11 -1 V
10 -1 V
8 -1 V
7 -1 V
6 0 V
4 -1 V
2 0 V
2 0 V
2 0 V
4 -1 V
6 0 V
7 -1 V
8 -1 V
10 -1 V
11 -1 V
11 -1 V
12 -1 V
12 -1 V
12 -1 V
11 -1 V
11 -1 V
10 -1 V
8 -1 V
8 -1 V
5 0 V
4 -1 V
3 0 V
1 0 V
2 0 V
4 -1 V
6 0 V
7 -1 V
9 -1 V
9 -1 V
11 -1 V
11 -1 V
12 -1 V
12 -1 V
12 -1 V
11 -1 V
11 -1 V
10 -1 V
9 -1 V
7 -1 V
6 0 V
4 -1 V
2 0 V
1 0 V
2 0 V
4 -1 V
6 0 V
7 -1 V
9 -1 V
10 -1 V
10 -1 V
12 -1 V
11 -1 V
12 -1 V
12 -1 V
12 -1 V
10 -1 V
10 -1 V
9 -1 V
7 -1 V
6 0 V
4 -1 V
2 0 V
1 0 V
2 0 V
4 -1 V
6 0 V
7 -1 V
9 -1 V
10 -1 V
10 -1 V
12 -1 V
12 -1 V
12 -1 V
11 -1 V
12 -1 V
11 -2 V
9 0 V
9 -1 V
7 -1 V
6 -1 V
4 0 V
2 0 V
1 0 V
3 -1 V
4 0 V
5 0 V
8 -1 V
8 -1 V
10 -1 V
11 -1 V
11 -1 V
12 -1 V
12 -1 V
stroke 3207 995 M
12 -2 V
11 -1 V
11 -1 V
9 -1 V
9 -1 V
7 0 V
6 -1 V
4 0 V
2 0 V
1 -1 V
3 0 V
4 0 V
5 -1 V
8 0 V
8 -1 V
10 -1 V
11 -1 V
11 -1 V
12 -2 V
12 -1 V
12 -1 V
11 -1 V
11 -1 V
10 -1 V
8 -1 V
8 -1 V
5 0 V
4 -1 V
3 0 V
1 0 V
2 0 V
4 0 V
6 -1 V
7 -1 V
9 -1 V
9 -1 V
11 -1 V
11 -1 V
12 -1 V
12 -1 V
12 -1 V
11 -1 V
11 -1 V
10 -1 V
8 -1 V
8 -1 V
5 0 V
4 -1 V
3 0 V
1 0 V
2 0 V
4 -1 V
6 0 V
7 -1 V
9 -1 V
9 -1 V
11 -1 V
12 -1 V
11 -1 V
12 -1 V
12 -2 V
12 -1 V
10 -1 V
10 -1 V
9 0 V
7 -1 V
6 -1 V
4 0 V
2 0 V
1 0 V
2 -1 V
4 0 V
6 -1 V
7 0 V
9 -1 V
10 -1 V
10 -1 V
12 -1 V
12 -2 V
12 -1 V
11 -1 V
12 -1 V
10 -1 V
10 -1 V
9 -1 V
7 -1 V
6 0 V
4 -1 V
2 -1 V
stroke
LTb
780 3683 N
780 640 L
3539 0 V
0 3043 V
-3539 0 V
Z stroke
1.000 UP
1.000 UL
LTb
stroke
grestore
end
showpage
  }}%
  \put(3182,1501){\makebox(0,0)[r]{\strut{}A$_-$}}%
  \put(3182,1701){\makebox(0,0)[r]{\strut{}A$_+$}}%
  \put(3182,1901){\makebox(0,0)[r]{\strut{}A$_0$}}%
  \put(3182,2101){\makebox(0,0)[r]{\strut{}V}}%
  \put(2549,3983){\makebox(0,0){\strut{}$\bar B_s\to D_s^*$}}%
  \put(2549,140){\makebox(0,0){\strut{}$q^2$ [$\rm GeV^2$]}}%
  \put(4319,440){\makebox(0,0){\strut{} 12}}%
  \put(3729,440){\makebox(0,0){\strut{} 10}}%
  \put(3139,440){\makebox(0,0){\strut{} 8}}%
  \put(2550,440){\makebox(0,0){\strut{} 6}}%
  \put(1960,440){\makebox(0,0){\strut{} 4}}%
  \put(1370,440){\makebox(0,0){\strut{} 2}}%
  \put(780,440){\makebox(0,0){\strut{} 0}}%
  \put(660,3683){\makebox(0,0)[r]{\strut{} 2.5}}%
  \put(660,3303){\makebox(0,0)[r]{\strut{} 2}}%
  \put(660,2922){\makebox(0,0)[r]{\strut{} 1.5}}%
  \put(660,2542){\makebox(0,0)[r]{\strut{} 1}}%
  \put(660,2162){\makebox(0,0)[r]{\strut{} 0.5}}%
  \put(660,1781){\makebox(0,0)[r]{\strut{} 0}}%
  \put(660,1401){\makebox(0,0)[r]{\strut{}-0.5}}%
  \put(660,1020){\makebox(0,0)[r]{\strut{}-1}}%
  \put(660,640){\makebox(0,0)[r]{\strut{}-1.5}}%
\end{picture}%
\endgroup
 

%% file: ff.1p0.1p1.tex
\begingroup%
\makeatletter%
\newcommand{\GNUPLOTspecial}{%
  \@sanitize\catcode`\%=14\relax\special}%
\setlength{\unitlength}{0.0500bp}%
\begin{picture}(7200,5040)(0,0)%
  {\GNUPLOTspecial{"
/gnudict 256 dict def
gnudict begin
%
%
/Color false def
/Blacktext true def
/Solid false def
/Dashlength 1 def
/Landscape false def
/Level1 false def
/Rounded false def
/ClipToBoundingBox false def
/SuppressPDFMark false def
/TransparentPatterns false def
/gnulinewidth 5.000 def
/userlinewidth gnulinewidth def
/Gamma 1.0 def
/BackgroundColor {-1.000 -1.000 -1.000} def
/vshift -66 def
/dl1 {
  10.0 Dashlength mul mul
  Rounded { currentlinewidth 0.75 mul sub dup 0 le { pop 0.01 } if } if
} def
/dl2 {
  10.0 Dashlength mul mul
  Rounded { currentlinewidth 0.75 mul add } if
} def
/hpt_ 31.5 def
/vpt_ 31.5 def
/hpt hpt_ def
/vpt vpt_ def
/doclip {
  ClipToBoundingBox {
    newpath 0 0 moveto 360 0 lineto 360 252 lineto 0 252 lineto closepath
    clip
  } if
} def
%
%
%
/M {moveto} bind def
/L {lineto} bind def
/R {rmoveto} bind def
/V {rlineto} bind def
/N {newpath moveto} bind def
/Z {closepath} bind def
/C {setrgbcolor} bind def
/f {rlineto fill} bind def
/g {setgray} bind def
/Gshow {show} def   
/vpt2 vpt 2 mul def
/hpt2 hpt 2 mul def
/Lshow {currentpoint stroke M 0 vshift R 
	Blacktext {gsave 0 setgray show grestore} {show} ifelse} def
/Rshow {currentpoint stroke M dup stringwidth pop neg vshift R
	Blacktext {gsave 0 setgray show grestore} {show} ifelse} def
/Cshow {currentpoint stroke M dup stringwidth pop -2 div vshift R 
	Blacktext {gsave 0 setgray show grestore} {show} ifelse} def
/UP {dup vpt_ mul /vpt exch def hpt_ mul /hpt exch def
  /hpt2 hpt 2 mul def /vpt2 vpt 2 mul def} def
/DL {Color {setrgbcolor Solid {pop []} if 0 setdash}
 {pop pop pop 0 setgray Solid {pop []} if 0 setdash} ifelse} def
/BL {stroke userlinewidth 2 mul setlinewidth
	Rounded {1 setlinejoin 1 setlinecap} if} def
/AL {stroke userlinewidth 2 div setlinewidth
	Rounded {1 setlinejoin 1 setlinecap} if} def
/UL {dup gnulinewidth mul /userlinewidth exch def
	dup 1 lt {pop 1} if 10 mul /udl exch def} def
/PL {stroke userlinewidth setlinewidth
	Rounded {1 setlinejoin 1 setlinecap} if} def
3.8 setmiterlimit
/LCw {1 1 1} def
/LCb {0 0 0} def
/LCa {0 0 0} def
/LC0 {1 0 0} def
/LC1 {0 1 0} def
/LC2 {0 0 1} def
/LC3 {1 0 1} def
/LC4 {0 1 1} def
/LC5 {1 1 0} def
/LC6 {0 0 0} def
/LC7 {1 0.3 0} def
/LC8 {0.5 0.5 0.5} def
/LTw {PL [] 1 setgray} def
/LTb {BL [] LCb DL} def
/LTa {AL [1 udl mul 2 udl mul] 0 setdash LCa setrgbcolor} def
/LT0 {PL [] LC0 DL} def
/LT1 {PL [4 dl1 2 dl2] LC1 DL} def
/LT2 {PL [2 dl1 3 dl2] LC2 DL} def
/LT3 {PL [1 dl1 1.5 dl2] LC3 DL} def
/LT4 {PL [6 dl1 2 dl2 1 dl1 2 dl2] LC4 DL} def
/LT5 {PL [3 dl1 3 dl2 1 dl1 3 dl2] LC5 DL} def
/LT6 {PL [2 dl1 2 dl2 2 dl1 6 dl2] LC6 DL} def
/LT7 {PL [1 dl1 2 dl2 6 dl1 2 dl2 1 dl1 2 dl2] LC7 DL} def
/LT8 {PL [2 dl1 2 dl2 2 dl1 2 dl2 2 dl1 2 dl2 2 dl1 4 dl2] LC8 DL} def
/Pnt {stroke [] 0 setdash gsave 1 setlinecap M 0 0 V stroke grestore} def
/Dia {stroke [] 0 setdash 2 copy vpt add M
  hpt neg vpt neg V hpt vpt neg V
  hpt vpt V hpt neg vpt V closepath stroke
  Pnt} def
/Pls {stroke [] 0 setdash vpt sub M 0 vpt2 V
  currentpoint stroke M
  hpt neg vpt neg R hpt2 0 V stroke
 } def
/Box {stroke [] 0 setdash 2 copy exch hpt sub exch vpt add M
  0 vpt2 neg V hpt2 0 V 0 vpt2 V
  hpt2 neg 0 V closepath stroke
  Pnt} def
/Crs {stroke [] 0 setdash exch hpt sub exch vpt add M
  hpt2 vpt2 neg V currentpoint stroke M
  hpt2 neg 0 R hpt2 vpt2 V stroke} def
/TriU {stroke [] 0 setdash 2 copy vpt 1.12 mul add M
  hpt neg vpt -1.62 mul V
  hpt 2 mul 0 V
  hpt neg vpt 1.62 mul V closepath stroke
  Pnt} def
/Star {2 copy Pls Crs} def
/BoxF {stroke [] 0 setdash exch hpt sub exch vpt add M
  0 vpt2 neg V hpt2 0 V 0 vpt2 V
  hpt2 neg 0 V closepath fill} def
/TriUF {stroke [] 0 setdash vpt 1.12 mul add M
  hpt neg vpt -1.62 mul V
  hpt 2 mul 0 V
  hpt neg vpt 1.62 mul V closepath fill} def
/TriD {stroke [] 0 setdash 2 copy vpt 1.12 mul sub M
  hpt neg vpt 1.62 mul V
  hpt 2 mul 0 V
  hpt neg vpt -1.62 mul V closepath stroke
  Pnt} def
/TriDF {stroke [] 0 setdash vpt 1.12 mul sub M
  hpt neg vpt 1.62 mul V
  hpt 2 mul 0 V
  hpt neg vpt -1.62 mul V closepath fill} def
/DiaF {stroke [] 0 setdash vpt add M
  hpt neg vpt neg V hpt vpt neg V
  hpt vpt V hpt neg vpt V closepath fill} def
/Pent {stroke [] 0 setdash 2 copy gsave
  translate 0 hpt M 4 {72 rotate 0 hpt L} repeat
  closepath stroke grestore Pnt} def
/PentF {stroke [] 0 setdash gsave
  translate 0 hpt M 4 {72 rotate 0 hpt L} repeat
  closepath fill grestore} def
/Circle {stroke [] 0 setdash 2 copy
  hpt 0 360 arc stroke Pnt} def
/CircleF {stroke [] 0 setdash hpt 0 360 arc fill} def
/C0 {BL [] 0 setdash 2 copy moveto vpt 90 450 arc} bind def
/C1 {BL [] 0 setdash 2 copy moveto
	2 copy vpt 0 90 arc closepath fill
	vpt 0 360 arc closepath} bind def
/C2 {BL [] 0 setdash 2 copy moveto
	2 copy vpt 90 180 arc closepath fill
	vpt 0 360 arc closepath} bind def
/C3 {BL [] 0 setdash 2 copy moveto
	2 copy vpt 0 180 arc closepath fill
	vpt 0 360 arc closepath} bind def
/C4 {BL [] 0 setdash 2 copy moveto
	2 copy vpt 180 270 arc closepath fill
	vpt 0 360 arc closepath} bind def
/C5 {BL [] 0 setdash 2 copy moveto
	2 copy vpt 0 90 arc
	2 copy moveto
	2 copy vpt 180 270 arc closepath fill
	vpt 0 360 arc} bind def
/C6 {BL [] 0 setdash 2 copy moveto
	2 copy vpt 90 270 arc closepath fill
	vpt 0 360 arc closepath} bind def
/C7 {BL [] 0 setdash 2 copy moveto
	2 copy vpt 0 270 arc closepath fill
	vpt 0 360 arc closepath} bind def
/C8 {BL [] 0 setdash 2 copy moveto
	2 copy vpt 270 360 arc closepath fill
	vpt 0 360 arc closepath} bind def
/C9 {BL [] 0 setdash 2 copy moveto
	2 copy vpt 270 450 arc closepath fill
	vpt 0 360 arc closepath} bind def
/C10 {BL [] 0 setdash 2 copy 2 copy moveto vpt 270 360 arc closepath fill
	2 copy moveto
	2 copy vpt 90 180 arc closepath fill
	vpt 0 360 arc closepath} bind def
/C11 {BL [] 0 setdash 2 copy moveto
	2 copy vpt 0 180 arc closepath fill
	2 copy moveto
	2 copy vpt 270 360 arc closepath fill
	vpt 0 360 arc closepath} bind def
/C12 {BL [] 0 setdash 2 copy moveto
	2 copy vpt 180 360 arc closepath fill
	vpt 0 360 arc closepath} bind def
/C13 {BL [] 0 setdash 2 copy moveto
	2 copy vpt 0 90 arc closepath fill
	2 copy moveto
	2 copy vpt 180 360 arc closepath fill
	vpt 0 360 arc closepath} bind def
/C14 {BL [] 0 setdash 2 copy moveto
	2 copy vpt 90 360 arc closepath fill
	vpt 0 360 arc} bind def
/C15 {BL [] 0 setdash 2 copy vpt 0 360 arc closepath fill
	vpt 0 360 arc closepath} bind def
/Rec {newpath 4 2 roll moveto 1 index 0 rlineto 0 exch rlineto
	neg 0 rlineto closepath} bind def
/Square {dup Rec} bind def
/Bsquare {vpt sub exch vpt sub exch vpt2 Square} bind def
/S0 {BL [] 0 setdash 2 copy moveto 0 vpt rlineto BL Bsquare} bind def
/S1 {BL [] 0 setdash 2 copy vpt Square fill Bsquare} bind def
/S2 {BL [] 0 setdash 2 copy exch vpt sub exch vpt Square fill Bsquare} bind def
/S3 {BL [] 0 setdash 2 copy exch vpt sub exch vpt2 vpt Rec fill Bsquare} bind def
/S4 {BL [] 0 setdash 2 copy exch vpt sub exch vpt sub vpt Square fill Bsquare} bind def
/S5 {BL [] 0 setdash 2 copy 2 copy vpt Square fill
	exch vpt sub exch vpt sub vpt Square fill Bsquare} bind def
/S6 {BL [] 0 setdash 2 copy exch vpt sub exch vpt sub vpt vpt2 Rec fill Bsquare} bind def
/S7 {BL [] 0 setdash 2 copy exch vpt sub exch vpt sub vpt vpt2 Rec fill
	2 copy vpt Square fill Bsquare} bind def
/S8 {BL [] 0 setdash 2 copy vpt sub vpt Square fill Bsquare} bind def
/S9 {BL [] 0 setdash 2 copy vpt sub vpt vpt2 Rec fill Bsquare} bind def
/S10 {BL [] 0 setdash 2 copy vpt sub vpt Square fill 2 copy exch vpt sub exch vpt Square fill
	Bsquare} bind def
/S11 {BL [] 0 setdash 2 copy vpt sub vpt Square fill 2 copy exch vpt sub exch vpt2 vpt Rec fill
	Bsquare} bind def
/S12 {BL [] 0 setdash 2 copy exch vpt sub exch vpt sub vpt2 vpt Rec fill Bsquare} bind def
/S13 {BL [] 0 setdash 2 copy exch vpt sub exch vpt sub vpt2 vpt Rec fill
	2 copy vpt Square fill Bsquare} bind def
/S14 {BL [] 0 setdash 2 copy exch vpt sub exch vpt sub vpt2 vpt Rec fill
	2 copy exch vpt sub exch vpt Square fill Bsquare} bind def
/S15 {BL [] 0 setdash 2 copy Bsquare fill Bsquare} bind def
/D0 {gsave translate 45 rotate 0 0 S0 stroke grestore} bind def
/D1 {gsave translate 45 rotate 0 0 S1 stroke grestore} bind def
/D2 {gsave translate 45 rotate 0 0 S2 stroke grestore} bind def
/D3 {gsave translate 45 rotate 0 0 S3 stroke grestore} bind def
/D4 {gsave translate 45 rotate 0 0 S4 stroke grestore} bind def
/D5 {gsave translate 45 rotate 0 0 S5 stroke grestore} bind def
/D6 {gsave translate 45 rotate 0 0 S6 stroke grestore} bind def
/D7 {gsave translate 45 rotate 0 0 S7 stroke grestore} bind def
/D8 {gsave translate 45 rotate 0 0 S8 stroke grestore} bind def
/D9 {gsave translate 45 rotate 0 0 S9 stroke grestore} bind def
/D10 {gsave translate 45 rotate 0 0 S10 stroke grestore} bind def
/D11 {gsave translate 45 rotate 0 0 S11 stroke grestore} bind def
/D12 {gsave translate 45 rotate 0 0 S12 stroke grestore} bind def
/D13 {gsave translate 45 rotate 0 0 S13 stroke grestore} bind def
/D14 {gsave translate 45 rotate 0 0 S14 stroke grestore} bind def
/D15 {gsave translate 45 rotate 0 0 S15 stroke grestore} bind def
/DiaE {stroke [] 0 setdash vpt add M
  hpt neg vpt neg V hpt vpt neg V
  hpt vpt V hpt neg vpt V closepath stroke} def
/BoxE {stroke [] 0 setdash exch hpt sub exch vpt add M
  0 vpt2 neg V hpt2 0 V 0 vpt2 V
  hpt2 neg 0 V closepath stroke} def
/TriUE {stroke [] 0 setdash vpt 1.12 mul add M
  hpt neg vpt -1.62 mul V
  hpt 2 mul 0 V
  hpt neg vpt 1.62 mul V closepath stroke} def
/TriDE {stroke [] 0 setdash vpt 1.12 mul sub M
  hpt neg vpt 1.62 mul V
  hpt 2 mul 0 V
  hpt neg vpt -1.62 mul V closepath stroke} def
/PentE {stroke [] 0 setdash gsave
  translate 0 hpt M 4 {72 rotate 0 hpt L} repeat
  closepath stroke grestore} def
/CircE {stroke [] 0 setdash 
  hpt 0 360 arc stroke} def
/Opaque {gsave closepath 1 setgray fill grestore 0 setgray closepath} def
/DiaW {stroke [] 0 setdash vpt add M
  hpt neg vpt neg V hpt vpt neg V
  hpt vpt V hpt neg vpt V Opaque stroke} def
/BoxW {stroke [] 0 setdash exch hpt sub exch vpt add M
  0 vpt2 neg V hpt2 0 V 0 vpt2 V
  hpt2 neg 0 V Opaque stroke} def
/TriUW {stroke [] 0 setdash vpt 1.12 mul add M
  hpt neg vpt -1.62 mul V
  hpt 2 mul 0 V
  hpt neg vpt 1.62 mul V Opaque stroke} def
/TriDW {stroke [] 0 setdash vpt 1.12 mul sub M
  hpt neg vpt 1.62 mul V
  hpt 2 mul 0 V
  hpt neg vpt -1.62 mul V Opaque stroke} def
/PentW {stroke [] 0 setdash gsave
  translate 0 hpt M 4 {72 rotate 0 hpt L} repeat
  Opaque stroke grestore} def
/CircW {stroke [] 0 setdash 
  hpt 0 360 arc Opaque stroke} def
/BoxFill {gsave Rec 1 setgray fill grestore} def
/Density {
  /Fillden exch def
  currentrgbcolor
  /ColB exch def /ColG exch def /ColR exch def
  /ColR ColR Fillden mul Fillden sub 1 add def
  /ColG ColG Fillden mul Fillden sub 1 add def
  /ColB ColB Fillden mul Fillden sub 1 add def
  ColR ColG ColB setrgbcolor} def
/BoxColFill {gsave Rec PolyFill} def
/PolyFill {gsave Density fill grestore grestore} def
/h {rlineto rlineto rlineto gsave closepath fill grestore} bind def
%
%
/PatternFill {gsave /PFa [ 9 2 roll ] def
  PFa 0 get PFa 2 get 2 div add PFa 1 get PFa 3 get 2 div add translate
  PFa 2 get -2 div PFa 3 get -2 div PFa 2 get PFa 3 get Rec
  gsave 1 setgray fill grestore clip
  currentlinewidth 0.5 mul setlinewidth
  /PFs PFa 2 get dup mul PFa 3 get dup mul add sqrt def
  0 0 M PFa 5 get rotate PFs -2 div dup translate
  0 1 PFs PFa 4 get div 1 add floor cvi
	{PFa 4 get mul 0 M 0 PFs V} for
  0 PFa 6 get ne {
	0 1 PFs PFa 4 get div 1 add floor cvi
	{PFa 4 get mul 0 2 1 roll M PFs 0 V} for
 } if
  stroke grestore} def
/languagelevel where
 {pop languagelevel} {1} ifelse
 2 lt
	{/InterpretLevel1 true def}
	{/InterpretLevel1 Level1 def}
 ifelse
%
%
/Level2PatternFill {
/Tile8x8 {/PaintType 2 /PatternType 1 /TilingType 1 /BBox [0 0 8 8] /XStep 8 /YStep 8}
	bind def
/KeepColor {currentrgbcolor [/Pattern /DeviceRGB] setcolorspace} bind def
<< Tile8x8
 /PaintProc {0.5 setlinewidth pop 0 0 M 8 8 L 0 8 M 8 0 L stroke} 
>> matrix makepattern
/Pat1 exch def
<< Tile8x8
 /PaintProc {0.5 setlinewidth pop 0 0 M 8 8 L 0 8 M 8 0 L stroke
	0 4 M 4 8 L 8 4 L 4 0 L 0 4 L stroke}
>> matrix makepattern
/Pat2 exch def
<< Tile8x8
 /PaintProc {0.5 setlinewidth pop 0 0 M 0 8 L
	8 8 L 8 0 L 0 0 L fill}
>> matrix makepattern
/Pat3 exch def
<< Tile8x8
 /PaintProc {0.5 setlinewidth pop -4 8 M 8 -4 L
	0 12 M 12 0 L stroke}
>> matrix makepattern
/Pat4 exch def
<< Tile8x8
 /PaintProc {0.5 setlinewidth pop -4 0 M 8 12 L
	0 -4 M 12 8 L stroke}
>> matrix makepattern
/Pat5 exch def
<< Tile8x8
 /PaintProc {0.5 setlinewidth pop -2 8 M 4 -4 L
	0 12 M 8 -4 L 4 12 M 10 0 L stroke}
>> matrix makepattern
/Pat6 exch def
<< Tile8x8
 /PaintProc {0.5 setlinewidth pop -2 0 M 4 12 L
	0 -4 M 8 12 L 4 -4 M 10 8 L stroke}
>> matrix makepattern
/Pat7 exch def
<< Tile8x8
 /PaintProc {0.5 setlinewidth pop 8 -2 M -4 4 L
	12 0 M -4 8 L 12 4 M 0 10 L stroke}
>> matrix makepattern
/Pat8 exch def
<< Tile8x8
 /PaintProc {0.5 setlinewidth pop 0 -2 M 12 4 L
	-4 0 M 12 8 L -4 4 M 8 10 L stroke}
>> matrix makepattern
/Pat9 exch def
/Pattern1 {PatternBgnd KeepColor Pat1 setpattern} bind def
/Pattern2 {PatternBgnd KeepColor Pat2 setpattern} bind def
/Pattern3 {PatternBgnd KeepColor Pat3 setpattern} bind def
/Pattern4 {PatternBgnd KeepColor Landscape {Pat5} {Pat4} ifelse setpattern} bind def
/Pattern5 {PatternBgnd KeepColor Landscape {Pat4} {Pat5} ifelse setpattern} bind def
/Pattern6 {PatternBgnd KeepColor Landscape {Pat9} {Pat6} ifelse setpattern} bind def
/Pattern7 {PatternBgnd KeepColor Landscape {Pat8} {Pat7} ifelse setpattern} bind def
} def
%
%
%
/PatternBgnd {
  TransparentPatterns {} {gsave 1 setgray fill grestore} ifelse
} def
%
%
/Level1PatternFill {
/Pattern1 {0.250 Density} bind def
/Pattern2 {0.500 Density} bind def
/Pattern3 {0.750 Density} bind def
/Pattern4 {0.125 Density} bind def
/Pattern5 {0.375 Density} bind def
/Pattern6 {0.625 Density} bind def
/Pattern7 {0.875 Density} bind def
} def
%
%
Level1 {Level1PatternFill} {Level2PatternFill} ifelse
/Symbol-Oblique /Symbol findfont [1 0 .167 1 0 0] makefont
dup length dict begin {1 index /FID eq {pop pop} {def} ifelse} forall
currentdict end definefont pop
Level1 SuppressPDFMark or 
{} {
/SDict 10 dict def
systemdict /pdfmark known not {
  userdict /pdfmark systemdict /cleartomark get put
} if
SDict begin [
  /Title (paper/ff.1p0.1p1.tex)
  /Subject (gnuplot plot)
  /Creator (gnuplot 4.6 patchlevel 0)
  /Author (conrado)
  /CreationDate (Tue Nov 12 01:22:49 2013)
  /DOCINFO pdfmark
end
} ifelse
end
gnudict begin
gsave
doclip
0 0 translate
0.050 0.050 scale
0 setgray
newpath
BackgroundColor 0 lt 3 1 roll 0 lt exch 0 lt or or not {BackgroundColor C 1.000 0 0 7200.00 5040.00 BoxColFill} if
1.000 UL
LTb
0 640 M
63 0 V
2816 0 R
-63 0 V
0 989 M
63 0 V
2816 0 R
-63 0 V
0 1338 M
63 0 V
2816 0 R
-63 0 V
0 1687 M
63 0 V
2816 0 R
-63 0 V
0 2035 M
63 0 V
2816 0 R
-63 0 V
0 2384 M
63 0 V
2816 0 R
-63 0 V
0 2733 M
63 0 V
2816 0 R
-63 0 V
0 3082 M
63 0 V
2816 0 R
-63 0 V
0 3431 M
63 0 V
2816 0 R
-63 0 V
0 640 M
0 63 V
0 3431 M
0 -63 V
320 640 M
0 63 V
0 2728 R
0 -63 V
640 640 M
0 63 V
0 2728 R
0 -63 V
960 640 M
0 63 V
0 2728 R
0 -63 V
1280 640 M
0 63 V
0 2728 R
0 -63 V
1599 640 M
0 63 V
0 2728 R
0 -63 V
1919 640 M
0 63 V
0 2728 R
0 -63 V
2239 640 M
0 63 V
0 2728 R
0 -63 V
2559 640 M
0 63 V
0 2728 R
0 -63 V
2879 640 M
0 63 V
0 2728 R
0 -63 V
stroke
LTa
0 2733 M
2879 0 V
stroke
LTb
0 3431 N
0 640 L
2879 0 V
0 2791 V
0 3431 L
Z stroke
LCb setrgbcolor
LTb
1.000 UP
1.000 UL
LTb
1.000 UL
LT0
LCb setrgbcolor
LT0
1928 2456 M
543 0 V
0 2684 M
2 0 V
4 0 V
5 0 V
6 0 V
8 0 V
8 0 V
9 0 V
10 0 V
10 0 V
11 -1 V
10 0 V
10 0 V
9 0 V
9 0 V
7 0 V
6 0 V
5 0 V
4 0 V
2 0 V
1 0 V
2 0 V
3 0 V
5 0 V
6 0 V
8 0 V
8 0 V
10 0 V
9 0 V
11 0 V
10 -1 V
10 0 V
10 0 V
9 0 V
9 0 V
7 0 V
6 0 V
5 0 V
4 0 V
2 0 V
1 0 V
2 0 V
3 0 V
5 0 V
7 0 V
7 0 V
8 0 V
10 0 V
10 0 V
10 -1 V
10 0 V
10 0 V
10 0 V
9 0 V
9 0 V
7 0 V
7 0 V
5 0 V
3 0 V
2 0 V
1 0 V
2 0 V
3 0 V
5 0 V
7 0 V
7 0 V
9 0 V
9 -1 V
10 0 V
10 0 V
10 0 V
11 0 V
9 0 V
10 0 V
8 0 V
8 0 V
6 0 V
5 0 V
3 0 V
2 0 V
1 0 V
2 0 V
4 0 V
5 0 V
6 0 V
7 0 V
9 -1 V
9 0 V
10 0 V
10 0 V
10 0 V
11 0 V
10 0 V
9 0 V
8 0 V
8 0 V
6 0 V
5 0 V
3 0 V
2 0 V
1 0 V
2 0 V
4 0 V
stroke 682 2679 M
5 0 V
6 -1 V
7 0 V
9 0 V
9 0 V
10 0 V
10 0 V
11 0 V
10 0 V
10 0 V
9 0 V
8 0 V
8 0 V
6 0 V
5 0 V
4 -1 V
2 0 V
1 0 V
2 0 V
3 0 V
5 0 V
6 0 V
8 0 V
8 0 V
9 0 V
10 0 V
10 0 V
11 0 V
10 0 V
10 0 V
9 0 V
9 0 V
7 -1 V
6 0 V
5 0 V
4 0 V
2 0 V
1 0 V
2 0 V
3 0 V
5 0 V
6 0 V
8 0 V
8 0 V
10 0 V
9 0 V
11 0 V
10 0 V
10 0 V
10 0 V
9 -1 V
9 0 V
7 0 V
7 0 V
5 0 V
3 0 V
2 0 V
1 0 V
2 0 V
3 0 V
5 0 V
7 0 V
7 0 V
8 0 V
10 0 V
10 0 V
10 0 V
10 0 V
10 -1 V
10 0 V
10 0 V
8 0 V
7 0 V
7 0 V
5 0 V
3 0 V
2 0 V
1 0 V
2 0 V
4 0 V
4 0 V
7 0 V
7 0 V
9 0 V
9 0 V
10 -1 V
10 0 V
10 0 V
11 0 V
9 0 V
10 0 V
8 0 V
8 0 V
6 0 V
5 0 V
3 0 V
2 0 V
1 0 V
2 0 V
4 0 V
5 0 V
6 0 V
7 0 V
9 -1 V
stroke 1385 2672 M
9 0 V
10 0 V
10 0 V
11 0 V
10 0 V
10 0 V
9 0 V
8 0 V
8 0 V
6 0 V
5 0 V
4 0 V
2 0 V
2 0 V
4 -1 V
5 0 V
6 0 V
8 0 V
8 0 V
9 0 V
10 0 V
10 0 V
11 0 V
10 0 V
10 0 V
9 0 V
9 0 V
7 -1 V
6 0 V
5 0 V
4 0 V
2 0 V
1 0 V
2 0 V
3 0 V
5 0 V
6 0 V
8 0 V
8 0 V
9 0 V
10 0 V
11 0 V
10 0 V
10 0 V
10 -1 V
9 0 V
9 0 V
7 0 V
6 0 V
5 0 V
4 0 V
2 0 V
1 0 V
2 0 V
3 0 V
5 0 V
6 0 V
8 0 V
8 0 V
10 0 V
10 0 V
10 -1 V
10 0 V
10 0 V
10 0 V
9 0 V
9 0 V
7 0 V
7 0 V
5 0 V
3 0 V
2 0 V
1 0 V
2 0 V
3 0 V
5 0 V
7 0 V
7 0 V
9 -1 V
9 0 V
10 0 V
10 0 V
10 0 V
10 0 V
10 0 V
10 0 V
8 0 V
8 0 V
6 0 V
5 0 V
3 -1 V
2 0 V
1 0 V
2 0 V
4 0 V
5 0 V
6 0 V
7 0 V
9 0 V
9 0 V
10 0 V
10 0 V
10 0 V
11 0 V
stroke 2111 2666 M
10 0 V
9 -1 V
8 0 V
8 0 V
6 0 V
5 0 V
3 0 V
2 0 V
1 0 V
2 0 V
4 0 V
5 0 V
6 0 V
7 0 V
9 0 V
9 0 V
10 0 V
10 0 V
11 -1 V
10 0 V
10 0 V
9 0 V
8 0 V
8 0 V
6 0 V
5 0 V
4 0 V
2 0 V
1 0 V
1 0 V
4 0 V
5 0 V
6 0 V
8 0 V
8 0 V
9 -1 V
10 0 V
10 0 V
11 0 V
10 0 V
10 0 V
9 0 V
9 0 V
7 0 V
6 0 V
5 0 V
4 0 V
2 0 V
1 0 V
2 -1 V
3 0 V
5 0 V
6 0 V
8 0 V
8 0 V
10 0 V
9 0 V
11 0 V
10 0 V
10 0 V
10 0 V
9 -1 V
9 0 V
7 0 V
7 0 V
4 0 V
4 0 V
2 0 V
1 0 V
2 0 V
3 0 V
5 0 V
7 0 V
7 0 V
8 0 V
10 0 V
10 0 V
10 0 V
10 -1 V
10 0 V
10 0 V
10 0 V
8 0 V
7 0 V
7 0 V
5 0 V
3 0 V
2 0 V
stroke
LT1
LCb setrgbcolor
LT1
1928 2256 M
543 0 V
0 3081 M
2 1 V
4 0 V
5 0 V
6 0 V
8 1 V
8 0 V
9 1 V
10 0 V
10 1 V
11 0 V
10 1 V
10 0 V
9 0 V
9 1 V
7 0 V
6 1 V
5 0 V
4 0 V
2 0 V
1 0 V
2 0 V
3 0 V
5 1 V
6 0 V
8 0 V
8 1 V
10 0 V
9 1 V
11 0 V
10 1 V
10 0 V
10 1 V
9 0 V
9 1 V
7 0 V
6 0 V
5 1 V
4 0 V
2 0 V
1 0 V
2 0 V
3 0 V
5 0 V
7 1 V
7 0 V
8 0 V
10 1 V
10 0 V
10 1 V
10 0 V
10 1 V
10 0 V
9 1 V
9 0 V
7 1 V
7 0 V
5 0 V
3 0 V
2 1 V
1 0 V
2 0 V
3 0 V
5 0 V
7 0 V
7 1 V
9 0 V
9 1 V
10 0 V
10 1 V
10 0 V
11 1 V
9 0 V
10 1 V
8 0 V
8 1 V
6 0 V
5 0 V
3 0 V
2 0 V
1 0 V
2 1 V
4 0 V
5 0 V
6 0 V
7 1 V
9 0 V
9 1 V
10 0 V
10 1 V
10 0 V
11 1 V
10 0 V
9 1 V
8 0 V
8 1 V
6 0 V
5 0 V
3 0 V
2 0 V
1 0 V
2 1 V
4 0 V
stroke 682 3116 M
5 0 V
6 0 V
7 1 V
9 0 V
9 1 V
10 0 V
10 1 V
11 0 V
10 1 V
10 0 V
9 1 V
8 0 V
8 1 V
6 0 V
5 0 V
4 0 V
2 0 V
1 0 V
2 1 V
3 0 V
5 0 V
6 0 V
8 1 V
8 0 V
9 1 V
10 0 V
10 1 V
11 0 V
10 1 V
10 0 V
9 1 V
9 0 V
7 1 V
6 0 V
5 0 V
4 0 V
2 1 V
1 0 V
2 0 V
3 0 V
5 0 V
6 0 V
8 1 V
8 0 V
10 1 V
9 0 V
11 1 V
10 0 V
10 1 V
10 1 V
9 0 V
9 0 V
7 1 V
7 0 V
5 0 V
3 1 V
2 0 V
1 0 V
2 0 V
3 0 V
5 0 V
7 1 V
7 0 V
8 1 V
10 0 V
10 1 V
10 0 V
10 1 V
10 0 V
10 1 V
10 0 V
8 1 V
7 0 V
7 0 V
5 1 V
3 0 V
2 0 V
1 0 V
2 0 V
4 0 V
4 1 V
7 0 V
7 0 V
9 1 V
9 0 V
10 1 V
10 0 V
10 1 V
11 1 V
9 0 V
10 1 V
8 0 V
8 0 V
6 1 V
5 0 V
3 0 V
2 0 V
1 0 V
2 1 V
4 0 V
5 0 V
6 0 V
7 1 V
9 0 V
stroke 1385 3153 M
9 1 V
10 0 V
10 1 V
11 0 V
10 1 V
10 0 V
9 1 V
8 0 V
8 1 V
6 0 V
5 0 V
4 1 V
2 0 V
2 0 V
4 0 V
5 0 V
6 1 V
8 0 V
8 1 V
9 0 V
10 1 V
10 0 V
11 1 V
10 0 V
10 1 V
9 0 V
9 1 V
7 0 V
6 1 V
5 0 V
4 0 V
2 0 V
1 0 V
2 0 V
3 1 V
5 0 V
6 0 V
8 1 V
8 0 V
9 1 V
10 0 V
11 1 V
10 0 V
10 1 V
10 1 V
9 0 V
9 1 V
7 0 V
6 0 V
5 1 V
4 0 V
2 0 V
1 0 V
2 0 V
3 0 V
5 1 V
6 0 V
8 0 V
8 1 V
10 0 V
10 1 V
10 0 V
10 1 V
10 1 V
10 0 V
9 1 V
9 0 V
7 1 V
7 0 V
5 0 V
3 0 V
2 1 V
1 0 V
2 0 V
3 0 V
5 0 V
7 1 V
7 0 V
9 0 V
9 1 V
10 1 V
10 0 V
10 1 V
10 0 V
10 1 V
10 0 V
8 1 V
8 0 V
6 1 V
5 0 V
3 0 V
2 0 V
1 0 V
2 0 V
4 1 V
5 0 V
6 0 V
7 1 V
9 0 V
9 1 V
10 0 V
10 1 V
10 0 V
11 1 V
stroke 2111 3194 M
10 1 V
9 0 V
8 1 V
8 0 V
6 0 V
5 1 V
3 0 V
2 0 V
1 0 V
2 0 V
4 0 V
5 1 V
6 0 V
7 0 V
9 1 V
9 1 V
10 0 V
10 1 V
11 0 V
10 1 V
10 0 V
9 1 V
8 0 V
8 1 V
6 0 V
5 1 V
4 0 V
2 0 V
1 0 V
1 0 V
4 0 V
5 1 V
6 0 V
8 0 V
8 1 V
9 0 V
10 1 V
10 1 V
11 0 V
10 1 V
10 0 V
9 1 V
9 0 V
7 1 V
6 0 V
5 1 V
4 0 V
2 0 V
1 0 V
2 0 V
3 0 V
5 1 V
6 0 V
8 0 V
8 1 V
10 0 V
9 1 V
11 1 V
10 0 V
10 1 V
10 0 V
9 1 V
9 0 V
7 1 V
7 0 V
4 1 V
4 0 V
2 0 V
1 0 V
2 0 V
3 0 V
5 1 V
7 0 V
7 0 V
8 1 V
10 0 V
10 1 V
10 1 V
10 0 V
10 1 V
10 0 V
10 1 V
8 0 V
7 1 V
7 0 V
5 1 V
3 0 V
2 0 V
stroke
LT2
LCb setrgbcolor
LT2
1928 2056 M
543 0 V
0 1531 M
2 0 V
4 -1 V
5 -1 V
6 -1 V
8 -1 V
8 -2 V
9 -2 V
10 -2 V
10 -2 V
11 -2 V
10 -2 V
10 -2 V
9 -2 V
9 -2 V
7 -1 V
6 -1 V
5 -1 V
4 -1 V
2 -1 V
1 0 V
2 0 V
3 -1 V
5 -1 V
6 -1 V
8 -2 V
8 -1 V
10 -2 V
9 -2 V
11 -2 V
10 -2 V
10 -3 V
10 -2 V
9 -1 V
9 -2 V
7 -2 V
6 -1 V
5 -1 V
4 -1 V
2 0 V
1 0 V
2 -1 V
3 0 V
5 -1 V
7 -2 V
7 -1 V
8 -2 V
10 -2 V
10 -2 V
10 -2 V
10 -2 V
10 -2 V
10 -2 V
9 -2 V
9 -2 V
7 -2 V
7 -1 V
5 -1 V
3 -1 V
2 0 V
1 0 V
2 -1 V
3 0 V
5 -1 V
7 -2 V
7 -1 V
9 -2 V
9 -2 V
10 -2 V
10 -2 V
10 -2 V
11 -2 V
9 -2 V
10 -2 V
8 -2 V
8 -2 V
6 -1 V
5 -1 V
3 -1 V
2 0 V
1 0 V
2 -1 V
4 0 V
5 -1 V
6 -2 V
7 -1 V
9 -2 V
9 -2 V
10 -2 V
10 -2 V
10 -2 V
11 -3 V
10 -2 V
9 -2 V
8 -1 V
8 -2 V
6 -1 V
5 -1 V
3 -1 V
2 -1 V
1 0 V
2 0 V
4 -1 V
stroke 682 1391 M
5 -1 V
6 -1 V
7 -2 V
9 -2 V
9 -2 V
10 -2 V
10 -2 V
11 -2 V
10 -2 V
10 -2 V
9 -2 V
8 -2 V
8 -2 V
6 -1 V
5 -1 V
4 -1 V
2 0 V
1 0 V
2 -1 V
3 -1 V
5 -1 V
6 -1 V
8 -2 V
8 -1 V
9 -2 V
10 -2 V
10 -3 V
11 -2 V
10 -2 V
10 -2 V
9 -2 V
9 -2 V
7 -2 V
6 -1 V
5 -1 V
4 -1 V
2 0 V
1 0 V
2 -1 V
3 -1 V
5 -1 V
6 -1 V
8 -2 V
8 -1 V
10 -2 V
9 -3 V
11 -2 V
10 -2 V
10 -2 V
10 -2 V
9 -2 V
9 -2 V
7 -2 V
7 -1 V
5 -1 V
3 -1 V
2 0 V
1 -1 V
2 0 V
3 -1 V
5 -1 V
7 -1 V
7 -2 V
8 -2 V
10 -2 V
10 -2 V
10 -2 V
10 -2 V
10 -2 V
10 -3 V
10 -2 V
8 -2 V
7 -1 V
7 -1 V
5 -2 V
3 0 V
2 -1 V
1 0 V
2 0 V
4 -1 V
4 -1 V
7 -2 V
7 -1 V
9 -2 V
9 -2 V
10 -2 V
10 -3 V
10 -2 V
11 -2 V
9 -2 V
10 -2 V
8 -2 V
8 -2 V
6 -1 V
5 -1 V
3 -1 V
2 0 V
1 -1 V
2 0 V
4 -1 V
5 -1 V
6 -1 V
7 -2 V
9 -2 V
stroke 1385 1239 M
9 -2 V
10 -2 V
10 -2 V
11 -2 V
10 -3 V
10 -2 V
9 -2 V
8 -2 V
8 -1 V
6 -2 V
5 -1 V
4 -1 V
2 0 V
2 -1 V
4 -1 V
5 -1 V
6 -1 V
8 -2 V
8 -2 V
9 -2 V
10 -2 V
10 -2 V
11 -2 V
10 -3 V
10 -2 V
9 -2 V
9 -2 V
7 -1 V
6 -2 V
5 -1 V
4 -1 V
2 0 V
1 0 V
2 -1 V
3 -1 V
5 -1 V
6 -1 V
8 -2 V
8 -2 V
9 -2 V
10 -2 V
11 -2 V
10 -2 V
10 -3 V
10 -2 V
9 -2 V
9 -2 V
7 -1 V
6 -2 V
5 -1 V
4 -1 V
2 0 V
1 0 V
2 -1 V
3 -1 V
5 -1 V
6 -1 V
8 -2 V
8 -2 V
10 -2 V
10 -2 V
10 -2 V
10 -2 V
10 -3 V
10 -2 V
9 -2 V
9 -2 V
7 -2 V
7 -1 V
5 -1 V
3 -1 V
2 0 V
1 0 V
2 -1 V
3 -1 V
5 -1 V
7 -1 V
7 -2 V
9 -2 V
9 -2 V
10 -2 V
10 -2 V
10 -3 V
10 -2 V
10 -2 V
10 -2 V
8 -2 V
8 -2 V
6 -1 V
5 -1 V
3 -1 V
2 -1 V
1 0 V
2 0 V
4 -1 V
5 -1 V
6 -1 V
7 -2 V
9 -2 V
9 -2 V
10 -2 V
10 -3 V
10 -2 V
11 -2 V
stroke 2111 1078 M
10 -2 V
9 -2 V
8 -2 V
8 -2 V
6 -1 V
5 -2 V
3 0 V
2 -1 V
1 0 V
2 0 V
4 -1 V
5 -1 V
6 -2 V
7 -1 V
9 -2 V
9 -2 V
10 -3 V
10 -2 V
11 -2 V
10 -2 V
10 -3 V
9 -2 V
8 -2 V
8 -1 V
6 -2 V
5 -1 V
4 -1 V
2 0 V
1 0 V
1 -1 V
4 0 V
5 -2 V
6 -1 V
8 -2 V
8 -1 V
9 -3 V
10 -2 V
10 -2 V
11 -2 V
10 -3 V
10 -2 V
9 -2 V
9 -2 V
7 -1 V
6 -2 V
5 -1 V
4 -1 V
2 0 V
1 0 V
2 -1 V
3 -1 V
5 -1 V
6 -1 V
8 -2 V
8 -2 V
10 -2 V
9 -2 V
11 -2 V
10 -2 V
10 -3 V
10 -2 V
9 -2 V
9 -2 V
7 -1 V
7 -2 V
4 -1 V
4 -1 V
2 0 V
1 0 V
2 -1 V
3 -1 V
5 -1 V
7 -1 V
7 -2 V
8 -2 V
10 -2 V
10 -2 V
10 -2 V
10 -2 V
10 -2 V
10 -2 V
10 -2 V
8 -2 V
7 -2 V
7 0 V
5 -2 V
3 3 V
2 2 V
stroke
LT3
LCb setrgbcolor
LT3
1928 1856 M
543 0 V
0 3086 M
2 0 V
4 0 V
5 0 V
6 1 V
8 0 V
8 1 V
9 0 V
10 1 V
10 0 V
11 1 V
10 0 V
10 1 V
9 1 V
9 0 V
7 1 V
6 0 V
5 0 V
4 0 V
2 1 V
1 0 V
2 0 V
3 0 V
5 0 V
6 1 V
8 0 V
8 0 V
10 1 V
9 1 V
11 0 V
10 1 V
10 0 V
10 1 V
9 1 V
9 0 V
7 0 V
6 1 V
5 0 V
4 0 V
2 0 V
1 1 V
2 0 V
3 0 V
5 0 V
7 1 V
7 0 V
8 0 V
10 1 V
10 1 V
10 0 V
10 1 V
10 0 V
10 1 V
9 1 V
9 0 V
7 0 V
7 1 V
5 0 V
3 0 V
2 0 V
1 1 V
2 0 V
3 0 V
5 0 V
7 1 V
7 0 V
9 0 V
9 1 V
10 1 V
10 0 V
10 1 V
11 0 V
9 1 V
10 1 V
8 0 V
8 1 V
6 0 V
5 0 V
3 0 V
2 1 V
1 0 V
2 0 V
4 0 V
5 0 V
6 1 V
7 0 V
9 1 V
9 0 V
10 1 V
10 0 V
10 1 V
11 0 V
10 1 V
9 1 V
8 0 V
8 1 V
6 0 V
5 0 V
3 0 V
2 1 V
1 0 V
2 0 V
4 0 V
stroke 682 3126 M
5 0 V
6 1 V
7 0 V
9 1 V
9 0 V
10 1 V
10 0 V
11 1 V
10 1 V
10 0 V
9 1 V
8 0 V
8 1 V
6 0 V
5 0 V
4 1 V
2 0 V
1 0 V
2 0 V
3 0 V
5 0 V
6 1 V
8 0 V
8 1 V
9 0 V
10 1 V
10 0 V
11 1 V
10 1 V
10 0 V
9 1 V
9 0 V
7 1 V
6 0 V
5 0 V
4 1 V
2 0 V
1 0 V
2 0 V
3 0 V
5 0 V
6 1 V
8 0 V
8 1 V
10 0 V
9 1 V
11 1 V
10 0 V
10 1 V
10 0 V
9 1 V
9 0 V
7 1 V
7 0 V
5 1 V
3 0 V
2 0 V
1 0 V
2 0 V
3 0 V
5 1 V
7 0 V
7 0 V
8 1 V
10 0 V
10 1 V
10 1 V
10 0 V
10 1 V
10 0 V
10 1 V
8 0 V
7 1 V
7 0 V
5 1 V
3 0 V
2 0 V
1 0 V
2 0 V
4 0 V
4 1 V
7 0 V
7 0 V
9 1 V
9 0 V
10 1 V
10 1 V
10 0 V
11 1 V
9 0 V
10 1 V
8 1 V
8 0 V
6 0 V
5 1 V
3 0 V
2 0 V
1 0 V
2 0 V
4 0 V
5 1 V
6 0 V
7 0 V
9 1 V
stroke 1385 3168 M
9 1 V
10 0 V
10 1 V
11 0 V
10 1 V
10 1 V
9 0 V
8 1 V
8 0 V
6 0 V
5 1 V
4 0 V
2 0 V
2 0 V
4 0 V
5 1 V
6 0 V
8 0 V
8 1 V
9 1 V
10 0 V
10 1 V
11 0 V
10 1 V
10 0 V
9 1 V
9 1 V
7 0 V
6 0 V
5 1 V
4 0 V
2 0 V
1 0 V
2 0 V
3 0 V
5 1 V
6 0 V
8 0 V
8 1 V
9 0 V
10 1 V
11 1 V
10 0 V
10 1 V
10 0 V
9 1 V
9 0 V
7 1 V
6 0 V
5 1 V
4 0 V
2 0 V
1 0 V
2 0 V
3 0 V
5 1 V
6 0 V
8 0 V
8 1 V
10 0 V
10 1 V
10 0 V
10 1 V
10 1 V
10 0 V
9 1 V
9 0 V
7 1 V
7 0 V
5 0 V
3 1 V
2 0 V
1 0 V
2 0 V
3 0 V
5 0 V
7 1 V
7 0 V
9 1 V
9 0 V
10 1 V
10 0 V
10 1 V
10 0 V
10 1 V
10 0 V
8 1 V
8 0 V
6 1 V
5 0 V
3 0 V
2 0 V
1 0 V
2 0 V
4 1 V
5 0 V
6 0 V
7 1 V
9 0 V
9 1 V
10 0 V
10 1 V
10 0 V
11 1 V
stroke 2111 3210 M
10 1 V
9 0 V
8 1 V
8 0 V
6 0 V
5 1 V
3 0 V
2 0 V
1 0 V
2 0 V
4 0 V
5 0 V
6 1 V
7 0 V
9 1 V
9 0 V
10 1 V
10 0 V
11 1 V
10 0 V
10 1 V
9 0 V
8 1 V
8 0 V
6 1 V
5 0 V
4 0 V
2 0 V
1 0 V
1 0 V
4 1 V
5 0 V
6 0 V
8 1 V
8 0 V
9 1 V
10 0 V
10 1 V
11 0 V
10 1 V
10 0 V
9 1 V
9 0 V
7 1 V
6 0 V
5 0 V
4 0 V
2 0 V
1 0 V
2 1 V
3 0 V
5 0 V
6 0 V
8 1 V
8 0 V
10 1 V
9 0 V
11 1 V
10 0 V
10 1 V
10 0 V
9 1 V
9 0 V
7 0 V
7 1 V
4 0 V
4 0 V
2 0 V
1 0 V
2 0 V
3 0 V
5 1 V
7 0 V
7 0 V
8 1 V
10 0 V
10 0 V
10 1 V
10 0 V
10 1 V
10 0 V
10 -1 V
8 1 V
7 0 V
7 -2 V
5 1 V
3 -9 V
2 -6 V
stroke
LTb
0 3431 N
0 640 L
2879 0 V
0 2791 V
0 3431 L
Z stroke
1.000 UP
1.000 UL
LTb
1.000 UL
LTb
4320 640 M
63 0 V
2816 0 R
-63 0 V
4320 950 M
63 0 V
2816 0 R
-63 0 V
4320 1260 M
63 0 V
2816 0 R
-63 0 V
4320 1570 M
63 0 V
2816 0 R
-63 0 V
4320 1880 M
63 0 V
2816 0 R
-63 0 V
4320 2191 M
63 0 V
2816 0 R
-63 0 V
4320 2501 M
63 0 V
2816 0 R
-63 0 V
4320 2811 M
63 0 V
2816 0 R
-63 0 V
4320 3121 M
63 0 V
2816 0 R
-63 0 V
4320 3431 M
63 0 V
2816 0 R
-63 0 V
4320 640 M
0 63 V
0 2728 R
0 -63 V
4640 640 M
0 63 V
0 2728 R
0 -63 V
4960 640 M
0 63 V
0 2728 R
0 -63 V
5280 640 M
0 63 V
0 2728 R
0 -63 V
5600 640 M
0 63 V
0 2728 R
0 -63 V
5919 640 M
0 63 V
0 2728 R
0 -63 V
6239 640 M
0 63 V
0 2728 R
0 -63 V
6559 640 M
0 63 V
0 2728 R
0 -63 V
6879 640 M
0 63 V
0 2728 R
0 -63 V
7199 640 M
0 63 V
0 2728 R
0 -63 V
stroke
LTa
4320 1570 M
2879 0 V
stroke
LTb
4320 3431 N
0 -2791 V
2879 0 V
0 2791 V
-2879 0 V
Z stroke
LCb setrgbcolor
LTb
1.000 UP
1.000 UL
LTb
1.000 UL
LT0
LCb setrgbcolor
LT0
6248 2680 M
543 0 V
4320 2936 M
2 1 V
4 0 V
4 1 V
6 1 V
7 1 V
8 1 V
9 2 V
10 2 V
9 1 V
10 2 V
10 1 V
9 2 V
9 1 V
8 2 V
7 1 V
6 1 V
5 1 V
3 0 V
2 1 V
1 0 V
2 0 V
3 1 V
5 0 V
6 1 V
7 2 V
8 1 V
8 1 V
10 2 V
9 2 V
10 1 V
10 2 V
9 1 V
9 2 V
8 1 V
7 1 V
6 1 V
5 1 V
3 1 V
2 0 V
1 0 V
2 0 V
3 1 V
5 1 V
6 1 V
7 1 V
8 1 V
9 2 V
9 1 V
10 2 V
10 2 V
9 1 V
10 2 V
8 1 V
8 2 V
7 1 V
6 1 V
5 1 V
3 0 V
2 0 V
1 1 V
2 0 V
3 0 V
5 1 V
6 1 V
7 1 V
8 2 V
9 1 V
9 2 V
10 1 V
10 2 V
9 2 V
10 1 V
9 2 V
8 1 V
7 1 V
6 1 V
4 1 V
4 0 V
2 1 V
2 0 V
4 1 V
4 1 V
6 1 V
7 1 V
8 1 V
9 1 V
10 2 V
9 2 V
10 1 V
10 2 V
9 1 V
9 2 V
8 1 V
7 1 V
6 1 V
5 1 V
3 1 V
2 0 V
1 0 V
2 1 V
3 0 V
5 1 V
stroke 4972 3045 M
6 1 V
7 1 V
8 1 V
8 2 V
10 1 V
9 2 V
10 2 V
10 1 V
9 2 V
9 1 V
8 2 V
7 1 V
6 1 V
5 0 V
3 1 V
2 0 V
1 0 V
2 1 V
3 0 V
5 1 V
6 1 V
7 1 V
8 2 V
9 1 V
9 2 V
10 1 V
10 2 V
9 1 V
10 2 V
8 1 V
8 2 V
7 1 V
6 1 V
5 1 V
3 0 V
2 1 V
1 0 V
2 0 V
3 1 V
5 0 V
6 1 V
7 1 V
8 2 V
9 1 V
9 2 V
10 1 V
10 2 V
9 2 V
10 1 V
9 2 V
8 1 V
7 1 V
6 1 V
4 1 V
4 0 V
2 1 V
2 0 V
4 1 V
4 0 V
6 1 V
7 2 V
8 1 V
9 1 V
9 2 V
10 1 V
10 2 V
10 2 V
9 1 V
9 2 V
8 1 V
7 1 V
6 1 V
5 1 V
3 0 V
2 1 V
1 0 V
2 0 V
3 1 V
5 0 V
5 1 V
8 1 V
8 2 V
8 1 V
10 2 V
9 1 V
10 2 V
10 2 V
9 1 V
9 1 V
8 2 V
7 1 V
6 1 V
5 1 V
3 0 V
2 0 V
1 1 V
2 0 V
3 0 V
5 1 V
6 1 V
7 1 V
8 2 V
9 1 V
9 2 V
stroke 5652 3157 M
10 1 V
10 2 V
9 1 V
10 2 V
8 1 V
8 2 V
7 1 V
6 1 V
5 0 V
3 1 V
2 0 V
1 0 V
2 1 V
3 0 V
5 1 V
6 1 V
7 1 V
8 1 V
9 2 V
9 1 V
10 2 V
10 1 V
9 2 V
10 1 V
9 2 V
8 1 V
7 1 V
6 1 V
4 1 V
4 0 V
2 1 V
2 0 V
4 1 V
4 0 V
6 1 V
7 2 V
8 1 V
9 1 V
9 2 V
10 1 V
10 2 V
10 1 V
9 2 V
9 1 V
8 2 V
7 1 V
6 1 V
5 0 V
3 1 V
2 0 V
1 0 V
2 1 V
3 0 V
5 1 V
5 1 V
8 1 V
8 1 V
8 2 V
10 1 V
9 2 V
10 1 V
10 2 V
9 1 V
9 1 V
8 2 V
7 1 V
6 1 V
5 0 V
3 1 V
2 0 V
1 0 V
2 1 V
3 0 V
5 1 V
6 1 V
7 1 V
8 1 V
9 2 V
9 1 V
10 2 V
9 1 V
10 2 V
10 1 V
8 1 V
8 2 V
7 1 V
6 1 V
5 0 V
3 1 V
2 0 V
1 0 V
2 1 V
3 0 V
5 1 V
6 1 V
7 1 V
8 1 V
9 2 V
9 1 V
10 1 V
10 2 V
9 1 V
10 2 V
9 1 V
stroke 6341 3265 M
8 1 V
7 2 V
6 0 V
4 1 V
4 1 V
2 0 V
2 0 V
4 1 V
4 1 V
6 1 V
7 1 V
8 1 V
9 1 V
9 2 V
10 1 V
10 2 V
10 1 V
9 1 V
9 2 V
8 1 V
7 1 V
6 1 V
5 1 V
3 0 V
2 0 V
1 1 V
2 0 V
3 0 V
5 1 V
5 1 V
8 1 V
8 1 V
8 1 V
10 2 V
9 1 V
10 2 V
10 1 V
9 2 V
9 1 V
8 1 V
7 1 V
6 1 V
5 1 V
3 0 V
2 1 V
1 0 V
2 0 V
3 0 V
5 1 V
6 1 V
7 1 V
8 1 V
9 2 V
9 1 V
10 1 V
9 2 V
10 1 V
10 2 V
8 1 V
8 1 V
7 1 V
6 1 V
5 1 V
3 0 V
2 0 V
1 0 V
2 1 V
3 0 V
5 1 V
6 1 V
7 1 V
8 1 V
9 1 V
9 2 V
10 1 V
10 1 V
9 2 V
10 1 V
9 1 V
8 2 V
7 1 V
6 0 V
4 1 V
4 1 V
2 0 V
stroke
LT1
LCb setrgbcolor
LT1
6248 2480 M
543 0 V
4320 835 M
2 0 V
4 1 V
4 0 V
6 1 V
7 1 V
8 0 V
9 1 V
10 2 V
9 1 V
10 1 V
10 1 V
9 1 V
9 1 V
8 1 V
7 1 V
6 0 V
5 1 V
3 0 V
2 1 V
1 0 V
2 0 V
3 0 V
5 1 V
6 1 V
7 0 V
8 1 V
8 1 V
10 2 V
9 1 V
10 1 V
10 1 V
9 1 V
9 1 V
8 1 V
7 1 V
6 1 V
5 0 V
3 1 V
2 0 V
1 0 V
2 0 V
3 1 V
5 0 V
6 1 V
7 1 V
8 1 V
9 1 V
9 1 V
10 2 V
10 1 V
9 1 V
10 1 V
8 1 V
8 1 V
7 1 V
6 1 V
5 1 V
3 0 V
2 0 V
1 0 V
2 1 V
3 0 V
5 1 V
6 1 V
7 0 V
8 2 V
9 1 V
9 1 V
10 1 V
10 1 V
9 2 V
10 1 V
9 1 V
8 1 V
7 1 V
6 1 V
4 1 V
4 0 V
2 0 V
0 1 V
2 0 V
4 0 V
4 1 V
6 1 V
7 1 V
8 1 V
9 1 V
10 1 V
9 2 V
10 1 V
10 1 V
9 2 V
9 1 V
8 1 V
7 1 V
6 1 V
5 0 V
3 1 V
2 0 V
1 0 V
2 1 V
3 0 V
stroke 4967 917 M
5 1 V
6 1 V
7 1 V
8 1 V
8 1 V
10 1 V
9 2 V
10 1 V
10 2 V
9 1 V
9 1 V
8 1 V
7 2 V
6 0 V
5 1 V
3 1 V
2 0 V
1 0 V
2 0 V
3 1 V
5 0 V
6 1 V
7 1 V
8 2 V
9 1 V
9 1 V
10 2 V
10 1 V
9 2 V
10 1 V
8 2 V
8 1 V
7 1 V
6 1 V
5 0 V
3 1 V
2 0 V
1 0 V
2 1 V
3 0 V
5 1 V
6 1 V
7 1 V
8 1 V
9 2 V
9 1 V
10 2 V
10 1 V
9 2 V
10 1 V
9 2 V
8 1 V
7 1 V
6 1 V
4 1 V
4 0 V
2 0 V
0 1 V
2 0 V
4 0 V
4 1 V
6 1 V
7 1 V
8 2 V
9 1 V
9 2 V
10 1 V
10 2 V
10 1 V
9 2 V
9 1 V
8 2 V
7 1 V
6 1 V
5 1 V
3 0 V
2 0 V
1 1 V
2 0 V
3 0 V
5 1 V
5 1 V
8 1 V
8 2 V
8 1 V
10 2 V
9 1 V
10 2 V
10 2 V
9 1 V
9 2 V
8 1 V
7 1 V
6 1 V
5 1 V
3 1 V
2 0 V
1 0 V
2 1 V
3 0 V
5 1 V
6 1 V
7 1 V
8 2 V
stroke 5634 1023 M
9 1 V
9 2 V
10 2 V
10 1 V
9 2 V
10 2 V
8 1 V
8 2 V
7 1 V
6 1 V
5 1 V
3 0 V
2 1 V
1 0 V
2 0 V
3 1 V
5 1 V
6 1 V
7 1 V
8 1 V
9 2 V
9 2 V
10 1 V
10 2 V
9 2 V
10 2 V
9 1 V
8 2 V
7 1 V
6 1 V
4 1 V
4 1 V
2 0 V
2 0 V
4 1 V
4 1 V
6 1 V
7 1 V
8 2 V
9 1 V
9 2 V
10 2 V
10 2 V
10 2 V
9 1 V
9 2 V
8 2 V
7 1 V
6 1 V
5 1 V
3 1 V
2 0 V
1 0 V
2 0 V
3 1 V
5 1 V
5 1 V
8 1 V
8 2 V
8 2 V
10 1 V
9 2 V
10 2 V
10 2 V
9 2 V
9 2 V
8 1 V
7 2 V
6 1 V
5 1 V
3 0 V
2 1 V
1 0 V
2 0 V
3 1 V
5 1 V
6 1 V
7 1 V
8 2 V
9 2 V
9 1 V
10 2 V
9 2 V
10 2 V
10 2 V
8 2 V
8 2 V
7 1 V
6 1 V
5 1 V
3 1 V
2 0 V
1 0 V
2 1 V
3 0 V
5 1 V
6 1 V
7 2 V
8 2 V
9 1 V
9 2 V
10 2 V
10 2 V
9 2 V
stroke 6322 1153 M
10 2 V
9 2 V
8 2 V
7 1 V
6 1 V
4 1 V
4 1 V
2 0 V
2 1 V
4 1 V
4 0 V
6 2 V
7 1 V
8 2 V
9 2 V
9 2 V
10 2 V
10 2 V
10 2 V
9 2 V
9 2 V
8 1 V
7 2 V
6 1 V
5 1 V
3 1 V
2 0 V
1 0 V
2 1 V
3 0 V
5 1 V
5 2 V
8 1 V
8 2 V
8 2 V
10 2 V
9 2 V
10 2 V
10 2 V
9 2 V
9 2 V
8 2 V
7 1 V
6 1 V
5 1 V
3 1 V
2 1 V
1 0 V
2 0 V
3 1 V
5 1 V
6 1 V
7 2 V
8 1 V
9 2 V
9 2 V
10 3 V
9 2 V
10 2 V
10 2 V
8 2 V
8 2 V
7 1 V
6 2 V
5 1 V
3 0 V
2 1 V
1 0 V
2 0 V
3 1 V
5 1 V
6 1 V
7 2 V
8 2 V
9 2 V
9 2 V
10 2 V
10 2 V
9 2 V
10 2 V
9 2 V
8 2 V
7 2 V
6 1 V
4 1 V
4 1 V
2 0 V
stroke
LT2
LCb setrgbcolor
LT2
6248 2280 M
543 0 V
4320 890 M
2 0 V
4 0 V
4 -1 V
6 0 V
7 0 V
8 -1 V
9 0 V
10 0 V
9 -1 V
10 0 V
10 -1 V
9 0 V
9 -1 V
8 0 V
7 0 V
6 -1 V
5 0 V
3 0 V
2 0 V
1 0 V
2 0 V
3 0 V
5 -1 V
6 0 V
7 0 V
8 -1 V
8 0 V
10 0 V
9 -1 V
10 0 V
10 -1 V
9 0 V
9 -1 V
8 0 V
7 0 V
6 -1 V
5 0 V
3 0 V
2 0 V
1 0 V
2 0 V
3 0 V
5 0 V
6 -1 V
7 0 V
8 0 V
9 -1 V
9 0 V
10 -1 V
10 0 V
9 -1 V
10 0 V
8 0 V
8 -1 V
7 0 V
6 0 V
5 0 V
3 -1 V
2 0 V
1 0 V
2 0 V
3 0 V
5 0 V
6 0 V
7 -1 V
8 0 V
9 0 V
9 -1 V
10 0 V
10 -1 V
9 0 V
10 0 V
9 -1 V
8 0 V
7 0 V
6 -1 V
4 0 V
4 0 V
2 0 V
2 0 V
4 0 V
4 -1 V
6 0 V
7 0 V
8 0 V
9 -1 V
10 0 V
9 -1 V
10 0 V
10 0 V
9 -1 V
9 0 V
8 0 V
7 -1 V
6 0 V
5 0 V
3 0 V
2 0 V
1 0 V
2 -1 V
3 0 V
5 0 V
stroke 4972 861 M
6 0 V
7 0 V
8 -1 V
8 0 V
10 0 V
9 -1 V
10 0 V
10 -1 V
9 0 V
9 0 V
8 -1 V
7 0 V
6 0 V
5 0 V
3 0 V
2 0 V
1 0 V
2 -1 V
3 0 V
5 0 V
6 0 V
7 0 V
8 -1 V
9 0 V
9 0 V
10 -1 V
10 0 V
9 0 V
10 -1 V
8 0 V
8 0 V
7 -1 V
6 0 V
5 0 V
3 0 V
2 0 V
1 0 V
2 0 V
3 0 V
5 -1 V
6 0 V
7 0 V
8 0 V
9 -1 V
9 0 V
10 0 V
10 -1 V
9 0 V
10 0 V
9 -1 V
8 0 V
7 0 V
6 0 V
4 -1 V
4 0 V
2 0 V
2 0 V
4 0 V
4 0 V
6 0 V
7 -1 V
8 0 V
9 0 V
9 -1 V
10 0 V
10 0 V
10 0 V
9 -1 V
9 0 V
8 0 V
7 -1 V
6 0 V
5 0 V
3 0 V
2 0 V
1 0 V
2 0 V
3 0 V
5 0 V
5 -1 V
8 0 V
8 0 V
8 0 V
10 -1 V
9 0 V
10 0 V
10 -1 V
9 0 V
9 0 V
8 0 V
7 -1 V
6 0 V
5 0 V
3 0 V
2 0 V
1 0 V
2 0 V
3 0 V
5 0 V
6 -1 V
7 0 V
8 0 V
9 0 V
9 -1 V
stroke 5652 837 M
10 0 V
10 0 V
9 -1 V
10 0 V
8 0 V
8 0 V
7 0 V
6 -1 V
5 0 V
3 0 V
2 0 V
1 0 V
2 0 V
3 0 V
5 0 V
6 0 V
7 -1 V
8 0 V
9 0 V
9 0 V
10 -1 V
10 0 V
9 0 V
10 0 V
9 -1 V
8 0 V
7 0 V
6 0 V
4 0 V
4 0 V
2 0 V
2 -1 V
4 0 V
4 0 V
6 0 V
7 0 V
8 0 V
9 0 V
9 -1 V
10 0 V
10 0 V
10 0 V
9 -1 V
9 0 V
8 0 V
7 0 V
6 0 V
5 -1 V
3 0 V
2 0 V
1 0 V
2 0 V
3 0 V
5 0 V
5 0 V
8 0 V
8 0 V
8 -1 V
10 0 V
9 0 V
10 0 V
10 -1 V
9 0 V
9 0 V
8 0 V
7 0 V
6 0 V
5 0 V
3 -1 V
2 0 V
1 0 V
2 0 V
3 0 V
5 0 V
6 0 V
7 0 V
8 0 V
9 0 V
9 -1 V
10 0 V
9 0 V
10 0 V
10 0 V
8 -1 V
8 0 V
7 0 V
6 0 V
5 0 V
3 0 V
2 0 V
1 0 V
2 0 V
3 0 V
5 0 V
6 -1 V
7 0 V
8 0 V
9 0 V
9 0 V
10 0 V
10 -1 V
9 0 V
10 0 V
9 0 V
stroke 6341 821 M
8 0 V
7 0 V
6 0 V
4 0 V
4 0 V
2 -1 V
2 0 V
4 0 V
4 0 V
6 0 V
7 0 V
8 0 V
9 0 V
9 0 V
10 0 V
10 -1 V
10 0 V
9 0 V
9 0 V
8 0 V
7 0 V
6 0 V
5 0 V
3 0 V
2 -1 V
1 0 V
2 0 V
3 0 V
5 0 V
5 0 V
8 0 V
8 0 V
8 0 V
10 0 V
9 0 V
10 0 V
10 -1 V
9 0 V
9 0 V
8 0 V
7 0 V
6 0 V
5 0 V
3 0 V
2 0 V
1 0 V
2 0 V
3 0 V
5 0 V
6 0 V
7 0 V
8 -1 V
9 0 V
9 0 V
10 0 V
9 0 V
10 0 V
10 0 V
8 0 V
8 0 V
7 0 V
6 0 V
5 0 V
3 -1 V
2 0 V
1 0 V
2 0 V
3 0 V
5 0 V
6 0 V
7 0 V
8 0 V
9 0 V
9 0 V
10 0 V
10 0 V
9 0 V
10 0 V
9 0 V
8 0 V
7 0 V
6 0 V
4 0 V
4 0 V
2 0 V
stroke
LT3
LCb setrgbcolor
LT3
6248 2080 M
543 0 V
4320 2774 M
2 0 V
4 1 V
4 1 V
6 0 V
7 1 V
8 1 V
9 2 V
10 1 V
9 1 V
10 1 V
10 2 V
9 1 V
9 1 V
8 1 V
7 1 V
6 1 V
5 1 V
3 0 V
2 0 V
1 0 V
2 1 V
3 0 V
5 1 V
6 0 V
7 1 V
8 1 V
8 2 V
10 1 V
9 1 V
10 1 V
10 2 V
9 1 V
9 1 V
8 1 V
7 1 V
6 1 V
5 1 V
3 0 V
2 0 V
1 0 V
2 1 V
3 0 V
5 1 V
6 0 V
7 1 V
8 1 V
9 2 V
9 1 V
10 1 V
10 1 V
9 2 V
10 1 V
8 1 V
8 1 V
7 1 V
6 1 V
5 1 V
3 0 V
2 0 V
1 0 V
2 1 V
3 0 V
5 1 V
6 0 V
7 1 V
8 1 V
9 2 V
9 1 V
10 1 V
10 1 V
9 2 V
10 1 V
9 1 V
8 1 V
7 1 V
6 1 V
4 0 V
4 1 V
2 0 V
2 0 V
4 1 V
4 0 V
6 1 V
7 1 V
8 1 V
9 1 V
10 2 V
9 1 V
10 1 V
10 1 V
9 2 V
9 1 V
8 1 V
7 1 V
6 0 V
5 1 V
3 0 V
2 1 V
1 0 V
2 0 V
3 1 V
5 0 V
stroke 4972 2860 M
6 1 V
7 1 V
8 1 V
8 1 V
10 1 V
9 1 V
10 2 V
10 1 V
9 1 V
9 1 V
8 1 V
7 1 V
6 1 V
5 1 V
3 0 V
2 0 V
1 0 V
2 1 V
3 0 V
5 1 V
6 0 V
7 1 V
8 1 V
9 1 V
9 2 V
10 1 V
10 1 V
9 1 V
10 2 V
8 1 V
8 1 V
7 1 V
6 0 V
5 1 V
3 0 V
2 1 V
1 0 V
2 0 V
3 0 V
5 1 V
6 1 V
7 1 V
8 1 V
9 1 V
9 1 V
10 1 V
10 1 V
9 2 V
10 1 V
9 1 V
8 1 V
7 1 V
6 1 V
4 0 V
4 1 V
2 0 V
2 0 V
4 1 V
4 0 V
6 1 V
7 1 V
8 1 V
9 1 V
9 1 V
10 1 V
10 1 V
10 2 V
9 1 V
9 1 V
8 1 V
7 1 V
6 1 V
5 0 V
3 1 V
2 0 V
1 0 V
2 0 V
3 0 V
5 1 V
5 1 V
8 1 V
8 1 V
8 1 V
10 1 V
9 1 V
10 1 V
10 1 V
9 2 V
9 1 V
8 1 V
7 1 V
6 0 V
5 1 V
3 0 V
2 0 V
1 1 V
2 0 V
3 0 V
5 1 V
6 1 V
7 0 V
8 1 V
9 1 V
9 2 V
stroke 5652 2946 M
10 1 V
10 1 V
9 1 V
10 1 V
8 1 V
8 1 V
7 1 V
6 1 V
5 0 V
3 1 V
2 0 V
1 0 V
2 0 V
3 1 V
5 0 V
6 1 V
7 1 V
8 1 V
9 1 V
9 1 V
10 1 V
10 1 V
9 1 V
10 2 V
9 1 V
8 0 V
7 1 V
6 1 V
4 1 V
4 0 V
2 0 V
2 1 V
4 0 V
4 0 V
6 1 V
7 1 V
8 1 V
9 1 V
9 1 V
10 1 V
10 1 V
10 1 V
9 2 V
9 1 V
8 0 V
7 1 V
6 1 V
5 1 V
3 0 V
2 0 V
1 0 V
2 0 V
3 1 V
5 0 V
5 1 V
8 1 V
8 1 V
8 1 V
10 1 V
9 1 V
10 1 V
10 1 V
9 1 V
9 1 V
8 1 V
7 1 V
6 1 V
5 0 V
3 1 V
2 0 V
1 0 V
2 0 V
3 0 V
5 1 V
6 1 V
7 0 V
8 1 V
9 1 V
9 1 V
10 1 V
9 1 V
10 2 V
10 1 V
8 1 V
8 0 V
7 1 V
6 1 V
5 0 V
3 1 V
2 0 V
1 0 V
2 0 V
3 1 V
5 0 V
6 1 V
7 1 V
8 0 V
9 1 V
9 1 V
10 2 V
10 1 V
9 1 V
10 1 V
9 1 V
stroke 6341 3025 M
8 0 V
7 1 V
6 1 V
4 0 V
4 1 V
2 0 V
2 0 V
4 1 V
4 0 V
6 1 V
7 0 V
8 1 V
9 1 V
9 1 V
10 1 V
10 1 V
10 1 V
9 1 V
9 1 V
8 1 V
7 1 V
6 0 V
5 1 V
3 0 V
2 0 V
1 1 V
2 0 V
3 0 V
5 1 V
5 0 V
8 1 V
8 1 V
8 1 V
10 1 V
9 1 V
10 1 V
10 1 V
9 1 V
9 0 V
8 1 V
7 1 V
6 1 V
5 0 V
3 0 V
2 1 V
1 0 V
2 0 V
3 0 V
5 1 V
6 0 V
7 1 V
8 1 V
9 1 V
9 0 V
10 1 V
9 1 V
10 1 V
10 1 V
8 1 V
8 1 V
7 1 V
6 0 V
5 1 V
3 0 V
2 0 V
1 0 V
2 0 V
3 1 V
5 0 V
6 1 V
7 0 V
8 1 V
9 1 V
9 1 V
10 1 V
10 1 V
9 1 V
10 1 V
9 0 V
8 1 V
7 1 V
6 0 V
4 1 V
4 -1 V
2 -1 V
stroke
LTb
4320 3431 N
0 -2791 V
2879 0 V
0 2791 V
-2879 0 V
Z stroke
1.000 UP
1.000 UL
LTb
stroke
grestore
end
showpage
  }}%
  \put(6128,2080){\makebox(0,0)[r]{\strut{}A$_-$}}%
  \put(6128,2280){\makebox(0,0)[r]{\strut{}A$_+$}}%
  \put(6128,2480){\makebox(0,0)[r]{\strut{}A$_0$}}%
  \put(6128,2680){\makebox(0,0)[r]{\strut{}V}}%
  \put(5759,3731){\makebox(0,0){\strut{}$\bar B_s\to c\bar s(^3P_1)$}}%
  \put(5759,140){\makebox(0,0){\strut{}$q^2$ [$\rm GeV^2$]}}%
  \put(7199,440){\makebox(0,0){\strut{} 9}}%
  \put(6879,440){\makebox(0,0){\strut{} 8}}%
  \put(6559,440){\makebox(0,0){\strut{} 7}}%
  \put(6239,440){\makebox(0,0){\strut{} 6}}%
  \put(5919,440){\makebox(0,0){\strut{} 5}}%
  \put(5600,440){\makebox(0,0){\strut{} 4}}%
  \put(5280,440){\makebox(0,0){\strut{} 3}}%
  \put(4960,440){\makebox(0,0){\strut{} 2}}%
  \put(4640,440){\makebox(0,0){\strut{} 1}}%
  \put(4320,440){\makebox(0,0){\strut{} 0}}%
  \put(4200,3431){\makebox(0,0)[r]{\strut{} 1.2}}%
  \put(4200,3121){\makebox(0,0)[r]{\strut{} 1}}%
  \put(4200,2811){\makebox(0,0)[r]{\strut{} 0.8}}%
  \put(4200,2501){\makebox(0,0)[r]{\strut{} 0.6}}%
  \put(4200,2191){\makebox(0,0)[r]{\strut{} 0.4}}%
  \put(4200,1880){\makebox(0,0)[r]{\strut{} 0.2}}%
  \put(4200,1570){\makebox(0,0)[r]{\strut{} 0}}%
  \put(4200,1260){\makebox(0,0)[r]{\strut{}-0.2}}%
  \put(4200,950){\makebox(0,0)[r]{\strut{}-0.4}}%
  \put(4200,640){\makebox(0,0)[r]{\strut{}-0.6}}%
  \put(1808,1856){\makebox(0,0)[r]{\strut{}A$_-$}}%
  \put(1808,2056){\makebox(0,0)[r]{\strut{}A$_+$}}%
  \put(1808,2256){\makebox(0,0)[r]{\strut{}A$_0$}}%
  \put(1808,2456){\makebox(0,0)[r]{\strut{}V}}%
  \put(1439,3731){\makebox(0,0){\strut{}$\bar B_s\to c\bar s(^1P_1)$}}%
  \put(1439,140){\makebox(0,0){\strut{}$q^2$ [$\rm GeV^2$]}}%
  \put(2879,440){\makebox(0,0){\strut{} 9}}%
  \put(2559,440){\makebox(0,0){\strut{} 8}}%
  \put(2239,440){\makebox(0,0){\strut{} 7}}%
  \put(1919,440){\makebox(0,0){\strut{} 6}}%
  \put(1599,440){\makebox(0,0){\strut{} 5}}%
  \put(1280,440){\makebox(0,0){\strut{} 4}}%
  \put(960,440){\makebox(0,0){\strut{} 3}}%
  \put(640,440){\makebox(0,0){\strut{} 2}}%
  \put(320,440){\makebox(0,0){\strut{} 1}}%
  \put(0,440){\makebox(0,0){\strut{} 0}}%
  \put(-120,3431){\makebox(0,0)[r]{\strut{} 0.4}}%
  \put(-120,3082){\makebox(0,0)[r]{\strut{} 0.2}}%
  \put(-120,2733){\makebox(0,0)[r]{\strut{} 0}}%
  \put(-120,2384){\makebox(0,0)[r]{\strut{}-0.2}}%
  \put(-120,2035){\makebox(0,0)[r]{\strut{}-0.4}}%
  \put(-120,1687){\makebox(0,0)[r]{\strut{}-0.6}}%
  \put(-120,1338){\makebox(0,0)[r]{\strut{}-0.8}}%
  \put(-120,989){\makebox(0,0)[r]{\strut{}-1}}%
  \put(-120,640){\makebox(0,0)[r]{\strut{}-1.2}}%
\end{picture}%
\endgroup
 

%% file: ff.2m.2p.tex
\begingroup%
\makeatletter%
\newcommand{\GNUPLOTspecial}{%
  \@sanitize\catcode`\%=14\relax\special}%
\setlength{\unitlength}{0.0500bp}%
\begin{picture}(7200,5040)(0,0)%
  {\GNUPLOTspecial{"
/gnudict 256 dict def
gnudict begin
%
%
/Color false def
/Blacktext true def
/Solid false def
/Dashlength 1 def
/Landscape false def
/Level1 false def
/Rounded false def
/ClipToBoundingBox false def
/SuppressPDFMark false def
/TransparentPatterns false def
/gnulinewidth 5.000 def
/userlinewidth gnulinewidth def
/Gamma 1.0 def
/BackgroundColor {-1.000 -1.000 -1.000} def
/vshift -66 def
/dl1 {
  10.0 Dashlength mul mul
  Rounded { currentlinewidth 0.75 mul sub dup 0 le { pop 0.01 } if } if
} def
/dl2 {
  10.0 Dashlength mul mul
  Rounded { currentlinewidth 0.75 mul add } if
} def
/hpt_ 31.5 def
/vpt_ 31.5 def
/hpt hpt_ def
/vpt vpt_ def
/doclip {
  ClipToBoundingBox {
    newpath 0 0 moveto 360 0 lineto 360 252 lineto 0 252 lineto closepath
    clip
  } if
} def
%
%
%
/M {moveto} bind def
/L {lineto} bind def
/R {rmoveto} bind def
/V {rlineto} bind def
/N {newpath moveto} bind def
/Z {closepath} bind def
/C {setrgbcolor} bind def
/f {rlineto fill} bind def
/g {setgray} bind def
/Gshow {show} def   
/vpt2 vpt 2 mul def
/hpt2 hpt 2 mul def
/Lshow {currentpoint stroke M 0 vshift R 
	Blacktext {gsave 0 setgray show grestore} {show} ifelse} def
/Rshow {currentpoint stroke M dup stringwidth pop neg vshift R
	Blacktext {gsave 0 setgray show grestore} {show} ifelse} def
/Cshow {currentpoint stroke M dup stringwidth pop -2 div vshift R 
	Blacktext {gsave 0 setgray show grestore} {show} ifelse} def
/UP {dup vpt_ mul /vpt exch def hpt_ mul /hpt exch def
  /hpt2 hpt 2 mul def /vpt2 vpt 2 mul def} def
/DL {Color {setrgbcolor Solid {pop []} if 0 setdash}
 {pop pop pop 0 setgray Solid {pop []} if 0 setdash} ifelse} def
/BL {stroke userlinewidth 2 mul setlinewidth
	Rounded {1 setlinejoin 1 setlinecap} if} def
/AL {stroke userlinewidth 2 div setlinewidth
	Rounded {1 setlinejoin 1 setlinecap} if} def
/UL {dup gnulinewidth mul /userlinewidth exch def
	dup 1 lt {pop 1} if 10 mul /udl exch def} def
/PL {stroke userlinewidth setlinewidth
	Rounded {1 setlinejoin 1 setlinecap} if} def
3.8 setmiterlimit
/LCw {1 1 1} def
/LCb {0 0 0} def
/LCa {0 0 0} def
/LC0 {1 0 0} def
/LC1 {0 1 0} def
/LC2 {0 0 1} def
/LC3 {1 0 1} def
/LC4 {0 1 1} def
/LC5 {1 1 0} def
/LC6 {0 0 0} def
/LC7 {1 0.3 0} def
/LC8 {0.5 0.5 0.5} def
/LTw {PL [] 1 setgray} def
/LTb {BL [] LCb DL} def
/LTa {AL [1 udl mul 2 udl mul] 0 setdash LCa setrgbcolor} def
/LT0 {PL [] LC0 DL} def
/LT1 {PL [4 dl1 2 dl2] LC1 DL} def
/LT2 {PL [2 dl1 3 dl2] LC2 DL} def
/LT3 {PL [1 dl1 1.5 dl2] LC3 DL} def
/LT4 {PL [6 dl1 2 dl2 1 dl1 2 dl2] LC4 DL} def
/LT5 {PL [3 dl1 3 dl2 1 dl1 3 dl2] LC5 DL} def
/LT6 {PL [2 dl1 2 dl2 2 dl1 6 dl2] LC6 DL} def
/LT7 {PL [1 dl1 2 dl2 6 dl1 2 dl2 1 dl1 2 dl2] LC7 DL} def
/LT8 {PL [2 dl1 2 dl2 2 dl1 2 dl2 2 dl1 2 dl2 2 dl1 4 dl2] LC8 DL} def
/Pnt {stroke [] 0 setdash gsave 1 setlinecap M 0 0 V stroke grestore} def
/Dia {stroke [] 0 setdash 2 copy vpt add M
  hpt neg vpt neg V hpt vpt neg V
  hpt vpt V hpt neg vpt V closepath stroke
  Pnt} def
/Pls {stroke [] 0 setdash vpt sub M 0 vpt2 V
  currentpoint stroke M
  hpt neg vpt neg R hpt2 0 V stroke
 } def
/Box {stroke [] 0 setdash 2 copy exch hpt sub exch vpt add M
  0 vpt2 neg V hpt2 0 V 0 vpt2 V
  hpt2 neg 0 V closepath stroke
  Pnt} def
/Crs {stroke [] 0 setdash exch hpt sub exch vpt add M
  hpt2 vpt2 neg V currentpoint stroke M
  hpt2 neg 0 R hpt2 vpt2 V stroke} def
/TriU {stroke [] 0 setdash 2 copy vpt 1.12 mul add M
  hpt neg vpt -1.62 mul V
  hpt 2 mul 0 V
  hpt neg vpt 1.62 mul V closepath stroke
  Pnt} def
/Star {2 copy Pls Crs} def
/BoxF {stroke [] 0 setdash exch hpt sub exch vpt add M
  0 vpt2 neg V hpt2 0 V 0 vpt2 V
  hpt2 neg 0 V closepath fill} def
/TriUF {stroke [] 0 setdash vpt 1.12 mul add M
  hpt neg vpt -1.62 mul V
  hpt 2 mul 0 V
  hpt neg vpt 1.62 mul V closepath fill} def
/TriD {stroke [] 0 setdash 2 copy vpt 1.12 mul sub M
  hpt neg vpt 1.62 mul V
  hpt 2 mul 0 V
  hpt neg vpt -1.62 mul V closepath stroke
  Pnt} def
/TriDF {stroke [] 0 setdash vpt 1.12 mul sub M
  hpt neg vpt 1.62 mul V
  hpt 2 mul 0 V
  hpt neg vpt -1.62 mul V closepath fill} def
/DiaF {stroke [] 0 setdash vpt add M
  hpt neg vpt neg V hpt vpt neg V
  hpt vpt V hpt neg vpt V closepath fill} def
/Pent {stroke [] 0 setdash 2 copy gsave
  translate 0 hpt M 4 {72 rotate 0 hpt L} repeat
  closepath stroke grestore Pnt} def
/PentF {stroke [] 0 setdash gsave
  translate 0 hpt M 4 {72 rotate 0 hpt L} repeat
  closepath fill grestore} def
/Circle {stroke [] 0 setdash 2 copy
  hpt 0 360 arc stroke Pnt} def
/CircleF {stroke [] 0 setdash hpt 0 360 arc fill} def
/C0 {BL [] 0 setdash 2 copy moveto vpt 90 450 arc} bind def
/C1 {BL [] 0 setdash 2 copy moveto
	2 copy vpt 0 90 arc closepath fill
	vpt 0 360 arc closepath} bind def
/C2 {BL [] 0 setdash 2 copy moveto
	2 copy vpt 90 180 arc closepath fill
	vpt 0 360 arc closepath} bind def
/C3 {BL [] 0 setdash 2 copy moveto
	2 copy vpt 0 180 arc closepath fill
	vpt 0 360 arc closepath} bind def
/C4 {BL [] 0 setdash 2 copy moveto
	2 copy vpt 180 270 arc closepath fill
	vpt 0 360 arc closepath} bind def
/C5 {BL [] 0 setdash 2 copy moveto
	2 copy vpt 0 90 arc
	2 copy moveto
	2 copy vpt 180 270 arc closepath fill
	vpt 0 360 arc} bind def
/C6 {BL [] 0 setdash 2 copy moveto
	2 copy vpt 90 270 arc closepath fill
	vpt 0 360 arc closepath} bind def
/C7 {BL [] 0 setdash 2 copy moveto
	2 copy vpt 0 270 arc closepath fill
	vpt 0 360 arc closepath} bind def
/C8 {BL [] 0 setdash 2 copy moveto
	2 copy vpt 270 360 arc closepath fill
	vpt 0 360 arc closepath} bind def
/C9 {BL [] 0 setdash 2 copy moveto
	2 copy vpt 270 450 arc closepath fill
	vpt 0 360 arc closepath} bind def
/C10 {BL [] 0 setdash 2 copy 2 copy moveto vpt 270 360 arc closepath fill
	2 copy moveto
	2 copy vpt 90 180 arc closepath fill
	vpt 0 360 arc closepath} bind def
/C11 {BL [] 0 setdash 2 copy moveto
	2 copy vpt 0 180 arc closepath fill
	2 copy moveto
	2 copy vpt 270 360 arc closepath fill
	vpt 0 360 arc closepath} bind def
/C12 {BL [] 0 setdash 2 copy moveto
	2 copy vpt 180 360 arc closepath fill
	vpt 0 360 arc closepath} bind def
/C13 {BL [] 0 setdash 2 copy moveto
	2 copy vpt 0 90 arc closepath fill
	2 copy moveto
	2 copy vpt 180 360 arc closepath fill
	vpt 0 360 arc closepath} bind def
/C14 {BL [] 0 setdash 2 copy moveto
	2 copy vpt 90 360 arc closepath fill
	vpt 0 360 arc} bind def
/C15 {BL [] 0 setdash 2 copy vpt 0 360 arc closepath fill
	vpt 0 360 arc closepath} bind def
/Rec {newpath 4 2 roll moveto 1 index 0 rlineto 0 exch rlineto
	neg 0 rlineto closepath} bind def
/Square {dup Rec} bind def
/Bsquare {vpt sub exch vpt sub exch vpt2 Square} bind def
/S0 {BL [] 0 setdash 2 copy moveto 0 vpt rlineto BL Bsquare} bind def
/S1 {BL [] 0 setdash 2 copy vpt Square fill Bsquare} bind def
/S2 {BL [] 0 setdash 2 copy exch vpt sub exch vpt Square fill Bsquare} bind def
/S3 {BL [] 0 setdash 2 copy exch vpt sub exch vpt2 vpt Rec fill Bsquare} bind def
/S4 {BL [] 0 setdash 2 copy exch vpt sub exch vpt sub vpt Square fill Bsquare} bind def
/S5 {BL [] 0 setdash 2 copy 2 copy vpt Square fill
	exch vpt sub exch vpt sub vpt Square fill Bsquare} bind def
/S6 {BL [] 0 setdash 2 copy exch vpt sub exch vpt sub vpt vpt2 Rec fill Bsquare} bind def
/S7 {BL [] 0 setdash 2 copy exch vpt sub exch vpt sub vpt vpt2 Rec fill
	2 copy vpt Square fill Bsquare} bind def
/S8 {BL [] 0 setdash 2 copy vpt sub vpt Square fill Bsquare} bind def
/S9 {BL [] 0 setdash 2 copy vpt sub vpt vpt2 Rec fill Bsquare} bind def
/S10 {BL [] 0 setdash 2 copy vpt sub vpt Square fill 2 copy exch vpt sub exch vpt Square fill
	Bsquare} bind def
/S11 {BL [] 0 setdash 2 copy vpt sub vpt Square fill 2 copy exch vpt sub exch vpt2 vpt Rec fill
	Bsquare} bind def
/S12 {BL [] 0 setdash 2 copy exch vpt sub exch vpt sub vpt2 vpt Rec fill Bsquare} bind def
/S13 {BL [] 0 setdash 2 copy exch vpt sub exch vpt sub vpt2 vpt Rec fill
	2 copy vpt Square fill Bsquare} bind def
/S14 {BL [] 0 setdash 2 copy exch vpt sub exch vpt sub vpt2 vpt Rec fill
	2 copy exch vpt sub exch vpt Square fill Bsquare} bind def
/S15 {BL [] 0 setdash 2 copy Bsquare fill Bsquare} bind def
/D0 {gsave translate 45 rotate 0 0 S0 stroke grestore} bind def
/D1 {gsave translate 45 rotate 0 0 S1 stroke grestore} bind def
/D2 {gsave translate 45 rotate 0 0 S2 stroke grestore} bind def
/D3 {gsave translate 45 rotate 0 0 S3 stroke grestore} bind def
/D4 {gsave translate 45 rotate 0 0 S4 stroke grestore} bind def
/D5 {gsave translate 45 rotate 0 0 S5 stroke grestore} bind def
/D6 {gsave translate 45 rotate 0 0 S6 stroke grestore} bind def
/D7 {gsave translate 45 rotate 0 0 S7 stroke grestore} bind def
/D8 {gsave translate 45 rotate 0 0 S8 stroke grestore} bind def
/D9 {gsave translate 45 rotate 0 0 S9 stroke grestore} bind def
/D10 {gsave translate 45 rotate 0 0 S10 stroke grestore} bind def
/D11 {gsave translate 45 rotate 0 0 S11 stroke grestore} bind def
/D12 {gsave translate 45 rotate 0 0 S12 stroke grestore} bind def
/D13 {gsave translate 45 rotate 0 0 S13 stroke grestore} bind def
/D14 {gsave translate 45 rotate 0 0 S14 stroke grestore} bind def
/D15 {gsave translate 45 rotate 0 0 S15 stroke grestore} bind def
/DiaE {stroke [] 0 setdash vpt add M
  hpt neg vpt neg V hpt vpt neg V
  hpt vpt V hpt neg vpt V closepath stroke} def
/BoxE {stroke [] 0 setdash exch hpt sub exch vpt add M
  0 vpt2 neg V hpt2 0 V 0 vpt2 V
  hpt2 neg 0 V closepath stroke} def
/TriUE {stroke [] 0 setdash vpt 1.12 mul add M
  hpt neg vpt -1.62 mul V
  hpt 2 mul 0 V
  hpt neg vpt 1.62 mul V closepath stroke} def
/TriDE {stroke [] 0 setdash vpt 1.12 mul sub M
  hpt neg vpt 1.62 mul V
  hpt 2 mul 0 V
  hpt neg vpt -1.62 mul V closepath stroke} def
/PentE {stroke [] 0 setdash gsave
  translate 0 hpt M 4 {72 rotate 0 hpt L} repeat
  closepath stroke grestore} def
/CircE {stroke [] 0 setdash 
  hpt 0 360 arc stroke} def
/Opaque {gsave closepath 1 setgray fill grestore 0 setgray closepath} def
/DiaW {stroke [] 0 setdash vpt add M
  hpt neg vpt neg V hpt vpt neg V
  hpt vpt V hpt neg vpt V Opaque stroke} def
/BoxW {stroke [] 0 setdash exch hpt sub exch vpt add M
  0 vpt2 neg V hpt2 0 V 0 vpt2 V
  hpt2 neg 0 V Opaque stroke} def
/TriUW {stroke [] 0 setdash vpt 1.12 mul add M
  hpt neg vpt -1.62 mul V
  hpt 2 mul 0 V
  hpt neg vpt 1.62 mul V Opaque stroke} def
/TriDW {stroke [] 0 setdash vpt 1.12 mul sub M
  hpt neg vpt 1.62 mul V
  hpt 2 mul 0 V
  hpt neg vpt -1.62 mul V Opaque stroke} def
/PentW {stroke [] 0 setdash gsave
  translate 0 hpt M 4 {72 rotate 0 hpt L} repeat
  Opaque stroke grestore} def
/CircW {stroke [] 0 setdash 
  hpt 0 360 arc Opaque stroke} def
/BoxFill {gsave Rec 1 setgray fill grestore} def
/Density {
  /Fillden exch def
  currentrgbcolor
  /ColB exch def /ColG exch def /ColR exch def
  /ColR ColR Fillden mul Fillden sub 1 add def
  /ColG ColG Fillden mul Fillden sub 1 add def
  /ColB ColB Fillden mul Fillden sub 1 add def
  ColR ColG ColB setrgbcolor} def
/BoxColFill {gsave Rec PolyFill} def
/PolyFill {gsave Density fill grestore grestore} def
/h {rlineto rlineto rlineto gsave closepath fill grestore} bind def
%
%
/PatternFill {gsave /PFa [ 9 2 roll ] def
  PFa 0 get PFa 2 get 2 div add PFa 1 get PFa 3 get 2 div add translate
  PFa 2 get -2 div PFa 3 get -2 div PFa 2 get PFa 3 get Rec
  gsave 1 setgray fill grestore clip
  currentlinewidth 0.5 mul setlinewidth
  /PFs PFa 2 get dup mul PFa 3 get dup mul add sqrt def
  0 0 M PFa 5 get rotate PFs -2 div dup translate
  0 1 PFs PFa 4 get div 1 add floor cvi
	{PFa 4 get mul 0 M 0 PFs V} for
  0 PFa 6 get ne {
	0 1 PFs PFa 4 get div 1 add floor cvi
	{PFa 4 get mul 0 2 1 roll M PFs 0 V} for
 } if
  stroke grestore} def
/languagelevel where
 {pop languagelevel} {1} ifelse
 2 lt
	{/InterpretLevel1 true def}
	{/InterpretLevel1 Level1 def}
 ifelse
%
%
/Level2PatternFill {
/Tile8x8 {/PaintType 2 /PatternType 1 /TilingType 1 /BBox [0 0 8 8] /XStep 8 /YStep 8}
	bind def
/KeepColor {currentrgbcolor [/Pattern /DeviceRGB] setcolorspace} bind def
<< Tile8x8
 /PaintProc {0.5 setlinewidth pop 0 0 M 8 8 L 0 8 M 8 0 L stroke} 
>> matrix makepattern
/Pat1 exch def
<< Tile8x8
 /PaintProc {0.5 setlinewidth pop 0 0 M 8 8 L 0 8 M 8 0 L stroke
	0 4 M 4 8 L 8 4 L 4 0 L 0 4 L stroke}
>> matrix makepattern
/Pat2 exch def
<< Tile8x8
 /PaintProc {0.5 setlinewidth pop 0 0 M 0 8 L
	8 8 L 8 0 L 0 0 L fill}
>> matrix makepattern
/Pat3 exch def
<< Tile8x8
 /PaintProc {0.5 setlinewidth pop -4 8 M 8 -4 L
	0 12 M 12 0 L stroke}
>> matrix makepattern
/Pat4 exch def
<< Tile8x8
 /PaintProc {0.5 setlinewidth pop -4 0 M 8 12 L
	0 -4 M 12 8 L stroke}
>> matrix makepattern
/Pat5 exch def
<< Tile8x8
 /PaintProc {0.5 setlinewidth pop -2 8 M 4 -4 L
	0 12 M 8 -4 L 4 12 M 10 0 L stroke}
>> matrix makepattern
/Pat6 exch def
<< Tile8x8
 /PaintProc {0.5 setlinewidth pop -2 0 M 4 12 L
	0 -4 M 8 12 L 4 -4 M 10 8 L stroke}
>> matrix makepattern
/Pat7 exch def
<< Tile8x8
 /PaintProc {0.5 setlinewidth pop 8 -2 M -4 4 L
	12 0 M -4 8 L 12 4 M 0 10 L stroke}
>> matrix makepattern
/Pat8 exch def
<< Tile8x8
 /PaintProc {0.5 setlinewidth pop 0 -2 M 12 4 L
	-4 0 M 12 8 L -4 4 M 8 10 L stroke}
>> matrix makepattern
/Pat9 exch def
/Pattern1 {PatternBgnd KeepColor Pat1 setpattern} bind def
/Pattern2 {PatternBgnd KeepColor Pat2 setpattern} bind def
/Pattern3 {PatternBgnd KeepColor Pat3 setpattern} bind def
/Pattern4 {PatternBgnd KeepColor Landscape {Pat5} {Pat4} ifelse setpattern} bind def
/Pattern5 {PatternBgnd KeepColor Landscape {Pat4} {Pat5} ifelse setpattern} bind def
/Pattern6 {PatternBgnd KeepColor Landscape {Pat9} {Pat6} ifelse setpattern} bind def
/Pattern7 {PatternBgnd KeepColor Landscape {Pat8} {Pat7} ifelse setpattern} bind def
} def
%
%
%
/PatternBgnd {
  TransparentPatterns {} {gsave 1 setgray fill grestore} ifelse
} def
%
%
/Level1PatternFill {
/Pattern1 {0.250 Density} bind def
/Pattern2 {0.500 Density} bind def
/Pattern3 {0.750 Density} bind def
/Pattern4 {0.125 Density} bind def
/Pattern5 {0.375 Density} bind def
/Pattern6 {0.625 Density} bind def
/Pattern7 {0.875 Density} bind def
} def
%
%
Level1 {Level1PatternFill} {Level2PatternFill} ifelse
/Symbol-Oblique /Symbol findfont [1 0 .167 1 0 0] makefont
dup length dict begin {1 index /FID eq {pop pop} {def} ifelse} forall
currentdict end definefont pop
Level1 SuppressPDFMark or 
{} {
/SDict 10 dict def
systemdict /pdfmark known not {
  userdict /pdfmark systemdict /cleartomark get put
} if
SDict begin [
  /Title (paper/ff.2m.2p.tex)
  /Subject (gnuplot plot)
  /Creator (gnuplot 4.6 patchlevel 0)
  /Author (conrado)
  /CreationDate (Sat Oct 12 12:39:34 2013)
  /DOCINFO pdfmark
end
} ifelse
end
gnudict begin
gsave
doclip
0 0 translate
0.050 0.050 scale
0 setgray
newpath
BackgroundColor 0 lt 3 1 roll 0 lt exch 0 lt or or not {BackgroundColor C 1.000 0 0 7200.00 5040.00 BoxColFill} if
1.000 UL
LTb
0 640 M
63 0 V
2816 0 R
-63 0 V
0 1198 M
63 0 V
2816 0 R
-63 0 V
0 1756 M
63 0 V
2816 0 R
-63 0 V
0 2315 M
63 0 V
2816 0 R
-63 0 V
0 2873 M
63 0 V
2816 0 R
-63 0 V
0 3431 M
63 0 V
2816 0 R
-63 0 V
0 640 M
0 63 V
0 3431 M
0 -63 V
411 640 M
0 63 V
0 2728 R
0 -63 V
823 640 M
0 63 V
0 2728 R
0 -63 V
1234 640 M
0 63 V
0 2728 R
0 -63 V
1645 640 M
0 63 V
0 2728 R
0 -63 V
2056 640 M
0 63 V
0 2728 R
0 -63 V
2468 640 M
0 63 V
0 2728 R
0 -63 V
2879 640 M
0 63 V
0 2728 R
0 -63 V
stroke
LTa
0 2873 M
2879 0 V
stroke
LTb
0 3431 N
0 640 L
2879 0 V
0 2791 V
0 3431 L
Z stroke
LCb setrgbcolor
LTb
1.000 UP
1.000 UL
LTb
1.000 UL
LT0
LCb setrgbcolor
LT0
1496 2317 M
543 0 V
0 684 M
2 1 V
4 1 V
5 1 V
6 2 V
8 2 V
8 2 V
9 3 V
10 3 V
10 3 V
11 3 V
10 2 V
10 3 V
9 3 V
8 2 V
8 2 V
6 2 V
5 2 V
3 1 V
2 0 V
1 1 V
2 0 V
4 1 V
5 2 V
6 1 V
7 3 V
9 2 V
9 3 V
10 3 V
10 3 V
10 3 V
10 3 V
10 3 V
10 3 V
8 2 V
7 2 V
7 2 V
5 2 V
3 1 V
2 0 V
1 1 V
2 0 V
3 1 V
5 2 V
7 2 V
7 2 V
8 3 V
10 3 V
10 3 V
10 3 V
10 3 V
10 3 V
10 3 V
9 3 V
9 3 V
7 2 V
6 2 V
5 2 V
4 1 V
2 1 V
1 0 V
2 1 V
3 1 V
5 1 V
6 3 V
8 2 V
8 3 V
9 3 V
10 3 V
10 3 V
11 4 V
10 3 V
10 3 V
9 3 V
9 3 V
7 3 V
6 2 V
5 2 V
4 1 V
2 0 V
0 1 V
2 0 V
4 2 V
5 1 V
6 2 V
8 3 V
8 3 V
9 3 V
10 3 V
10 4 V
11 3 V
10 4 V
10 3 V
9 4 V
8 3 V
8 2 V
6 2 V
5 2 V
3 1 V
2 1 V
1 0 V
2 1 V
4 1 V
stroke 680 898 M
5 2 V
6 2 V
7 3 V
9 3 V
9 3 V
10 4 V
10 4 V
10 3 V
11 4 V
9 4 V
10 3 V
8 3 V
7 3 V
7 2 V
5 2 V
3 1 V
2 1 V
1 0 V
2 1 V
3 2 V
5 1 V
7 3 V
7 2 V
8 4 V
10 3 V
10 4 V
10 4 V
10 4 V
10 4 V
10 3 V
9 4 V
9 3 V
7 3 V
6 2 V
5 2 V
4 2 V
2 0 V
1 1 V
2 1 V
3 1 V
5 2 V
6 2 V
8 3 V
8 3 V
9 4 V
10 4 V
11 4 V
10 4 V
10 4 V
10 4 V
9 4 V
9 3 V
7 3 V
6 3 V
5 2 V
4 1 V
2 1 V
1 0 V
1 1 V
4 2 V
5 2 V
6 2 V
8 3 V
8 4 V
9 3 V
10 4 V
10 5 V
11 4 V
10 4 V
10 4 V
9 4 V
8 4 V
8 3 V
6 2 V
5 3 V
3 1 V
2 1 V
1 0 V
2 1 V
4 2 V
5 2 V
6 2 V
7 3 V
9 4 V
9 4 V
10 4 V
10 5 V
10 4 V
11 5 V
9 4 V
10 4 V
8 3 V
7 4 V
7 2 V
5 3 V
3 1 V
2 1 V
1 0 V
2 1 V
3 2 V
5 2 V
7 3 V
7 3 V
9 4 V
stroke 1381 1179 M
9 4 V
10 4 V
10 5 V
10 5 V
10 4 V
10 5 V
9 4 V
9 4 V
7 3 V
6 3 V
5 2 V
4 2 V
2 1 V
1 0 V
2 1 V
3 2 V
5 2 V
6 3 V
8 3 V
8 4 V
10 4 V
9 5 V
11 5 V
10 4 V
10 5 V
10 5 V
9 4 V
9 4 V
7 4 V
6 3 V
5 2 V
4 2 V
2 1 V
1 0 V
2 1 V
3 2 V
5 2 V
6 3 V
8 3 V
8 5 V
9 4 V
10 5 V
10 5 V
11 5 V
10 5 V
10 4 V
9 5 V
8 4 V
8 4 V
6 3 V
5 2 V
3 2 V
2 1 V
1 0 V
2 1 V
4 2 V
5 2 V
6 4 V
7 3 V
9 4 V
9 5 V
10 5 V
10 5 V
10 5 V
11 5 V
9 5 V
10 5 V
8 4 V
7 4 V
7 3 V
5 3 V
3 2 V
2 1 V
1 0 V
2 1 V
4 2 V
4 2 V
7 4 V
7 3 V
9 5 V
9 4 V
10 6 V
10 5 V
10 5 V
10 6 V
10 5 V
9 5 V
9 4 V
7 4 V
7 3 V
4 3 V
4 2 V
2 1 V
1 0 V
2 1 V
3 2 V
5 3 V
6 3 V
8 4 V
8 4 V
10 5 V
9 6 V
11 5 V
10 6 V
stroke 2094 1530 M
10 5 V
10 5 V
9 6 V
9 4 V
7 4 V
6 4 V
5 2 V
4 2 V
2 1 V
1 1 V
2 1 V
3 2 V
5 2 V
6 4 V
8 4 V
8 5 V
9 5 V
10 5 V
10 6 V
11 6 V
10 5 V
10 6 V
9 5 V
8 5 V
8 4 V
6 3 V
5 3 V
3 2 V
2 1 V
1 1 V
2 1 V
4 2 V
5 3 V
6 3 V
7 4 V
9 5 V
9 5 V
10 6 V
10 6 V
10 6 V
11 6 V
9 5 V
10 6 V
8 5 V
8 4 V
6 4 V
5 2 V
3 3 V
2 1 V
1 0 V
2 1 V
4 2 V
4 3 V
7 4 V
7 4 V
9 5 V
9 6 V
10 6 V
10 6 V
10 6 V
10 6 V
10 6 V
9 5 V
9 5 V
7 5 V
7 3 V
4 3 V
4 2 V
2 2 V
1 0 V
2 1 V
3 2 V
5 3 V
6 4 V
8 5 V
8 5 V
10 5 V
9 6 V
11 7 V
10 6 V
10 6 V
10 6 V
9 6 V
9 5 V
7 5 V
6 4 V
5 3 V
4 2 V
stroke
LT1
LCb setrgbcolor
LT1
1496 2117 M
543 0 V
0 2967 M
2 0 V
4 0 V
5 0 V
6 0 V
8 0 V
8 0 V
9 0 V
10 0 V
10 0 V
11 0 V
10 1 V
10 0 V
9 0 V
8 0 V
8 0 V
6 0 V
5 0 V
3 0 V
2 0 V
1 0 V
2 0 V
4 0 V
5 0 V
6 0 V
7 0 V
9 0 V
9 0 V
10 0 V
10 0 V
10 0 V
10 0 V
10 1 V
10 0 V
8 0 V
7 0 V
7 0 V
5 0 V
3 0 V
2 0 V
1 0 V
2 0 V
3 0 V
5 0 V
7 0 V
7 0 V
8 0 V
10 0 V
10 0 V
10 0 V
10 0 V
10 0 V
10 0 V
9 1 V
9 0 V
7 0 V
6 0 V
5 0 V
4 0 V
2 0 V
1 0 V
2 0 V
3 0 V
5 0 V
6 0 V
8 0 V
8 0 V
9 0 V
10 0 V
10 0 V
11 0 V
10 0 V
10 0 V
9 0 V
9 0 V
7 1 V
6 0 V
5 0 V
4 0 V
2 0 V
2 0 V
4 0 V
5 0 V
6 0 V
8 0 V
8 0 V
9 0 V
10 0 V
10 0 V
11 0 V
10 0 V
10 0 V
9 0 V
8 0 V
8 0 V
6 0 V
5 0 V
3 0 V
2 0 V
1 0 V
2 1 V
4 0 V
5 0 V
stroke 685 2972 M
6 0 V
7 0 V
9 0 V
9 0 V
10 0 V
10 0 V
10 0 V
11 0 V
9 0 V
10 0 V
8 0 V
7 0 V
7 0 V
5 0 V
3 0 V
2 0 V
1 0 V
2 0 V
3 0 V
5 0 V
7 0 V
7 0 V
8 1 V
10 0 V
10 0 V
10 0 V
10 0 V
10 0 V
10 0 V
9 0 V
9 0 V
7 0 V
6 0 V
5 0 V
4 0 V
2 0 V
1 0 V
2 0 V
3 0 V
5 0 V
6 0 V
8 0 V
8 0 V
9 0 V
10 0 V
11 1 V
10 0 V
10 0 V
10 0 V
9 0 V
9 0 V
7 0 V
6 0 V
5 0 V
4 0 V
2 0 V
1 0 V
1 0 V
4 0 V
5 0 V
6 0 V
8 0 V
8 0 V
9 0 V
10 0 V
10 0 V
11 0 V
10 0 V
10 1 V
9 0 V
8 0 V
8 0 V
6 0 V
5 0 V
3 0 V
2 0 V
1 0 V
2 0 V
4 0 V
5 0 V
6 0 V
7 0 V
9 0 V
9 0 V
10 0 V
10 0 V
10 0 V
11 0 V
9 0 V
10 0 V
8 0 V
7 0 V
7 0 V
5 0 V
3 1 V
2 0 V
1 0 V
2 0 V
3 0 V
5 0 V
7 0 V
7 0 V
9 0 V
9 0 V
stroke 1390 2976 M
10 0 V
10 0 V
10 0 V
10 0 V
10 0 V
9 0 V
9 0 V
7 0 V
6 0 V
5 0 V
4 0 V
2 0 V
1 0 V
2 0 V
3 0 V
5 0 V
6 0 V
8 0 V
8 0 V
10 0 V
9 1 V
11 0 V
10 0 V
10 0 V
10 0 V
9 0 V
9 0 V
7 0 V
6 0 V
5 0 V
4 0 V
2 0 V
1 0 V
2 0 V
3 0 V
5 0 V
6 0 V
8 0 V
8 0 V
9 0 V
10 0 V
10 0 V
11 0 V
10 0 V
10 0 V
9 0 V
8 1 V
8 0 V
6 0 V
5 0 V
3 0 V
2 0 V
1 0 V
2 0 V
4 0 V
5 0 V
6 0 V
7 0 V
9 0 V
9 0 V
10 0 V
10 0 V
10 0 V
11 0 V
9 0 V
10 0 V
8 0 V
7 0 V
7 0 V
5 0 V
3 0 V
2 0 V
1 0 V
2 0 V
4 0 V
4 0 V
7 0 V
7 0 V
9 0 V
9 1 V
10 0 V
10 0 V
10 0 V
10 0 V
10 0 V
9 0 V
9 0 V
7 0 V
7 0 V
4 0 V
4 0 V
2 0 V
1 0 V
2 0 V
3 0 V
5 0 V
6 0 V
8 0 V
8 0 V
10 0 V
9 0 V
11 0 V
10 0 V
10 0 V
stroke 2104 2979 M
10 0 V
9 0 V
9 0 V
7 0 V
6 0 V
5 1 V
4 0 V
2 0 V
1 0 V
2 0 V
3 0 V
5 0 V
6 0 V
8 0 V
8 0 V
9 0 V
10 0 V
10 0 V
11 0 V
10 0 V
10 0 V
9 0 V
8 0 V
8 0 V
6 0 V
5 0 V
3 0 V
2 0 V
1 0 V
2 0 V
4 0 V
5 0 V
6 0 V
7 0 V
9 0 V
9 0 V
10 0 V
10 0 V
10 0 V
11 0 V
9 0 V
10 0 V
8 1 V
8 0 V
6 0 V
5 0 V
3 0 V
2 0 V
1 0 V
2 0 V
4 0 V
4 0 V
7 0 V
7 0 V
9 0 V
9 0 V
10 0 V
10 0 V
10 0 V
10 0 V
10 0 V
9 0 V
9 0 V
7 0 V
7 0 V
4 0 V
4 0 V
2 0 V
1 0 V
2 0 V
3 0 V
5 0 V
6 0 V
8 0 V
8 0 V
10 0 V
9 0 V
11 0 V
10 0 V
10 0 V
10 0 V
9 0 V
9 0 V
7 0 V
6 1 V
5 -2 V
4 3 V
stroke
LT2
LCb setrgbcolor
LT2
1496 1917 M
543 0 V
0 2687 M
2 0 V
4 0 V
5 0 V
6 0 V
8 0 V
8 0 V
9 0 V
10 -1 V
10 0 V
11 0 V
10 0 V
10 -1 V
9 0 V
8 0 V
8 0 V
6 0 V
5 0 V
3 0 V
2 0 V
1 0 V
2 0 V
4 -1 V
5 0 V
6 0 V
7 0 V
9 0 V
9 0 V
10 -1 V
10 0 V
10 0 V
10 0 V
10 0 V
10 -1 V
8 0 V
7 0 V
7 0 V
5 0 V
3 0 V
2 0 V
1 0 V
2 0 V
3 0 V
5 0 V
7 -1 V
7 0 V
8 0 V
10 0 V
10 0 V
10 -1 V
10 0 V
10 0 V
10 0 V
9 0 V
9 -1 V
7 0 V
6 0 V
5 0 V
4 0 V
2 0 V
1 0 V
2 0 V
3 0 V
5 0 V
6 0 V
8 -1 V
8 0 V
9 0 V
10 0 V
10 0 V
11 -1 V
10 0 V
10 0 V
9 0 V
9 0 V
7 -1 V
6 0 V
5 0 V
4 0 V
2 0 V
2 0 V
4 0 V
5 0 V
6 0 V
8 0 V
8 -1 V
9 0 V
10 0 V
10 0 V
11 0 V
10 -1 V
10 0 V
9 0 V
8 0 V
8 0 V
6 -1 V
5 0 V
3 0 V
2 0 V
1 0 V
2 0 V
4 0 V
5 0 V
stroke 685 2673 M
6 0 V
7 0 V
9 0 V
9 -1 V
10 0 V
10 0 V
10 0 V
11 -1 V
9 0 V
10 0 V
8 0 V
7 0 V
7 0 V
5 -1 V
3 0 V
2 0 V
1 0 V
2 0 V
3 0 V
5 0 V
7 0 V
7 0 V
8 0 V
10 -1 V
10 0 V
10 0 V
10 0 V
10 0 V
10 -1 V
9 0 V
9 0 V
7 0 V
6 0 V
5 0 V
4 0 V
2 0 V
1 0 V
2 -1 V
3 0 V
5 0 V
6 0 V
8 0 V
8 0 V
9 0 V
10 -1 V
11 0 V
10 0 V
10 0 V
10 0 V
9 -1 V
9 0 V
7 0 V
6 0 V
5 0 V
4 0 V
2 0 V
1 0 V
1 0 V
4 0 V
5 -1 V
6 0 V
8 0 V
8 0 V
9 0 V
10 0 V
10 -1 V
11 0 V
10 0 V
10 0 V
9 -1 V
8 0 V
8 0 V
6 0 V
5 0 V
3 0 V
2 0 V
1 0 V
2 0 V
4 0 V
5 0 V
6 -1 V
7 0 V
9 0 V
9 0 V
10 0 V
10 -1 V
10 0 V
11 0 V
9 0 V
10 0 V
8 -1 V
7 0 V
7 0 V
5 0 V
3 0 V
2 0 V
1 0 V
2 0 V
3 0 V
5 0 V
7 0 V
7 -1 V
9 0 V
9 0 V
stroke 1390 2658 M
10 0 V
10 0 V
10 -1 V
10 0 V
10 0 V
9 0 V
9 0 V
7 -1 V
6 0 V
5 0 V
4 0 V
2 0 V
1 0 V
2 0 V
3 0 V
5 0 V
6 0 V
8 0 V
8 -1 V
10 0 V
9 0 V
11 0 V
10 0 V
10 -1 V
10 0 V
9 0 V
9 0 V
7 0 V
6 -1 V
5 0 V
4 0 V
2 0 V
1 0 V
2 0 V
3 0 V
5 0 V
6 0 V
8 0 V
8 0 V
9 -1 V
10 0 V
10 0 V
11 0 V
10 -1 V
10 0 V
9 0 V
8 0 V
8 0 V
6 0 V
5 0 V
3 -1 V
2 0 V
1 0 V
2 0 V
4 0 V
5 0 V
6 0 V
7 0 V
9 0 V
9 -1 V
10 0 V
10 0 V
10 0 V
11 0 V
9 -1 V
10 0 V
8 0 V
7 0 V
7 0 V
5 0 V
3 0 V
2 0 V
1 0 V
2 -1 V
4 0 V
4 0 V
7 0 V
7 0 V
9 0 V
9 0 V
10 -1 V
10 0 V
10 0 V
10 0 V
10 0 V
9 -1 V
9 0 V
7 0 V
7 0 V
4 0 V
4 0 V
2 0 V
1 0 V
2 0 V
3 0 V
5 -1 V
6 0 V
8 0 V
8 0 V
10 0 V
9 0 V
11 -1 V
10 0 V
10 0 V
stroke 2104 2643 M
10 0 V
9 0 V
9 -1 V
7 0 V
6 0 V
5 0 V
4 0 V
2 0 V
1 0 V
2 0 V
3 0 V
5 0 V
6 0 V
8 -1 V
8 0 V
9 0 V
10 0 V
10 0 V
11 -1 V
10 0 V
10 0 V
9 0 V
8 0 V
8 -1 V
6 0 V
5 0 V
3 0 V
2 0 V
1 0 V
2 0 V
4 0 V
5 0 V
6 0 V
7 0 V
9 -1 V
9 0 V
10 0 V
10 0 V
10 0 V
11 -1 V
9 0 V
10 0 V
8 0 V
8 0 V
6 0 V
5 -1 V
3 0 V
2 0 V
1 0 V
2 0 V
4 0 V
4 0 V
7 0 V
7 0 V
9 0 V
9 -1 V
10 0 V
10 0 V
10 0 V
10 0 V
10 -1 V
9 0 V
9 0 V
7 0 V
7 0 V
4 0 V
4 0 V
2 0 V
1 0 V
2 -1 V
3 0 V
5 0 V
6 0 V
8 0 V
8 0 V
10 0 V
9 -1 V
11 0 V
10 0 V
10 0 V
10 0 V
9 -1 V
9 0 V
7 1 V
6 -2 V
5 4 V
4 -9 V
stroke
LT3
LCb setrgbcolor
LT3
1496 1717 M
543 0 V
0 2676 M
2 0 V
4 0 V
5 0 V
6 -1 V
8 0 V
8 0 V
9 0 V
10 0 V
10 -1 V
11 0 V
10 0 V
10 0 V
9 0 V
8 -1 V
8 0 V
6 0 V
5 0 V
3 0 V
2 0 V
1 0 V
2 0 V
4 0 V
5 0 V
6 -1 V
7 0 V
9 0 V
9 0 V
10 0 V
10 -1 V
10 0 V
10 0 V
10 0 V
10 -1 V
8 0 V
7 0 V
7 0 V
5 0 V
3 0 V
2 0 V
1 0 V
2 0 V
3 0 V
5 -1 V
7 0 V
7 0 V
8 0 V
10 0 V
10 -1 V
10 0 V
10 0 V
10 0 V
10 -1 V
9 0 V
9 0 V
7 0 V
6 0 V
5 0 V
4 0 V
2 0 V
1 -1 V
2 0 V
3 0 V
5 0 V
6 0 V
8 0 V
8 0 V
9 0 V
10 -1 V
10 0 V
11 0 V
10 0 V
10 -1 V
9 0 V
9 0 V
7 0 V
6 0 V
5 -1 V
4 0 V
2 0 V
2 0 V
4 0 V
5 0 V
6 0 V
8 0 V
8 0 V
9 -1 V
10 0 V
10 0 V
11 0 V
10 -1 V
10 0 V
9 0 V
8 0 V
8 0 V
6 -1 V
5 0 V
3 0 V
2 0 V
1 0 V
2 0 V
4 0 V
5 0 V
stroke 685 2660 M
6 0 V
7 0 V
9 -1 V
9 0 V
10 0 V
10 0 V
10 -1 V
11 0 V
9 0 V
10 0 V
8 0 V
7 -1 V
7 0 V
5 0 V
3 0 V
2 0 V
1 0 V
2 0 V
3 0 V
5 0 V
7 0 V
7 -1 V
8 0 V
10 0 V
10 0 V
10 0 V
10 -1 V
10 0 V
10 0 V
9 0 V
9 -1 V
7 0 V
6 0 V
5 0 V
4 0 V
2 0 V
1 0 V
2 0 V
3 0 V
5 0 V
6 -1 V
8 0 V
8 0 V
9 0 V
10 0 V
11 -1 V
10 0 V
10 0 V
10 0 V
9 -1 V
9 0 V
7 0 V
6 0 V
5 0 V
4 0 V
2 0 V
1 0 V
1 0 V
4 -1 V
5 0 V
6 0 V
8 0 V
8 0 V
9 0 V
10 -1 V
10 0 V
11 0 V
10 0 V
10 -1 V
9 0 V
8 0 V
8 0 V
6 0 V
5 0 V
3 -1 V
2 0 V
1 0 V
2 0 V
4 0 V
5 0 V
6 0 V
7 0 V
9 0 V
9 -1 V
10 0 V
10 0 V
10 0 V
11 -1 V
9 0 V
10 0 V
8 0 V
7 0 V
7 -1 V
5 0 V
3 0 V
2 0 V
1 0 V
2 0 V
3 0 V
5 0 V
7 0 V
7 0 V
9 -1 V
9 0 V
stroke 1390 2643 M
10 0 V
10 0 V
10 -1 V
10 0 V
10 0 V
9 0 V
9 0 V
7 -1 V
6 0 V
5 0 V
4 0 V
2 0 V
1 0 V
2 0 V
3 0 V
5 0 V
6 0 V
8 -1 V
8 0 V
10 0 V
9 0 V
11 0 V
10 -1 V
10 0 V
10 0 V
9 0 V
9 -1 V
7 0 V
6 0 V
5 0 V
4 0 V
2 0 V
1 0 V
2 0 V
3 0 V
5 0 V
6 -1 V
8 0 V
8 0 V
9 0 V
10 0 V
10 -1 V
11 0 V
10 0 V
10 0 V
9 -1 V
8 0 V
8 0 V
6 0 V
5 0 V
3 0 V
2 0 V
1 0 V
2 -1 V
4 0 V
5 0 V
6 0 V
7 0 V
9 0 V
9 0 V
10 -1 V
10 0 V
10 0 V
11 0 V
9 -1 V
10 0 V
8 0 V
7 0 V
7 0 V
5 -1 V
3 0 V
2 0 V
1 0 V
2 0 V
4 0 V
4 0 V
7 0 V
7 0 V
9 0 V
9 -1 V
10 0 V
10 0 V
10 0 V
10 -1 V
10 0 V
9 0 V
9 0 V
7 0 V
7 -1 V
4 0 V
4 0 V
2 0 V
1 0 V
2 0 V
3 0 V
5 0 V
6 0 V
8 0 V
8 -1 V
10 0 V
9 0 V
11 0 V
10 -1 V
10 0 V
stroke 2104 2626 M
10 0 V
9 0 V
9 -1 V
7 0 V
6 0 V
5 0 V
4 0 V
2 0 V
1 0 V
2 0 V
3 0 V
5 0 V
6 -1 V
8 0 V
8 0 V
9 0 V
10 0 V
10 -1 V
11 0 V
10 0 V
10 0 V
9 -1 V
8 0 V
8 0 V
6 0 V
5 0 V
3 0 V
2 0 V
1 0 V
2 0 V
4 0 V
5 -1 V
6 0 V
7 0 V
9 0 V
9 0 V
10 -1 V
10 0 V
10 0 V
11 0 V
9 0 V
10 -1 V
8 0 V
8 0 V
6 0 V
5 0 V
3 0 V
2 0 V
1 -1 V
2 0 V
4 0 V
4 0 V
7 0 V
7 0 V
9 0 V
9 0 V
10 -1 V
10 0 V
10 0 V
10 0 V
10 -1 V
9 0 V
9 0 V
7 0 V
7 0 V
4 -1 V
4 0 V
2 0 V
1 0 V
2 0 V
3 0 V
5 0 V
6 0 V
8 0 V
8 0 V
10 -1 V
9 0 V
11 0 V
10 0 V
10 -1 V
10 0 V
9 0 V
9 0 V
7 0 V
6 0 V
5 0 V
4 1 V
stroke
LTb
0 3431 N
0 640 L
2879 0 V
0 2791 V
0 3431 L
Z stroke
1.000 UP
1.000 UL
LTb
1.000 UL
LTb
4320 640 M
63 0 V
2816 0 R
-63 0 V
4320 873 M
63 0 V
2816 0 R
-63 0 V
4320 1105 M
63 0 V
2816 0 R
-63 0 V
4320 1338 M
63 0 V
2816 0 R
-63 0 V
4320 1570 M
63 0 V
2816 0 R
-63 0 V
4320 1803 M
63 0 V
2816 0 R
-63 0 V
4320 2035 M
63 0 V
2816 0 R
-63 0 V
4320 2268 M
63 0 V
2816 0 R
-63 0 V
4320 2501 M
63 0 V
2816 0 R
-63 0 V
4320 2733 M
63 0 V
2816 0 R
-63 0 V
4320 2966 M
63 0 V
2816 0 R
-63 0 V
4320 3198 M
63 0 V
2816 0 R
-63 0 V
4320 3431 M
63 0 V
2816 0 R
-63 0 V
4320 640 M
0 63 V
0 2728 R
0 -63 V
4680 640 M
0 63 V
0 2728 R
0 -63 V
5040 640 M
0 63 V
0 2728 R
0 -63 V
5400 640 M
0 63 V
0 2728 R
0 -63 V
5760 640 M
0 63 V
0 2728 R
0 -63 V
6119 640 M
0 63 V
0 2728 R
0 -63 V
6479 640 M
0 63 V
0 2728 R
0 -63 V
6839 640 M
0 63 V
0 2728 R
0 -63 V
7199 640 M
0 63 V
0 2728 R
0 -63 V
stroke
LTa
4320 873 M
2879 0 V
stroke
LTb
4320 3431 N
0 -2791 V
2879 0 V
0 2791 V
-2879 0 V
Z stroke
LCb setrgbcolor
LTb
1.000 UP
1.000 UL
LTb
1.000 UL
LT0
LCb setrgbcolor
LT0
6248 2456 M
543 0 V
4320 2834 M
3 0 V
3 1 V
5 1 V
7 1 V
8 2 V
8 2 V
10 2 V
10 2 V
11 3 V
11 2 V
10 2 V
11 2 V
9 3 V
9 1 V
8 2 V
6 2 V
5 1 V
4 0 V
2 1 V
1 0 V
2 1 V
4 0 V
5 1 V
6 2 V
8 2 V
9 1 V
10 3 V
10 2 V
10 2 V
11 2 V
11 3 V
10 2 V
10 2 V
9 2 V
7 2 V
7 1 V
5 1 V
4 1 V
2 0 V
1 1 V
2 0 V
3 1 V
5 1 V
7 1 V
8 2 V
8 2 V
10 2 V
10 2 V
11 3 V
11 2 V
10 2 V
11 3 V
9 2 V
9 2 V
8 1 V
6 2 V
5 1 V
4 1 V
2 0 V
1 0 V
2 1 V
4 1 V
5 1 V
6 1 V
8 2 V
9 2 V
10 2 V
10 2 V
11 2 V
10 3 V
11 2 V
10 2 V
10 2 V
9 2 V
7 2 V
7 1 V
5 2 V
4 0 V
2 1 V
1 0 V
2 0 V
3 1 V
6 1 V
6 2 V
8 1 V
9 2 V
9 2 V
10 3 V
11 2 V
11 2 V
11 3 V
10 2 V
9 2 V
9 2 V
8 2 V
6 1 V
6 1 V
3 1 V
2 0 V
1 1 V
2 0 V
4 1 V
stroke 5029 2988 M
5 1 V
7 1 V
7 2 V
9 2 V
10 2 V
10 2 V
11 3 V
10 2 V
11 2 V
10 2 V
10 3 V
9 1 V
8 2 V
6 2 V
5 1 V
4 0 V
2 1 V
1 0 V
2 1 V
3 0 V
6 1 V
6 2 V
8 2 V
9 2 V
9 2 V
11 2 V
10 2 V
11 2 V
11 3 V
10 2 V
10 2 V
8 2 V
8 2 V
7 1 V
5 1 V
3 1 V
2 0 V
1 1 V
2 0 V
4 1 V
5 1 V
7 1 V
7 2 V
9 2 V
10 2 V
10 2 V
11 3 V
11 2 V
10 2 V
10 2 V
10 3 V
9 1 V
8 2 V
6 2 V
5 1 V
4 0 V
2 1 V
1 0 V
2 1 V
4 0 V
5 1 V
6 2 V
8 1 V
9 2 V
9 2 V
11 3 V
10 2 V
11 2 V
11 3 V
10 2 V
10 2 V
8 2 V
8 1 V
7 2 V
5 1 V
3 1 V
3 0 V
3 1 V
3 1 V
5 1 V
7 1 V
8 2 V
8 2 V
10 2 V
10 2 V
11 2 V
11 3 V
10 2 V
11 2 V
9 2 V
9 2 V
8 2 V
6 1 V
5 1 V
4 1 V
2 0 V
1 1 V
2 0 V
4 1 V
5 1 V
6 1 V
8 2 V
9 2 V
10 2 V
stroke 5770 3148 M
10 2 V
10 2 V
11 3 V
11 2 V
10 2 V
10 2 V
9 2 V
7 2 V
7 1 V
5 1 V
4 1 V
2 0 V
1 1 V
2 0 V
3 1 V
5 1 V
7 1 V
8 2 V
8 2 V
10 2 V
10 2 V
11 2 V
11 3 V
10 2 V
11 2 V
9 2 V
9 2 V
8 2 V
6 1 V
5 1 V
4 1 V
2 0 V
1 1 V
2 0 V
4 1 V
5 1 V
6 1 V
8 2 V
9 2 V
10 2 V
10 2 V
11 2 V
10 2 V
11 3 V
10 2 V
10 2 V
9 2 V
7 1 V
7 2 V
5 1 V
4 1 V
2 0 V
1 0 V
2 1 V
3 0 V
6 1 V
6 2 V
8 1 V
9 2 V
9 2 V
10 2 V
11 3 V
11 2 V
10 2 V
11 2 V
9 2 V
9 2 V
8 2 V
6 1 V
6 1 V
3 1 V
2 0 V
1 1 V
2 0 V
4 1 V
5 1 V
7 1 V
7 2 V
9 2 V
10 2 V
10 2 V
11 2 V
10 2 V
11 2 V
10 3 V
10 2 V
9 1 V
7 2 V
7 1 V
5 1 V
4 1 V
2 1 V
1 0 V
2 0 V
3 1 V
6 1 V
6 1 V
8 2 V
9 2 V
9 2 V
11 2 V
10 2 V
11 2 V
11 2 V
stroke 6515 3304 M
10 3 V
9 2 V
9 1 V
8 2 V
7 1 V
5 1 V
3 1 V
2 1 V
1 0 V
2 0 V
4 1 V
5 1 V
7 1 V
7 2 V
9 2 V
10 2 V
10 2 V
11 2 V
11 2 V
10 2 V
10 2 V
10 2 V
9 2 V
8 2 V
6 1 V
5 1 V
4 1 V
2 0 V
1 0 V
2 1 V
4 0 V
5 1 V
6 2 V
8 1 V
9 2 V
9 2 V
11 2 V
10 2 V
11 2 V
11 3 V
10 2 V
10 2 V
8 1 V
8 2 V
7 1 V
5 1 V
3 1 V
2 0 V
1 1 V
3 0 V
3 1 V
5 1 V
7 1 V
8 2 V
8 1 V
10 2 V
10 2 V
11 2 V
11 2 V
10 3 V
11 2 V
9 2 V
9 1 V
8 2 V
6 1 V
5 1 V
4 1 V
2 0 V
1 0 V
2 1 V
4 1 V
5 1 V
6 1 V
8 1 V
9 2 V
10 2 V
10 2 V
10 2 V
11 2 V
11 2 V
10 2 V
10 2 V
9 2 V
7 1 V
7 2 V
5 1 V
4 0 V
2 1 V
stroke
LT1
LCb setrgbcolor
LT1
6248 2256 M
543 0 V
4320 840 M
3 0 V
3 0 V
5 0 V
7 0 V
8 0 V
8 0 V
10 0 V
10 0 V
11 0 V
11 0 V
10 0 V
11 0 V
9 -1 V
9 0 V
8 0 V
6 0 V
5 0 V
4 0 V
2 0 V
1 0 V
2 0 V
4 0 V
5 0 V
6 0 V
8 0 V
9 0 V
10 0 V
10 0 V
10 0 V
11 0 V
11 0 V
10 0 V
10 0 V
9 0 V
7 0 V
7 0 V
5 0 V
4 0 V
2 0 V
1 0 V
2 0 V
3 0 V
5 0 V
7 -1 V
8 0 V
8 0 V
10 0 V
10 0 V
11 0 V
11 0 V
10 0 V
11 0 V
9 0 V
9 0 V
8 0 V
6 0 V
5 0 V
4 0 V
2 0 V
1 0 V
2 0 V
4 0 V
5 0 V
6 0 V
8 0 V
9 0 V
10 0 V
10 0 V
11 0 V
10 -1 V
11 0 V
10 0 V
10 0 V
9 0 V
7 0 V
7 0 V
5 0 V
4 0 V
2 0 V
1 0 V
2 0 V
3 0 V
6 0 V
6 0 V
8 0 V
9 0 V
9 0 V
10 0 V
11 0 V
11 0 V
11 0 V
10 0 V
9 0 V
9 -1 V
8 0 V
6 0 V
6 0 V
3 0 V
2 0 V
1 0 V
2 0 V
4 0 V
stroke 5029 836 M
5 0 V
7 0 V
7 0 V
9 0 V
10 0 V
10 0 V
11 0 V
10 0 V
11 0 V
10 0 V
10 0 V
9 0 V
8 0 V
6 0 V
5 0 V
4 0 V
2 0 V
1 0 V
2 0 V
3 0 V
6 0 V
6 -1 V
8 0 V
9 0 V
9 0 V
11 0 V
10 0 V
11 0 V
11 0 V
10 0 V
10 0 V
8 0 V
8 0 V
7 0 V
5 0 V
3 0 V
2 0 V
1 0 V
2 0 V
4 0 V
5 0 V
7 0 V
7 0 V
9 0 V
10 0 V
10 0 V
11 -1 V
11 0 V
10 0 V
10 0 V
10 0 V
9 0 V
8 0 V
6 0 V
5 0 V
4 0 V
2 0 V
1 0 V
2 0 V
4 0 V
5 0 V
6 0 V
8 0 V
9 0 V
9 0 V
11 0 V
10 0 V
11 0 V
11 0 V
10 -1 V
10 0 V
8 0 V
8 0 V
7 0 V
5 0 V
3 0 V
3 0 V
3 0 V
3 0 V
5 0 V
7 0 V
8 0 V
8 0 V
10 0 V
10 0 V
11 0 V
11 0 V
10 0 V
11 0 V
9 0 V
9 0 V
8 0 V
6 0 V
5 -1 V
4 0 V
2 0 V
1 0 V
2 0 V
4 0 V
5 0 V
6 0 V
8 0 V
9 0 V
10 0 V
stroke 5770 832 M
10 0 V
10 0 V
11 0 V
11 0 V
10 0 V
10 0 V
9 0 V
7 0 V
7 0 V
5 0 V
4 0 V
2 0 V
1 0 V
2 0 V
3 0 V
5 0 V
7 0 V
8 -1 V
8 0 V
10 0 V
10 0 V
11 0 V
11 0 V
10 0 V
11 0 V
9 0 V
9 0 V
8 0 V
6 0 V
5 0 V
4 0 V
2 0 V
1 0 V
2 0 V
4 0 V
5 0 V
6 0 V
8 0 V
9 0 V
10 0 V
10 0 V
11 -1 V
10 0 V
11 0 V
10 0 V
10 0 V
9 0 V
7 0 V
7 0 V
5 0 V
4 0 V
2 0 V
1 0 V
2 0 V
3 0 V
6 0 V
6 0 V
8 0 V
9 0 V
9 0 V
10 0 V
11 0 V
11 0 V
10 -1 V
11 0 V
9 0 V
9 0 V
8 0 V
6 0 V
6 0 V
3 0 V
2 0 V
1 0 V
2 0 V
4 0 V
5 0 V
7 0 V
7 0 V
9 0 V
10 0 V
10 0 V
11 0 V
10 0 V
11 0 V
10 0 V
10 0 V
9 -1 V
7 0 V
7 0 V
5 0 V
4 0 V
2 0 V
1 0 V
2 0 V
3 0 V
6 0 V
6 0 V
8 0 V
9 0 V
9 0 V
11 0 V
10 0 V
11 0 V
11 0 V
stroke 6515 828 M
10 0 V
9 0 V
9 0 V
8 0 V
7 0 V
5 0 V
3 -1 V
2 0 V
1 0 V
2 0 V
4 0 V
5 0 V
7 0 V
7 0 V
9 0 V
10 0 V
10 0 V
11 0 V
11 0 V
10 0 V
10 0 V
10 0 V
9 0 V
8 0 V
6 0 V
5 0 V
4 0 V
2 0 V
1 0 V
2 0 V
4 0 V
5 0 V
6 0 V
8 -1 V
9 0 V
9 0 V
11 0 V
10 0 V
11 0 V
11 0 V
10 0 V
10 0 V
8 0 V
8 0 V
7 0 V
5 0 V
3 0 V
2 0 V
1 0 V
3 0 V
3 0 V
5 0 V
7 0 V
8 0 V
8 0 V
10 -1 V
10 0 V
11 0 V
11 0 V
10 0 V
11 0 V
9 0 V
9 0 V
8 0 V
6 0 V
5 0 V
4 0 V
2 0 V
1 0 V
2 0 V
4 0 V
5 0 V
6 0 V
8 0 V
9 0 V
10 0 V
10 0 V
10 -1 V
11 0 V
11 0 V
10 0 V
10 0 V
9 0 V
7 0 V
7 0 V
5 0 V
4 0 V
2 -1 V
stroke
LT2
LCb setrgbcolor
LT2
6248 2056 M
543 0 V
4320 907 M
3 0 V
3 0 V
5 0 V
7 0 V
8 0 V
8 0 V
10 0 V
10 0 V
11 0 V
11 0 V
10 0 V
11 0 V
9 0 V
9 0 V
8 0 V
6 1 V
5 0 V
4 0 V
2 0 V
1 0 V
2 0 V
4 0 V
5 0 V
6 0 V
8 0 V
9 0 V
10 0 V
10 0 V
10 0 V
11 0 V
11 0 V
10 0 V
10 0 V
9 0 V
7 0 V
7 0 V
5 0 V
4 0 V
2 0 V
1 0 V
2 0 V
3 0 V
5 0 V
7 0 V
8 0 V
8 1 V
10 0 V
10 0 V
11 0 V
11 0 V
10 0 V
11 0 V
9 0 V
9 0 V
8 0 V
6 0 V
5 0 V
4 0 V
2 0 V
1 0 V
2 0 V
4 0 V
5 0 V
6 0 V
8 0 V
9 0 V
10 0 V
10 0 V
11 0 V
10 1 V
11 0 V
10 0 V
10 0 V
9 0 V
7 0 V
7 0 V
5 0 V
4 0 V
2 0 V
1 0 V
2 0 V
3 0 V
6 0 V
6 0 V
8 0 V
9 0 V
9 0 V
10 0 V
11 0 V
11 0 V
11 0 V
10 0 V
9 1 V
9 0 V
8 0 V
6 0 V
6 0 V
3 0 V
2 0 V
1 0 V
2 0 V
4 0 V
stroke 5029 911 M
5 0 V
7 0 V
7 0 V
9 0 V
10 0 V
10 0 V
11 0 V
10 0 V
11 0 V
10 0 V
10 0 V
9 0 V
8 0 V
6 0 V
5 0 V
4 0 V
2 0 V
1 1 V
2 0 V
3 0 V
6 0 V
6 0 V
8 0 V
9 0 V
9 0 V
11 0 V
10 0 V
11 0 V
11 0 V
10 0 V
10 0 V
8 0 V
8 0 V
7 0 V
5 0 V
3 0 V
2 0 V
1 0 V
2 0 V
4 0 V
5 0 V
7 0 V
7 0 V
9 1 V
10 0 V
10 0 V
11 0 V
11 0 V
10 0 V
10 0 V
10 0 V
9 0 V
8 0 V
6 0 V
5 0 V
4 0 V
2 0 V
1 0 V
2 0 V
4 0 V
5 0 V
6 0 V
8 0 V
9 0 V
9 0 V
11 0 V
10 1 V
11 0 V
11 0 V
10 0 V
10 0 V
8 0 V
8 0 V
7 0 V
5 0 V
3 0 V
3 0 V
3 0 V
3 0 V
5 0 V
7 0 V
8 0 V
8 0 V
10 0 V
10 0 V
11 0 V
11 0 V
10 0 V
11 1 V
9 0 V
9 0 V
8 0 V
6 0 V
5 0 V
4 0 V
2 0 V
1 0 V
2 0 V
4 0 V
5 0 V
6 0 V
8 0 V
9 0 V
10 0 V
stroke 5770 915 M
10 0 V
10 0 V
11 0 V
11 0 V
10 0 V
10 0 V
9 0 V
7 1 V
7 0 V
5 0 V
4 0 V
2 0 V
1 0 V
2 0 V
3 0 V
5 0 V
7 0 V
8 0 V
8 0 V
10 0 V
10 0 V
11 0 V
11 0 V
10 0 V
11 0 V
9 0 V
9 0 V
8 0 V
6 0 V
5 0 V
4 0 V
2 0 V
1 0 V
2 0 V
4 1 V
5 0 V
6 0 V
8 0 V
9 0 V
10 0 V
10 0 V
11 0 V
10 0 V
11 0 V
10 0 V
10 0 V
9 0 V
7 0 V
7 0 V
5 0 V
4 0 V
2 0 V
1 0 V
2 0 V
3 0 V
6 0 V
6 0 V
8 0 V
9 1 V
9 0 V
10 0 V
11 0 V
11 0 V
10 0 V
11 0 V
9 0 V
9 0 V
8 0 V
6 0 V
6 0 V
3 0 V
2 0 V
1 0 V
2 0 V
4 0 V
5 0 V
7 0 V
7 0 V
9 0 V
10 0 V
10 1 V
11 0 V
10 0 V
11 0 V
10 0 V
10 0 V
9 0 V
7 0 V
7 0 V
5 0 V
4 0 V
2 0 V
1 0 V
2 0 V
3 0 V
6 0 V
6 0 V
8 0 V
9 0 V
9 0 V
11 0 V
10 0 V
11 1 V
11 0 V
stroke 6515 920 M
10 0 V
9 0 V
9 0 V
8 0 V
7 0 V
5 0 V
3 0 V
2 0 V
1 0 V
2 0 V
4 0 V
5 0 V
7 0 V
7 0 V
9 0 V
10 0 V
10 0 V
11 0 V
11 0 V
10 0 V
10 1 V
10 0 V
9 0 V
8 0 V
6 0 V
5 0 V
4 0 V
2 0 V
1 0 V
2 0 V
4 0 V
5 0 V
6 0 V
8 0 V
9 0 V
9 0 V
11 0 V
10 0 V
11 0 V
11 0 V
10 0 V
10 0 V
8 1 V
8 0 V
7 0 V
5 0 V
3 0 V
2 0 V
1 0 V
3 0 V
3 0 V
5 0 V
7 0 V
8 0 V
8 0 V
10 0 V
10 0 V
11 0 V
11 0 V
10 0 V
11 0 V
9 0 V
9 0 V
8 0 V
6 1 V
5 0 V
4 0 V
2 0 V
1 0 V
2 0 V
4 0 V
5 0 V
6 0 V
8 0 V
9 0 V
10 0 V
10 0 V
10 0 V
11 0 V
11 0 V
10 0 V
10 0 V
9 0 V
7 0 V
7 0 V
5 1 V
4 -1 V
2 4 V
stroke
LT3
LCb setrgbcolor
LT3
6248 1856 M
543 0 V
4320 907 M
3 0 V
3 0 V
5 0 V
7 0 V
8 1 V
8 0 V
10 0 V
10 0 V
11 0 V
11 0 V
10 0 V
11 0 V
9 0 V
9 0 V
8 0 V
6 0 V
5 0 V
4 0 V
2 0 V
1 0 V
2 0 V
4 0 V
5 0 V
6 0 V
8 0 V
9 0 V
10 0 V
10 0 V
10 1 V
11 0 V
11 0 V
10 0 V
10 0 V
9 0 V
7 0 V
7 0 V
5 0 V
4 0 V
2 0 V
1 0 V
2 0 V
3 0 V
5 0 V
7 0 V
8 0 V
8 0 V
10 0 V
10 0 V
11 0 V
11 0 V
10 0 V
11 0 V
9 1 V
9 0 V
8 0 V
6 0 V
5 0 V
4 0 V
2 0 V
1 0 V
2 0 V
4 0 V
5 0 V
6 0 V
8 0 V
9 0 V
10 0 V
10 0 V
11 0 V
10 0 V
11 0 V
10 0 V
10 0 V
9 0 V
7 0 V
7 0 V
5 0 V
4 1 V
2 0 V
1 0 V
2 0 V
3 0 V
6 0 V
6 0 V
8 0 V
9 0 V
9 0 V
10 0 V
11 0 V
11 0 V
11 0 V
10 0 V
9 0 V
9 0 V
8 0 V
6 0 V
6 0 V
3 0 V
2 0 V
1 0 V
2 0 V
4 0 V
stroke 5029 911 M
5 0 V
7 0 V
7 0 V
9 1 V
10 0 V
10 0 V
11 0 V
10 0 V
11 0 V
10 0 V
10 0 V
9 0 V
8 0 V
6 0 V
5 0 V
4 0 V
2 0 V
1 0 V
2 0 V
3 0 V
6 0 V
6 0 V
8 0 V
9 0 V
9 0 V
11 0 V
10 1 V
11 0 V
11 0 V
10 0 V
10 0 V
8 0 V
8 0 V
7 0 V
5 0 V
3 0 V
2 0 V
1 0 V
2 0 V
4 0 V
5 0 V
7 0 V
7 0 V
9 0 V
10 0 V
10 0 V
11 0 V
11 0 V
10 0 V
10 1 V
10 0 V
9 0 V
8 0 V
6 0 V
5 0 V
4 0 V
2 0 V
1 0 V
2 0 V
4 0 V
5 0 V
6 0 V
8 0 V
9 0 V
9 0 V
11 0 V
10 0 V
11 0 V
11 0 V
10 0 V
10 0 V
8 0 V
8 1 V
7 0 V
5 0 V
3 0 V
3 0 V
3 0 V
3 0 V
5 0 V
7 0 V
8 0 V
8 0 V
10 0 V
10 0 V
11 0 V
11 0 V
10 0 V
11 0 V
9 0 V
9 0 V
8 0 V
6 0 V
5 0 V
4 0 V
2 0 V
1 0 V
2 1 V
4 0 V
5 0 V
6 0 V
8 0 V
9 0 V
10 0 V
stroke 5770 916 M
10 0 V
10 0 V
11 0 V
11 0 V
10 0 V
10 0 V
9 0 V
7 0 V
7 0 V
5 0 V
4 0 V
2 0 V
1 0 V
2 0 V
3 0 V
5 0 V
7 0 V
8 0 V
8 1 V
10 0 V
10 0 V
11 0 V
11 0 V
10 0 V
11 0 V
9 0 V
9 0 V
8 0 V
6 0 V
5 0 V
4 0 V
2 0 V
1 0 V
2 0 V
4 0 V
5 0 V
6 0 V
8 0 V
9 0 V
10 0 V
10 1 V
11 0 V
10 0 V
11 0 V
10 0 V
10 0 V
9 0 V
7 0 V
7 0 V
5 0 V
4 0 V
2 0 V
1 0 V
2 0 V
3 0 V
6 0 V
6 0 V
8 0 V
9 0 V
9 0 V
10 0 V
11 0 V
11 1 V
10 0 V
11 0 V
9 0 V
9 0 V
8 0 V
6 0 V
6 0 V
3 0 V
2 0 V
1 0 V
2 0 V
4 0 V
5 0 V
7 0 V
7 0 V
9 0 V
10 0 V
10 0 V
11 0 V
10 0 V
11 0 V
10 1 V
10 0 V
9 0 V
7 0 V
7 0 V
5 0 V
4 0 V
2 0 V
1 0 V
2 0 V
3 0 V
6 0 V
6 0 V
8 0 V
9 0 V
9 0 V
11 0 V
10 0 V
11 0 V
11 0 V
stroke 6515 920 M
10 0 V
9 0 V
9 1 V
8 0 V
7 0 V
5 0 V
3 0 V
2 0 V
1 0 V
2 0 V
4 0 V
5 0 V
7 0 V
7 0 V
9 0 V
10 0 V
10 0 V
11 0 V
11 0 V
10 0 V
10 0 V
10 0 V
9 0 V
8 0 V
6 1 V
5 0 V
4 0 V
2 0 V
1 0 V
2 0 V
4 0 V
5 0 V
6 0 V
8 0 V
9 0 V
9 0 V
11 0 V
10 0 V
11 0 V
11 0 V
10 0 V
10 0 V
8 0 V
8 0 V
7 0 V
5 0 V
3 0 V
2 0 V
1 0 V
3 0 V
3 1 V
5 0 V
7 0 V
8 0 V
8 0 V
10 0 V
10 0 V
11 0 V
11 0 V
10 0 V
11 0 V
9 0 V
9 0 V
8 0 V
6 0 V
5 0 V
4 0 V
2 0 V
1 0 V
2 0 V
4 0 V
5 0 V
6 0 V
8 1 V
9 0 V
10 0 V
10 0 V
10 0 V
11 0 V
11 0 V
10 0 V
10 0 V
9 0 V
7 0 V
7 0 V
5 0 V
4 0 V
2 0 V
stroke
LTb
4320 3431 N
0 -2791 V
2879 0 V
0 2791 V
-2879 0 V
Z stroke
1.000 UP
1.000 UL
LTb
stroke
grestore
end
showpage
  }}%
  \put(6128,1856){\makebox(0,0)[r]{\strut{}T$_4$ [GeV$^{-2}$]}}%
  \put(6128,2056){\makebox(0,0)[r]{\strut{}T$_3$ [GeV$^{-2}$]}}%
  \put(6128,2256){\makebox(0,0)[r]{\strut{}T$_2$ [GeV$^{-2}$]}}%
  \put(6128,2456){\makebox(0,0)[r]{\strut{}T$_1$}}%
  \put(5759,3731){\makebox(0,0){\strut{}$\bar B_s\to D_{s2}^*(2573)$}}%
  \put(5759,140){\makebox(0,0){\strut{}$q^2$ [$\rm GeV^2$]}}%
  \put(7199,440){\makebox(0,0){\strut{} 8}}%
  \put(6839,440){\makebox(0,0){\strut{} 7}}%
  \put(6479,440){\makebox(0,0){\strut{} 6}}%
  \put(6119,440){\makebox(0,0){\strut{} 5}}%
  \put(5760,440){\makebox(0,0){\strut{} 4}}%
  \put(5400,440){\makebox(0,0){\strut{} 3}}%
  \put(5040,440){\makebox(0,0){\strut{} 2}}%
  \put(4680,440){\makebox(0,0){\strut{} 1}}%
  \put(4320,440){\makebox(0,0){\strut{} 0}}%
  \put(4200,3431){\makebox(0,0)[r]{\strut{} 1.1}}%
  \put(4200,3198){\makebox(0,0)[r]{\strut{} 1}}%
  \put(4200,2966){\makebox(0,0)[r]{\strut{} 0.9}}%
  \put(4200,2733){\makebox(0,0)[r]{\strut{} 0.8}}%
  \put(4200,2501){\makebox(0,0)[r]{\strut{} 0.7}}%
  \put(4200,2268){\makebox(0,0)[r]{\strut{} 0.6}}%
  \put(4200,2035){\makebox(0,0)[r]{\strut{} 0.5}}%
  \put(4200,1803){\makebox(0,0)[r]{\strut{} 0.4}}%
  \put(4200,1570){\makebox(0,0)[r]{\strut{} 0.3}}%
  \put(4200,1338){\makebox(0,0)[r]{\strut{} 0.2}}%
  \put(4200,1105){\makebox(0,0)[r]{\strut{} 0.1}}%
  \put(4200,873){\makebox(0,0)[r]{\strut{} 0}}%
  \put(4200,640){\makebox(0,0)[r]{\strut{}-0.1}}%
  \put(1376,1717){\makebox(0,0)[r]{\strut{}T$_4$ [GeV$^{-2}$]}}%
  \put(1376,1917){\makebox(0,0)[r]{\strut{}T$_3$ [GeV$^{-2}$]}}%
  \put(1376,2117){\makebox(0,0)[r]{\strut{}T$_2$ [GeV$^{-2}$]}}%
  \put(1376,2317){\makebox(0,0)[r]{\strut{}T$_1$}}%
  \put(1439,3731){\makebox(0,0){\strut{}$\bar B_s\to c\bar s (2^-)$}}%
  \put(1439,140){\makebox(0,0){\strut{}$q^2$ [$\rm GeV^2$]}}%
  \put(2879,440){\makebox(0,0){\strut{} 7}}%
  \put(2468,440){\makebox(0,0){\strut{} 6}}%
  \put(2056,440){\makebox(0,0){\strut{} 5}}%
  \put(1645,440){\makebox(0,0){\strut{} 4}}%
  \put(1234,440){\makebox(0,0){\strut{} 3}}%
  \put(823,440){\makebox(0,0){\strut{} 2}}%
  \put(411,440){\makebox(0,0){\strut{} 1}}%
  \put(0,440){\makebox(0,0){\strut{} 0}}%
  \put(-120,3431){\makebox(0,0)[r]{\strut{} 0.05}}%
  \put(-120,2873){\makebox(0,0)[r]{\strut{} 0}}%
  \put(-120,2315){\makebox(0,0)[r]{\strut{}-0.05}}%
  \put(-120,1756){\makebox(0,0)[r]{\strut{}-0.1}}%
  \put(-120,1198){\makebox(0,0)[r]{\strut{}-0.15}}%
  \put(-120,640){\makebox(0,0)[r]{\strut{}-0.2}}%
\end{picture}%
\endgroup
 

%% file: dg.0m.0p.tex
\begingroup%
\makeatletter%
\newcommand{\GNUPLOTspecial}{%
  \@sanitize\catcode`\%=14\relax\special}%
\setlength{\unitlength}{0.0500bp}%
\begin{picture}(7200,5040)(0,0)%
  {\GNUPLOTspecial{"
/gnudict 256 dict def
gnudict begin
%
%
/Color false def
/Blacktext true def
/Solid false def
/Dashlength 1 def
/Landscape false def
/Level1 false def
/Rounded false def
/ClipToBoundingBox false def
/SuppressPDFMark false def
/TransparentPatterns false def
/gnulinewidth 5.000 def
/userlinewidth gnulinewidth def
/Gamma 1.0 def
/BackgroundColor {-1.000 -1.000 -1.000} def
/vshift -66 def
/dl1 {
  10.0 Dashlength mul mul
  Rounded { currentlinewidth 0.75 mul sub dup 0 le { pop 0.01 } if } if
} def
/dl2 {
  10.0 Dashlength mul mul
  Rounded { currentlinewidth 0.75 mul add } if
} def
/hpt_ 31.5 def
/vpt_ 31.5 def
/hpt hpt_ def
/vpt vpt_ def
/doclip {
  ClipToBoundingBox {
    newpath 0 0 moveto 360 0 lineto 360 252 lineto 0 252 lineto closepath
    clip
  } if
} def
%
%
%
/M {moveto} bind def
/L {lineto} bind def
/R {rmoveto} bind def
/V {rlineto} bind def
/N {newpath moveto} bind def
/Z {closepath} bind def
/C {setrgbcolor} bind def
/f {rlineto fill} bind def
/g {setgray} bind def
/Gshow {show} def   
/vpt2 vpt 2 mul def
/hpt2 hpt 2 mul def
/Lshow {currentpoint stroke M 0 vshift R 
	Blacktext {gsave 0 setgray show grestore} {show} ifelse} def
/Rshow {currentpoint stroke M dup stringwidth pop neg vshift R
	Blacktext {gsave 0 setgray show grestore} {show} ifelse} def
/Cshow {currentpoint stroke M dup stringwidth pop -2 div vshift R 
	Blacktext {gsave 0 setgray show grestore} {show} ifelse} def
/UP {dup vpt_ mul /vpt exch def hpt_ mul /hpt exch def
  /hpt2 hpt 2 mul def /vpt2 vpt 2 mul def} def
/DL {Color {setrgbcolor Solid {pop []} if 0 setdash}
 {pop pop pop 0 setgray Solid {pop []} if 0 setdash} ifelse} def
/BL {stroke userlinewidth 2 mul setlinewidth
	Rounded {1 setlinejoin 1 setlinecap} if} def
/AL {stroke userlinewidth 2 div setlinewidth
	Rounded {1 setlinejoin 1 setlinecap} if} def
/UL {dup gnulinewidth mul /userlinewidth exch def
	dup 1 lt {pop 1} if 10 mul /udl exch def} def
/PL {stroke userlinewidth setlinewidth
	Rounded {1 setlinejoin 1 setlinecap} if} def
3.8 setmiterlimit
/LCw {1 1 1} def
/LCb {0 0 0} def
/LCa {0 0 0} def
/LC0 {1 0 0} def
/LC1 {0 1 0} def
/LC2 {0 0 1} def
/LC3 {1 0 1} def
/LC4 {0 1 1} def
/LC5 {1 1 0} def
/LC6 {0 0 0} def
/LC7 {1 0.3 0} def
/LC8 {0.5 0.5 0.5} def
/LTw {PL [] 1 setgray} def
/LTb {BL [] LCb DL} def
/LTa {AL [1 udl mul 2 udl mul] 0 setdash LCa setrgbcolor} def
/LT0 {PL [] LC0 DL} def
/LT1 {PL [4 dl1 2 dl2] LC1 DL} def
/LT2 {PL [2 dl1 3 dl2] LC2 DL} def
/LT3 {PL [1 dl1 1.5 dl2] LC3 DL} def
/LT4 {PL [6 dl1 2 dl2 1 dl1 2 dl2] LC4 DL} def
/LT5 {PL [3 dl1 3 dl2 1 dl1 3 dl2] LC5 DL} def
/LT6 {PL [2 dl1 2 dl2 2 dl1 6 dl2] LC6 DL} def
/LT7 {PL [1 dl1 2 dl2 6 dl1 2 dl2 1 dl1 2 dl2] LC7 DL} def
/LT8 {PL [2 dl1 2 dl2 2 dl1 2 dl2 2 dl1 2 dl2 2 dl1 4 dl2] LC8 DL} def
/Pnt {stroke [] 0 setdash gsave 1 setlinecap M 0 0 V stroke grestore} def
/Dia {stroke [] 0 setdash 2 copy vpt add M
  hpt neg vpt neg V hpt vpt neg V
  hpt vpt V hpt neg vpt V closepath stroke
  Pnt} def
/Pls {stroke [] 0 setdash vpt sub M 0 vpt2 V
  currentpoint stroke M
  hpt neg vpt neg R hpt2 0 V stroke
 } def
/Box {stroke [] 0 setdash 2 copy exch hpt sub exch vpt add M
  0 vpt2 neg V hpt2 0 V 0 vpt2 V
  hpt2 neg 0 V closepath stroke
  Pnt} def
/Crs {stroke [] 0 setdash exch hpt sub exch vpt add M
  hpt2 vpt2 neg V currentpoint stroke M
  hpt2 neg 0 R hpt2 vpt2 V stroke} def
/TriU {stroke [] 0 setdash 2 copy vpt 1.12 mul add M
  hpt neg vpt -1.62 mul V
  hpt 2 mul 0 V
  hpt neg vpt 1.62 mul V closepath stroke
  Pnt} def
/Star {2 copy Pls Crs} def
/BoxF {stroke [] 0 setdash exch hpt sub exch vpt add M
  0 vpt2 neg V hpt2 0 V 0 vpt2 V
  hpt2 neg 0 V closepath fill} def
/TriUF {stroke [] 0 setdash vpt 1.12 mul add M
  hpt neg vpt -1.62 mul V
  hpt 2 mul 0 V
  hpt neg vpt 1.62 mul V closepath fill} def
/TriD {stroke [] 0 setdash 2 copy vpt 1.12 mul sub M
  hpt neg vpt 1.62 mul V
  hpt 2 mul 0 V
  hpt neg vpt -1.62 mul V closepath stroke
  Pnt} def
/TriDF {stroke [] 0 setdash vpt 1.12 mul sub M
  hpt neg vpt 1.62 mul V
  hpt 2 mul 0 V
  hpt neg vpt -1.62 mul V closepath fill} def
/DiaF {stroke [] 0 setdash vpt add M
  hpt neg vpt neg V hpt vpt neg V
  hpt vpt V hpt neg vpt V closepath fill} def
/Pent {stroke [] 0 setdash 2 copy gsave
  translate 0 hpt M 4 {72 rotate 0 hpt L} repeat
  closepath stroke grestore Pnt} def
/PentF {stroke [] 0 setdash gsave
  translate 0 hpt M 4 {72 rotate 0 hpt L} repeat
  closepath fill grestore} def
/Circle {stroke [] 0 setdash 2 copy
  hpt 0 360 arc stroke Pnt} def
/CircleF {stroke [] 0 setdash hpt 0 360 arc fill} def
/C0 {BL [] 0 setdash 2 copy moveto vpt 90 450 arc} bind def
/C1 {BL [] 0 setdash 2 copy moveto
	2 copy vpt 0 90 arc closepath fill
	vpt 0 360 arc closepath} bind def
/C2 {BL [] 0 setdash 2 copy moveto
	2 copy vpt 90 180 arc closepath fill
	vpt 0 360 arc closepath} bind def
/C3 {BL [] 0 setdash 2 copy moveto
	2 copy vpt 0 180 arc closepath fill
	vpt 0 360 arc closepath} bind def
/C4 {BL [] 0 setdash 2 copy moveto
	2 copy vpt 180 270 arc closepath fill
	vpt 0 360 arc closepath} bind def
/C5 {BL [] 0 setdash 2 copy moveto
	2 copy vpt 0 90 arc
	2 copy moveto
	2 copy vpt 180 270 arc closepath fill
	vpt 0 360 arc} bind def
/C6 {BL [] 0 setdash 2 copy moveto
	2 copy vpt 90 270 arc closepath fill
	vpt 0 360 arc closepath} bind def
/C7 {BL [] 0 setdash 2 copy moveto
	2 copy vpt 0 270 arc closepath fill
	vpt 0 360 arc closepath} bind def
/C8 {BL [] 0 setdash 2 copy moveto
	2 copy vpt 270 360 arc closepath fill
	vpt 0 360 arc closepath} bind def
/C9 {BL [] 0 setdash 2 copy moveto
	2 copy vpt 270 450 arc closepath fill
	vpt 0 360 arc closepath} bind def
/C10 {BL [] 0 setdash 2 copy 2 copy moveto vpt 270 360 arc closepath fill
	2 copy moveto
	2 copy vpt 90 180 arc closepath fill
	vpt 0 360 arc closepath} bind def
/C11 {BL [] 0 setdash 2 copy moveto
	2 copy vpt 0 180 arc closepath fill
	2 copy moveto
	2 copy vpt 270 360 arc closepath fill
	vpt 0 360 arc closepath} bind def
/C12 {BL [] 0 setdash 2 copy moveto
	2 copy vpt 180 360 arc closepath fill
	vpt 0 360 arc closepath} bind def
/C13 {BL [] 0 setdash 2 copy moveto
	2 copy vpt 0 90 arc closepath fill
	2 copy moveto
	2 copy vpt 180 360 arc closepath fill
	vpt 0 360 arc closepath} bind def
/C14 {BL [] 0 setdash 2 copy moveto
	2 copy vpt 90 360 arc closepath fill
	vpt 0 360 arc} bind def
/C15 {BL [] 0 setdash 2 copy vpt 0 360 arc closepath fill
	vpt 0 360 arc closepath} bind def
/Rec {newpath 4 2 roll moveto 1 index 0 rlineto 0 exch rlineto
	neg 0 rlineto closepath} bind def
/Square {dup Rec} bind def
/Bsquare {vpt sub exch vpt sub exch vpt2 Square} bind def
/S0 {BL [] 0 setdash 2 copy moveto 0 vpt rlineto BL Bsquare} bind def
/S1 {BL [] 0 setdash 2 copy vpt Square fill Bsquare} bind def
/S2 {BL [] 0 setdash 2 copy exch vpt sub exch vpt Square fill Bsquare} bind def
/S3 {BL [] 0 setdash 2 copy exch vpt sub exch vpt2 vpt Rec fill Bsquare} bind def
/S4 {BL [] 0 setdash 2 copy exch vpt sub exch vpt sub vpt Square fill Bsquare} bind def
/S5 {BL [] 0 setdash 2 copy 2 copy vpt Square fill
	exch vpt sub exch vpt sub vpt Square fill Bsquare} bind def
/S6 {BL [] 0 setdash 2 copy exch vpt sub exch vpt sub vpt vpt2 Rec fill Bsquare} bind def
/S7 {BL [] 0 setdash 2 copy exch vpt sub exch vpt sub vpt vpt2 Rec fill
	2 copy vpt Square fill Bsquare} bind def
/S8 {BL [] 0 setdash 2 copy vpt sub vpt Square fill Bsquare} bind def
/S9 {BL [] 0 setdash 2 copy vpt sub vpt vpt2 Rec fill Bsquare} bind def
/S10 {BL [] 0 setdash 2 copy vpt sub vpt Square fill 2 copy exch vpt sub exch vpt Square fill
	Bsquare} bind def
/S11 {BL [] 0 setdash 2 copy vpt sub vpt Square fill 2 copy exch vpt sub exch vpt2 vpt Rec fill
	Bsquare} bind def
/S12 {BL [] 0 setdash 2 copy exch vpt sub exch vpt sub vpt2 vpt Rec fill Bsquare} bind def
/S13 {BL [] 0 setdash 2 copy exch vpt sub exch vpt sub vpt2 vpt Rec fill
	2 copy vpt Square fill Bsquare} bind def
/S14 {BL [] 0 setdash 2 copy exch vpt sub exch vpt sub vpt2 vpt Rec fill
	2 copy exch vpt sub exch vpt Square fill Bsquare} bind def
/S15 {BL [] 0 setdash 2 copy Bsquare fill Bsquare} bind def
/D0 {gsave translate 45 rotate 0 0 S0 stroke grestore} bind def
/D1 {gsave translate 45 rotate 0 0 S1 stroke grestore} bind def
/D2 {gsave translate 45 rotate 0 0 S2 stroke grestore} bind def
/D3 {gsave translate 45 rotate 0 0 S3 stroke grestore} bind def
/D4 {gsave translate 45 rotate 0 0 S4 stroke grestore} bind def
/D5 {gsave translate 45 rotate 0 0 S5 stroke grestore} bind def
/D6 {gsave translate 45 rotate 0 0 S6 stroke grestore} bind def
/D7 {gsave translate 45 rotate 0 0 S7 stroke grestore} bind def
/D8 {gsave translate 45 rotate 0 0 S8 stroke grestore} bind def
/D9 {gsave translate 45 rotate 0 0 S9 stroke grestore} bind def
/D10 {gsave translate 45 rotate 0 0 S10 stroke grestore} bind def
/D11 {gsave translate 45 rotate 0 0 S11 stroke grestore} bind def
/D12 {gsave translate 45 rotate 0 0 S12 stroke grestore} bind def
/D13 {gsave translate 45 rotate 0 0 S13 stroke grestore} bind def
/D14 {gsave translate 45 rotate 0 0 S14 stroke grestore} bind def
/D15 {gsave translate 45 rotate 0 0 S15 stroke grestore} bind def
/DiaE {stroke [] 0 setdash vpt add M
  hpt neg vpt neg V hpt vpt neg V
  hpt vpt V hpt neg vpt V closepath stroke} def
/BoxE {stroke [] 0 setdash exch hpt sub exch vpt add M
  0 vpt2 neg V hpt2 0 V 0 vpt2 V
  hpt2 neg 0 V closepath stroke} def
/TriUE {stroke [] 0 setdash vpt 1.12 mul add M
  hpt neg vpt -1.62 mul V
  hpt 2 mul 0 V
  hpt neg vpt 1.62 mul V closepath stroke} def
/TriDE {stroke [] 0 setdash vpt 1.12 mul sub M
  hpt neg vpt 1.62 mul V
  hpt 2 mul 0 V
  hpt neg vpt -1.62 mul V closepath stroke} def
/PentE {stroke [] 0 setdash gsave
  translate 0 hpt M 4 {72 rotate 0 hpt L} repeat
  closepath stroke grestore} def
/CircE {stroke [] 0 setdash 
  hpt 0 360 arc stroke} def
/Opaque {gsave closepath 1 setgray fill grestore 0 setgray closepath} def
/DiaW {stroke [] 0 setdash vpt add M
  hpt neg vpt neg V hpt vpt neg V
  hpt vpt V hpt neg vpt V Opaque stroke} def
/BoxW {stroke [] 0 setdash exch hpt sub exch vpt add M
  0 vpt2 neg V hpt2 0 V 0 vpt2 V
  hpt2 neg 0 V Opaque stroke} def
/TriUW {stroke [] 0 setdash vpt 1.12 mul add M
  hpt neg vpt -1.62 mul V
  hpt 2 mul 0 V
  hpt neg vpt 1.62 mul V Opaque stroke} def
/TriDW {stroke [] 0 setdash vpt 1.12 mul sub M
  hpt neg vpt 1.62 mul V
  hpt 2 mul 0 V
  hpt neg vpt -1.62 mul V Opaque stroke} def
/PentW {stroke [] 0 setdash gsave
  translate 0 hpt M 4 {72 rotate 0 hpt L} repeat
  Opaque stroke grestore} def
/CircW {stroke [] 0 setdash 
  hpt 0 360 arc Opaque stroke} def
/BoxFill {gsave Rec 1 setgray fill grestore} def
/Density {
  /Fillden exch def
  currentrgbcolor
  /ColB exch def /ColG exch def /ColR exch def
  /ColR ColR Fillden mul Fillden sub 1 add def
  /ColG ColG Fillden mul Fillden sub 1 add def
  /ColB ColB Fillden mul Fillden sub 1 add def
  ColR ColG ColB setrgbcolor} def
/BoxColFill {gsave Rec PolyFill} def
/PolyFill {gsave Density fill grestore grestore} def
/h {rlineto rlineto rlineto gsave closepath fill grestore} bind def
%
%
/PatternFill {gsave /PFa [ 9 2 roll ] def
  PFa 0 get PFa 2 get 2 div add PFa 1 get PFa 3 get 2 div add translate
  PFa 2 get -2 div PFa 3 get -2 div PFa 2 get PFa 3 get Rec
  gsave 1 setgray fill grestore clip
  currentlinewidth 0.5 mul setlinewidth
  /PFs PFa 2 get dup mul PFa 3 get dup mul add sqrt def
  0 0 M PFa 5 get rotate PFs -2 div dup translate
  0 1 PFs PFa 4 get div 1 add floor cvi
	{PFa 4 get mul 0 M 0 PFs V} for
  0 PFa 6 get ne {
	0 1 PFs PFa 4 get div 1 add floor cvi
	{PFa 4 get mul 0 2 1 roll M PFs 0 V} for
 } if
  stroke grestore} def
/languagelevel where
 {pop languagelevel} {1} ifelse
 2 lt
	{/InterpretLevel1 true def}
	{/InterpretLevel1 Level1 def}
 ifelse
%
%
/Level2PatternFill {
/Tile8x8 {/PaintType 2 /PatternType 1 /TilingType 1 /BBox [0 0 8 8] /XStep 8 /YStep 8}
	bind def
/KeepColor {currentrgbcolor [/Pattern /DeviceRGB] setcolorspace} bind def
<< Tile8x8
 /PaintProc {0.5 setlinewidth pop 0 0 M 8 8 L 0 8 M 8 0 L stroke} 
>> matrix makepattern
/Pat1 exch def
<< Tile8x8
 /PaintProc {0.5 setlinewidth pop 0 0 M 8 8 L 0 8 M 8 0 L stroke
	0 4 M 4 8 L 8 4 L 4 0 L 0 4 L stroke}
>> matrix makepattern
/Pat2 exch def
<< Tile8x8
 /PaintProc {0.5 setlinewidth pop 0 0 M 0 8 L
	8 8 L 8 0 L 0 0 L fill}
>> matrix makepattern
/Pat3 exch def
<< Tile8x8
 /PaintProc {0.5 setlinewidth pop -4 8 M 8 -4 L
	0 12 M 12 0 L stroke}
>> matrix makepattern
/Pat4 exch def
<< Tile8x8
 /PaintProc {0.5 setlinewidth pop -4 0 M 8 12 L
	0 -4 M 12 8 L stroke}
>> matrix makepattern
/Pat5 exch def
<< Tile8x8
 /PaintProc {0.5 setlinewidth pop -2 8 M 4 -4 L
	0 12 M 8 -4 L 4 12 M 10 0 L stroke}
>> matrix makepattern
/Pat6 exch def
<< Tile8x8
 /PaintProc {0.5 setlinewidth pop -2 0 M 4 12 L
	0 -4 M 8 12 L 4 -4 M 10 8 L stroke}
>> matrix makepattern
/Pat7 exch def
<< Tile8x8
 /PaintProc {0.5 setlinewidth pop 8 -2 M -4 4 L
	12 0 M -4 8 L 12 4 M 0 10 L stroke}
>> matrix makepattern
/Pat8 exch def
<< Tile8x8
 /PaintProc {0.5 setlinewidth pop 0 -2 M 12 4 L
	-4 0 M 12 8 L -4 4 M 8 10 L stroke}
>> matrix makepattern
/Pat9 exch def
/Pattern1 {PatternBgnd KeepColor Pat1 setpattern} bind def
/Pattern2 {PatternBgnd KeepColor Pat2 setpattern} bind def
/Pattern3 {PatternBgnd KeepColor Pat3 setpattern} bind def
/Pattern4 {PatternBgnd KeepColor Landscape {Pat5} {Pat4} ifelse setpattern} bind def
/Pattern5 {PatternBgnd KeepColor Landscape {Pat4} {Pat5} ifelse setpattern} bind def
/Pattern6 {PatternBgnd KeepColor Landscape {Pat9} {Pat6} ifelse setpattern} bind def
/Pattern7 {PatternBgnd KeepColor Landscape {Pat8} {Pat7} ifelse setpattern} bind def
} def
%
%
%
/PatternBgnd {
  TransparentPatterns {} {gsave 1 setgray fill grestore} ifelse
} def
%
%
/Level1PatternFill {
/Pattern1 {0.250 Density} bind def
/Pattern2 {0.500 Density} bind def
/Pattern3 {0.750 Density} bind def
/Pattern4 {0.125 Density} bind def
/Pattern5 {0.375 Density} bind def
/Pattern6 {0.625 Density} bind def
/Pattern7 {0.875 Density} bind def
} def
%
%
Level1 {Level1PatternFill} {Level2PatternFill} ifelse
/Symbol-Oblique /Symbol findfont [1 0 .167 1 0 0] makefont
dup length dict begin {1 index /FID eq {pop pop} {def} ifelse} forall
currentdict end definefont pop
Level1 SuppressPDFMark or 
{} {
/SDict 10 dict def
systemdict /pdfmark known not {
  userdict /pdfmark systemdict /cleartomark get put
} if
SDict begin [
  /Title (paper/dg.0m.0p.tex)
  /Subject (gnuplot plot)
  /Creator (gnuplot 4.6 patchlevel 0)
  /Author (conrado)
  /CreationDate (Tue Oct 15 09:50:31 2013)
  /DOCINFO pdfmark
end
} ifelse
end
gnudict begin
gsave
doclip
0 0 translate
0.050 0.050 scale
0 setgray
newpath
BackgroundColor 0 lt 3 1 roll 0 lt exch 0 lt or or not {BackgroundColor C 1.000 0 0 7200.00 5040.00 BoxColFill} if
1.000 UL
LTb
0 640 M
63 0 V
2816 0 R
-63 0 V
0 950 M
63 0 V
2816 0 R
-63 0 V
0 1260 M
63 0 V
2816 0 R
-63 0 V
0 1570 M
63 0 V
2816 0 R
-63 0 V
0 1880 M
63 0 V
2816 0 R
-63 0 V
0 2191 M
63 0 V
2816 0 R
-63 0 V
0 2501 M
63 0 V
2816 0 R
-63 0 V
0 2811 M
63 0 V
2816 0 R
-63 0 V
0 3121 M
63 0 V
2816 0 R
-63 0 V
0 3431 M
63 0 V
2816 0 R
-63 0 V
0 640 M
0 63 V
0 3431 M
0 -63 V
480 640 M
0 63 V
0 2728 R
0 -63 V
960 640 M
0 63 V
0 2728 R
0 -63 V
1440 640 M
0 63 V
0 2728 R
0 -63 V
1919 640 M
0 63 V
0 2728 R
0 -63 V
2399 640 M
0 63 V
0 2728 R
0 -63 V
2879 640 M
0 63 V
0 2728 R
0 -63 V
stroke
LTa
0 640 M
2879 0 V
stroke
LTb
0 3431 N
0 640 L
2879 0 V
0 2791 V
0 3431 L
Z stroke
LCb setrgbcolor
LTb
LCb setrgbcolor
LTb
1.000 UP
1.000 UL
LTb
1.000 UL
LT0
LCb setrgbcolor
LT0
1928 2336 M
543 0 V
0 3356 M
2 -2 V
4 -3 V
5 -4 V
7 -6 V
7 -7 V
9 -8 V
9 -8 V
10 -9 V
11 -10 V
11 -9 V
10 -10 V
10 -9 V
10 -9 V
8 -8 V
8 -7 V
6 -6 V
5 -5 V
4 -3 V
2 -2 V
1 -1 V
2 -2 V
4 -3 V
5 -5 V
6 -6 V
8 -7 V
8 -8 V
10 -9 V
10 -9 V
10 -10 V
11 -10 V
11 -10 V
10 -10 V
9 -9 V
9 -9 V
7 -7 V
7 -6 V
5 -5 V
4 -4 V
2 -2 V
1 0 V
2 -2 V
3 -4 V
5 -5 V
7 -6 V
7 -8 V
9 -8 V
9 -10 V
11 -10 V
10 -10 V
11 -10 V
10 -11 V
10 -10 V
10 -10 V
8 -8 V
8 -8 V
6 -6 V
6 -6 V
3 -3 V
2 -2 V
1 -1 V
2 -2 V
4 -4 V
5 -5 V
6 -7 V
8 -7 V
8 -9 V
10 -10 V
10 -10 V
11 -11 V
10 -11 V
11 -11 V
10 -10 V
9 -10 V
9 -9 V
8 -8 V
6 -7 V
5 -5 V
4 -4 V
2 -2 V
1 -1 V
2 -2 V
3 -4 V
5 -5 V
7 -7 V
7 -8 V
9 -9 V
10 -11 V
10 -10 V
10 -11 V
11 -12 V
10 -11 V
10 -11 V
10 -10 V
8 -9 V
8 -8 V
7 -7 V
5 -6 V
3 -4 V
2 -2 V
1 -1 V
2 -2 V
4 -4 V
stroke 699 2664 M
5 -6 V
6 -7 V
8 -8 V
9 -9 V
9 -11 V
10 -11 V
11 -11 V
10 -12 V
11 -11 V
10 -11 V
9 -11 V
9 -9 V
8 -9 V
6 -7 V
5 -6 V
4 -4 V
2 -2 V
1 -1 V
2 -2 V
3 -4 V
5 -6 V
7 -7 V
7 -8 V
9 -10 V
10 -11 V
10 -11 V
10 -12 V
11 -12 V
10 -11 V
10 -12 V
10 -10 V
9 -10 V
7 -9 V
7 -7 V
5 -6 V
3 -4 V
2 -2 V
1 -1 V
2 -2 V
4 -4 V
5 -6 V
6 -7 V
8 -9 V
9 -10 V
9 -11 V
10 -11 V
11 -12 V
10 -12 V
11 -12 V
10 -12 V
9 -11 V
9 -9 V
8 -9 V
6 -8 V
5 -5 V
4 -4 V
2 -3 V
1 -1 V
2 -2 V
3 -4 V
5 -6 V
7 -7 V
7 -9 V
9 -10 V
10 -11 V
10 -12 V
10 -12 V
11 -12 V
10 -12 V
10 -12 V
10 -11 V
9 -10 V
7 -8 V
7 -8 V
5 -6 V
3 -4 V
2 -2 V
1 -1 V
2 -3 V
4 -4 V
5 -5 V
6 -8 V
8 -9 V
9 -10 V
9 -11 V
10 -11 V
11 -12 V
10 -13 V
11 -12 V
10 -11 V
9 -11 V
9 -10 V
8 -9 V
6 -8 V
5 -6 V
4 -4 V
2 -2 V
1 -1 V
2 -2 V
3 -5 V
5 -5 V
7 -8 V
8 -9 V
8 -10 V
stroke 1419 1847 M
10 -11 V
10 -11 V
10 -12 V
11 -12 V
10 -12 V
10 -12 V
10 -11 V
9 -10 V
7 -9 V
7 -7 V
5 -6 V
3 -4 V
2 -2 V
1 -1 V
2 -3 V
4 -4 V
5 -6 V
6 -7 V
8 -9 V
9 -9 V
9 -11 V
10 -12 V
11 -12 V
10 -12 V
11 -12 V
10 -11 V
9 -11 V
9 -10 V
8 -8 V
6 -8 V
5 -5 V
4 -4 V
2 -3 V
1 -1 V
2 -2 V
3 -4 V
6 -6 V
6 -7 V
8 -8 V
8 -10 V
10 -11 V
10 -11 V
10 -12 V
11 -11 V
10 -12 V
11 -11 V
9 -11 V
9 -9 V
7 -9 V
7 -7 V
5 -5 V
3 -4 V
2 -2 V
1 -2 V
2 -2 V
4 -4 V
5 -5 V
6 -7 V
8 -8 V
9 -10 V
9 -10 V
10 -11 V
11 -11 V
10 -12 V
11 -11 V
10 -11 V
10 -10 V
8 -9 V
8 -8 V
6 -7 V
5 -5 V
4 -4 V
2 -2 V
1 -1 V
2 -2 V
4 -4 V
5 -5 V
6 -7 V
8 -8 V
8 -9 V
10 -10 V
10 -10 V
10 -11 V
11 -11 V
10 -11 V
11 -10 V
9 -10 V
9 -8 V
7 -8 V
7 -6 V
5 -6 V
3 -3 V
2 -2 V
1 -1 V
2 -2 V
4 -4 V
5 -5 V
7 -6 V
7 -7 V
9 -9 V
9 -9 V
10 -10 V
11 -10 V
11 -10 V
stroke 2153 1053 M
10 -10 V
10 -10 V
10 -9 V
8 -8 V
8 -7 V
6 -6 V
5 -4 V
4 -4 V
2 -2 V
1 0 V
2 -2 V
4 -4 V
5 -4 V
6 -6 V
8 -7 V
8 -8 V
10 -8 V
10 -9 V
10 -9 V
11 -9 V
11 -9 V
10 -9 V
9 -8 V
9 -7 V
7 -7 V
7 -5 V
5 -4 V
4 -3 V
2 -2 V
1 -1 V
2 -1 V
3 -3 V
5 -4 V
7 -5 V
7 -7 V
9 -6 V
9 -8 V
11 -8 V
10 -8 V
11 -8 V
10 -7 V
10 -8 V
10 -7 V
8 -6 V
8 -5 V
6 -5 V
5 -3 V
4 -3 V
2 -1 V
1 -1 V
2 -1 V
4 -3 V
5 -3 V
6 -4 V
8 -5 V
8 -6 V
10 -6 V
10 -6 V
11 -7 V
10 -6 V
11 -6 V
10 -6 V
9 -5 V
9 -5 V
8 -4 V
6 -3 V
5 -3 V
4 -2 V
2 -1 V
1 0 V
2 -1 V
3 -2 V
5 -2 V
7 -3 V
7 -4 V
9 -4 V
9 -4 V
11 -4 V
10 -4 V
11 -3 V
10 -4 V
10 -3 V
10 -2 V
8 -2 V
8 -2 V
6 -1 V
6 -1 V
3 0 V
2 0 V
stroke
LT1
LCb setrgbcolor
LT1
1928 2136 M
543 0 V
758 640 M
1 0 V
3 0 V
4 1 V
4 1 V
6 2 V
6 3 V
7 4 V
7 5 V
8 6 V
8 7 V
7 7 V
8 8 V
6 8 V
7 8 V
5 7 V
5 6 V
4 5 V
2 3 V
2 2 V
0 1 V
2 2 V
3 4 V
3 5 V
5 7 V
5 8 V
7 9 V
7 11 V
7 11 V
8 12 V
7 12 V
8 13 V
7 11 V
7 12 V
6 10 V
6 9 V
5 7 V
3 6 V
3 5 V
1 2 V
1 1 V
2 3 V
2 4 V
4 6 V
5 7 V
5 9 V
6 11 V
7 11 V
8 11 V
7 13 V
8 12 V
8 12 V
7 11 V
7 11 V
6 9 V
6 9 V
4 7 V
4 5 V
3 4 V
1 2 V
1 1 V
1 2 V
3 4 V
4 6 V
4 6 V
6 8 V
6 9 V
7 10 V
7 10 V
8 11 V
8 10 V
7 10 V
8 10 V
6 8 V
7 8 V
5 7 V
5 6 V
4 4 V
2 3 V
2 2 V
0 1 V
2 2 V
2 3 V
4 4 V
5 6 V
5 6 V
7 7 V
7 8 V
7 8 V
8 8 V
7 8 V
8 7 V
7 8 V
7 6 V
6 6 V
6 5 V
5 5 V
3 3 V
3 2 V
1 2 V
1 0 V
2 1 V
2 3 V
stroke 1265 1282 M
4 3 V
5 4 V
5 5 V
6 5 V
7 5 V
8 6 V
7 6 V
8 5 V
8 6 V
7 5 V
7 4 V
6 5 V
6 3 V
4 3 V
4 2 V
3 2 V
1 1 V
1 0 V
1 1 V
3 1 V
3 3 V
5 2 V
6 3 V
6 4 V
7 3 V
7 4 V
8 4 V
8 3 V
7 3 V
8 4 V
6 2 V
7 3 V
5 2 V
5 1 V
4 2 V
2 1 V
2 0 V
2 1 V
2 1 V
4 1 V
5 1 V
5 2 V
7 2 V
7 2 V
7 1 V
8 2 V
7 2 V
8 1 V
7 2 V
7 1 V
6 1 V
6 1 V
5 0 V
3 1 V
3 0 V
1 0 V
1 0 V
2 1 V
2 0 V
4 0 V
4 1 V
6 0 V
6 0 V
7 1 V
8 0 V
7 0 V
8 0 V
8 0 V
7 0 V
7 0 V
6 0 V
6 0 V
4 -1 V
4 0 V
3 0 V
1 0 V
1 0 V
1 0 V
3 -1 V
3 0 V
5 0 V
6 -1 V
6 -1 V
7 -1 V
7 -1 V
8 -1 V
8 -1 V
7 -1 V
8 -2 V
6 -1 V
7 -1 V
5 -2 V
5 -1 V
4 -1 V
2 0 V
2 0 V
0 -1 V
2 0 V
2 -1 V
4 -1 V
5 -1 V
5 -1 V
7 -2 V
7 -2 V
stroke 1796 1397 M
7 -2 V
8 -2 V
7 -3 V
8 -2 V
7 -2 V
7 -3 V
6 -2 V
6 -2 V
5 -2 V
3 -1 V
3 -1 V
1 -1 V
1 0 V
2 0 V
2 -1 V
4 -2 V
4 -2 V
6 -2 V
6 -2 V
7 -3 V
8 -3 V
7 -3 V
8 -4 V
7 -3 V
8 -3 V
7 -3 V
6 -3 V
6 -3 V
4 -2 V
4 -2 V
3 -1 V
1 -1 V
1 0 V
1 -1 V
3 -1 V
3 -2 V
5 -2 V
6 -3 V
6 -3 V
7 -4 V
7 -4 V
8 -4 V
8 -4 V
7 -4 V
7 -4 V
7 -4 V
7 -3 V
5 -3 V
5 -3 V
4 -2 V
2 -2 V
2 0 V
0 -1 V
2 -1 V
2 -1 V
4 -2 V
5 -3 V
5 -3 V
7 -4 V
7 -4 V
7 -5 V
7 -4 V
8 -5 V
8 -5 V
7 -4 V
7 -5 V
6 -4 V
6 -3 V
5 -3 V
3 -3 V
3 -1 V
1 -1 V
1 -1 V
1 -1 V
3 -1 V
4 -3 V
4 -3 V
6 -4 V
6 -4 V
7 -4 V
8 -5 V
7 -6 V
8 -5 V
7 -5 V
8 -5 V
7 -5 V
6 -4 V
6 -4 V
4 -4 V
4 -2 V
3 -2 V
1 -1 V
1 -1 V
1 -1 V
3 -1 V
3 -3 V
5 -3 V
6 -4 V
6 -5 V
7 -5 V
7 -5 V
8 -6 V
8 -6 V
7 -5 V
stroke 2329 1102 M
7 -6 V
7 -5 V
7 -5 V
5 -4 V
5 -3 V
4 -3 V
2 -2 V
2 -1 V
0 -1 V
2 -1 V
2 -2 V
4 -3 V
5 -3 V
5 -5 V
7 -5 V
7 -5 V
7 -6 V
7 -6 V
8 -6 V
8 -6 V
7 -6 V
7 -5 V
6 -5 V
6 -5 V
5 -3 V
3 -3 V
3 -3 V
1 -1 V
1 0 V
1 -1 V
3 -3 V
4 -3 V
4 -3 V
6 -5 V
6 -5 V
7 -6 V
8 -6 V
7 -6 V
8 -7 V
7 -6 V
8 -6 V
7 -6 V
6 -6 V
6 -5 V
4 -4 V
4 -3 V
2 -2 V
2 -1 V
1 -1 V
1 -1 V
3 -3 V
3 -3 V
5 -4 V
6 -5 V
6 -6 V
7 -6 V
7 -6 V
8 -7 V
7 -8 V
8 -7 V
7 -7 V
7 -6 V
7 -7 V
5 -5 V
5 -5 V
4 -3 V
2 -3 V
2 -2 V
2 -2 V
2 -3 V
4 -3 V
5 -5 V
5 -6 V
7 -7 V
6 -8 V
8 -9 V
7 -9 V
8 -10 V
8 -10 V
7 -11 V
7 -11 V
6 -12 V
6 -11 V
5 -12 V
3 -12 V
3 -12 V
1 -12 V
stroke
LTb
0 3431 N
0 640 L
2879 0 V
0 2791 V
0 3431 L
Z stroke
1.000 UP
1.000 UL
LTb
1.000 UL
LTb
4320 640 M
63 0 V
2816 0 R
-63 0 V
4320 919 M
63 0 V
2816 0 R
-63 0 V
4320 1198 M
63 0 V
2816 0 R
-63 0 V
4320 1477 M
63 0 V
2816 0 R
-63 0 V
4320 1756 M
63 0 V
2816 0 R
-63 0 V
4320 2036 M
63 0 V
2816 0 R
-63 0 V
4320 2315 M
63 0 V
2816 0 R
-63 0 V
4320 2594 M
63 0 V
2816 0 R
-63 0 V
4320 2873 M
63 0 V
2816 0 R
-63 0 V
4320 3152 M
63 0 V
2816 0 R
-63 0 V
4320 3431 M
63 0 V
2816 0 R
-63 0 V
4320 640 M
0 63 V
0 2728 R
0 -63 V
4608 640 M
0 63 V
0 2728 R
0 -63 V
4896 640 M
0 63 V
0 2728 R
0 -63 V
5184 640 M
0 63 V
0 2728 R
0 -63 V
5472 640 M
0 63 V
0 2728 R
0 -63 V
5760 640 M
0 63 V
0 2728 R
0 -63 V
6047 640 M
0 63 V
0 2728 R
0 -63 V
6335 640 M
0 63 V
0 2728 R
0 -63 V
6623 640 M
0 63 V
0 2728 R
0 -63 V
6911 640 M
0 63 V
0 2728 R
0 -63 V
7199 640 M
0 63 V
0 2728 R
0 -63 V
stroke
LTa
4320 640 M
2879 0 V
stroke
LTb
4320 3431 N
0 -2791 V
2879 0 V
0 2791 V
-2879 0 V
Z stroke
LCb setrgbcolor
LTb
LCb setrgbcolor
LTb
1.000 UP
1.000 UL
LTb
1.000 UL
LT0
LCb setrgbcolor
LT0
6248 2615 M
543 0 V
4320 3216 M
2 -1 V
4 -3 V
5 -5 V
6 -5 V
7 -7 V
9 -7 V
9 -9 V
10 -8 V
10 -10 V
10 -9 V
10 -9 V
10 -9 V
9 -8 V
8 -8 V
8 -7 V
6 -6 V
5 -4 V
3 -3 V
2 -2 V
1 -1 V
2 -2 V
4 -3 V
5 -5 V
6 -6 V
7 -6 V
9 -8 V
9 -9 V
9 -9 V
11 -10 V
10 -9 V
10 -10 V
10 -9 V
9 -9 V
8 -8 V
8 -7 V
6 -6 V
5 -5 V
3 -3 V
2 -2 V
1 -1 V
2 -2 V
4 -3 V
4 -5 V
7 -6 V
7 -7 V
8 -9 V
10 -9 V
9 -9 V
10 -10 V
11 -10 V
10 -10 V
10 -10 V
9 -9 V
8 -8 V
7 -8 V
7 -6 V
5 -5 V
3 -3 V
2 -2 V
1 -1 V
2 -2 V
3 -4 V
5 -5 V
6 -6 V
8 -7 V
8 -9 V
9 -9 V
10 -10 V
10 -10 V
10 -11 V
11 -10 V
9 -10 V
10 -10 V
8 -8 V
7 -8 V
7 -6 V
4 -5 V
4 -4 V
2 -2 V
1 -1 V
2 -2 V
3 -4 V
5 -5 V
6 -6 V
8 -8 V
8 -9 V
9 -9 V
10 -10 V
10 -11 V
10 -11 V
10 -11 V
10 -10 V
9 -10 V
9 -8 V
7 -8 V
6 -7 V
5 -5 V
4 -4 V
2 -2 V
1 -1 V
2 -2 V
3 -4 V
stroke 4995 2552 M
5 -5 V
6 -6 V
7 -8 V
9 -9 V
9 -10 V
10 -11 V
10 -11 V
10 -11 V
10 -11 V
10 -10 V
9 -10 V
9 -9 V
7 -8 V
6 -7 V
5 -5 V
4 -4 V
1 -2 V
1 -1 V
2 -2 V
4 -4 V
5 -5 V
6 -7 V
7 -8 V
9 -9 V
9 -10 V
10 -11 V
10 -11 V
10 -11 V
10 -12 V
10 -10 V
9 -10 V
8 -10 V
8 -8 V
6 -7 V
5 -5 V
3 -4 V
2 -2 V
1 -1 V
2 -2 V
4 -4 V
4 -6 V
7 -7 V
7 -8 V
8 -9 V
10 -10 V
9 -11 V
11 -11 V
10 -12 V
10 -11 V
10 -11 V
9 -10 V
8 -10 V
8 -8 V
6 -7 V
5 -5 V
3 -4 V
2 -2 V
1 -1 V
2 -3 V
3 -3 V
5 -6 V
7 -7 V
7 -8 V
8 -10 V
9 -10 V
10 -11 V
10 -11 V
11 -12 V
10 -11 V
9 -11 V
10 -10 V
8 -10 V
7 -8 V
7 -7 V
4 -6 V
4 -3 V
2 -3 V
1 -1 V
2 -2 V
3 -4 V
5 -5 V
6 -7 V
8 -9 V
8 -9 V
9 -10 V
10 -11 V
10 -12 V
10 -11 V
11 -12 V
9 -11 V
9 -10 V
9 -9 V
7 -9 V
6 -7 V
5 -5 V
4 -4 V
2 -2 V
1 -1 V
2 -3 V
3 -3 V
5 -6 V
6 -7 V
8 -8 V
8 -10 V
stroke 5691 1780 M
9 -10 V
10 -11 V
10 -11 V
10 -12 V
10 -11 V
10 -11 V
9 -10 V
9 -10 V
7 -8 V
6 -7 V
5 -5 V
4 -4 V
2 -2 V
0 -1 V
2 -2 V
4 -4 V
5 -6 V
6 -7 V
7 -8 V
9 -9 V
9 -10 V
10 -11 V
10 -11 V
10 -12 V
10 -11 V
10 -11 V
9 -10 V
8 -9 V
8 -8 V
6 -7 V
5 -5 V
3 -4 V
2 -2 V
1 -1 V
2 -2 V
4 -4 V
5 -6 V
6 -6 V
7 -8 V
9 -10 V
9 -10 V
9 -10 V
11 -11 V
10 -11 V
10 -11 V
10 -11 V
9 -10 V
8 -9 V
8 -7 V
6 -7 V
5 -5 V
3 -4 V
2 -2 V
1 -1 V
2 -2 V
4 -4 V
4 -5 V
7 -7 V
7 -7 V
8 -9 V
10 -10 V
9 -10 V
10 -11 V
11 -11 V
10 -10 V
10 -10 V
9 -10 V
8 -9 V
7 -7 V
7 -7 V
5 -5 V
3 -3 V
2 -2 V
1 -1 V
2 -2 V
3 -4 V
5 -5 V
6 -6 V
8 -8 V
8 -8 V
9 -9 V
10 -10 V
10 -10 V
10 -10 V
11 -11 V
9 -9 V
10 -9 V
8 -9 V
7 -7 V
7 -6 V
4 -5 V
4 -3 V
2 -2 V
1 -1 V
2 -2 V
3 -3 V
5 -5 V
6 -6 V
8 -7 V
8 -8 V
9 -8 V
10 -10 V
10 -9 V
10 -10 V
stroke 6399 1032 M
10 -9 V
10 -9 V
9 -9 V
9 -7 V
7 -7 V
6 -6 V
5 -4 V
4 -3 V
2 -2 V
1 -1 V
2 -1 V
3 -4 V
5 -4 V
6 -5 V
7 -7 V
9 -7 V
9 -8 V
10 -9 V
10 -8 V
10 -9 V
10 -9 V
10 -8 V
9 -8 V
9 -6 V
7 -6 V
6 -5 V
5 -4 V
4 -3 V
1 -2 V
1 -1 V
2 -1 V
4 -3 V
5 -4 V
6 -5 V
7 -5 V
9 -7 V
9 -7 V
10 -7 V
10 -8 V
10 -8 V
10 -7 V
10 -7 V
9 -7 V
8 -6 V
8 -5 V
6 -4 V
5 -3 V
3 -3 V
2 -1 V
1 -1 V
2 -1 V
4 -2 V
5 -3 V
6 -5 V
7 -4 V
8 -6 V
10 -5 V
9 -6 V
11 -7 V
10 -6 V
10 -5 V
10 -6 V
9 -5 V
8 -5 V
8 -3 V
6 -4 V
5 -2 V
3 -2 V
2 -1 V
1 0 V
2 -1 V
3 -2 V
5 -2 V
7 -3 V
7 -3 V
8 -4 V
9 -4 V
10 -4 V
10 -3 V
11 -4 V
10 -3 V
9 -3 V
10 -3 V
8 -2 V
7 -1 V
7 -1 V
5 -1 V
3 0 V
2 0 V
stroke
LT1
LCb setrgbcolor
LT1
6248 2415 M
543 0 V
5229 640 M
2 0 V
2 0 V
3 0 V
4 1 V
5 0 V
6 2 V
6 1 V
6 2 V
7 2 V
6 3 V
7 3 V
7 3 V
6 4 V
5 3 V
5 3 V
4 2 V
3 2 V
3 2 V
1 0 V
1 1 V
1 1 V
2 1 V
3 2 V
5 3 V
4 3 V
6 4 V
6 5 V
6 5 V
7 5 V
7 5 V
7 5 V
6 5 V
6 5 V
6 5 V
4 4 V
5 3 V
3 2 V
2 2 V
1 1 V
1 1 V
1 1 V
3 2 V
3 3 V
4 3 V
5 4 V
5 4 V
6 5 V
7 6 V
6 5 V
7 6 V
7 5 V
6 5 V
6 5 V
6 5 V
5 3 V
4 4 V
3 2 V
2 2 V
2 1 V
0 1 V
2 1 V
2 1 V
3 3 V
4 3 V
5 4 V
6 4 V
6 4 V
6 5 V
7 5 V
7 5 V
6 5 V
7 4 V
6 4 V
5 4 V
5 3 V
4 3 V
4 2 V
2 2 V
1 0 V
1 1 V
1 1 V
2 1 V
4 2 V
4 3 V
4 3 V
6 3 V
6 4 V
7 4 V
6 4 V
7 3 V
7 4 V
6 4 V
6 3 V
6 3 V
5 2 V
4 2 V
3 2 V
2 1 V
1 1 V
1 0 V
1 0 V
3 2 V
stroke 5675 928 M
3 1 V
4 2 V
5 2 V
5 3 V
6 2 V
7 3 V
7 3 V
6 3 V
7 2 V
6 3 V
7 2 V
5 2 V
5 1 V
4 2 V
3 1 V
3 1 V
1 0 V
2 1 V
2 0 V
3 1 V
4 2 V
5 1 V
6 2 V
6 1 V
6 2 V
7 2 V
7 1 V
6 2 V
7 1 V
6 1 V
5 2 V
5 0 V
4 1 V
4 1 V
2 0 V
1 0 V
1 0 V
1 1 V
2 0 V
4 1 V
4 0 V
5 1 V
5 1 V
6 0 V
7 1 V
6 1 V
7 1 V
7 0 V
6 1 V
6 0 V
6 0 V
5 1 V
4 0 V
3 0 V
2 0 V
2 0 V
2 0 V
2 0 V
3 0 V
4 0 V
5 0 V
6 0 V
6 0 V
6 0 V
7 0 V
6 0 V
7 -1 V
7 0 V
6 0 V
5 -1 V
5 0 V
4 0 V
3 0 V
3 -1 V
1 0 V
1 0 V
1 0 V
2 0 V
3 0 V
4 -1 V
5 0 V
6 -1 V
6 -1 V
6 0 V
7 -1 V
7 -1 V
7 -1 V
6 -1 V
6 -1 V
6 -1 V
4 -1 V
5 -1 V
3 0 V
2 -1 V
1 0 V
1 0 V
1 0 V
3 -1 V
3 0 V
4 -1 V
5 -1 V
5 -1 V
6 -2 V
7 -1 V
stroke 6147 969 M
6 -1 V
7 -2 V
7 -2 V
6 -1 V
6 -2 V
6 -1 V
5 -1 V
4 -1 V
3 -1 V
2 -1 V
2 0 V
2 -1 V
2 0 V
3 -1 V
4 -1 V
5 -2 V
6 -1 V
6 -2 V
6 -2 V
7 -2 V
7 -2 V
6 -2 V
7 -2 V
6 -2 V
5 -2 V
5 -1 V
4 -2 V
4 -1 V
2 0 V
1 -1 V
1 0 V
1 0 V
2 -1 V
4 -1 V
4 -2 V
4 -1 V
6 -2 V
6 -2 V
7 -3 V
6 -2 V
7 -2 V
7 -3 V
6 -2 V
6 -2 V
6 -2 V
5 -2 V
4 -2 V
3 -1 V
2 -1 V
1 0 V
1 -1 V
1 0 V
3 -1 V
3 -1 V
4 -2 V
5 -2 V
5 -2 V
6 -2 V
7 -3 V
7 -2 V
6 -3 V
7 -3 V
6 -2 V
7 -3 V
5 -2 V
5 -2 V
4 -2 V
3 -1 V
3 -1 V
1 0 V
0 -1 V
2 0 V
2 -1 V
3 -2 V
4 -1 V
5 -2 V
6 -3 V
6 -2 V
6 -3 V
7 -3 V
7 -3 V
6 -2 V
7 -3 V
6 -3 V
5 -2 V
5 -2 V
4 -2 V
4 -1 V
2 -1 V
1 -1 V
1 0 V
1 -1 V
2 -1 V
4 -1 V
4 -2 V
5 -2 V
5 -2 V
6 -3 V
7 -3 V
6 -3 V
7 -3 V
7 -3 V
6 -2 V
6 -3 V
stroke 6621 796 M
6 -3 V
5 -2 V
4 -1 V
3 -2 V
2 -1 V
2 0 V
0 -1 V
1 0 V
3 -1 V
3 -2 V
4 -1 V
5 -3 V
6 -2 V
6 -3 V
6 -2 V
7 -3 V
6 -3 V
7 -3 V
7 -3 V
6 -3 V
5 -2 V
5 -2 V
4 -2 V
3 -1 V
3 -1 V
1 -1 V
1 0 V
1 0 V
2 -1 V
3 -2 V
4 -2 V
5 -2 V
6 -2 V
6 -3 V
6 -2 V
7 -3 V
7 -3 V
7 -3 V
6 -2 V
6 -3 V
6 -2 V
4 -2 V
5 -2 V
3 -1 V
2 -1 V
1 0 V
1 -1 V
1 0 V
3 -1 V
3 -1 V
4 -2 V
5 -2 V
5 -2 V
6 -2 V
7 -3 V
6 -3 V
7 -2 V
7 -3 V
6 -2 V
6 -2 V
6 -3 V
5 -1 V
4 -2 V
3 -1 V
2 -1 V
2 0 V
0 -1 V
2 0 V
2 -1 V
3 -1 V
4 -2 V
5 -1 V
6 -2 V
6 -2 V
6 -3 V
7 -2 V
7 -3 V
6 -2 V
7 -2 V
6 -3 V
5 -2 V
5 -2 V
4 -2 V
4 -2 V
2 -2 V
1 -2 V
stroke
LTb
4320 3431 N
0 -2791 V
2879 0 V
0 2791 V
-2879 0 V
Z stroke
1.000 UP
1.000 UL
LTb
stroke
grestore
end
showpage
  }}%
  \put(6128,2415){\makebox(0,0)[r]{\strut{}$l=\tau $}}%
  \put(6128,2615){\makebox(0,0)[r]{\strut{}$l=e$}}%
  \put(5759,3731){\makebox(0,0){\strut{}$\bar B_s\to D_{s0}^*(2371) l\nu_l$}}%
  \put(5759,140){\makebox(0,0){\strut{}$q^2$ [$\rm GeV^2$]}}%
  \put(3380,2035){%
  \special{ps: gsave currentpoint currentpoint translate
630 rotate neg exch neg exch translate}%
  \makebox(0,0){\strut{}$d\Gamma/dq^2$ [$|V_{cb}|^2\ 10^{-13} \rm GeV^{-1}$]}%
  \special{ps: currentpoint grestore moveto}%
  }%
  \put(7199,440){\makebox(0,0){\strut{} 10}}%
  \put(6911,440){\makebox(0,0){\strut{} 9}}%
  \put(6623,440){\makebox(0,0){\strut{} 8}}%
  \put(6335,440){\makebox(0,0){\strut{} 7}}%
  \put(6047,440){\makebox(0,0){\strut{} 6}}%
  \put(5760,440){\makebox(0,0){\strut{} 5}}%
  \put(5472,440){\makebox(0,0){\strut{} 4}}%
  \put(5184,440){\makebox(0,0){\strut{} 3}}%
  \put(4896,440){\makebox(0,0){\strut{} 2}}%
  \put(4608,440){\makebox(0,0){\strut{} 1}}%
  \put(4320,440){\makebox(0,0){\strut{} 0}}%
  \put(4200,3431){\makebox(0,0)[r]{\strut{} 0.1}}%
  \put(4200,3152){\makebox(0,0)[r]{\strut{} 0.09}}%
  \put(4200,2873){\makebox(0,0)[r]{\strut{} 0.08}}%
  \put(4200,2594){\makebox(0,0)[r]{\strut{} 0.07}}%
  \put(4200,2315){\makebox(0,0)[r]{\strut{} 0.06}}%
  \put(4200,2036){\makebox(0,0)[r]{\strut{} 0.05}}%
  \put(4200,1756){\makebox(0,0)[r]{\strut{} 0.04}}%
  \put(4200,1477){\makebox(0,0)[r]{\strut{} 0.03}}%
  \put(4200,1198){\makebox(0,0)[r]{\strut{} 0.02}}%
  \put(4200,919){\makebox(0,0)[r]{\strut{} 0.01}}%
  \put(4200,640){\makebox(0,0)[r]{\strut{} 0}}%
  \put(1808,2136){\makebox(0,0)[r]{\strut{}$l= \tau $}}%
  \put(1808,2336){\makebox(0,0)[r]{\strut{}$l=e$}}%
  \put(1439,3731){\makebox(0,0){\strut{}$\bar B_s\to D_s l\nu_l$}}%
  \put(1439,140){\makebox(0,0){\strut{}$q^2$ [$\rm GeV^2$]}}%
  \put(-940,2035){%
  \special{ps: gsave currentpoint currentpoint translate
630 rotate neg exch neg exch translate}%
  \makebox(0,0){\strut{}$d\Gamma/dq^2$ [$|V_{cb}|^2\ 10^{-13} \rm GeV^{-1}$]}%
  \special{ps: currentpoint grestore moveto}%
  }%
  \put(2879,440){\makebox(0,0){\strut{} 12}}%
  \put(2399,440){\makebox(0,0){\strut{} 10}}%
  \put(1919,440){\makebox(0,0){\strut{} 8}}%
  \put(1440,440){\makebox(0,0){\strut{} 6}}%
  \put(960,440){\makebox(0,0){\strut{} 4}}%
  \put(480,440){\makebox(0,0){\strut{} 2}}%
  \put(0,440){\makebox(0,0){\strut{} 0}}%
  \put(-120,3431){\makebox(0,0)[r]{\strut{} 0.45}}%
  \put(-120,3121){\makebox(0,0)[r]{\strut{} 0.4}}%
  \put(-120,2811){\makebox(0,0)[r]{\strut{} 0.35}}%
  \put(-120,2501){\makebox(0,0)[r]{\strut{} 0.3}}%
  \put(-120,2191){\makebox(0,0)[r]{\strut{} 0.25}}%
  \put(-120,1880){\makebox(0,0)[r]{\strut{} 0.2}}%
  \put(-120,1570){\makebox(0,0)[r]{\strut{} 0.15}}%
  \put(-120,1260){\makebox(0,0)[r]{\strut{} 0.1}}%
  \put(-120,950){\makebox(0,0)[r]{\strut{} 0.05}}%
  \put(-120,640){\makebox(0,0)[r]{\strut{} 0}}%
\end{picture}%
\endgroup
 

%% file: dg.1m1m.tex
\begingroup%
\makeatletter%
\newcommand{\GNUPLOTspecial}{%
  \@sanitize\catcode`\%=14\relax\special}%
\setlength{\unitlength}{0.0500bp}%
\begin{picture}(7200,4032)(0,0)%
  {\GNUPLOTspecial{"
/gnudict 256 dict def
gnudict begin
%
%
/Color false def
/Blacktext true def
/Solid false def
/Dashlength 1 def
/Landscape false def
/Level1 false def
/Rounded false def
/ClipToBoundingBox false def
/SuppressPDFMark false def
/TransparentPatterns false def
/gnulinewidth 5.000 def
/userlinewidth gnulinewidth def
/Gamma 1.0 def
/BackgroundColor {-1.000 -1.000 -1.000} def
/vshift -66 def
/dl1 {
  10.0 Dashlength mul mul
  Rounded { currentlinewidth 0.75 mul sub dup 0 le { pop 0.01 } if } if
} def
/dl2 {
  10.0 Dashlength mul mul
  Rounded { currentlinewidth 0.75 mul add } if
} def
/hpt_ 31.5 def
/vpt_ 31.5 def
/hpt hpt_ def
/vpt vpt_ def
/doclip {
  ClipToBoundingBox {
    newpath 0 0 moveto 360 0 lineto 360 201 lineto 0 201 lineto closepath
    clip
  } if
} def
%
%
%
/M {moveto} bind def
/L {lineto} bind def
/R {rmoveto} bind def
/V {rlineto} bind def
/N {newpath moveto} bind def
/Z {closepath} bind def
/C {setrgbcolor} bind def
/f {rlineto fill} bind def
/g {setgray} bind def
/Gshow {show} def   
/vpt2 vpt 2 mul def
/hpt2 hpt 2 mul def
/Lshow {currentpoint stroke M 0 vshift R 
	Blacktext {gsave 0 setgray show grestore} {show} ifelse} def
/Rshow {currentpoint stroke M dup stringwidth pop neg vshift R
	Blacktext {gsave 0 setgray show grestore} {show} ifelse} def
/Cshow {currentpoint stroke M dup stringwidth pop -2 div vshift R 
	Blacktext {gsave 0 setgray show grestore} {show} ifelse} def
/UP {dup vpt_ mul /vpt exch def hpt_ mul /hpt exch def
  /hpt2 hpt 2 mul def /vpt2 vpt 2 mul def} def
/DL {Color {setrgbcolor Solid {pop []} if 0 setdash}
 {pop pop pop 0 setgray Solid {pop []} if 0 setdash} ifelse} def
/BL {stroke userlinewidth 2 mul setlinewidth
	Rounded {1 setlinejoin 1 setlinecap} if} def
/AL {stroke userlinewidth 2 div setlinewidth
	Rounded {1 setlinejoin 1 setlinecap} if} def
/UL {dup gnulinewidth mul /userlinewidth exch def
	dup 1 lt {pop 1} if 10 mul /udl exch def} def
/PL {stroke userlinewidth setlinewidth
	Rounded {1 setlinejoin 1 setlinecap} if} def
3.8 setmiterlimit
/LCw {1 1 1} def
/LCb {0 0 0} def
/LCa {0 0 0} def
/LC0 {1 0 0} def
/LC1 {0 1 0} def
/LC2 {0 0 1} def
/LC3 {1 0 1} def
/LC4 {0 1 1} def
/LC5 {1 1 0} def
/LC6 {0 0 0} def
/LC7 {1 0.3 0} def
/LC8 {0.5 0.5 0.5} def
/LTw {PL [] 1 setgray} def
/LTb {BL [] LCb DL} def
/LTa {AL [1 udl mul 2 udl mul] 0 setdash LCa setrgbcolor} def
/LT0 {PL [] LC0 DL} def
/LT1 {PL [4 dl1 2 dl2] LC1 DL} def
/LT2 {PL [2 dl1 3 dl2] LC2 DL} def
/LT3 {PL [1 dl1 1.5 dl2] LC3 DL} def
/LT4 {PL [6 dl1 2 dl2 1 dl1 2 dl2] LC4 DL} def
/LT5 {PL [3 dl1 3 dl2 1 dl1 3 dl2] LC5 DL} def
/LT6 {PL [2 dl1 2 dl2 2 dl1 6 dl2] LC6 DL} def
/LT7 {PL [1 dl1 2 dl2 6 dl1 2 dl2 1 dl1 2 dl2] LC7 DL} def
/LT8 {PL [2 dl1 2 dl2 2 dl1 2 dl2 2 dl1 2 dl2 2 dl1 4 dl2] LC8 DL} def
/Pnt {stroke [] 0 setdash gsave 1 setlinecap M 0 0 V stroke grestore} def
/Dia {stroke [] 0 setdash 2 copy vpt add M
  hpt neg vpt neg V hpt vpt neg V
  hpt vpt V hpt neg vpt V closepath stroke
  Pnt} def
/Pls {stroke [] 0 setdash vpt sub M 0 vpt2 V
  currentpoint stroke M
  hpt neg vpt neg R hpt2 0 V stroke
 } def
/Box {stroke [] 0 setdash 2 copy exch hpt sub exch vpt add M
  0 vpt2 neg V hpt2 0 V 0 vpt2 V
  hpt2 neg 0 V closepath stroke
  Pnt} def
/Crs {stroke [] 0 setdash exch hpt sub exch vpt add M
  hpt2 vpt2 neg V currentpoint stroke M
  hpt2 neg 0 R hpt2 vpt2 V stroke} def
/TriU {stroke [] 0 setdash 2 copy vpt 1.12 mul add M
  hpt neg vpt -1.62 mul V
  hpt 2 mul 0 V
  hpt neg vpt 1.62 mul V closepath stroke
  Pnt} def
/Star {2 copy Pls Crs} def
/BoxF {stroke [] 0 setdash exch hpt sub exch vpt add M
  0 vpt2 neg V hpt2 0 V 0 vpt2 V
  hpt2 neg 0 V closepath fill} def
/TriUF {stroke [] 0 setdash vpt 1.12 mul add M
  hpt neg vpt -1.62 mul V
  hpt 2 mul 0 V
  hpt neg vpt 1.62 mul V closepath fill} def
/TriD {stroke [] 0 setdash 2 copy vpt 1.12 mul sub M
  hpt neg vpt 1.62 mul V
  hpt 2 mul 0 V
  hpt neg vpt -1.62 mul V closepath stroke
  Pnt} def
/TriDF {stroke [] 0 setdash vpt 1.12 mul sub M
  hpt neg vpt 1.62 mul V
  hpt 2 mul 0 V
  hpt neg vpt -1.62 mul V closepath fill} def
/DiaF {stroke [] 0 setdash vpt add M
  hpt neg vpt neg V hpt vpt neg V
  hpt vpt V hpt neg vpt V closepath fill} def
/Pent {stroke [] 0 setdash 2 copy gsave
  translate 0 hpt M 4 {72 rotate 0 hpt L} repeat
  closepath stroke grestore Pnt} def
/PentF {stroke [] 0 setdash gsave
  translate 0 hpt M 4 {72 rotate 0 hpt L} repeat
  closepath fill grestore} def
/Circle {stroke [] 0 setdash 2 copy
  hpt 0 360 arc stroke Pnt} def
/CircleF {stroke [] 0 setdash hpt 0 360 arc fill} def
/C0 {BL [] 0 setdash 2 copy moveto vpt 90 450 arc} bind def
/C1 {BL [] 0 setdash 2 copy moveto
	2 copy vpt 0 90 arc closepath fill
	vpt 0 360 arc closepath} bind def
/C2 {BL [] 0 setdash 2 copy moveto
	2 copy vpt 90 180 arc closepath fill
	vpt 0 360 arc closepath} bind def
/C3 {BL [] 0 setdash 2 copy moveto
	2 copy vpt 0 180 arc closepath fill
	vpt 0 360 arc closepath} bind def
/C4 {BL [] 0 setdash 2 copy moveto
	2 copy vpt 180 270 arc closepath fill
	vpt 0 360 arc closepath} bind def
/C5 {BL [] 0 setdash 2 copy moveto
	2 copy vpt 0 90 arc
	2 copy moveto
	2 copy vpt 180 270 arc closepath fill
	vpt 0 360 arc} bind def
/C6 {BL [] 0 setdash 2 copy moveto
	2 copy vpt 90 270 arc closepath fill
	vpt 0 360 arc closepath} bind def
/C7 {BL [] 0 setdash 2 copy moveto
	2 copy vpt 0 270 arc closepath fill
	vpt 0 360 arc closepath} bind def
/C8 {BL [] 0 setdash 2 copy moveto
	2 copy vpt 270 360 arc closepath fill
	vpt 0 360 arc closepath} bind def
/C9 {BL [] 0 setdash 2 copy moveto
	2 copy vpt 270 450 arc closepath fill
	vpt 0 360 arc closepath} bind def
/C10 {BL [] 0 setdash 2 copy 2 copy moveto vpt 270 360 arc closepath fill
	2 copy moveto
	2 copy vpt 90 180 arc closepath fill
	vpt 0 360 arc closepath} bind def
/C11 {BL [] 0 setdash 2 copy moveto
	2 copy vpt 0 180 arc closepath fill
	2 copy moveto
	2 copy vpt 270 360 arc closepath fill
	vpt 0 360 arc closepath} bind def
/C12 {BL [] 0 setdash 2 copy moveto
	2 copy vpt 180 360 arc closepath fill
	vpt 0 360 arc closepath} bind def
/C13 {BL [] 0 setdash 2 copy moveto
	2 copy vpt 0 90 arc closepath fill
	2 copy moveto
	2 copy vpt 180 360 arc closepath fill
	vpt 0 360 arc closepath} bind def
/C14 {BL [] 0 setdash 2 copy moveto
	2 copy vpt 90 360 arc closepath fill
	vpt 0 360 arc} bind def
/C15 {BL [] 0 setdash 2 copy vpt 0 360 arc closepath fill
	vpt 0 360 arc closepath} bind def
/Rec {newpath 4 2 roll moveto 1 index 0 rlineto 0 exch rlineto
	neg 0 rlineto closepath} bind def
/Square {dup Rec} bind def
/Bsquare {vpt sub exch vpt sub exch vpt2 Square} bind def
/S0 {BL [] 0 setdash 2 copy moveto 0 vpt rlineto BL Bsquare} bind def
/S1 {BL [] 0 setdash 2 copy vpt Square fill Bsquare} bind def
/S2 {BL [] 0 setdash 2 copy exch vpt sub exch vpt Square fill Bsquare} bind def
/S3 {BL [] 0 setdash 2 copy exch vpt sub exch vpt2 vpt Rec fill Bsquare} bind def
/S4 {BL [] 0 setdash 2 copy exch vpt sub exch vpt sub vpt Square fill Bsquare} bind def
/S5 {BL [] 0 setdash 2 copy 2 copy vpt Square fill
	exch vpt sub exch vpt sub vpt Square fill Bsquare} bind def
/S6 {BL [] 0 setdash 2 copy exch vpt sub exch vpt sub vpt vpt2 Rec fill Bsquare} bind def
/S7 {BL [] 0 setdash 2 copy exch vpt sub exch vpt sub vpt vpt2 Rec fill
	2 copy vpt Square fill Bsquare} bind def
/S8 {BL [] 0 setdash 2 copy vpt sub vpt Square fill Bsquare} bind def
/S9 {BL [] 0 setdash 2 copy vpt sub vpt vpt2 Rec fill Bsquare} bind def
/S10 {BL [] 0 setdash 2 copy vpt sub vpt Square fill 2 copy exch vpt sub exch vpt Square fill
	Bsquare} bind def
/S11 {BL [] 0 setdash 2 copy vpt sub vpt Square fill 2 copy exch vpt sub exch vpt2 vpt Rec fill
	Bsquare} bind def
/S12 {BL [] 0 setdash 2 copy exch vpt sub exch vpt sub vpt2 vpt Rec fill Bsquare} bind def
/S13 {BL [] 0 setdash 2 copy exch vpt sub exch vpt sub vpt2 vpt Rec fill
	2 copy vpt Square fill Bsquare} bind def
/S14 {BL [] 0 setdash 2 copy exch vpt sub exch vpt sub vpt2 vpt Rec fill
	2 copy exch vpt sub exch vpt Square fill Bsquare} bind def
/S15 {BL [] 0 setdash 2 copy Bsquare fill Bsquare} bind def
/D0 {gsave translate 45 rotate 0 0 S0 stroke grestore} bind def
/D1 {gsave translate 45 rotate 0 0 S1 stroke grestore} bind def
/D2 {gsave translate 45 rotate 0 0 S2 stroke grestore} bind def
/D3 {gsave translate 45 rotate 0 0 S3 stroke grestore} bind def
/D4 {gsave translate 45 rotate 0 0 S4 stroke grestore} bind def
/D5 {gsave translate 45 rotate 0 0 S5 stroke grestore} bind def
/D6 {gsave translate 45 rotate 0 0 S6 stroke grestore} bind def
/D7 {gsave translate 45 rotate 0 0 S7 stroke grestore} bind def
/D8 {gsave translate 45 rotate 0 0 S8 stroke grestore} bind def
/D9 {gsave translate 45 rotate 0 0 S9 stroke grestore} bind def
/D10 {gsave translate 45 rotate 0 0 S10 stroke grestore} bind def
/D11 {gsave translate 45 rotate 0 0 S11 stroke grestore} bind def
/D12 {gsave translate 45 rotate 0 0 S12 stroke grestore} bind def
/D13 {gsave translate 45 rotate 0 0 S13 stroke grestore} bind def
/D14 {gsave translate 45 rotate 0 0 S14 stroke grestore} bind def
/D15 {gsave translate 45 rotate 0 0 S15 stroke grestore} bind def
/DiaE {stroke [] 0 setdash vpt add M
  hpt neg vpt neg V hpt vpt neg V
  hpt vpt V hpt neg vpt V closepath stroke} def
/BoxE {stroke [] 0 setdash exch hpt sub exch vpt add M
  0 vpt2 neg V hpt2 0 V 0 vpt2 V
  hpt2 neg 0 V closepath stroke} def
/TriUE {stroke [] 0 setdash vpt 1.12 mul add M
  hpt neg vpt -1.62 mul V
  hpt 2 mul 0 V
  hpt neg vpt 1.62 mul V closepath stroke} def
/TriDE {stroke [] 0 setdash vpt 1.12 mul sub M
  hpt neg vpt 1.62 mul V
  hpt 2 mul 0 V
  hpt neg vpt -1.62 mul V closepath stroke} def
/PentE {stroke [] 0 setdash gsave
  translate 0 hpt M 4 {72 rotate 0 hpt L} repeat
  closepath stroke grestore} def
/CircE {stroke [] 0 setdash 
  hpt 0 360 arc stroke} def
/Opaque {gsave closepath 1 setgray fill grestore 0 setgray closepath} def
/DiaW {stroke [] 0 setdash vpt add M
  hpt neg vpt neg V hpt vpt neg V
  hpt vpt V hpt neg vpt V Opaque stroke} def
/BoxW {stroke [] 0 setdash exch hpt sub exch vpt add M
  0 vpt2 neg V hpt2 0 V 0 vpt2 V
  hpt2 neg 0 V Opaque stroke} def
/TriUW {stroke [] 0 setdash vpt 1.12 mul add M
  hpt neg vpt -1.62 mul V
  hpt 2 mul 0 V
  hpt neg vpt 1.62 mul V Opaque stroke} def
/TriDW {stroke [] 0 setdash vpt 1.12 mul sub M
  hpt neg vpt 1.62 mul V
  hpt 2 mul 0 V
  hpt neg vpt -1.62 mul V Opaque stroke} def
/PentW {stroke [] 0 setdash gsave
  translate 0 hpt M 4 {72 rotate 0 hpt L} repeat
  Opaque stroke grestore} def
/CircW {stroke [] 0 setdash 
  hpt 0 360 arc Opaque stroke} def
/BoxFill {gsave Rec 1 setgray fill grestore} def
/Density {
  /Fillden exch def
  currentrgbcolor
  /ColB exch def /ColG exch def /ColR exch def
  /ColR ColR Fillden mul Fillden sub 1 add def
  /ColG ColG Fillden mul Fillden sub 1 add def
  /ColB ColB Fillden mul Fillden sub 1 add def
  ColR ColG ColB setrgbcolor} def
/BoxColFill {gsave Rec PolyFill} def
/PolyFill {gsave Density fill grestore grestore} def
/h {rlineto rlineto rlineto gsave closepath fill grestore} bind def
%
%
/PatternFill {gsave /PFa [ 9 2 roll ] def
  PFa 0 get PFa 2 get 2 div add PFa 1 get PFa 3 get 2 div add translate
  PFa 2 get -2 div PFa 3 get -2 div PFa 2 get PFa 3 get Rec
  gsave 1 setgray fill grestore clip
  currentlinewidth 0.5 mul setlinewidth
  /PFs PFa 2 get dup mul PFa 3 get dup mul add sqrt def
  0 0 M PFa 5 get rotate PFs -2 div dup translate
  0 1 PFs PFa 4 get div 1 add floor cvi
	{PFa 4 get mul 0 M 0 PFs V} for
  0 PFa 6 get ne {
	0 1 PFs PFa 4 get div 1 add floor cvi
	{PFa 4 get mul 0 2 1 roll M PFs 0 V} for
 } if
  stroke grestore} def
/languagelevel where
 {pop languagelevel} {1} ifelse
 2 lt
	{/InterpretLevel1 true def}
	{/InterpretLevel1 Level1 def}
 ifelse
%
%
/Level2PatternFill {
/Tile8x8 {/PaintType 2 /PatternType 1 /TilingType 1 /BBox [0 0 8 8] /XStep 8 /YStep 8}
	bind def
/KeepColor {currentrgbcolor [/Pattern /DeviceRGB] setcolorspace} bind def
<< Tile8x8
 /PaintProc {0.5 setlinewidth pop 0 0 M 8 8 L 0 8 M 8 0 L stroke} 
>> matrix makepattern
/Pat1 exch def
<< Tile8x8
 /PaintProc {0.5 setlinewidth pop 0 0 M 8 8 L 0 8 M 8 0 L stroke
	0 4 M 4 8 L 8 4 L 4 0 L 0 4 L stroke}
>> matrix makepattern
/Pat2 exch def
<< Tile8x8
 /PaintProc {0.5 setlinewidth pop 0 0 M 0 8 L
	8 8 L 8 0 L 0 0 L fill}
>> matrix makepattern
/Pat3 exch def
<< Tile8x8
 /PaintProc {0.5 setlinewidth pop -4 8 M 8 -4 L
	0 12 M 12 0 L stroke}
>> matrix makepattern
/Pat4 exch def
<< Tile8x8
 /PaintProc {0.5 setlinewidth pop -4 0 M 8 12 L
	0 -4 M 12 8 L stroke}
>> matrix makepattern
/Pat5 exch def
<< Tile8x8
 /PaintProc {0.5 setlinewidth pop -2 8 M 4 -4 L
	0 12 M 8 -4 L 4 12 M 10 0 L stroke}
>> matrix makepattern
/Pat6 exch def
<< Tile8x8
 /PaintProc {0.5 setlinewidth pop -2 0 M 4 12 L
	0 -4 M 8 12 L 4 -4 M 10 8 L stroke}
>> matrix makepattern
/Pat7 exch def
<< Tile8x8
 /PaintProc {0.5 setlinewidth pop 8 -2 M -4 4 L
	12 0 M -4 8 L 12 4 M 0 10 L stroke}
>> matrix makepattern
/Pat8 exch def
<< Tile8x8
 /PaintProc {0.5 setlinewidth pop 0 -2 M 12 4 L
	-4 0 M 12 8 L -4 4 M 8 10 L stroke}
>> matrix makepattern
/Pat9 exch def
/Pattern1 {PatternBgnd KeepColor Pat1 setpattern} bind def
/Pattern2 {PatternBgnd KeepColor Pat2 setpattern} bind def
/Pattern3 {PatternBgnd KeepColor Pat3 setpattern} bind def
/Pattern4 {PatternBgnd KeepColor Landscape {Pat5} {Pat4} ifelse setpattern} bind def
/Pattern5 {PatternBgnd KeepColor Landscape {Pat4} {Pat5} ifelse setpattern} bind def
/Pattern6 {PatternBgnd KeepColor Landscape {Pat9} {Pat6} ifelse setpattern} bind def
/Pattern7 {PatternBgnd KeepColor Landscape {Pat8} {Pat7} ifelse setpattern} bind def
} def
%
%
%
/PatternBgnd {
  TransparentPatterns {} {gsave 1 setgray fill grestore} ifelse
} def
%
%
/Level1PatternFill {
/Pattern1 {0.250 Density} bind def
/Pattern2 {0.500 Density} bind def
/Pattern3 {0.750 Density} bind def
/Pattern4 {0.125 Density} bind def
/Pattern5 {0.375 Density} bind def
/Pattern6 {0.625 Density} bind def
/Pattern7 {0.875 Density} bind def
} def
%
%
Level1 {Level1PatternFill} {Level2PatternFill} ifelse
/Symbol-Oblique /Symbol findfont [1 0 .167 1 0 0] makefont
dup length dict begin {1 index /FID eq {pop pop} {def} ifelse} forall
currentdict end definefont pop
Level1 SuppressPDFMark or 
{} {
/SDict 10 dict def
systemdict /pdfmark known not {
  userdict /pdfmark systemdict /cleartomark get put
} if
SDict begin [
  /Title (paper/dg.1m1m.tex)
  /Subject (gnuplot plot)
  /Creator (gnuplot 4.6 patchlevel 0)
  /Author (conrado)
  /CreationDate (Tue Oct 15 09:50:37 2013)
  /DOCINFO pdfmark
end
} ifelse
end
gnudict begin
gsave
doclip
0 0 translate
0.050 0.050 scale
0 setgray
newpath
BackgroundColor 0 lt 3 1 roll 0 lt exch 0 lt or or not {BackgroundColor C 1.000 0 0 7200.00 4032.00 BoxColFill} if
1.000 UL
LTb
980 640 M
63 0 V
5796 0 R
-63 0 V
980 1338 M
63 0 V
5796 0 R
-63 0 V
980 2036 M
63 0 V
5796 0 R
-63 0 V
980 2733 M
63 0 V
5796 0 R
-63 0 V
980 3431 M
63 0 V
5796 0 R
-63 0 V
980 640 M
0 63 V
0 2728 R
0 -63 V
1957 640 M
0 63 V
0 2728 R
0 -63 V
2933 640 M
0 63 V
0 2728 R
0 -63 V
3910 640 M
0 63 V
0 2728 R
0 -63 V
4886 640 M
0 63 V
0 2728 R
0 -63 V
5863 640 M
0 63 V
0 2728 R
0 -63 V
6839 640 M
0 63 V
0 2728 R
0 -63 V
stroke
LTa
980 640 M
5859 0 V
stroke
LTb
980 3431 N
980 640 L
5859 0 V
0 2791 V
-5859 0 V
Z stroke
LCb setrgbcolor
LTb
LCb setrgbcolor
LTb
1.000 UP
1.000 UL
LTb
1.000 UL
LT0
LCb setrgbcolor
LT0
3832 2475 M
543 0 V
981 1945 M
4 2 V
6 5 V
10 6 V
12 8 V
14 10 V
16 11 V
18 12 V
19 13 V
19 13 V
20 14 V
20 13 V
19 13 V
17 12 V
17 10 V
14 10 V
12 8 V
9 6 V
7 5 V
4 2 V
1 1 V
4 3 V
7 4 V
9 6 V
12 8 V
15 10 V
16 10 V
18 12 V
18 12 V
20 13 V
20 13 V
19 13 V
19 12 V
18 11 V
16 11 V
14 9 V
12 8 V
10 6 V
6 4 V
4 2 V
2 1 V
4 3 V
6 4 V
10 6 V
12 8 V
14 9 V
16 10 V
18 11 V
19 12 V
20 12 V
19 12 V
20 12 V
19 12 V
17 11 V
17 9 V
14 9 V
12 7 V
9 6 V
7 4 V
4 2 V
2 1 V
3 3 V
7 4 V
10 5 V
11 8 V
15 8 V
16 10 V
18 10 V
19 11 V
19 12 V
20 11 V
19 12 V
19 11 V
18 10 V
16 9 V
14 8 V
12 7 V
10 5 V
7 4 V
3 2 V
2 1 V
4 2 V
7 4 V
9 5 V
12 7 V
14 8 V
17 9 V
17 10 V
19 10 V
20 11 V
19 10 V
20 11 V
19 10 V
18 9 V
16 9 V
14 7 V
12 7 V
9 5 V
7 3 V
4 2 V
2 1 V
3 2 V
7 3 V
stroke 2284 2743 M
10 5 V
12 6 V
14 8 V
16 8 V
18 9 V
19 9 V
19 10 V
20 10 V
20 9 V
18 9 V
18 9 V
16 7 V
15 7 V
12 6 V
9 4 V
7 3 V
4 2 V
1 1 V
4 2 V
7 3 V
9 4 V
12 6 V
14 6 V
17 8 V
17 7 V
19 9 V
20 8 V
20 9 V
19 9 V
19 8 V
18 7 V
16 7 V
14 6 V
12 5 V
10 4 V
6 2 V
4 2 V
2 1 V
4 1 V
6 3 V
10 4 V
12 4 V
14 6 V
16 6 V
18 7 V
19 8 V
19 7 V
20 7 V
20 7 V
19 7 V
17 7 V
16 5 V
15 5 V
12 5 V
9 3 V
7 2 V
4 1 V
1 1 V
4 1 V
7 2 V
9 4 V
12 4 V
15 4 V
16 5 V
18 6 V
18 6 V
20 6 V
20 6 V
19 6 V
19 5 V
18 5 V
16 5 V
14 4 V
12 3 V
10 3 V
6 1 V
4 1 V
2 1 V
4 1 V
6 1 V
10 3 V
12 3 V
14 4 V
16 4 V
18 4 V
19 4 V
19 5 V
20 4 V
20 4 V
19 4 V
17 4 V
17 3 V
14 3 V
12 2 V
9 2 V
7 1 V
4 0 V
2 1 V
3 0 V
7 2 V
9 1 V
12 2 V
15 3 V
16 2 V
stroke 3629 3219 M
18 3 V
19 2 V
19 3 V
20 3 V
19 2 V
19 2 V
18 2 V
16 2 V
14 1 V
12 1 V
10 1 V
7 1 V
3 0 V
2 0 V
4 1 V
7 0 V
9 1 V
12 1 V
14 1 V
16 0 V
18 1 V
19 1 V
20 1 V
19 0 V
20 1 V
19 0 V
18 0 V
16 0 V
14 0 V
12 0 V
9 -1 V
7 0 V
4 0 V
2 0 V
3 0 V
7 0 V
10 -1 V
12 0 V
14 -1 V
16 -1 V
18 -1 V
19 -1 V
19 -2 V
20 -1 V
20 -2 V
18 -2 V
18 -3 V
16 -2 V
15 -1 V
12 -2 V
9 -2 V
7 -1 V
3 0 V
2 0 V
4 -1 V
7 -1 V
9 -2 V
12 -2 V
14 -2 V
17 -3 V
17 -4 V
19 -3 V
20 -5 V
20 -4 V
19 -5 V
19 -4 V
18 -5 V
16 -4 V
14 -4 V
12 -4 V
10 -2 V
6 -2 V
4 -1 V
2 -1 V
4 -1 V
6 -2 V
10 -3 V
12 -4 V
14 -5 V
16 -5 V
18 -6 V
19 -7 V
19 -7 V
20 -8 V
20 -8 V
19 -7 V
17 -8 V
16 -7 V
15 -6 V
12 -6 V
9 -4 V
7 -3 V
4 -2 V
1 -1 V
4 -2 V
7 -3 V
9 -5 V
12 -6 V
15 -7 V
16 -8 V
17 -10 V
19 -10 V
20 -11 V
20 -12 V
stroke 4998 2993 M
19 -11 V
19 -12 V
18 -11 V
16 -10 V
14 -10 V
12 -8 V
10 -6 V
6 -5 V
4 -2 V
2 -2 V
4 -2 V
6 -5 V
10 -7 V
12 -8 V
14 -11 V
16 -12 V
18 -13 V
19 -15 V
19 -16 V
20 -16 V
20 -17 V
19 -16 V
17 -16 V
17 -15 V
14 -13 V
12 -12 V
9 -9 V
7 -6 V
4 -4 V
2 -2 V
3 -4 V
7 -6 V
9 -10 V
12 -12 V
15 -15 V
16 -17 V
18 -19 V
18 -21 V
20 -23 V
20 -23 V
19 -24 V
19 -23 V
18 -23 V
16 -21 V
14 -19 V
12 -16 V
10 -14 V
7 -9 V
3 -5 V
2 -3 V
4 -5 V
7 -10 V
9 -14 V
12 -17 V
14 -22 V
16 -25 V
18 -29 V
19 -31 V
20 -33 V
19 -35 V
20 -36 V
19 -36 V
18 -34 V
16 -33 V
14 -30 V
12 -27 V
9 -21 V
7 -15 V
4 -8 V
2 -4 V
3 -9 V
7 -16 V
10 -23 V
12 -29 V
14 -37 V
16 -43 V
18 -50 V
19 -56 V
19 -63 V
20 -68 V
20 -75 V
18 -79 V
18 -85 V
16 -89 V
15 -92 V
12 -96 V
9 -99 V
7 -100 V
3 -102 V
stroke
LT1
LCb setrgbcolor
LT1
3832 2275 M
543 0 V
2522 640 M
3 0 V
4 0 V
7 0 V
8 1 V
10 1 V
12 2 V
12 2 V
14 3 V
13 3 V
14 4 V
14 5 V
13 5 V
13 4 V
11 5 V
10 4 V
8 4 V
7 3 V
5 3 V
2 1 V
2 0 V
2 2 V
5 2 V
7 3 V
8 5 V
10 5 V
11 6 V
13 7 V
13 7 V
14 9 V
14 8 V
13 8 V
14 9 V
12 8 V
11 7 V
10 6 V
9 6 V
6 4 V
5 4 V
3 1 V
1 1 V
3 2 V
4 3 V
7 5 V
8 5 V
10 7 V
12 8 V
12 9 V
14 9 V
13 10 V
14 10 V
14 10 V
13 9 V
13 9 V
11 8 V
10 8 V
8 6 V
7 5 V
5 3 V
2 2 V
2 1 V
2 2 V
5 3 V
7 5 V
8 6 V
10 7 V
11 9 V
13 9 V
13 10 V
14 10 V
14 10 V
13 10 V
14 9 V
12 9 V
11 9 V
10 7 V
9 6 V
6 5 V
5 3 V
3 2 V
1 1 V
3 2 V
4 3 V
7 5 V
8 6 V
10 7 V
12 8 V
12 9 V
14 9 V
13 10 V
14 10 V
14 9 V
13 10 V
13 8 V
11 8 V
10 7 V
8 6 V
7 4 V
5 4 V
2 1 V
2 1 V
2 2 V
5 3 V
stroke 3437 1191 M
7 5 V
8 5 V
10 7 V
11 8 V
13 8 V
13 9 V
14 9 V
14 9 V
13 8 V
14 9 V
12 8 V
11 7 V
10 6 V
9 6 V
6 4 V
5 3 V
3 1 V
1 1 V
3 2 V
4 3 V
7 4 V
8 5 V
10 6 V
12 7 V
12 8 V
13 7 V
14 9 V
14 8 V
14 8 V
13 7 V
13 8 V
11 6 V
10 6 V
8 4 V
7 4 V
5 3 V
2 1 V
2 1 V
2 1 V
5 3 V
7 4 V
8 4 V
10 6 V
11 6 V
13 6 V
13 7 V
14 8 V
14 7 V
13 7 V
14 6 V
12 7 V
11 5 V
10 5 V
9 4 V
6 4 V
5 2 V
3 1 V
1 1 V
3 1 V
4 2 V
7 3 V
8 4 V
10 5 V
12 5 V
12 6 V
13 6 V
14 6 V
14 6 V
14 6 V
13 6 V
13 5 V
11 5 V
10 4 V
8 4 V
7 3 V
5 1 V
2 2 V
2 0 V
2 1 V
5 2 V
6 3 V
9 3 V
10 4 V
11 4 V
13 5 V
13 5 V
14 5 V
14 5 V
13 5 V
14 5 V
12 4 V
11 4 V
10 3 V
9 3 V
6 2 V
5 2 V
3 1 V
1 0 V
3 1 V
4 1 V
7 2 V
8 3 V
10 3 V
12 4 V
stroke 4381 1665 M
12 3 V
13 4 V
14 4 V
14 4 V
14 4 V
13 3 V
12 3 V
12 3 V
10 3 V
8 2 V
7 2 V
5 1 V
2 0 V
2 1 V
2 0 V
5 1 V
6 2 V
9 2 V
10 2 V
11 2 V
13 3 V
13 3 V
14 2 V
14 3 V
13 3 V
14 2 V
12 2 V
11 2 V
10 1 V
9 2 V
6 1 V
5 0 V
3 1 V
1 0 V
3 0 V
4 1 V
7 1 V
8 1 V
10 1 V
12 2 V
12 1 V
13 1 V
14 2 V
14 1 V
14 1 V
13 1 V
12 1 V
12 1 V
10 0 V
8 1 V
7 0 V
5 0 V
2 0 V
1 0 V
3 0 V
5 1 V
6 0 V
9 0 V
10 0 V
11 1 V
13 0 V
13 0 V
14 0 V
14 -1 V
13 0 V
14 0 V
12 -1 V
11 -1 V
10 0 V
9 -1 V
6 0 V
5 0 V
3 -1 V
1 0 V
3 0 V
4 0 V
7 -1 V
8 -1 V
10 -1 V
12 -1 V
12 -1 V
13 -2 V
14 -1 V
14 -2 V
14 -2 V
13 -3 V
12 -2 V
12 -2 V
10 -1 V
8 -2 V
7 -1 V
5 -1 V
2 -1 V
1 0 V
3 -1 V
5 -1 V
6 -1 V
9 -2 V
10 -2 V
11 -3 V
13 -3 V
13 -4 V
14 -3 V
14 -4 V
stroke 5342 1700 M
13 -4 V
14 -5 V
12 -4 V
11 -3 V
10 -4 V
9 -3 V
6 -2 V
5 -2 V
3 -1 V
1 0 V
3 -1 V
4 -2 V
7 -2 V
8 -4 V
10 -4 V
12 -4 V
12 -6 V
13 -5 V
14 -7 V
14 -6 V
14 -7 V
13 -6 V
12 -7 V
12 -6 V
10 -5 V
8 -5 V
7 -4 V
5 -2 V
2 -2 V
1 0 V
3 -2 V
5 -3 V
6 -4 V
9 -5 V
10 -6 V
11 -7 V
13 -8 V
13 -9 V
14 -10 V
14 -10 V
13 -10 V
14 -10 V
12 -10 V
11 -9 V
10 -8 V
9 -8 V
6 -5 V
5 -5 V
3 -2 V
1 -1 V
3 -2 V
4 -5 V
7 -6 V
8 -8 V
10 -9 V
12 -11 V
12 -13 V
13 -14 V
14 -15 V
14 -16 V
14 -16 V
13 -16 V
12 -16 V
12 -15 V
10 -14 V
8 -11 V
7 -10 V
5 -7 V
2 -4 V
1 -2 V
3 -4 V
5 -7 V
6 -11 V
9 -13 V
10 -17 V
11 -20 V
13 -23 V
13 -26 V
14 -30 V
14 -32 V
13 -34 V
14 -38 V
12 -40 V
11 -41 V
10 -44 V
9 -46 V
6 -46 V
5 -48 V
3 -48 V
stroke
LTb
980 3431 N
980 640 L
5859 0 V
0 2791 V
-5859 0 V
Z stroke
1.000 UP
1.000 UL
LTb
stroke
grestore
end
showpage
  }}%
  \put(3712,2275){\makebox(0,0)[r]{\strut{}$l= \tau $}}%
  \put(3712,2475){\makebox(0,0)[r]{\strut{}$l=e$}}%
  \put(3909,3731){\makebox(0,0){\strut{}$\bar B_s\to D_s^* l\nu_l$}}%
  \put(3909,140){\makebox(0,0){\strut{}$q^2$ [$\rm GeV^2$]}}%
  \put(160,2035){%
  \special{ps: gsave currentpoint currentpoint translate
630 rotate neg exch neg exch translate}%
  \makebox(0,0){\strut{}$d\Gamma/dq^2$ [$|V_{cb}|^2\ 10^{-13} \rm GeV^{-1}$]}%
  \special{ps: currentpoint grestore moveto}%
  }%
  \put(6839,440){\makebox(0,0){\strut{} 12}}%
  \put(5863,440){\makebox(0,0){\strut{} 10}}%
  \put(4886,440){\makebox(0,0){\strut{} 8}}%
  \put(3910,440){\makebox(0,0){\strut{} 6}}%
  \put(2933,440){\makebox(0,0){\strut{} 4}}%
  \put(1957,440){\makebox(0,0){\strut{} 2}}%
  \put(980,440){\makebox(0,0){\strut{} 0}}%
  \put(860,3431){\makebox(0,0)[r]{\strut{} 0.8}}%
  \put(860,2733){\makebox(0,0)[r]{\strut{} 0.6}}%
  \put(860,2036){\makebox(0,0)[r]{\strut{} 0.4}}%
  \put(860,1338){\makebox(0,0)[r]{\strut{} 0.2}}%
  \put(860,640){\makebox(0,0)[r]{\strut{} 0}}%
\end{picture}%
\endgroup
 

%% file: dg.1ps0s1.tex
\begingroup%
\makeatletter%
\newcommand{\GNUPLOTspecial}{%
  \@sanitize\catcode`\%=14\relax\special}%
\setlength{\unitlength}{0.0500bp}%
\begin{picture}(7200,5040)(0,0)%
  {\GNUPLOTspecial{"
/gnudict 256 dict def
gnudict begin
%
%
/Color false def
/Blacktext true def
/Solid false def
/Dashlength 1 def
/Landscape false def
/Level1 false def
/Rounded false def
/ClipToBoundingBox false def
/SuppressPDFMark false def
/TransparentPatterns false def
/gnulinewidth 5.000 def
/userlinewidth gnulinewidth def
/Gamma 1.0 def
/BackgroundColor {-1.000 -1.000 -1.000} def
/vshift -66 def
/dl1 {
  10.0 Dashlength mul mul
  Rounded { currentlinewidth 0.75 mul sub dup 0 le { pop 0.01 } if } if
} def
/dl2 {
  10.0 Dashlength mul mul
  Rounded { currentlinewidth 0.75 mul add } if
} def
/hpt_ 31.5 def
/vpt_ 31.5 def
/hpt hpt_ def
/vpt vpt_ def
/doclip {
  ClipToBoundingBox {
    newpath 0 0 moveto 360 0 lineto 360 252 lineto 0 252 lineto closepath
    clip
  } if
} def
%
%
%
/M {moveto} bind def
/L {lineto} bind def
/R {rmoveto} bind def
/V {rlineto} bind def
/N {newpath moveto} bind def
/Z {closepath} bind def
/C {setrgbcolor} bind def
/f {rlineto fill} bind def
/g {setgray} bind def
/Gshow {show} def   
/vpt2 vpt 2 mul def
/hpt2 hpt 2 mul def
/Lshow {currentpoint stroke M 0 vshift R 
	Blacktext {gsave 0 setgray show grestore} {show} ifelse} def
/Rshow {currentpoint stroke M dup stringwidth pop neg vshift R
	Blacktext {gsave 0 setgray show grestore} {show} ifelse} def
/Cshow {currentpoint stroke M dup stringwidth pop -2 div vshift R 
	Blacktext {gsave 0 setgray show grestore} {show} ifelse} def
/UP {dup vpt_ mul /vpt exch def hpt_ mul /hpt exch def
  /hpt2 hpt 2 mul def /vpt2 vpt 2 mul def} def
/DL {Color {setrgbcolor Solid {pop []} if 0 setdash}
 {pop pop pop 0 setgray Solid {pop []} if 0 setdash} ifelse} def
/BL {stroke userlinewidth 2 mul setlinewidth
	Rounded {1 setlinejoin 1 setlinecap} if} def
/AL {stroke userlinewidth 2 div setlinewidth
	Rounded {1 setlinejoin 1 setlinecap} if} def
/UL {dup gnulinewidth mul /userlinewidth exch def
	dup 1 lt {pop 1} if 10 mul /udl exch def} def
/PL {stroke userlinewidth setlinewidth
	Rounded {1 setlinejoin 1 setlinecap} if} def
3.8 setmiterlimit
/LCw {1 1 1} def
/LCb {0 0 0} def
/LCa {0 0 0} def
/LC0 {1 0 0} def
/LC1 {0 1 0} def
/LC2 {0 0 1} def
/LC3 {1 0 1} def
/LC4 {0 1 1} def
/LC5 {1 1 0} def
/LC6 {0 0 0} def
/LC7 {1 0.3 0} def
/LC8 {0.5 0.5 0.5} def
/LTw {PL [] 1 setgray} def
/LTb {BL [] LCb DL} def
/LTa {AL [1 udl mul 2 udl mul] 0 setdash LCa setrgbcolor} def
/LT0 {PL [] LC0 DL} def
/LT1 {PL [4 dl1 2 dl2] LC1 DL} def
/LT2 {PL [2 dl1 3 dl2] LC2 DL} def
/LT3 {PL [1 dl1 1.5 dl2] LC3 DL} def
/LT4 {PL [6 dl1 2 dl2 1 dl1 2 dl2] LC4 DL} def
/LT5 {PL [3 dl1 3 dl2 1 dl1 3 dl2] LC5 DL} def
/LT6 {PL [2 dl1 2 dl2 2 dl1 6 dl2] LC6 DL} def
/LT7 {PL [1 dl1 2 dl2 6 dl1 2 dl2 1 dl1 2 dl2] LC7 DL} def
/LT8 {PL [2 dl1 2 dl2 2 dl1 2 dl2 2 dl1 2 dl2 2 dl1 4 dl2] LC8 DL} def
/Pnt {stroke [] 0 setdash gsave 1 setlinecap M 0 0 V stroke grestore} def
/Dia {stroke [] 0 setdash 2 copy vpt add M
  hpt neg vpt neg V hpt vpt neg V
  hpt vpt V hpt neg vpt V closepath stroke
  Pnt} def
/Pls {stroke [] 0 setdash vpt sub M 0 vpt2 V
  currentpoint stroke M
  hpt neg vpt neg R hpt2 0 V stroke
 } def
/Box {stroke [] 0 setdash 2 copy exch hpt sub exch vpt add M
  0 vpt2 neg V hpt2 0 V 0 vpt2 V
  hpt2 neg 0 V closepath stroke
  Pnt} def
/Crs {stroke [] 0 setdash exch hpt sub exch vpt add M
  hpt2 vpt2 neg V currentpoint stroke M
  hpt2 neg 0 R hpt2 vpt2 V stroke} def
/TriU {stroke [] 0 setdash 2 copy vpt 1.12 mul add M
  hpt neg vpt -1.62 mul V
  hpt 2 mul 0 V
  hpt neg vpt 1.62 mul V closepath stroke
  Pnt} def
/Star {2 copy Pls Crs} def
/BoxF {stroke [] 0 setdash exch hpt sub exch vpt add M
  0 vpt2 neg V hpt2 0 V 0 vpt2 V
  hpt2 neg 0 V closepath fill} def
/TriUF {stroke [] 0 setdash vpt 1.12 mul add M
  hpt neg vpt -1.62 mul V
  hpt 2 mul 0 V
  hpt neg vpt 1.62 mul V closepath fill} def
/TriD {stroke [] 0 setdash 2 copy vpt 1.12 mul sub M
  hpt neg vpt 1.62 mul V
  hpt 2 mul 0 V
  hpt neg vpt -1.62 mul V closepath stroke
  Pnt} def
/TriDF {stroke [] 0 setdash vpt 1.12 mul sub M
  hpt neg vpt 1.62 mul V
  hpt 2 mul 0 V
  hpt neg vpt -1.62 mul V closepath fill} def
/DiaF {stroke [] 0 setdash vpt add M
  hpt neg vpt neg V hpt vpt neg V
  hpt vpt V hpt neg vpt V closepath fill} def
/Pent {stroke [] 0 setdash 2 copy gsave
  translate 0 hpt M 4 {72 rotate 0 hpt L} repeat
  closepath stroke grestore Pnt} def
/PentF {stroke [] 0 setdash gsave
  translate 0 hpt M 4 {72 rotate 0 hpt L} repeat
  closepath fill grestore} def
/Circle {stroke [] 0 setdash 2 copy
  hpt 0 360 arc stroke Pnt} def
/CircleF {stroke [] 0 setdash hpt 0 360 arc fill} def
/C0 {BL [] 0 setdash 2 copy moveto vpt 90 450 arc} bind def
/C1 {BL [] 0 setdash 2 copy moveto
	2 copy vpt 0 90 arc closepath fill
	vpt 0 360 arc closepath} bind def
/C2 {BL [] 0 setdash 2 copy moveto
	2 copy vpt 90 180 arc closepath fill
	vpt 0 360 arc closepath} bind def
/C3 {BL [] 0 setdash 2 copy moveto
	2 copy vpt 0 180 arc closepath fill
	vpt 0 360 arc closepath} bind def
/C4 {BL [] 0 setdash 2 copy moveto
	2 copy vpt 180 270 arc closepath fill
	vpt 0 360 arc closepath} bind def
/C5 {BL [] 0 setdash 2 copy moveto
	2 copy vpt 0 90 arc
	2 copy moveto
	2 copy vpt 180 270 arc closepath fill
	vpt 0 360 arc} bind def
/C6 {BL [] 0 setdash 2 copy moveto
	2 copy vpt 90 270 arc closepath fill
	vpt 0 360 arc closepath} bind def
/C7 {BL [] 0 setdash 2 copy moveto
	2 copy vpt 0 270 arc closepath fill
	vpt 0 360 arc closepath} bind def
/C8 {BL [] 0 setdash 2 copy moveto
	2 copy vpt 270 360 arc closepath fill
	vpt 0 360 arc closepath} bind def
/C9 {BL [] 0 setdash 2 copy moveto
	2 copy vpt 270 450 arc closepath fill
	vpt 0 360 arc closepath} bind def
/C10 {BL [] 0 setdash 2 copy 2 copy moveto vpt 270 360 arc closepath fill
	2 copy moveto
	2 copy vpt 90 180 arc closepath fill
	vpt 0 360 arc closepath} bind def
/C11 {BL [] 0 setdash 2 copy moveto
	2 copy vpt 0 180 arc closepath fill
	2 copy moveto
	2 copy vpt 270 360 arc closepath fill
	vpt 0 360 arc closepath} bind def
/C12 {BL [] 0 setdash 2 copy moveto
	2 copy vpt 180 360 arc closepath fill
	vpt 0 360 arc closepath} bind def
/C13 {BL [] 0 setdash 2 copy moveto
	2 copy vpt 0 90 arc closepath fill
	2 copy moveto
	2 copy vpt 180 360 arc closepath fill
	vpt 0 360 arc closepath} bind def
/C14 {BL [] 0 setdash 2 copy moveto
	2 copy vpt 90 360 arc closepath fill
	vpt 0 360 arc} bind def
/C15 {BL [] 0 setdash 2 copy vpt 0 360 arc closepath fill
	vpt 0 360 arc closepath} bind def
/Rec {newpath 4 2 roll moveto 1 index 0 rlineto 0 exch rlineto
	neg 0 rlineto closepath} bind def
/Square {dup Rec} bind def
/Bsquare {vpt sub exch vpt sub exch vpt2 Square} bind def
/S0 {BL [] 0 setdash 2 copy moveto 0 vpt rlineto BL Bsquare} bind def
/S1 {BL [] 0 setdash 2 copy vpt Square fill Bsquare} bind def
/S2 {BL [] 0 setdash 2 copy exch vpt sub exch vpt Square fill Bsquare} bind def
/S3 {BL [] 0 setdash 2 copy exch vpt sub exch vpt2 vpt Rec fill Bsquare} bind def
/S4 {BL [] 0 setdash 2 copy exch vpt sub exch vpt sub vpt Square fill Bsquare} bind def
/S5 {BL [] 0 setdash 2 copy 2 copy vpt Square fill
	exch vpt sub exch vpt sub vpt Square fill Bsquare} bind def
/S6 {BL [] 0 setdash 2 copy exch vpt sub exch vpt sub vpt vpt2 Rec fill Bsquare} bind def
/S7 {BL [] 0 setdash 2 copy exch vpt sub exch vpt sub vpt vpt2 Rec fill
	2 copy vpt Square fill Bsquare} bind def
/S8 {BL [] 0 setdash 2 copy vpt sub vpt Square fill Bsquare} bind def
/S9 {BL [] 0 setdash 2 copy vpt sub vpt vpt2 Rec fill Bsquare} bind def
/S10 {BL [] 0 setdash 2 copy vpt sub vpt Square fill 2 copy exch vpt sub exch vpt Square fill
	Bsquare} bind def
/S11 {BL [] 0 setdash 2 copy vpt sub vpt Square fill 2 copy exch vpt sub exch vpt2 vpt Rec fill
	Bsquare} bind def
/S12 {BL [] 0 setdash 2 copy exch vpt sub exch vpt sub vpt2 vpt Rec fill Bsquare} bind def
/S13 {BL [] 0 setdash 2 copy exch vpt sub exch vpt sub vpt2 vpt Rec fill
	2 copy vpt Square fill Bsquare} bind def
/S14 {BL [] 0 setdash 2 copy exch vpt sub exch vpt sub vpt2 vpt Rec fill
	2 copy exch vpt sub exch vpt Square fill Bsquare} bind def
/S15 {BL [] 0 setdash 2 copy Bsquare fill Bsquare} bind def
/D0 {gsave translate 45 rotate 0 0 S0 stroke grestore} bind def
/D1 {gsave translate 45 rotate 0 0 S1 stroke grestore} bind def
/D2 {gsave translate 45 rotate 0 0 S2 stroke grestore} bind def
/D3 {gsave translate 45 rotate 0 0 S3 stroke grestore} bind def
/D4 {gsave translate 45 rotate 0 0 S4 stroke grestore} bind def
/D5 {gsave translate 45 rotate 0 0 S5 stroke grestore} bind def
/D6 {gsave translate 45 rotate 0 0 S6 stroke grestore} bind def
/D7 {gsave translate 45 rotate 0 0 S7 stroke grestore} bind def
/D8 {gsave translate 45 rotate 0 0 S8 stroke grestore} bind def
/D9 {gsave translate 45 rotate 0 0 S9 stroke grestore} bind def
/D10 {gsave translate 45 rotate 0 0 S10 stroke grestore} bind def
/D11 {gsave translate 45 rotate 0 0 S11 stroke grestore} bind def
/D12 {gsave translate 45 rotate 0 0 S12 stroke grestore} bind def
/D13 {gsave translate 45 rotate 0 0 S13 stroke grestore} bind def
/D14 {gsave translate 45 rotate 0 0 S14 stroke grestore} bind def
/D15 {gsave translate 45 rotate 0 0 S15 stroke grestore} bind def
/DiaE {stroke [] 0 setdash vpt add M
  hpt neg vpt neg V hpt vpt neg V
  hpt vpt V hpt neg vpt V closepath stroke} def
/BoxE {stroke [] 0 setdash exch hpt sub exch vpt add M
  0 vpt2 neg V hpt2 0 V 0 vpt2 V
  hpt2 neg 0 V closepath stroke} def
/TriUE {stroke [] 0 setdash vpt 1.12 mul add M
  hpt neg vpt -1.62 mul V
  hpt 2 mul 0 V
  hpt neg vpt 1.62 mul V closepath stroke} def
/TriDE {stroke [] 0 setdash vpt 1.12 mul sub M
  hpt neg vpt 1.62 mul V
  hpt 2 mul 0 V
  hpt neg vpt -1.62 mul V closepath stroke} def
/PentE {stroke [] 0 setdash gsave
  translate 0 hpt M 4 {72 rotate 0 hpt L} repeat
  closepath stroke grestore} def
/CircE {stroke [] 0 setdash 
  hpt 0 360 arc stroke} def
/Opaque {gsave closepath 1 setgray fill grestore 0 setgray closepath} def
/DiaW {stroke [] 0 setdash vpt add M
  hpt neg vpt neg V hpt vpt neg V
  hpt vpt V hpt neg vpt V Opaque stroke} def
/BoxW {stroke [] 0 setdash exch hpt sub exch vpt add M
  0 vpt2 neg V hpt2 0 V 0 vpt2 V
  hpt2 neg 0 V Opaque stroke} def
/TriUW {stroke [] 0 setdash vpt 1.12 mul add M
  hpt neg vpt -1.62 mul V
  hpt 2 mul 0 V
  hpt neg vpt 1.62 mul V Opaque stroke} def
/TriDW {stroke [] 0 setdash vpt 1.12 mul sub M
  hpt neg vpt 1.62 mul V
  hpt 2 mul 0 V
  hpt neg vpt -1.62 mul V Opaque stroke} def
/PentW {stroke [] 0 setdash gsave
  translate 0 hpt M 4 {72 rotate 0 hpt L} repeat
  Opaque stroke grestore} def
/CircW {stroke [] 0 setdash 
  hpt 0 360 arc Opaque stroke} def
/BoxFill {gsave Rec 1 setgray fill grestore} def
/Density {
  /Fillden exch def
  currentrgbcolor
  /ColB exch def /ColG exch def /ColR exch def
  /ColR ColR Fillden mul Fillden sub 1 add def
  /ColG ColG Fillden mul Fillden sub 1 add def
  /ColB ColB Fillden mul Fillden sub 1 add def
  ColR ColG ColB setrgbcolor} def
/BoxColFill {gsave Rec PolyFill} def
/PolyFill {gsave Density fill grestore grestore} def
/h {rlineto rlineto rlineto gsave closepath fill grestore} bind def
%
%
/PatternFill {gsave /PFa [ 9 2 roll ] def
  PFa 0 get PFa 2 get 2 div add PFa 1 get PFa 3 get 2 div add translate
  PFa 2 get -2 div PFa 3 get -2 div PFa 2 get PFa 3 get Rec
  gsave 1 setgray fill grestore clip
  currentlinewidth 0.5 mul setlinewidth
  /PFs PFa 2 get dup mul PFa 3 get dup mul add sqrt def
  0 0 M PFa 5 get rotate PFs -2 div dup translate
  0 1 PFs PFa 4 get div 1 add floor cvi
	{PFa 4 get mul 0 M 0 PFs V} for
  0 PFa 6 get ne {
	0 1 PFs PFa 4 get div 1 add floor cvi
	{PFa 4 get mul 0 2 1 roll M PFs 0 V} for
 } if
  stroke grestore} def
/languagelevel where
 {pop languagelevel} {1} ifelse
 2 lt
	{/InterpretLevel1 true def}
	{/InterpretLevel1 Level1 def}
 ifelse
%
%
/Level2PatternFill {
/Tile8x8 {/PaintType 2 /PatternType 1 /TilingType 1 /BBox [0 0 8 8] /XStep 8 /YStep 8}
	bind def
/KeepColor {currentrgbcolor [/Pattern /DeviceRGB] setcolorspace} bind def
<< Tile8x8
 /PaintProc {0.5 setlinewidth pop 0 0 M 8 8 L 0 8 M 8 0 L stroke} 
>> matrix makepattern
/Pat1 exch def
<< Tile8x8
 /PaintProc {0.5 setlinewidth pop 0 0 M 8 8 L 0 8 M 8 0 L stroke
	0 4 M 4 8 L 8 4 L 4 0 L 0 4 L stroke}
>> matrix makepattern
/Pat2 exch def
<< Tile8x8
 /PaintProc {0.5 setlinewidth pop 0 0 M 0 8 L
	8 8 L 8 0 L 0 0 L fill}
>> matrix makepattern
/Pat3 exch def
<< Tile8x8
 /PaintProc {0.5 setlinewidth pop -4 8 M 8 -4 L
	0 12 M 12 0 L stroke}
>> matrix makepattern
/Pat4 exch def
<< Tile8x8
 /PaintProc {0.5 setlinewidth pop -4 0 M 8 12 L
	0 -4 M 12 8 L stroke}
>> matrix makepattern
/Pat5 exch def
<< Tile8x8
 /PaintProc {0.5 setlinewidth pop -2 8 M 4 -4 L
	0 12 M 8 -4 L 4 12 M 10 0 L stroke}
>> matrix makepattern
/Pat6 exch def
<< Tile8x8
 /PaintProc {0.5 setlinewidth pop -2 0 M 4 12 L
	0 -4 M 8 12 L 4 -4 M 10 8 L stroke}
>> matrix makepattern
/Pat7 exch def
<< Tile8x8
 /PaintProc {0.5 setlinewidth pop 8 -2 M -4 4 L
	12 0 M -4 8 L 12 4 M 0 10 L stroke}
>> matrix makepattern
/Pat8 exch def
<< Tile8x8
 /PaintProc {0.5 setlinewidth pop 0 -2 M 12 4 L
	-4 0 M 12 8 L -4 4 M 8 10 L stroke}
>> matrix makepattern
/Pat9 exch def
/Pattern1 {PatternBgnd KeepColor Pat1 setpattern} bind def
/Pattern2 {PatternBgnd KeepColor Pat2 setpattern} bind def
/Pattern3 {PatternBgnd KeepColor Pat3 setpattern} bind def
/Pattern4 {PatternBgnd KeepColor Landscape {Pat5} {Pat4} ifelse setpattern} bind def
/Pattern5 {PatternBgnd KeepColor Landscape {Pat4} {Pat5} ifelse setpattern} bind def
/Pattern6 {PatternBgnd KeepColor Landscape {Pat9} {Pat6} ifelse setpattern} bind def
/Pattern7 {PatternBgnd KeepColor Landscape {Pat8} {Pat7} ifelse setpattern} bind def
} def
%
%
%
/PatternBgnd {
  TransparentPatterns {} {gsave 1 setgray fill grestore} ifelse
} def
%
%
/Level1PatternFill {
/Pattern1 {0.250 Density} bind def
/Pattern2 {0.500 Density} bind def
/Pattern3 {0.750 Density} bind def
/Pattern4 {0.125 Density} bind def
/Pattern5 {0.375 Density} bind def
/Pattern6 {0.625 Density} bind def
/Pattern7 {0.875 Density} bind def
} def
%
%
Level1 {Level1PatternFill} {Level2PatternFill} ifelse
/Symbol-Oblique /Symbol findfont [1 0 .167 1 0 0] makefont
dup length dict begin {1 index /FID eq {pop pop} {def} ifelse} forall
currentdict end definefont pop
Level1 SuppressPDFMark or 
{} {
/SDict 10 dict def
systemdict /pdfmark known not {
  userdict /pdfmark systemdict /cleartomark get put
} if
SDict begin [
  /Title (paper/dg.1ps0s1.tex)
  /Subject (gnuplot plot)
  /Creator (gnuplot 4.6 patchlevel 0)
  /Author (conrado)
  /CreationDate (Tue Nov 12 01:29:07 2013)
  /DOCINFO pdfmark
end
} ifelse
end
gnudict begin
gsave
doclip
0 0 translate
0.050 0.050 scale
0 setgray
newpath
BackgroundColor 0 lt 3 1 roll 0 lt exch 0 lt or or not {BackgroundColor C 1.000 0 0 7200.00 5040.00 BoxColFill} if
1.000 UL
LTb
0 640 M
63 0 V
2816 0 R
-63 0 V
0 919 M
63 0 V
2816 0 R
-63 0 V
0 1198 M
63 0 V
2816 0 R
-63 0 V
0 1477 M
63 0 V
2816 0 R
-63 0 V
0 1756 M
63 0 V
2816 0 R
-63 0 V
0 2036 M
63 0 V
2816 0 R
-63 0 V
0 2315 M
63 0 V
2816 0 R
-63 0 V
0 2594 M
63 0 V
2816 0 R
-63 0 V
0 2873 M
63 0 V
2816 0 R
-63 0 V
0 3152 M
63 0 V
2816 0 R
-63 0 V
0 3431 M
63 0 V
2816 0 R
-63 0 V
0 640 M
0 63 V
0 3431 M
0 -63 V
320 640 M
0 63 V
0 2728 R
0 -63 V
640 640 M
0 63 V
0 2728 R
0 -63 V
960 640 M
0 63 V
0 2728 R
0 -63 V
1280 640 M
0 63 V
0 2728 R
0 -63 V
1599 640 M
0 63 V
0 2728 R
0 -63 V
1919 640 M
0 63 V
0 2728 R
0 -63 V
2239 640 M
0 63 V
0 2728 R
0 -63 V
2559 640 M
0 63 V
0 2728 R
0 -63 V
2879 640 M
0 63 V
0 2728 R
0 -63 V
stroke
LTa
0 640 M
2879 0 V
stroke
LTb
0 3431 N
0 640 L
2879 0 V
0 2791 V
0 3431 L
Z stroke
LCb setrgbcolor
LTb
LCb setrgbcolor
LTb
1.000 UP
1.000 UL
LTb
1.000 UL
LT0
LCb setrgbcolor
LT0
1928 2615 M
543 0 V
0 3248 M
2 -3 V
4 -5 V
5 -8 V
6 -9 V
8 -12 V
8 -13 V
9 -14 V
10 -15 V
10 -15 V
11 -16 V
10 -16 V
10 -15 V
9 -14 V
9 -13 V
7 -11 V
6 -9 V
5 -8 V
4 -5 V
2 -3 V
1 -1 V
2 -3 V
3 -6 V
5 -7 V
6 -10 V
8 -11 V
8 -12 V
10 -14 V
9 -15 V
11 -15 V
10 -16 V
10 -15 V
10 -15 V
9 -13 V
9 -13 V
7 -11 V
6 -9 V
5 -7 V
4 -6 V
2 -3 V
1 -1 V
2 -3 V
3 -5 V
5 -7 V
7 -9 V
7 -11 V
8 -13 V
10 -13 V
10 -15 V
10 -15 V
10 -15 V
10 -14 V
10 -15 V
9 -13 V
9 -12 V
7 -11 V
7 -9 V
5 -7 V
3 -5 V
2 -3 V
1 -1 V
2 -3 V
3 -5 V
5 -7 V
7 -9 V
7 -11 V
9 -12 V
9 -13 V
10 -14 V
10 -14 V
10 -15 V
11 -14 V
9 -14 V
10 -13 V
8 -12 V
8 -10 V
6 -9 V
5 -7 V
3 -4 V
2 -3 V
1 -1 V
2 -3 V
4 -5 V
5 -7 V
6 -9 V
7 -10 V
9 -11 V
9 -13 V
10 -14 V
10 -13 V
10 -15 V
11 -13 V
10 -14 V
9 -12 V
8 -12 V
8 -9 V
6 -9 V
5 -6 V
3 -5 V
2 -3 V
1 -1 V
2 -3 V
4 -4 V
stroke 682 2264 M
5 -7 V
6 -8 V
7 -10 V
9 -11 V
9 -12 V
10 -13 V
10 -14 V
11 -13 V
10 -13 V
10 -13 V
9 -12 V
8 -11 V
8 -9 V
6 -8 V
5 -7 V
4 -4 V
2 -3 V
1 -1 V
2 -2 V
3 -5 V
5 -6 V
6 -8 V
8 -9 V
8 -11 V
9 -12 V
10 -12 V
10 -13 V
11 -13 V
10 -12 V
10 -12 V
9 -12 V
9 -10 V
7 -9 V
6 -8 V
5 -6 V
4 -4 V
2 -2 V
1 -1 V
2 -3 V
3 -4 V
5 -6 V
6 -8 V
8 -8 V
8 -11 V
10 -11 V
9 -11 V
11 -12 V
10 -12 V
10 -12 V
10 -12 V
9 -11 V
9 -9 V
7 -9 V
7 -7 V
5 -6 V
3 -4 V
2 -2 V
1 -1 V
2 -2 V
3 -4 V
5 -6 V
7 -7 V
7 -8 V
8 -10 V
10 -10 V
10 -11 V
10 -12 V
10 -11 V
10 -11 V
10 -11 V
10 -10 V
8 -9 V
7 -8 V
7 -7 V
5 -5 V
3 -4 V
2 -2 V
1 -1 V
2 -2 V
4 -4 V
4 -5 V
7 -7 V
7 -8 V
9 -8 V
9 -10 V
10 -10 V
10 -11 V
10 -11 V
11 -10 V
9 -10 V
10 -9 V
8 -9 V
8 -7 V
6 -7 V
5 -4 V
3 -4 V
2 -2 V
1 -1 V
2 -2 V
4 -3 V
5 -5 V
6 -6 V
7 -7 V
9 -9 V
stroke 1385 1449 M
9 -9 V
10 -9 V
10 -10 V
11 -10 V
10 -9 V
10 -9 V
9 -9 V
8 -8 V
8 -7 V
6 -6 V
5 -4 V
4 -3 V
2 -2 V
0 -1 V
2 -2 V
4 -3 V
5 -4 V
6 -6 V
8 -7 V
8 -7 V
9 -8 V
10 -9 V
10 -9 V
11 -9 V
10 -9 V
10 -8 V
9 -8 V
9 -7 V
7 -6 V
6 -6 V
5 -4 V
4 -3 V
2 -1 V
1 -1 V
2 -2 V
3 -2 V
5 -5 V
6 -5 V
8 -6 V
8 -7 V
9 -7 V
10 -8 V
11 -8 V
10 -8 V
10 -8 V
10 -8 V
9 -7 V
9 -6 V
7 -6 V
6 -4 V
5 -4 V
4 -3 V
2 -1 V
1 -1 V
2 -1 V
3 -3 V
5 -3 V
6 -5 V
8 -6 V
8 -6 V
10 -6 V
10 -7 V
10 -8 V
10 -7 V
10 -7 V
10 -7 V
9 -6 V
9 -6 V
7 -5 V
7 -4 V
5 -3 V
3 -2 V
2 -2 V
1 0 V
2 -2 V
3 -2 V
5 -3 V
7 -4 V
7 -5 V
9 -5 V
9 -6 V
10 -7 V
10 -6 V
10 -6 V
10 -7 V
10 -5 V
10 -6 V
8 -5 V
8 -4 V
6 -4 V
5 -3 V
3 -2 V
2 -1 V
1 -1 V
2 -1 V
4 -2 V
5 -3 V
6 -3 V
7 -4 V
9 -5 V
9 -5 V
10 -6 V
10 -5 V
10 -6 V
stroke 2100 912 M
11 -5 V
10 -5 V
9 -5 V
8 -4 V
8 -4 V
6 -3 V
5 -3 V
3 -2 V
2 -1 V
1 0 V
2 -1 V
4 -2 V
5 -2 V
6 -3 V
7 -4 V
9 -4 V
9 -4 V
10 -5 V
10 -5 V
11 -5 V
10 -4 V
10 -5 V
9 -4 V
8 -4 V
8 -3 V
6 -3 V
5 -2 V
4 -1 V
2 -1 V
1 -1 V
1 0 V
4 -2 V
5 -2 V
6 -3 V
8 -3 V
8 -3 V
9 -4 V
10 -4 V
10 -4 V
11 -4 V
10 -4 V
10 -4 V
9 -4 V
9 -3 V
7 -3 V
6 -2 V
5 -2 V
4 -1 V
2 -1 V
1 0 V
2 -1 V
3 -1 V
5 -2 V
6 -2 V
8 -3 V
8 -3 V
10 -4 V
9 -3 V
11 -4 V
10 -4 V
10 -3 V
10 -4 V
9 -3 V
9 -3 V
7 -3 V
7 -2 V
4 -2 V
4 -1 V
2 -1 V
1 0 V
2 -1 V
3 -1 V
5 -2 V
7 -2 V
7 -3 V
8 -3 V
10 -3 V
10 -4 V
10 -4 V
10 -5 V
10 -4 V
10 -5 V
10 -5 V
8 -4 V
7 -5 V
7 -5 V
5 -5 V
3 -5 V
2 -5 V
stroke
LT1
LCb setrgbcolor
LT1
1928 2415 M
543 0 V
1010 640 M
1 0 V
3 0 V
3 0 V
4 0 V
4 1 V
6 1 V
5 1 V
7 1 V
6 1 V
6 2 V
7 2 V
6 2 V
6 2 V
5 2 V
5 2 V
4 2 V
3 1 V
2 1 V
1 1 V
1 0 V
1 0 V
2 1 V
3 2 V
4 2 V
5 2 V
5 2 V
6 3 V
6 3 V
7 4 V
6 3 V
7 3 V
6 4 V
6 3 V
5 3 V
5 3 V
3 2 V
4 2 V
2 1 V
1 0 V
1 1 V
1 0 V
2 2 V
3 1 V
4 3 V
5 2 V
5 3 V
6 4 V
6 3 V
6 4 V
7 3 V
6 4 V
6 3 V
6 4 V
6 3 V
4 2 V
4 2 V
3 2 V
2 1 V
2 1 V
1 1 V
3 1 V
3 2 V
4 2 V
4 2 V
6 3 V
5 3 V
7 3 V
6 3 V
7 4 V
6 3 V
6 3 V
6 2 V
5 3 V
5 2 V
4 2 V
3 1 V
2 1 V
1 1 V
1 0 V
1 1 V
2 0 V
3 2 V
4 2 V
5 2 V
5 2 V
6 2 V
6 3 V
7 2 V
6 3 V
7 2 V
6 2 V
6 3 V
5 1 V
5 2 V
4 1 V
3 1 V
2 1 V
1 1 V
1 0 V
1 0 V
2 1 V
3 1 V
stroke 1440 830 M
4 1 V
5 1 V
5 2 V
6 2 V
6 2 V
6 1 V
7 2 V
6 2 V
6 1 V
6 2 V
6 1 V
4 1 V
4 1 V
3 0 V
2 1 V
2 0 V
2 1 V
2 0 V
3 1 V
4 0 V
4 1 V
6 1 V
6 1 V
6 1 V
6 1 V
7 1 V
6 1 V
6 1 V
6 1 V
5 0 V
5 1 V
4 0 V
3 1 V
2 0 V
1 0 V
1 0 V
1 0 V
2 0 V
4 1 V
3 0 V
5 1 V
5 0 V
6 0 V
6 1 V
7 0 V
6 0 V
7 1 V
6 0 V
6 0 V
5 0 V
5 0 V
4 0 V
3 0 V
2 0 V
1 0 V
1 0 V
1 0 V
2 0 V
3 0 V
4 0 V
5 0 V
5 0 V
6 0 V
6 0 V
6 0 V
7 -1 V
6 0 V
7 0 V
5 -1 V
6 0 V
4 0 V
4 -1 V
3 0 V
3 0 V
1 0 V
2 0 V
2 0 V
3 -1 V
4 0 V
4 0 V
6 -1 V
6 0 V
6 -1 V
6 -1 V
7 -1 V
6 0 V
6 -1 V
6 -1 V
5 0 V
5 -1 V
4 -1 V
3 0 V
2 0 V
1 -1 V
1 0 V
1 0 V
2 0 V
4 -1 V
3 0 V
5 -1 V
5 -1 V
6 -1 V
6 -1 V
7 -1 V
stroke 1896 848 M
6 -1 V
7 -1 V
6 -1 V
6 -1 V
5 -1 V
5 -1 V
4 -1 V
3 0 V
2 -1 V
1 0 V
1 0 V
1 0 V
2 -1 V
3 0 V
4 -1 V
5 -1 V
5 -1 V
6 -1 V
6 -1 V
7 -2 V
6 -1 V
6 -2 V
7 -1 V
5 -1 V
6 -1 V
4 -1 V
4 -1 V
3 -1 V
3 0 V
1 -1 V
2 0 V
2 -1 V
3 0 V
4 -1 V
5 -1 V
5 -2 V
6 -1 V
6 -1 V
6 -2 V
7 -2 V
6 -1 V
6 -2 V
6 -1 V
5 -1 V
5 -2 V
4 -1 V
3 0 V
2 -1 V
2 0 V
1 -1 V
3 0 V
3 -1 V
4 -1 V
4 -1 V
6 -2 V
5 -1 V
6 -2 V
7 -1 V
6 -2 V
7 -2 V
6 -1 V
6 -2 V
5 -1 V
5 -1 V
4 -2 V
3 0 V
2 -1 V
1 0 V
1 0 V
1 -1 V
2 0 V
3 -1 V
4 -1 V
5 -1 V
5 -2 V
6 -1 V
6 -2 V
7 -2 V
6 -2 V
6 -1 V
7 -2 V
5 -1 V
6 -2 V
4 -1 V
4 -1 V
3 -1 V
3 -1 V
1 0 V
2 0 V
2 -1 V
3 -1 V
4 -1 V
5 -1 V
5 -2 V
6 -1 V
6 -2 V
6 -2 V
7 -1 V
6 -2 V
6 -2 V
6 -1 V
5 -2 V
5 -1 V
4 -1 V
stroke 2358 736 M
3 -1 V
2 0 V
2 -1 V
1 0 V
3 -1 V
3 -1 V
4 -1 V
4 -1 V
6 -1 V
5 -2 V
7 -1 V
6 -2 V
6 -2 V
7 -2 V
6 -1 V
6 -2 V
5 -1 V
5 -1 V
4 -1 V
3 -1 V
2 -1 V
1 0 V
1 0 V
1 0 V
2 -1 V
3 -1 V
4 -1 V
5 -1 V
5 -1 V
6 -2 V
6 -1 V
7 -2 V
6 -2 V
7 -1 V
6 -2 V
6 -1 V
5 -2 V
4 -1 V
4 -1 V
4 0 V
2 -1 V
1 0 V
1 0 V
1 -1 V
2 0 V
3 -1 V
4 -1 V
5 -1 V
5 -1 V
6 -2 V
6 -1 V
6 -2 V
7 -1 V
6 -2 V
6 -2 V
6 -1 V
5 -1 V
5 -1 V
4 -1 V
3 -1 V
2 -1 V
2 0 V
1 0 V
3 -1 V
3 -1 V
4 -1 V
4 -1 V
6 -1 V
5 -2 V
7 -1 V
6 -2 V
6 -2 V
7 -2 V
6 -1 V
6 -2 V
5 -2 V
5 -2 V
4 -2 V
3 -2 V
2 -2 V
1 -2 V
stroke
LTb
0 3431 N
0 640 L
2879 0 V
0 2791 V
0 3431 L
Z stroke
1.000 UP
1.000 UL
LTb
1.000 UL
LTb
4320 640 M
63 0 V
2816 0 R
-63 0 V
4320 989 M
63 0 V
2816 0 R
-63 0 V
4320 1338 M
63 0 V
2816 0 R
-63 0 V
4320 1687 M
63 0 V
2816 0 R
-63 0 V
4320 2036 M
63 0 V
2816 0 R
-63 0 V
4320 2384 M
63 0 V
2816 0 R
-63 0 V
4320 2733 M
63 0 V
2816 0 R
-63 0 V
4320 3082 M
63 0 V
2816 0 R
-63 0 V
4320 3431 M
63 0 V
2816 0 R
-63 0 V
4320 640 M
0 63 V
0 2728 R
0 -63 V
4640 640 M
0 63 V
0 2728 R
0 -63 V
4960 640 M
0 63 V
0 2728 R
0 -63 V
5280 640 M
0 63 V
0 2728 R
0 -63 V
5600 640 M
0 63 V
0 2728 R
0 -63 V
5919 640 M
0 63 V
0 2728 R
0 -63 V
6239 640 M
0 63 V
0 2728 R
0 -63 V
6559 640 M
0 63 V
0 2728 R
0 -63 V
6879 640 M
0 63 V
0 2728 R
0 -63 V
7199 640 M
0 63 V
0 2728 R
0 -63 V
stroke
LTa
4320 640 M
2879 0 V
stroke
LTb
4320 3431 N
0 -2791 V
2879 0 V
0 2791 V
-2879 0 V
Z stroke
LCb setrgbcolor
LTb
LCb setrgbcolor
LTb
1.000 UP
1.000 UL
LTb
1.000 UL
LT0
LCb setrgbcolor
LT0
5672 1777 M
543 0 V
4320 660 M
2 8 V
4 15 V
4 21 V
6 27 V
7 31 V
8 36 V
9 38 V
10 41 V
9 42 V
10 42 V
10 41 V
9 39 V
9 37 V
8 33 V
7 29 V
6 24 V
5 19 V
3 13 V
2 8 V
1 4 V
2 7 V
3 13 V
5 19 V
6 24 V
7 27 V
8 32 V
8 34 V
10 36 V
9 36 V
10 37 V
10 36 V
9 35 V
9 32 V
8 29 V
7 25 V
6 21 V
5 16 V
3 12 V
2 7 V
1 3 V
2 6 V
3 12 V
5 16 V
6 20 V
7 25 V
8 27 V
9 29 V
9 31 V
10 32 V
10 31 V
9 31 V
10 30 V
8 27 V
8 25 V
7 21 V
6 18 V
5 14 V
3 10 V
2 6 V
1 2 V
2 6 V
3 9 V
5 14 V
6 17 V
7 21 V
8 23 V
9 24 V
9 26 V
10 27 V
10 26 V
9 26 V
10 24 V
9 23 V
8 20 V
7 18 V
6 15 V
4 11 V
4 9 V
2 4 V
0 2 V
2 5 V
4 8 V
4 11 V
6 14 V
7 17 V
8 18 V
9 20 V
10 21 V
9 21 V
10 22 V
10 20 V
9 19 V
9 18 V
8 16 V
7 14 V
6 12 V
5 9 V
3 6 V
2 4 V
1 2 V
2 3 V
3 6 V
stroke 4967 2733 M
5 9 V
6 11 V
7 13 V
8 14 V
8 15 V
10 16 V
9 16 V
10 16 V
10 16 V
9 14 V
9 13 V
8 12 V
7 10 V
6 9 V
5 6 V
3 5 V
2 2 V
1 1 V
2 3 V
3 4 V
5 7 V
6 7 V
7 9 V
8 10 V
9 11 V
9 11 V
10 11 V
10 11 V
9 10 V
10 10 V
8 8 V
8 8 V
7 7 V
6 5 V
5 4 V
3 3 V
2 1 V
1 1 V
2 2 V
3 2 V
5 4 V
6 5 V
7 5 V
8 6 V
9 6 V
9 6 V
10 6 V
10 6 V
9 5 V
10 5 V
9 4 V
8 4 V
7 3 V
6 2 V
4 2 V
4 1 V
2 1 V
2 0 V
4 1 V
4 2 V
6 2 V
7 1 V
8 2 V
9 2 V
9 1 V
10 2 V
10 1 V
10 0 V
9 0 V
9 0 V
8 0 V
7 -1 V
6 -1 V
5 0 V
3 -1 V
2 0 V
1 0 V
2 0 V
3 -1 V
5 -1 V
5 -1 V
8 -2 V
8 -2 V
8 -2 V
10 -3 V
9 -3 V
10 -4 V
10 -4 V
9 -5 V
9 -4 V
8 -4 V
7 -4 V
6 -3 V
5 -3 V
3 -2 V
2 -1 V
1 -1 V
2 -1 V
3 -2 V
5 -3 V
6 -4 V
7 -5 V
8 -5 V
9 -7 V
stroke 5643 3081 M
9 -7 V
10 -8 V
10 -8 V
9 -8 V
10 -9 V
8 -8 V
8 -7 V
7 -7 V
6 -6 V
5 -5 V
3 -3 V
2 -2 V
1 -1 V
2 -2 V
3 -4 V
5 -5 V
6 -6 V
7 -8 V
8 -9 V
9 -10 V
9 -11 V
10 -12 V
10 -12 V
9 -12 V
10 -12 V
9 -12 V
8 -11 V
7 -10 V
6 -8 V
4 -6 V
4 -5 V
2 -3 V
0 -1 V
2 -3 V
4 -4 V
4 -7 V
6 -9 V
7 -11 V
8 -12 V
9 -13 V
9 -15 V
10 -15 V
10 -16 V
10 -15 V
9 -16 V
9 -15 V
8 -13 V
7 -12 V
6 -11 V
5 -8 V
3 -6 V
2 -3 V
1 -2 V
2 -3 V
3 -6 V
5 -8 V
5 -11 V
8 -13 V
8 -15 V
8 -16 V
10 -17 V
9 -19 V
10 -19 V
10 -18 V
9 -19 V
9 -17 V
8 -16 V
7 -14 V
6 -12 V
5 -10 V
3 -7 V
2 -4 V
1 -1 V
2 -4 V
3 -7 V
5 -10 V
6 -12 V
7 -15 V
8 -17 V
9 -18 V
9 -20 V
10 -21 V
9 -22 V
10 -21 V
10 -20 V
8 -20 V
8 -18 V
7 -16 V
6 -13 V
5 -11 V
3 -7 V
2 -4 V
1 -2 V
2 -5 V
3 -7 V
5 -11 V
6 -13 V
7 -17 V
8 -18 V
9 -20 V
9 -22 V
10 -23 V
10 -23 V
9 -23 V
stroke 6322 1932 M
10 -22 V
9 -21 V
8 -19 V
7 -17 V
6 -14 V
4 -11 V
4 -8 V
2 -5 V
0 -2 V
2 -4 V
4 -8 V
4 -12 V
6 -14 V
7 -17 V
8 -19 V
9 -22 V
9 -22 V
10 -24 V
10 -24 V
10 -23 V
9 -23 V
9 -22 V
8 -19 V
7 -17 V
6 -15 V
5 -11 V
3 -8 V
2 -5 V
1 -2 V
2 -5 V
3 -8 V
5 -11 V
5 -15 V
8 -17 V
8 -19 V
8 -22 V
10 -22 V
9 -24 V
10 -23 V
10 -23 V
9 -23 V
9 -21 V
8 -19 V
7 -16 V
6 -14 V
5 -11 V
3 -8 V
2 -5 V
1 -2 V
2 -4 V
3 -8 V
5 -11 V
6 -14 V
7 -16 V
8 -18 V
9 -20 V
9 -22 V
10 -21 V
9 -22 V
10 -22 V
10 -20 V
8 -19 V
8 -17 V
7 -15 V
6 -13 V
5 -10 V
3 -7 V
2 -4 V
1 -1 V
2 -4 V
3 -7 V
5 -10 V
6 -12 V
7 -14 V
8 -16 V
9 -17 V
9 -19 V
10 -18 V
10 -19 V
9 -18 V
10 -17 V
9 -16 V
8 -16 V
7 -13 V
6 -13 V
4 -12 V
4 -11 V
2 -10 V
stroke
LT1
LCb setrgbcolor
LT1
5672 1577 M
543 0 V
5330 640 M
1 0 V
2 0 V
3 0 V
4 0 V
4 1 V
5 0 V
5 1 V
6 1 V
6 1 V
6 2 V
6 2 V
5 1 V
6 2 V
5 2 V
4 2 V
3 1 V
3 1 V
2 1 V
1 1 V
1 0 V
1 1 V
2 0 V
3 2 V
4 2 V
4 2 V
5 2 V
5 3 V
6 3 V
6 4 V
6 3 V
5 4 V
6 3 V
5 4 V
5 3 V
5 3 V
3 2 V
3 2 V
2 1 V
1 1 V
1 1 V
1 0 V
2 2 V
3 2 V
3 2 V
5 3 V
4 4 V
6 4 V
5 4 V
6 5 V
6 4 V
6 5 V
6 4 V
5 4 V
5 4 V
4 4 V
4 3 V
3 2 V
2 2 V
1 0 V
0 1 V
2 1 V
2 1 V
2 3 V
4 3 V
4 3 V
5 4 V
6 5 V
5 4 V
6 5 V
6 5 V
6 5 V
6 5 V
5 4 V
5 4 V
4 4 V
4 3 V
3 2 V
2 2 V
1 0 V
0 1 V
1 1 V
2 1 V
3 3 V
4 3 V
4 3 V
5 4 V
5 5 V
6 4 V
6 5 V
6 5 V
6 5 V
5 4 V
6 4 V
5 4 V
4 4 V
3 2 V
3 3 V
2 1 V
1 1 V
1 1 V
1 0 V
2 2 V
stroke 5722 893 M
3 2 V
4 3 V
4 3 V
5 4 V
5 4 V
6 4 V
6 4 V
6 5 V
5 4 V
6 4 V
5 4 V
5 3 V
5 3 V
3 3 V
3 2 V
2 1 V
1 1 V
1 0 V
1 1 V
2 1 V
3 2 V
3 3 V
5 3 V
4 3 V
6 3 V
5 4 V
6 4 V
6 3 V
6 4 V
6 3 V
5 3 V
5 3 V
4 3 V
4 2 V
3 1 V
2 1 V
1 1 V
2 1 V
2 1 V
2 2 V
4 2 V
4 2 V
5 2 V
6 3 V
5 3 V
6 3 V
6 3 V
6 2 V
6 3 V
5 2 V
5 3 V
4 1 V
4 2 V
3 1 V
2 1 V
1 0 V
0 1 V
1 0 V
2 1 V
3 1 V
4 1 V
4 2 V
5 2 V
5 2 V
6 2 V
6 2 V
6 2 V
6 1 V
5 2 V
6 2 V
5 1 V
4 1 V
3 1 V
3 1 V
2 0 V
1 1 V
1 0 V
1 0 V
2 0 V
3 1 V
4 1 V
4 1 V
5 1 V
5 1 V
6 1 V
6 1 V
6 1 V
5 1 V
6 0 V
5 1 V
5 1 V
5 0 V
3 0 V
3 1 V
2 0 V
1 0 V
1 0 V
1 0 V
2 0 V
3 0 V
3 1 V
5 0 V
4 0 V
6 0 V
stroke 6132 1071 M
5 0 V
6 0 V
6 0 V
6 0 V
6 0 V
5 0 V
5 -1 V
4 0 V
4 0 V
3 0 V
2 0 V
1 -1 V
2 0 V
2 0 V
2 0 V
4 -1 V
4 0 V
5 -1 V
6 0 V
5 -1 V
6 -1 V
6 -1 V
6 -1 V
6 -1 V
5 -1 V
5 -1 V
4 -1 V
4 -1 V
3 -1 V
2 0 V
1 -1 V
1 0 V
2 -1 V
3 0 V
4 -1 V
4 -1 V
5 -2 V
5 -1 V
6 -2 V
6 -2 V
6 -2 V
6 -2 V
5 -2 V
6 -2 V
5 -1 V
4 -2 V
3 -1 V
3 -2 V
2 0 V
1 -1 V
1 0 V
1 0 V
2 -1 V
3 -1 V
4 -2 V
4 -2 V
5 -2 V
5 -2 V
6 -3 V
6 -2 V
6 -3 V
5 -3 V
6 -3 V
5 -2 V
5 -3 V
5 -2 V
3 -2 V
3 -1 V
2 -1 V
1 -1 V
1 0 V
1 -1 V
2 -1 V
3 -1 V
3 -2 V
5 -3 V
4 -2 V
6 -3 V
5 -4 V
6 -3 V
6 -4 V
6 -3 V
6 -4 V
5 -3 V
5 -3 V
4 -3 V
4 -2 V
3 -2 V
2 -1 V
1 -1 V
2 -1 V
2 -1 V
2 -2 V
4 -2 V
4 -3 V
5 -3 V
6 -4 V
5 -4 V
6 -4 V
6 -4 V
6 -4 V
6 -4 V
5 -4 V
5 -3 V
stroke 6560 906 M
4 -3 V
4 -3 V
3 -2 V
2 -1 V
1 -1 V
0 -1 V
1 -1 V
2 -1 V
3 -2 V
4 -3 V
4 -3 V
5 -4 V
5 -4 V
6 -4 V
6 -4 V
6 -5 V
6 -4 V
5 -5 V
6 -4 V
5 -4 V
4 -3 V
3 -3 V
3 -2 V
2 -2 V
1 -1 V
1 0 V
1 -1 V
2 -2 V
3 -2 V
4 -3 V
4 -3 V
5 -4 V
5 -4 V
6 -5 V
6 -4 V
6 -5 V
5 -5 V
6 -4 V
5 -5 V
5 -4 V
5 -3 V
3 -3 V
3 -2 V
2 -2 V
1 -1 V
1 0 V
1 -1 V
2 -2 V
3 -2 V
3 -3 V
5 -3 V
4 -4 V
6 -5 V
5 -4 V
6 -5 V
6 -5 V
6 -4 V
6 -5 V
5 -4 V
5 -4 V
4 -3 V
4 -3 V
3 -2 V
2 -2 V
1 -1 V
2 -1 V
2 -2 V
2 -2 V
4 -3 V
4 -3 V
5 -4 V
6 -4 V
5 -4 V
6 -5 V
6 -5 V
6 -4 V
6 -5 V
5 -4 V
5 -5 V
4 -4 V
4 -4 V
3 -3 V
2 -4 V
1 -3 V
stroke
LTb
4320 3431 N
0 -2791 V
2879 0 V
0 2791 V
-2879 0 V
Z stroke
1.000 UP
1.000 UL
LTb
stroke
grestore
end
showpage
  }}%
  \put(5552,1577){\makebox(0,0)[r]{\strut{}$l=\tau $}}%
  \put(5552,1777){\makebox(0,0)[r]{\strut{}$l=e$}}%
  \put(5759,3731){\makebox(0,0){\strut{}$\bar B_s\to c\bar s(^1P_1) l\nu_l$}}%
  \put(5759,140){\makebox(0,0){\strut{}$q^2$ [$\rm GeV^2$]}}%
  \put(3260,2035){%
  \special{ps: gsave currentpoint currentpoint translate
630 rotate neg exch neg exch translate}%
  \makebox(0,0){\strut{}$d\Gamma/dq^2$ [$|V_{cb}|^2\ 10^{-13} \rm GeV^{-1}$]}%
  \special{ps: currentpoint grestore moveto}%
  }%
  \put(7199,440){\makebox(0,0){\strut{} 9}}%
  \put(6879,440){\makebox(0,0){\strut{} 8}}%
  \put(6559,440){\makebox(0,0){\strut{} 7}}%
  \put(6239,440){\makebox(0,0){\strut{} 6}}%
  \put(5919,440){\makebox(0,0){\strut{} 5}}%
  \put(5600,440){\makebox(0,0){\strut{} 4}}%
  \put(5280,440){\makebox(0,0){\strut{} 3}}%
  \put(4960,440){\makebox(0,0){\strut{} 2}}%
  \put(4640,440){\makebox(0,0){\strut{} 1}}%
  \put(4320,440){\makebox(0,0){\strut{} 0}}%
  \put(4200,3431){\makebox(0,0)[r]{\strut{} 0.04}}%
  \put(4200,3082){\makebox(0,0)[r]{\strut{} 0.035}}%
  \put(4200,2733){\makebox(0,0)[r]{\strut{} 0.03}}%
  \put(4200,2384){\makebox(0,0)[r]{\strut{} 0.025}}%
  \put(4200,2036){\makebox(0,0)[r]{\strut{} 0.02}}%
  \put(4200,1687){\makebox(0,0)[r]{\strut{} 0.015}}%
  \put(4200,1338){\makebox(0,0)[r]{\strut{} 0.01}}%
  \put(4200,989){\makebox(0,0)[r]{\strut{} 0.005}}%
  \put(4200,640){\makebox(0,0)[r]{\strut{} 0}}%
  \put(1808,2415){\makebox(0,0)[r]{\strut{}$l= \tau $}}%
  \put(1808,2615){\makebox(0,0)[r]{\strut{}$l=e$}}%
  \put(1439,3731){\makebox(0,0){\strut{}$\bar B_s\to c\bar s(^1P_1) l\nu_l$}}%
  \put(1439,140){\makebox(0,0){\strut{}$q^2$ [$\rm GeV^2$]}}%
  \put(-940,2035){%
  \special{ps: gsave currentpoint currentpoint translate
630 rotate neg exch neg exch translate}%
  \makebox(0,0){\strut{}$d\Gamma/dq^2$ [$|V_{cb}|^2\ 10^{-13} \rm GeV^{-1}$]}%
  \special{ps: currentpoint grestore moveto}%
  }%
  \put(2879,440){\makebox(0,0){\strut{} 9}}%
  \put(2559,440){\makebox(0,0){\strut{} 8}}%
  \put(2239,440){\makebox(0,0){\strut{} 7}}%
  \put(1919,440){\makebox(0,0){\strut{} 6}}%
  \put(1599,440){\makebox(0,0){\strut{} 5}}%
  \put(1280,440){\makebox(0,0){\strut{} 4}}%
  \put(960,440){\makebox(0,0){\strut{} 3}}%
  \put(640,440){\makebox(0,0){\strut{} 2}}%
  \put(320,440){\makebox(0,0){\strut{} 1}}%
  \put(0,440){\makebox(0,0){\strut{} 0}}%
  \put(-120,3431){\makebox(0,0)[r]{\strut{} 0.2}}%
  \put(-120,3152){\makebox(0,0)[r]{\strut{} 0.18}}%
  \put(-120,2873){\makebox(0,0)[r]{\strut{} 0.16}}%
  \put(-120,2594){\makebox(0,0)[r]{\strut{} 0.14}}%
  \put(-120,2315){\makebox(0,0)[r]{\strut{} 0.12}}%
  \put(-120,2036){\makebox(0,0)[r]{\strut{} 0.1}}%
  \put(-120,1756){\makebox(0,0)[r]{\strut{} 0.08}}%
  \put(-120,1477){\makebox(0,0)[r]{\strut{} 0.06}}%
  \put(-120,1198){\makebox(0,0)[r]{\strut{} 0.04}}%
  \put(-120,919){\makebox(0,0)[r]{\strut{} 0.02}}%
  \put(-120,640){\makebox(0,0)[r]{\strut{} 0}}%
\end{picture}%
\endgroup
 

%% file: dg.2m.2p.tex
\begingroup%
\makeatletter%
\newcommand{\GNUPLOTspecial}{%
  \@sanitize\catcode`\%=14\relax\special}%
\setlength{\unitlength}{0.0500bp}%
\begin{picture}(7200,5040)(0,0)%
  {\GNUPLOTspecial{"
/gnudict 256 dict def
gnudict begin
%
%
/Color false def
/Blacktext true def
/Solid false def
/Dashlength 1 def
/Landscape false def
/Level1 false def
/Rounded false def
/ClipToBoundingBox false def
/SuppressPDFMark false def
/TransparentPatterns false def
/gnulinewidth 5.000 def
/userlinewidth gnulinewidth def
/Gamma 1.0 def
/BackgroundColor {-1.000 -1.000 -1.000} def
/vshift -66 def
/dl1 {
  10.0 Dashlength mul mul
  Rounded { currentlinewidth 0.75 mul sub dup 0 le { pop 0.01 } if } if
} def
/dl2 {
  10.0 Dashlength mul mul
  Rounded { currentlinewidth 0.75 mul add } if
} def
/hpt_ 31.5 def
/vpt_ 31.5 def
/hpt hpt_ def
/vpt vpt_ def
/doclip {
  ClipToBoundingBox {
    newpath 0 0 moveto 360 0 lineto 360 252 lineto 0 252 lineto closepath
    clip
  } if
} def
%
%
%
/M {moveto} bind def
/L {lineto} bind def
/R {rmoveto} bind def
/V {rlineto} bind def
/N {newpath moveto} bind def
/Z {closepath} bind def
/C {setrgbcolor} bind def
/f {rlineto fill} bind def
/g {setgray} bind def
/Gshow {show} def   
/vpt2 vpt 2 mul def
/hpt2 hpt 2 mul def
/Lshow {currentpoint stroke M 0 vshift R 
	Blacktext {gsave 0 setgray show grestore} {show} ifelse} def
/Rshow {currentpoint stroke M dup stringwidth pop neg vshift R
	Blacktext {gsave 0 setgray show grestore} {show} ifelse} def
/Cshow {currentpoint stroke M dup stringwidth pop -2 div vshift R 
	Blacktext {gsave 0 setgray show grestore} {show} ifelse} def
/UP {dup vpt_ mul /vpt exch def hpt_ mul /hpt exch def
  /hpt2 hpt 2 mul def /vpt2 vpt 2 mul def} def
/DL {Color {setrgbcolor Solid {pop []} if 0 setdash}
 {pop pop pop 0 setgray Solid {pop []} if 0 setdash} ifelse} def
/BL {stroke userlinewidth 2 mul setlinewidth
	Rounded {1 setlinejoin 1 setlinecap} if} def
/AL {stroke userlinewidth 2 div setlinewidth
	Rounded {1 setlinejoin 1 setlinecap} if} def
/UL {dup gnulinewidth mul /userlinewidth exch def
	dup 1 lt {pop 1} if 10 mul /udl exch def} def
/PL {stroke userlinewidth setlinewidth
	Rounded {1 setlinejoin 1 setlinecap} if} def
3.8 setmiterlimit
/LCw {1 1 1} def
/LCb {0 0 0} def
/LCa {0 0 0} def
/LC0 {1 0 0} def
/LC1 {0 1 0} def
/LC2 {0 0 1} def
/LC3 {1 0 1} def
/LC4 {0 1 1} def
/LC5 {1 1 0} def
/LC6 {0 0 0} def
/LC7 {1 0.3 0} def
/LC8 {0.5 0.5 0.5} def
/LTw {PL [] 1 setgray} def
/LTb {BL [] LCb DL} def
/LTa {AL [1 udl mul 2 udl mul] 0 setdash LCa setrgbcolor} def
/LT0 {PL [] LC0 DL} def
/LT1 {PL [4 dl1 2 dl2] LC1 DL} def
/LT2 {PL [2 dl1 3 dl2] LC2 DL} def
/LT3 {PL [1 dl1 1.5 dl2] LC3 DL} def
/LT4 {PL [6 dl1 2 dl2 1 dl1 2 dl2] LC4 DL} def
/LT5 {PL [3 dl1 3 dl2 1 dl1 3 dl2] LC5 DL} def
/LT6 {PL [2 dl1 2 dl2 2 dl1 6 dl2] LC6 DL} def
/LT7 {PL [1 dl1 2 dl2 6 dl1 2 dl2 1 dl1 2 dl2] LC7 DL} def
/LT8 {PL [2 dl1 2 dl2 2 dl1 2 dl2 2 dl1 2 dl2 2 dl1 4 dl2] LC8 DL} def
/Pnt {stroke [] 0 setdash gsave 1 setlinecap M 0 0 V stroke grestore} def
/Dia {stroke [] 0 setdash 2 copy vpt add M
  hpt neg vpt neg V hpt vpt neg V
  hpt vpt V hpt neg vpt V closepath stroke
  Pnt} def
/Pls {stroke [] 0 setdash vpt sub M 0 vpt2 V
  currentpoint stroke M
  hpt neg vpt neg R hpt2 0 V stroke
 } def
/Box {stroke [] 0 setdash 2 copy exch hpt sub exch vpt add M
  0 vpt2 neg V hpt2 0 V 0 vpt2 V
  hpt2 neg 0 V closepath stroke
  Pnt} def
/Crs {stroke [] 0 setdash exch hpt sub exch vpt add M
  hpt2 vpt2 neg V currentpoint stroke M
  hpt2 neg 0 R hpt2 vpt2 V stroke} def
/TriU {stroke [] 0 setdash 2 copy vpt 1.12 mul add M
  hpt neg vpt -1.62 mul V
  hpt 2 mul 0 V
  hpt neg vpt 1.62 mul V closepath stroke
  Pnt} def
/Star {2 copy Pls Crs} def
/BoxF {stroke [] 0 setdash exch hpt sub exch vpt add M
  0 vpt2 neg V hpt2 0 V 0 vpt2 V
  hpt2 neg 0 V closepath fill} def
/TriUF {stroke [] 0 setdash vpt 1.12 mul add M
  hpt neg vpt -1.62 mul V
  hpt 2 mul 0 V
  hpt neg vpt 1.62 mul V closepath fill} def
/TriD {stroke [] 0 setdash 2 copy vpt 1.12 mul sub M
  hpt neg vpt 1.62 mul V
  hpt 2 mul 0 V
  hpt neg vpt -1.62 mul V closepath stroke
  Pnt} def
/TriDF {stroke [] 0 setdash vpt 1.12 mul sub M
  hpt neg vpt 1.62 mul V
  hpt 2 mul 0 V
  hpt neg vpt -1.62 mul V closepath fill} def
/DiaF {stroke [] 0 setdash vpt add M
  hpt neg vpt neg V hpt vpt neg V
  hpt vpt V hpt neg vpt V closepath fill} def
/Pent {stroke [] 0 setdash 2 copy gsave
  translate 0 hpt M 4 {72 rotate 0 hpt L} repeat
  closepath stroke grestore Pnt} def
/PentF {stroke [] 0 setdash gsave
  translate 0 hpt M 4 {72 rotate 0 hpt L} repeat
  closepath fill grestore} def
/Circle {stroke [] 0 setdash 2 copy
  hpt 0 360 arc stroke Pnt} def
/CircleF {stroke [] 0 setdash hpt 0 360 arc fill} def
/C0 {BL [] 0 setdash 2 copy moveto vpt 90 450 arc} bind def
/C1 {BL [] 0 setdash 2 copy moveto
	2 copy vpt 0 90 arc closepath fill
	vpt 0 360 arc closepath} bind def
/C2 {BL [] 0 setdash 2 copy moveto
	2 copy vpt 90 180 arc closepath fill
	vpt 0 360 arc closepath} bind def
/C3 {BL [] 0 setdash 2 copy moveto
	2 copy vpt 0 180 arc closepath fill
	vpt 0 360 arc closepath} bind def
/C4 {BL [] 0 setdash 2 copy moveto
	2 copy vpt 180 270 arc closepath fill
	vpt 0 360 arc closepath} bind def
/C5 {BL [] 0 setdash 2 copy moveto
	2 copy vpt 0 90 arc
	2 copy moveto
	2 copy vpt 180 270 arc closepath fill
	vpt 0 360 arc} bind def
/C6 {BL [] 0 setdash 2 copy moveto
	2 copy vpt 90 270 arc closepath fill
	vpt 0 360 arc closepath} bind def
/C7 {BL [] 0 setdash 2 copy moveto
	2 copy vpt 0 270 arc closepath fill
	vpt 0 360 arc closepath} bind def
/C8 {BL [] 0 setdash 2 copy moveto
	2 copy vpt 270 360 arc closepath fill
	vpt 0 360 arc closepath} bind def
/C9 {BL [] 0 setdash 2 copy moveto
	2 copy vpt 270 450 arc closepath fill
	vpt 0 360 arc closepath} bind def
/C10 {BL [] 0 setdash 2 copy 2 copy moveto vpt 270 360 arc closepath fill
	2 copy moveto
	2 copy vpt 90 180 arc closepath fill
	vpt 0 360 arc closepath} bind def
/C11 {BL [] 0 setdash 2 copy moveto
	2 copy vpt 0 180 arc closepath fill
	2 copy moveto
	2 copy vpt 270 360 arc closepath fill
	vpt 0 360 arc closepath} bind def
/C12 {BL [] 0 setdash 2 copy moveto
	2 copy vpt 180 360 arc closepath fill
	vpt 0 360 arc closepath} bind def
/C13 {BL [] 0 setdash 2 copy moveto
	2 copy vpt 0 90 arc closepath fill
	2 copy moveto
	2 copy vpt 180 360 arc closepath fill
	vpt 0 360 arc closepath} bind def
/C14 {BL [] 0 setdash 2 copy moveto
	2 copy vpt 90 360 arc closepath fill
	vpt 0 360 arc} bind def
/C15 {BL [] 0 setdash 2 copy vpt 0 360 arc closepath fill
	vpt 0 360 arc closepath} bind def
/Rec {newpath 4 2 roll moveto 1 index 0 rlineto 0 exch rlineto
	neg 0 rlineto closepath} bind def
/Square {dup Rec} bind def
/Bsquare {vpt sub exch vpt sub exch vpt2 Square} bind def
/S0 {BL [] 0 setdash 2 copy moveto 0 vpt rlineto BL Bsquare} bind def
/S1 {BL [] 0 setdash 2 copy vpt Square fill Bsquare} bind def
/S2 {BL [] 0 setdash 2 copy exch vpt sub exch vpt Square fill Bsquare} bind def
/S3 {BL [] 0 setdash 2 copy exch vpt sub exch vpt2 vpt Rec fill Bsquare} bind def
/S4 {BL [] 0 setdash 2 copy exch vpt sub exch vpt sub vpt Square fill Bsquare} bind def
/S5 {BL [] 0 setdash 2 copy 2 copy vpt Square fill
	exch vpt sub exch vpt sub vpt Square fill Bsquare} bind def
/S6 {BL [] 0 setdash 2 copy exch vpt sub exch vpt sub vpt vpt2 Rec fill Bsquare} bind def
/S7 {BL [] 0 setdash 2 copy exch vpt sub exch vpt sub vpt vpt2 Rec fill
	2 copy vpt Square fill Bsquare} bind def
/S8 {BL [] 0 setdash 2 copy vpt sub vpt Square fill Bsquare} bind def
/S9 {BL [] 0 setdash 2 copy vpt sub vpt vpt2 Rec fill Bsquare} bind def
/S10 {BL [] 0 setdash 2 copy vpt sub vpt Square fill 2 copy exch vpt sub exch vpt Square fill
	Bsquare} bind def
/S11 {BL [] 0 setdash 2 copy vpt sub vpt Square fill 2 copy exch vpt sub exch vpt2 vpt Rec fill
	Bsquare} bind def
/S12 {BL [] 0 setdash 2 copy exch vpt sub exch vpt sub vpt2 vpt Rec fill Bsquare} bind def
/S13 {BL [] 0 setdash 2 copy exch vpt sub exch vpt sub vpt2 vpt Rec fill
	2 copy vpt Square fill Bsquare} bind def
/S14 {BL [] 0 setdash 2 copy exch vpt sub exch vpt sub vpt2 vpt Rec fill
	2 copy exch vpt sub exch vpt Square fill Bsquare} bind def
/S15 {BL [] 0 setdash 2 copy Bsquare fill Bsquare} bind def
/D0 {gsave translate 45 rotate 0 0 S0 stroke grestore} bind def
/D1 {gsave translate 45 rotate 0 0 S1 stroke grestore} bind def
/D2 {gsave translate 45 rotate 0 0 S2 stroke grestore} bind def
/D3 {gsave translate 45 rotate 0 0 S3 stroke grestore} bind def
/D4 {gsave translate 45 rotate 0 0 S4 stroke grestore} bind def
/D5 {gsave translate 45 rotate 0 0 S5 stroke grestore} bind def
/D6 {gsave translate 45 rotate 0 0 S6 stroke grestore} bind def
/D7 {gsave translate 45 rotate 0 0 S7 stroke grestore} bind def
/D8 {gsave translate 45 rotate 0 0 S8 stroke grestore} bind def
/D9 {gsave translate 45 rotate 0 0 S9 stroke grestore} bind def
/D10 {gsave translate 45 rotate 0 0 S10 stroke grestore} bind def
/D11 {gsave translate 45 rotate 0 0 S11 stroke grestore} bind def
/D12 {gsave translate 45 rotate 0 0 S12 stroke grestore} bind def
/D13 {gsave translate 45 rotate 0 0 S13 stroke grestore} bind def
/D14 {gsave translate 45 rotate 0 0 S14 stroke grestore} bind def
/D15 {gsave translate 45 rotate 0 0 S15 stroke grestore} bind def
/DiaE {stroke [] 0 setdash vpt add M
  hpt neg vpt neg V hpt vpt neg V
  hpt vpt V hpt neg vpt V closepath stroke} def
/BoxE {stroke [] 0 setdash exch hpt sub exch vpt add M
  0 vpt2 neg V hpt2 0 V 0 vpt2 V
  hpt2 neg 0 V closepath stroke} def
/TriUE {stroke [] 0 setdash vpt 1.12 mul add M
  hpt neg vpt -1.62 mul V
  hpt 2 mul 0 V
  hpt neg vpt 1.62 mul V closepath stroke} def
/TriDE {stroke [] 0 setdash vpt 1.12 mul sub M
  hpt neg vpt 1.62 mul V
  hpt 2 mul 0 V
  hpt neg vpt -1.62 mul V closepath stroke} def
/PentE {stroke [] 0 setdash gsave
  translate 0 hpt M 4 {72 rotate 0 hpt L} repeat
  closepath stroke grestore} def
/CircE {stroke [] 0 setdash 
  hpt 0 360 arc stroke} def
/Opaque {gsave closepath 1 setgray fill grestore 0 setgray closepath} def
/DiaW {stroke [] 0 setdash vpt add M
  hpt neg vpt neg V hpt vpt neg V
  hpt vpt V hpt neg vpt V Opaque stroke} def
/BoxW {stroke [] 0 setdash exch hpt sub exch vpt add M
  0 vpt2 neg V hpt2 0 V 0 vpt2 V
  hpt2 neg 0 V Opaque stroke} def
/TriUW {stroke [] 0 setdash vpt 1.12 mul add M
  hpt neg vpt -1.62 mul V
  hpt 2 mul 0 V
  hpt neg vpt 1.62 mul V Opaque stroke} def
/TriDW {stroke [] 0 setdash vpt 1.12 mul sub M
  hpt neg vpt 1.62 mul V
  hpt 2 mul 0 V
  hpt neg vpt -1.62 mul V Opaque stroke} def
/PentW {stroke [] 0 setdash gsave
  translate 0 hpt M 4 {72 rotate 0 hpt L} repeat
  Opaque stroke grestore} def
/CircW {stroke [] 0 setdash 
  hpt 0 360 arc Opaque stroke} def
/BoxFill {gsave Rec 1 setgray fill grestore} def
/Density {
  /Fillden exch def
  currentrgbcolor
  /ColB exch def /ColG exch def /ColR exch def
  /ColR ColR Fillden mul Fillden sub 1 add def
  /ColG ColG Fillden mul Fillden sub 1 add def
  /ColB ColB Fillden mul Fillden sub 1 add def
  ColR ColG ColB setrgbcolor} def
/BoxColFill {gsave Rec PolyFill} def
/PolyFill {gsave Density fill grestore grestore} def
/h {rlineto rlineto rlineto gsave closepath fill grestore} bind def
%
%
/PatternFill {gsave /PFa [ 9 2 roll ] def
  PFa 0 get PFa 2 get 2 div add PFa 1 get PFa 3 get 2 div add translate
  PFa 2 get -2 div PFa 3 get -2 div PFa 2 get PFa 3 get Rec
  gsave 1 setgray fill grestore clip
  currentlinewidth 0.5 mul setlinewidth
  /PFs PFa 2 get dup mul PFa 3 get dup mul add sqrt def
  0 0 M PFa 5 get rotate PFs -2 div dup translate
  0 1 PFs PFa 4 get div 1 add floor cvi
	{PFa 4 get mul 0 M 0 PFs V} for
  0 PFa 6 get ne {
	0 1 PFs PFa 4 get div 1 add floor cvi
	{PFa 4 get mul 0 2 1 roll M PFs 0 V} for
 } if
  stroke grestore} def
/languagelevel where
 {pop languagelevel} {1} ifelse
 2 lt
	{/InterpretLevel1 true def}
	{/InterpretLevel1 Level1 def}
 ifelse
%
%
/Level2PatternFill {
/Tile8x8 {/PaintType 2 /PatternType 1 /TilingType 1 /BBox [0 0 8 8] /XStep 8 /YStep 8}
	bind def
/KeepColor {currentrgbcolor [/Pattern /DeviceRGB] setcolorspace} bind def
<< Tile8x8
 /PaintProc {0.5 setlinewidth pop 0 0 M 8 8 L 0 8 M 8 0 L stroke} 
>> matrix makepattern
/Pat1 exch def
<< Tile8x8
 /PaintProc {0.5 setlinewidth pop 0 0 M 8 8 L 0 8 M 8 0 L stroke
	0 4 M 4 8 L 8 4 L 4 0 L 0 4 L stroke}
>> matrix makepattern
/Pat2 exch def
<< Tile8x8
 /PaintProc {0.5 setlinewidth pop 0 0 M 0 8 L
	8 8 L 8 0 L 0 0 L fill}
>> matrix makepattern
/Pat3 exch def
<< Tile8x8
 /PaintProc {0.5 setlinewidth pop -4 8 M 8 -4 L
	0 12 M 12 0 L stroke}
>> matrix makepattern
/Pat4 exch def
<< Tile8x8
 /PaintProc {0.5 setlinewidth pop -4 0 M 8 12 L
	0 -4 M 12 8 L stroke}
>> matrix makepattern
/Pat5 exch def
<< Tile8x8
 /PaintProc {0.5 setlinewidth pop -2 8 M 4 -4 L
	0 12 M 8 -4 L 4 12 M 10 0 L stroke}
>> matrix makepattern
/Pat6 exch def
<< Tile8x8
 /PaintProc {0.5 setlinewidth pop -2 0 M 4 12 L
	0 -4 M 8 12 L 4 -4 M 10 8 L stroke}
>> matrix makepattern
/Pat7 exch def
<< Tile8x8
 /PaintProc {0.5 setlinewidth pop 8 -2 M -4 4 L
	12 0 M -4 8 L 12 4 M 0 10 L stroke}
>> matrix makepattern
/Pat8 exch def
<< Tile8x8
 /PaintProc {0.5 setlinewidth pop 0 -2 M 12 4 L
	-4 0 M 12 8 L -4 4 M 8 10 L stroke}
>> matrix makepattern
/Pat9 exch def
/Pattern1 {PatternBgnd KeepColor Pat1 setpattern} bind def
/Pattern2 {PatternBgnd KeepColor Pat2 setpattern} bind def
/Pattern3 {PatternBgnd KeepColor Pat3 setpattern} bind def
/Pattern4 {PatternBgnd KeepColor Landscape {Pat5} {Pat4} ifelse setpattern} bind def
/Pattern5 {PatternBgnd KeepColor Landscape {Pat4} {Pat5} ifelse setpattern} bind def
/Pattern6 {PatternBgnd KeepColor Landscape {Pat9} {Pat6} ifelse setpattern} bind def
/Pattern7 {PatternBgnd KeepColor Landscape {Pat8} {Pat7} ifelse setpattern} bind def
} def
%
%
%
/PatternBgnd {
  TransparentPatterns {} {gsave 1 setgray fill grestore} ifelse
} def
%
%
/Level1PatternFill {
/Pattern1 {0.250 Density} bind def
/Pattern2 {0.500 Density} bind def
/Pattern3 {0.750 Density} bind def
/Pattern4 {0.125 Density} bind def
/Pattern5 {0.375 Density} bind def
/Pattern6 {0.625 Density} bind def
/Pattern7 {0.875 Density} bind def
} def
%
%
Level1 {Level1PatternFill} {Level2PatternFill} ifelse
/Symbol-Oblique /Symbol findfont [1 0 .167 1 0 0] makefont
dup length dict begin {1 index /FID eq {pop pop} {def} ifelse} forall
currentdict end definefont pop
Level1 SuppressPDFMark or 
{} {
/SDict 10 dict def
systemdict /pdfmark known not {
  userdict /pdfmark systemdict /cleartomark get put
} if
SDict begin [
  /Title (paper/dg.2m.2p.tex)
  /Subject (gnuplot plot)
  /Creator (gnuplot 4.6 patchlevel 0)
  /Author (conrado)
  /CreationDate (Tue Oct 15 10:07:52 2013)
  /DOCINFO pdfmark
end
} ifelse
end
gnudict begin
gsave
doclip
0 0 translate
0.050 0.050 scale
0 setgray
newpath
BackgroundColor 0 lt 3 1 roll 0 lt exch 0 lt or or not {BackgroundColor C 1.000 0 0 7200.00 5040.00 BoxColFill} if
1.000 UL
LTb
0 640 M
63 0 V
2816 0 R
-63 0 V
0 1105 M
63 0 V
2816 0 R
-63 0 V
0 1570 M
63 0 V
2816 0 R
-63 0 V
0 2036 M
63 0 V
2816 0 R
-63 0 V
0 2501 M
63 0 V
2816 0 R
-63 0 V
0 2966 M
63 0 V
2816 0 R
-63 0 V
0 3431 M
63 0 V
2816 0 R
-63 0 V
0 640 M
0 63 V
0 3431 M
0 -63 V
411 640 M
0 63 V
0 2728 R
0 -63 V
823 640 M
0 63 V
0 2728 R
0 -63 V
1234 640 M
0 63 V
0 2728 R
0 -63 V
1645 640 M
0 63 V
0 2728 R
0 -63 V
2056 640 M
0 63 V
0 2728 R
0 -63 V
2468 640 M
0 63 V
0 2728 R
0 -63 V
2879 640 M
0 63 V
0 2728 R
0 -63 V
stroke
LTa
0 640 M
2879 0 V
stroke
LTb
0 3431 N
0 640 L
2879 0 V
0 2791 V
0 3431 L
Z stroke
LCb setrgbcolor
LTb
LCb setrgbcolor
LTb
1.000 UP
1.000 UL
LTb
1.000 UL
LT0
LCb setrgbcolor
LT0
1208 1498 M
543 0 V
0 698 M
2 13 V
4 23 V
5 32 V
6 41 V
8 48 V
8 53 V
9 58 V
10 60 V
10 62 V
11 61 V
10 59 V
10 57 V
9 52 V
8 47 V
8 40 V
6 34 V
5 26 V
3 19 V
2 10 V
1 5 V
2 10 V
4 19 V
5 25 V
6 32 V
7 38 V
9 42 V
9 45 V
10 47 V
10 48 V
10 48 V
10 46 V
10 43 V
10 40 V
8 36 V
7 31 V
7 26 V
5 20 V
3 14 V
2 8 V
1 4 V
2 7 V
3 14 V
5 20 V
7 24 V
7 28 V
8 31 V
10 34 V
10 35 V
10 36 V
10 35 V
10 34 V
10 31 V
9 29 V
9 26 V
7 23 V
6 18 V
5 14 V
4 10 V
2 6 V
1 2 V
2 6 V
3 10 V
5 13 V
6 17 V
8 20 V
8 22 V
9 23 V
10 25 V
10 24 V
11 23 V
10 23 V
10 21 V
9 19 V
9 17 V
7 15 V
6 11 V
5 10 V
4 6 V
2 3 V
0 2 V
2 4 V
4 6 V
5 8 V
6 11 V
8 12 V
8 13 V
9 14 V
10 15 V
10 14 V
11 13 V
10 13 V
10 12 V
9 10 V
8 9 V
8 7 V
6 6 V
5 5 V
3 3 V
2 2 V
1 1 V
2 1 V
4 3 V
stroke 680 3069 M
5 4 V
6 5 V
7 6 V
9 5 V
9 6 V
10 6 V
10 5 V
10 5 V
11 4 V
9 3 V
10 3 V
8 2 V
7 1 V
7 1 V
5 1 V
3 0 V
2 0 V
1 1 V
2 0 V
3 0 V
5 0 V
7 0 V
7 0 V
8 -1 V
10 -1 V
10 -2 V
10 -2 V
10 -3 V
10 -3 V
10 -4 V
9 -4 V
9 -4 V
7 -3 V
6 -4 V
5 -2 V
4 -2 V
2 -2 V
1 0 V
2 -1 V
3 -3 V
5 -3 V
6 -4 V
8 -5 V
8 -6 V
9 -7 V
10 -8 V
11 -8 V
10 -9 V
10 -10 V
10 -9 V
9 -10 V
9 -8 V
7 -8 V
6 -7 V
5 -5 V
4 -4 V
2 -3 V
1 -1 V
1 -2 V
4 -4 V
5 -6 V
6 -7 V
8 -9 V
8 -11 V
9 -11 V
10 -13 V
10 -14 V
11 -14 V
10 -14 V
10 -14 V
9 -13 V
8 -13 V
8 -11 V
6 -9 V
5 -8 V
3 -5 V
2 -3 V
1 -2 V
2 -3 V
4 -5 V
5 -8 V
6 -10 V
7 -12 V
9 -14 V
9 -15 V
10 -17 V
10 -17 V
10 -18 V
11 -17 V
9 -18 V
10 -16 V
8 -15 V
7 -14 V
7 -11 V
5 -9 V
3 -7 V
2 -3 V
1 -2 V
2 -4 V
3 -6 V
5 -10 V
7 -11 V
7 -14 V
9 -16 V
stroke 1381 2496 M
9 -18 V
10 -19 V
10 -20 V
10 -20 V
10 -20 V
10 -19 V
9 -19 V
9 -17 V
7 -15 V
6 -12 V
5 -10 V
4 -7 V
2 -4 V
1 -2 V
2 -4 V
3 -7 V
5 -10 V
6 -13 V
8 -15 V
8 -18 V
10 -19 V
9 -20 V
11 -21 V
10 -21 V
10 -22 V
10 -20 V
9 -19 V
9 -18 V
7 -15 V
6 -14 V
5 -10 V
4 -7 V
2 -4 V
1 -2 V
2 -4 V
3 -8 V
5 -10 V
6 -13 V
8 -16 V
8 -17 V
9 -20 V
10 -20 V
10 -21 V
11 -22 V
10 -21 V
10 -21 V
9 -19 V
8 -17 V
8 -15 V
6 -13 V
5 -10 V
3 -8 V
2 -4 V
1 -2 V
2 -4 V
4 -7 V
5 -10 V
6 -12 V
7 -16 V
9 -17 V
9 -18 V
10 -20 V
10 -21 V
10 -20 V
11 -20 V
9 -20 V
10 -18 V
8 -16 V
7 -15 V
7 -12 V
5 -9 V
3 -7 V
2 -4 V
1 -1 V
2 -4 V
4 -7 V
4 -9 V
7 -12 V
7 -14 V
9 -16 V
9 -17 V
10 -18 V
10 -19 V
10 -18 V
10 -19 V
10 -17 V
9 -16 V
9 -15 V
7 -13 V
7 -11 V
4 -8 V
4 -6 V
2 -3 V
1 -2 V
2 -3 V
3 -6 V
5 -8 V
6 -11 V
8 -12 V
8 -14 V
10 -15 V
9 -15 V
11 -16 V
10 -16 V
stroke 2094 1107 M
10 -16 V
10 -15 V
9 -13 V
9 -13 V
7 -10 V
6 -9 V
5 -7 V
4 -5 V
2 -3 V
1 -1 V
2 -3 V
3 -5 V
5 -7 V
6 -8 V
8 -10 V
8 -11 V
9 -12 V
10 -13 V
10 -13 V
11 -13 V
10 -12 V
10 -11 V
9 -11 V
8 -10 V
8 -8 V
6 -7 V
5 -5 V
3 -4 V
2 -2 V
1 -1 V
2 -2 V
4 -4 V
5 -5 V
6 -6 V
7 -7 V
9 -9 V
9 -9 V
10 -9 V
10 -9 V
10 -9 V
11 -8 V
9 -9 V
10 -7 V
8 -6 V
8 -6 V
6 -4 V
5 -4 V
3 -2 V
2 -2 V
1 0 V
2 -2 V
4 -2 V
4 -3 V
7 -4 V
7 -5 V
9 -5 V
9 -6 V
10 -5 V
10 -6 V
10 -5 V
10 -5 V
10 -4 V
9 -4 V
9 -4 V
7 -3 V
7 -2 V
4 -2 V
4 -1 V
2 0 V
1 -1 V
2 0 V
3 -2 V
5 -1 V
6 -2 V
8 -2 V
8 -2 V
10 -2 V
9 -3 V
11 -2 V
10 -1 V
10 -2 V
10 -1 V
9 -1 V
9 -1 V
7 0 V
6 -1 V
5 0 V
4 0 V
2 0 V
stroke
LT1
LCb setrgbcolor
LT1
1208 1298 M
543 0 V
1299 640 M
1 0 V
2 0 V
2 0 V
3 0 V
4 0 V
5 1 V
4 0 V
5 0 V
6 1 V
5 0 V
5 1 V
5 1 V
5 0 V
5 1 V
3 1 V
4 0 V
2 1 V
2 0 V
1 0 V
1 0 V
1 0 V
1 1 V
3 0 V
3 1 V
4 1 V
4 1 V
5 1 V
5 1 V
6 1 V
5 2 V
5 1 V
5 1 V
5 1 V
4 2 V
4 1 V
4 0 V
2 1 V
2 1 V
1 0 V
1 0 V
2 1 V
3 1 V
3 0 V
4 2 V
4 1 V
5 1 V
5 2 V
5 1 V
6 2 V
5 1 V
5 2 V
5 1 V
4 2 V
4 1 V
3 1 V
3 1 V
2 0 V
1 0 V
0 1 V
1 0 V
2 0 V
3 1 V
3 1 V
4 1 V
4 2 V
5 1 V
5 2 V
5 1 V
6 2 V
5 1 V
5 2 V
5 1 V
4 1 V
4 1 V
3 1 V
3 1 V
2 0 V
1 1 V
1 0 V
2 1 V
2 0 V
4 1 V
4 1 V
4 1 V
5 2 V
5 1 V
5 1 V
5 2 V
6 1 V
5 1 V
5 2 V
4 1 V
4 0 V
3 1 V
3 1 V
1 0 V
1 0 V
1 1 V
1 0 V
2 0 V
2 1 V
3 0 V
stroke 1656 724 M
4 1 V
5 1 V
4 1 V
6 1 V
5 1 V
5 1 V
6 1 V
5 1 V
4 1 V
5 1 V
4 1 V
3 0 V
2 1 V
2 0 V
1 0 V
1 0 V
1 0 V
2 1 V
2 0 V
3 0 V
4 1 V
5 1 V
4 0 V
5 1 V
6 1 V
5 0 V
5 1 V
5 1 V
5 0 V
5 1 V
3 0 V
4 0 V
2 0 V
2 1 V
1 0 V
1 0 V
1 0 V
1 0 V
3 0 V
3 0 V
4 1 V
4 0 V
5 0 V
5 1 V
6 0 V
5 0 V
5 1 V
5 0 V
5 0 V
4 0 V
4 0 V
4 0 V
2 0 V
2 0 V
1 0 V
1 0 V
2 0 V
3 0 V
3 0 V
4 1 V
4 0 V
5 0 V
5 -1 V
5 0 V
6 0 V
5 0 V
5 0 V
5 0 V
4 0 V
4 0 V
3 -1 V
3 0 V
2 0 V
1 0 V
1 0 V
2 0 V
3 0 V
3 0 V
4 0 V
4 -1 V
5 0 V
5 0 V
5 -1 V
6 0 V
5 -1 V
5 0 V
5 -1 V
4 0 V
4 0 V
3 -1 V
3 0 V
2 0 V
1 0 V
1 0 V
2 -1 V
2 0 V
4 0 V
4 -1 V
4 0 V
5 -1 V
5 0 V
5 -1 V
5 -1 V
6 0 V
stroke 2040 737 M
5 -1 V
5 -1 V
4 0 V
4 -1 V
3 0 V
3 -1 V
1 0 V
1 0 V
1 0 V
1 0 V
2 -1 V
2 0 V
3 -1 V
4 0 V
5 -1 V
4 -1 V
6 0 V
5 -1 V
5 -1 V
6 -1 V
5 -1 V
4 -1 V
5 -1 V
4 0 V
3 -1 V
2 0 V
2 -1 V
1 0 V
1 0 V
1 0 V
2 0 V
2 -1 V
3 -1 V
4 0 V
5 -1 V
4 -1 V
5 -1 V
6 -1 V
5 -1 V
5 -1 V
5 -1 V
5 -1 V
5 -1 V
3 -1 V
4 0 V
2 -1 V
2 0 V
1 0 V
1 0 V
1 -1 V
1 0 V
3 0 V
3 -1 V
4 -1 V
4 -1 V
5 -1 V
5 -1 V
6 -1 V
5 -1 V
5 -1 V
5 -1 V
5 -1 V
4 -1 V
4 -1 V
4 0 V
2 -1 V
2 0 V
1 0 V
1 -1 V
2 0 V
3 -1 V
3 0 V
4 -1 V
4 -1 V
5 -1 V
5 -1 V
5 -1 V
6 -1 V
5 -1 V
5 -1 V
5 -1 V
4 -1 V
4 -1 V
3 0 V
3 -1 V
2 0 V
1 0 V
1 -1 V
2 0 V
3 0 V
3 -1 V
4 -1 V
4 -1 V
5 0 V
5 -1 V
5 -1 V
6 -1 V
5 -1 V
5 -1 V
5 -1 V
4 -1 V
4 -1 V
3 0 V
3 -1 V
stroke 2413 667 M
2 0 V
1 0 V
1 0 V
2 -1 V
2 0 V
4 -1 V
4 0 V
4 -1 V
5 -1 V
5 -1 V
5 0 V
5 -1 V
6 -1 V
5 -1 V
5 -1 V
4 0 V
4 -1 V
3 0 V
3 -1 V
1 0 V
1 0 V
1 0 V
1 0 V
2 0 V
2 -1 V
3 0 V
4 -1 V
5 0 V
4 -1 V
6 -1 V
5 0 V
5 -1 V
5 -1 V
6 0 V
4 -1 V
5 0 V
4 0 V
3 -1 V
2 0 V
2 0 V
1 0 V
1 0 V
1 -1 V
2 0 V
2 0 V
3 0 V
4 -1 V
5 0 V
4 0 V
5 -1 V
6 0 V
5 -1 V
5 0 V
5 0 V
5 -1 V
5 0 V
3 0 V
4 0 V
2 -1 V
2 0 V
1 0 V
1 0 V
1 0 V
1 0 V
3 0 V
3 0 V
4 0 V
4 -1 V
5 0 V
5 0 V
6 0 V
5 0 V
5 -1 V
5 0 V
5 0 V
4 0 V
4 0 V
4 0 V
2 0 V
2 0 V
1 0 V
stroke
LTb
0 3431 N
0 640 L
2879 0 V
0 2791 V
0 3431 L
Z stroke
1.000 UP
1.000 UL
LTb
1.000 UL
LTb
4320 640 M
63 0 V
2816 0 R
-63 0 V
4320 950 M
63 0 V
2816 0 R
-63 0 V
4320 1260 M
63 0 V
2816 0 R
-63 0 V
4320 1570 M
63 0 V
2816 0 R
-63 0 V
4320 1880 M
63 0 V
2816 0 R
-63 0 V
4320 2191 M
63 0 V
2816 0 R
-63 0 V
4320 2501 M
63 0 V
2816 0 R
-63 0 V
4320 2811 M
63 0 V
2816 0 R
-63 0 V
4320 3121 M
63 0 V
2816 0 R
-63 0 V
4320 3431 M
63 0 V
2816 0 R
-63 0 V
4320 640 M
0 63 V
0 2728 R
0 -63 V
4680 640 M
0 63 V
0 2728 R
0 -63 V
5040 640 M
0 63 V
0 2728 R
0 -63 V
5400 640 M
0 63 V
0 2728 R
0 -63 V
5760 640 M
0 63 V
0 2728 R
0 -63 V
6119 640 M
0 63 V
0 2728 R
0 -63 V
6479 640 M
0 63 V
0 2728 R
0 -63 V
6839 640 M
0 63 V
0 2728 R
0 -63 V
7199 640 M
0 63 V
0 2728 R
0 -63 V
stroke
LTa
4320 640 M
2879 0 V
stroke
LTb
4320 3431 N
0 -2791 V
2879 0 V
0 2791 V
-2879 0 V
Z stroke
LCb setrgbcolor
LTb
LCb setrgbcolor
LTb
1.000 UP
1.000 UL
LTb
1.000 UL
LT0
LCb setrgbcolor
LT0
5528 1917 M
543 0 V
4320 3232 M
3 2 V
3 2 V
5 4 V
7 5 V
8 5 V
8 7 V
10 6 V
10 7 V
11 7 V
11 7 V
10 6 V
11 7 V
9 5 V
9 5 V
8 4 V
6 4 V
5 3 V
4 2 V
2 1 V
1 0 V
2 1 V
4 2 V
5 3 V
6 3 V
8 4 V
9 4 V
10 4 V
10 5 V
10 4 V
11 4 V
11 5 V
10 3 V
10 4 V
9 3 V
7 2 V
7 2 V
5 2 V
4 1 V
2 1 V
1 0 V
2 0 V
3 1 V
5 2 V
7 2 V
8 1 V
8 3 V
10 2 V
10 2 V
11 2 V
11 2 V
10 1 V
11 2 V
9 1 V
9 1 V
8 1 V
6 0 V
5 1 V
4 0 V
2 0 V
1 0 V
2 0 V
4 0 V
5 1 V
6 0 V
8 0 V
9 0 V
10 0 V
10 0 V
11 -1 V
10 0 V
11 -1 V
10 -1 V
10 -1 V
9 -1 V
7 -1 V
7 -1 V
5 -1 V
4 0 V
2 -1 V
1 0 V
2 0 V
3 -1 V
6 -1 V
6 -1 V
8 -1 V
9 -2 V
9 -3 V
10 -2 V
11 -3 V
11 -3 V
11 -3 V
10 -3 V
9 -3 V
9 -3 V
8 -3 V
6 -2 V
6 -2 V
3 -2 V
2 -1 V
1 0 V
2 -1 V
4 -1 V
stroke 5029 3353 M
5 -2 V
7 -3 V
7 -3 V
9 -4 V
10 -4 V
10 -5 V
11 -5 V
10 -5 V
11 -6 V
10 -5 V
10 -6 V
9 -4 V
8 -5 V
6 -4 V
5 -3 V
4 -2 V
2 -1 V
1 -1 V
2 -1 V
3 -2 V
6 -3 V
6 -4 V
8 -5 V
9 -6 V
9 -6 V
11 -7 V
10 -7 V
11 -8 V
11 -8 V
10 -7 V
10 -7 V
8 -7 V
8 -6 V
7 -5 V
5 -4 V
3 -3 V
2 -2 V
1 -1 V
2 -1 V
4 -3 V
5 -5 V
7 -5 V
7 -6 V
9 -8 V
10 -8 V
10 -9 V
11 -10 V
11 -9 V
10 -10 V
10 -10 V
10 -9 V
9 -8 V
8 -8 V
6 -6 V
5 -5 V
4 -4 V
2 -2 V
1 -1 V
2 -2 V
4 -3 V
5 -6 V
6 -6 V
8 -8 V
9 -9 V
9 -10 V
11 -11 V
10 -12 V
11 -11 V
11 -12 V
10 -12 V
10 -10 V
8 -10 V
8 -9 V
7 -8 V
5 -6 V
3 -4 V
3 -2 V
0 -2 V
3 -2 V
3 -4 V
5 -6 V
7 -8 V
8 -9 V
8 -11 V
10 -12 V
10 -13 V
11 -13 V
11 -13 V
10 -14 V
11 -13 V
9 -12 V
9 -12 V
8 -10 V
6 -9 V
5 -6 V
4 -5 V
2 -3 V
1 -1 V
2 -3 V
4 -5 V
5 -7 V
6 -9 V
8 -10 V
9 -12 V
stroke 5760 2684 M
10 -13 V
10 -15 V
10 -14 V
11 -16 V
11 -15 V
10 -14 V
10 -14 V
9 -13 V
7 -11 V
7 -10 V
5 -7 V
4 -6 V
2 -3 V
1 -1 V
2 -3 V
3 -5 V
5 -8 V
7 -10 V
8 -11 V
8 -13 V
10 -15 V
10 -16 V
11 -16 V
11 -16 V
10 -17 V
11 -16 V
9 -15 V
9 -13 V
8 -13 V
6 -10 V
5 -8 V
4 -6 V
2 -3 V
1 -2 V
2 -3 V
4 -6 V
5 -8 V
6 -10 V
8 -13 V
9 -14 V
10 -15 V
10 -17 V
11 -17 V
10 -18 V
11 -17 V
10 -17 V
10 -16 V
9 -15 V
7 -13 V
7 -11 V
5 -8 V
4 -6 V
2 -4 V
1 -1 V
2 -4 V
3 -6 V
6 -8 V
6 -11 V
8 -13 V
9 -15 V
9 -17 V
10 -17 V
11 -18 V
11 -18 V
10 -19 V
11 -17 V
9 -17 V
9 -15 V
8 -13 V
6 -11 V
6 -9 V
3 -6 V
2 -4 V
1 -2 V
2 -3 V
4 -6 V
5 -9 V
7 -12 V
7 -13 V
9 -15 V
10 -17 V
10 -18 V
11 -18 V
10 -19 V
11 -18 V
10 -18 V
10 -17 V
9 -15 V
7 -14 V
7 -11 V
5 -9 V
4 -6 V
2 -4 V
1 -1 V
2 -4 V
3 -6 V
6 -9 V
6 -12 V
8 -13 V
9 -15 V
9 -17 V
11 -18 V
10 -18 V
11 -19 V
stroke 6504 1473 M
11 -18 V
10 -17 V
9 -17 V
9 -15 V
8 -13 V
7 -11 V
5 -9 V
3 -6 V
2 -4 V
1 -1 V
2 -4 V
4 -6 V
5 -9 V
7 -11 V
7 -13 V
9 -15 V
10 -16 V
10 -17 V
11 -18 V
11 -17 V
10 -18 V
10 -17 V
10 -16 V
9 -14 V
8 -12 V
6 -11 V
5 -8 V
4 -6 V
2 -3 V
1 -2 V
2 -3 V
4 -6 V
5 -8 V
6 -10 V
8 -12 V
9 -14 V
9 -15 V
11 -16 V
10 -16 V
11 -16 V
11 -16 V
10 -16 V
10 -14 V
8 -12 V
8 -12 V
7 -9 V
5 -7 V
3 -5 V
2 -3 V
1 -2 V
3 -3 V
3 -5 V
5 -7 V
7 -9 V
8 -10 V
8 -12 V
10 -13 V
10 -13 V
11 -14 V
11 -14 V
10 -13 V
11 -12 V
9 -11 V
9 -11 V
8 -8 V
6 -8 V
5 -5 V
4 -4 V
2 -2 V
1 -1 V
2 -2 V
4 -4 V
5 -6 V
6 -6 V
8 -8 V
9 -8 V
10 -9 V
10 -9 V
10 -9 V
11 -8 V
11 -8 V
10 -7 V
10 -5 V
9 -5 V
7 -3 V
7 -3 V
5 -1 V
4 -1 V
2 0 V
stroke
LT1
LCb setrgbcolor
LT1
5528 1717 M
543 0 V
5456 640 M
2 0 V
2 0 V
3 0 V
4 0 V
4 1 V
6 1 V
5 0 V
6 2 V
7 1 V
6 2 V
7 2 V
6 2 V
5 2 V
6 2 V
4 2 V
4 1 V
3 2 V
2 1 V
2 0 V
1 1 V
3 1 V
3 1 V
4 2 V
4 2 V
5 3 V
6 3 V
6 3 V
7 4 V
6 4 V
6 3 V
6 4 V
6 3 V
5 3 V
5 3 V
4 3 V
3 2 V
2 1 V
1 1 V
1 0 V
1 1 V
2 1 V
3 2 V
4 3 V
5 3 V
5 3 V
6 4 V
6 4 V
6 4 V
7 4 V
6 4 V
6 4 V
6 4 V
5 4 V
5 3 V
4 2 V
3 2 V
2 2 V
1 1 V
1 0 V
1 1 V
2 1 V
3 2 V
4 3 V
5 3 V
5 3 V
6 4 V
6 4 V
6 4 V
6 4 V
7 4 V
6 4 V
6 4 V
5 3 V
5 3 V
3 2 V
3 2 V
3 1 V
1 1 V
0 1 V
2 0 V
2 2 V
3 1 V
4 3 V
4 2 V
6 4 V
5 3 V
6 3 V
7 4 V
6 4 V
7 3 V
6 4 V
5 3 V
6 2 V
4 3 V
4 2 V
3 1 V
2 2 V
2 0 V
1 1 V
3 1 V
3 2 V
4 2 V
stroke 5886 869 M
4 2 V
5 2 V
6 3 V
6 3 V
7 3 V
6 3 V
6 3 V
6 2 V
6 3 V
5 2 V
5 2 V
4 1 V
3 2 V
2 0 V
1 1 V
1 0 V
1 1 V
2 0 V
3 2 V
4 1 V
5 2 V
5 2 V
6 2 V
6 2 V
6 2 V
7 2 V
6 2 V
6 2 V
6 2 V
5 1 V
5 1 V
4 1 V
3 1 V
2 1 V
1 0 V
1 0 V
1 1 V
2 0 V
3 1 V
4 1 V
5 1 V
5 1 V
6 1 V
6 2 V
6 1 V
6 1 V
7 1 V
6 2 V
6 1 V
5 0 V
5 1 V
3 1 V
3 0 V
3 0 V
1 0 V
0 1 V
2 0 V
2 0 V
3 0 V
4 1 V
4 0 V
6 1 V
5 0 V
6 1 V
7 0 V
6 1 V
7 0 V
6 1 V
5 0 V
6 0 V
4 0 V
4 0 V
3 0 V
2 0 V
2 0 V
1 0 V
3 0 V
3 0 V
3 0 V
5 0 V
5 0 V
6 0 V
6 0 V
7 0 V
6 -1 V
6 0 V
6 -1 V
6 0 V
5 0 V
5 -1 V
4 0 V
3 0 V
2 -1 V
1 0 V
1 0 V
1 0 V
2 0 V
3 -1 V
4 0 V
5 -1 V
5 0 V
6 -1 V
6 -1 V
6 -1 V
stroke 6332 941 M
7 -1 V
6 -1 V
6 -1 V
6 -1 V
5 -1 V
5 -1 V
4 -1 V
3 0 V
2 -1 V
1 0 V
1 0 V
1 0 V
2 -1 V
3 0 V
4 -1 V
5 -1 V
5 -2 V
6 -1 V
6 -1 V
6 -2 V
6 -2 V
7 -1 V
6 -2 V
6 -1 V
5 -2 V
4 -1 V
4 -1 V
3 -1 V
3 -1 V
1 0 V
2 -1 V
2 0 V
3 -1 V
4 -1 V
4 -2 V
6 -2 V
5 -1 V
6 -2 V
7 -2 V
6 -3 V
7 -2 V
6 -2 V
5 -2 V
6 -2 V
4 -1 V
4 -2 V
3 -1 V
2 -1 V
2 0 V
1 -1 V
3 -1 V
3 -1 V
3 -1 V
5 -2 V
5 -2 V
6 -2 V
6 -3 V
7 -2 V
6 -3 V
6 -2 V
6 -3 V
6 -2 V
5 -2 V
5 -2 V
4 -2 V
3 -1 V
2 -1 V
1 -1 V
1 0 V
1 -1 V
2 0 V
3 -2 V
4 -2 V
5 -2 V
5 -2 V
6 -2 V
6 -3 V
6 -3 V
7 -3 V
6 -3 V
6 -2 V
6 -3 V
5 -3 V
5 -2 V
4 -1 V
3 -2 V
2 -1 V
1 0 V
1 -1 V
1 0 V
2 -1 V
3 -2 V
4 -2 V
5 -2 V
5 -2 V
6 -3 V
6 -3 V
6 -3 V
6 -3 V
7 -3 V
6 -3 V
6 -3 V
5 -2 V
4 -3 V
stroke 6785 777 M
4 -2 V
3 -1 V
3 -1 V
1 -1 V
2 -1 V
2 -1 V
3 -1 V
4 -2 V
4 -2 V
6 -3 V
5 -3 V
6 -3 V
7 -3 V
6 -3 V
7 -3 V
6 -3 V
5 -3 V
6 -2 V
4 -2 V
4 -2 V
3 -2 V
2 -1 V
2 0 V
0 -1 V
1 0 V
3 -1 V
3 -2 V
3 -1 V
5 -3 V
5 -2 V
6 -3 V
6 -3 V
7 -3 V
6 -3 V
6 -3 V
6 -2 V
6 -3 V
5 -2 V
5 -2 V
4 -2 V
3 -1 V
2 -1 V
1 -1 V
1 0 V
1 -1 V
2 -1 V
3 -1 V
4 -2 V
5 -2 V
5 -2 V
6 -2 V
6 -3 V
6 -2 V
7 -3 V
6 -2 V
6 -3 V
6 -2 V
5 -2 V
5 -2 V
4 -1 V
3 -1 V
2 -1 V
1 0 V
1 0 V
1 -1 V
2 0 V
3 -1 V
4 -2 V
5 -1 V
5 -2 V
6 -2 V
6 -1 V
6 -2 V
6 -2 V
7 -1 V
6 -1 V
6 -2 V
5 0 V
4 -1 V
4 -1 V
3 0 V
3 0 V
1 0 V
stroke
LTb
4320 3431 N
0 -2791 V
2879 0 V
0 2791 V
-2879 0 V
Z stroke
1.000 UP
1.000 UL
LTb
stroke
grestore
end
showpage
  }}%
  \put(5408,1717){\makebox(0,0)[r]{\strut{}$l=\tau $}}%
  \put(5408,1917){\makebox(0,0)[r]{\strut{}$l=e$}}%
  \put(5759,3731){\makebox(0,0){\strut{}$\bar B_s\to D_{s2}^*(2573) l\nu_l$}}%
  \put(5759,140){\makebox(0,0){\strut{}$q^2$ [$\rm GeV^2$]}}%
  \put(3380,2035){%
  \special{ps: gsave currentpoint currentpoint translate
630 rotate neg exch neg exch translate}%
  \makebox(0,0){\strut{}$d\Gamma/dq^2$ [$|V_{cb}|^2\ 10^{-13} \rm GeV^{-1}$]}%
  \special{ps: currentpoint grestore moveto}%
  }%
  \put(7199,440){\makebox(0,0){\strut{} 8}}%
  \put(6839,440){\makebox(0,0){\strut{} 7}}%
  \put(6479,440){\makebox(0,0){\strut{} 6}}%
  \put(6119,440){\makebox(0,0){\strut{} 5}}%
  \put(5760,440){\makebox(0,0){\strut{} 4}}%
  \put(5400,440){\makebox(0,0){\strut{} 3}}%
  \put(5040,440){\makebox(0,0){\strut{} 2}}%
  \put(4680,440){\makebox(0,0){\strut{} 1}}%
  \put(4320,440){\makebox(0,0){\strut{} 0}}%
  \put(4200,3431){\makebox(0,0)[r]{\strut{} 0.09}}%
  \put(4200,3121){\makebox(0,0)[r]{\strut{} 0.08}}%
  \put(4200,2811){\makebox(0,0)[r]{\strut{} 0.07}}%
  \put(4200,2501){\makebox(0,0)[r]{\strut{} 0.06}}%
  \put(4200,2191){\makebox(0,0)[r]{\strut{} 0.05}}%
  \put(4200,1880){\makebox(0,0)[r]{\strut{} 0.04}}%
  \put(4200,1570){\makebox(0,0)[r]{\strut{} 0.03}}%
  \put(4200,1260){\makebox(0,0)[r]{\strut{} 0.02}}%
  \put(4200,950){\makebox(0,0)[r]{\strut{} 0.01}}%
  \put(4200,640){\makebox(0,0)[r]{\strut{} 0}}%
  \put(1088,1298){\makebox(0,0)[r]{\strut{}$l= \tau $}}%
  \put(1088,1498){\makebox(0,0)[r]{\strut{}$l=e$}}%
  \put(1439,3731){\makebox(0,0){\strut{}$\bar B_s\to c\bar s (2^-) l\nu_l$}}%
  \put(1439,140){\makebox(0,0){\strut{}$q^2$ [$\rm GeV^2$]}}%
  \put(-1180,2035){%
  \special{ps: gsave currentpoint currentpoint translate
630 rotate neg exch neg exch translate}%
  \makebox(0,0){\strut{}$d\Gamma/dq^2$ [$|V_{cb}|^2\ 10^{-13} \rm GeV^{-1}$]}%
  \special{ps: currentpoint grestore moveto}%
  }%
  \put(2879,440){\makebox(0,0){\strut{} 7}}%
  \put(2468,440){\makebox(0,0){\strut{} 6}}%
  \put(2056,440){\makebox(0,0){\strut{} 5}}%
  \put(1645,440){\makebox(0,0){\strut{} 4}}%
  \put(1234,440){\makebox(0,0){\strut{} 3}}%
  \put(823,440){\makebox(0,0){\strut{} 2}}%
  \put(411,440){\makebox(0,0){\strut{} 1}}%
  \put(0,440){\makebox(0,0){\strut{} 0}}%
  \put(-120,3431){\makebox(0,0)[r]{\strut{} 0.003}}%
  \put(-120,2966){\makebox(0,0)[r]{\strut{} 0.0025}}%
  \put(-120,2501){\makebox(0,0)[r]{\strut{} 0.002}}%
  \put(-120,2036){\makebox(0,0)[r]{\strut{} 0.0015}}%
  \put(-120,1570){\makebox(0,0)[r]{\strut{} 0.001}}%
  \put(-120,1105){\makebox(0,0)[r]{\strut{} 0.0005}}%
  \put(-120,640){\makebox(0,0)[r]{\strut{} 0}}%
\end{picture}%
\endgroup
 

%% file: hqs1.tex
\begingroup%
\makeatletter%
\newcommand{\GNUPLOTspecial}{%
  \@sanitize\catcode`\%=14\relax\special}%
\setlength{\unitlength}{0.0500bp}%
\begin{picture}(7200,5040)(0,0)%
  {\GNUPLOTspecial{"
/gnudict 256 dict def
gnudict begin
%
%
/Color false def
/Blacktext true def
/Solid false def
/Dashlength 1 def
/Landscape false def
/Level1 false def
/Rounded false def
/ClipToBoundingBox false def
/SuppressPDFMark false def
/TransparentPatterns false def
/gnulinewidth 5.000 def
/userlinewidth gnulinewidth def
/Gamma 1.0 def
/BackgroundColor {-1.000 -1.000 -1.000} def
/vshift -66 def
/dl1 {
  10.0 Dashlength mul mul
  Rounded { currentlinewidth 0.75 mul sub dup 0 le { pop 0.01 } if } if
} def
/dl2 {
  10.0 Dashlength mul mul
  Rounded { currentlinewidth 0.75 mul add } if
} def
/hpt_ 31.5 def
/vpt_ 31.5 def
/hpt hpt_ def
/vpt vpt_ def
/doclip {
  ClipToBoundingBox {
    newpath 0 0 moveto 360 0 lineto 360 252 lineto 0 252 lineto closepath
    clip
  } if
} def
%
%
%
/M {moveto} bind def
/L {lineto} bind def
/R {rmoveto} bind def
/V {rlineto} bind def
/N {newpath moveto} bind def
/Z {closepath} bind def
/C {setrgbcolor} bind def
/f {rlineto fill} bind def
/g {setgray} bind def
/Gshow {show} def   
/vpt2 vpt 2 mul def
/hpt2 hpt 2 mul def
/Lshow {currentpoint stroke M 0 vshift R 
	Blacktext {gsave 0 setgray show grestore} {show} ifelse} def
/Rshow {currentpoint stroke M dup stringwidth pop neg vshift R
	Blacktext {gsave 0 setgray show grestore} {show} ifelse} def
/Cshow {currentpoint stroke M dup stringwidth pop -2 div vshift R 
	Blacktext {gsave 0 setgray show grestore} {show} ifelse} def
/UP {dup vpt_ mul /vpt exch def hpt_ mul /hpt exch def
  /hpt2 hpt 2 mul def /vpt2 vpt 2 mul def} def
/DL {Color {setrgbcolor Solid {pop []} if 0 setdash}
 {pop pop pop 0 setgray Solid {pop []} if 0 setdash} ifelse} def
/BL {stroke userlinewidth 2 mul setlinewidth
	Rounded {1 setlinejoin 1 setlinecap} if} def
/AL {stroke userlinewidth 2 div setlinewidth
	Rounded {1 setlinejoin 1 setlinecap} if} def
/UL {dup gnulinewidth mul /userlinewidth exch def
	dup 1 lt {pop 1} if 10 mul /udl exch def} def
/PL {stroke userlinewidth setlinewidth
	Rounded {1 setlinejoin 1 setlinecap} if} def
3.8 setmiterlimit
/LCw {1 1 1} def
/LCb {0 0 0} def
/LCa {0 0 0} def
/LC0 {1 0 0} def
/LC1 {0 1 0} def
/LC2 {0 0 1} def
/LC3 {1 0 1} def
/LC4 {0 1 1} def
/LC5 {1 1 0} def
/LC6 {0 0 0} def
/LC7 {1 0.3 0} def
/LC8 {0.5 0.5 0.5} def
/LTw {PL [] 1 setgray} def
/LTb {BL [] LCb DL} def
/LTa {AL [1 udl mul 2 udl mul] 0 setdash LCa setrgbcolor} def
/LT0 {PL [] LC0 DL} def
/LT1 {PL [4 dl1 2 dl2] LC1 DL} def
/LT2 {PL [2 dl1 3 dl2] LC2 DL} def
/LT3 {PL [1 dl1 1.5 dl2] LC3 DL} def
/LT4 {PL [6 dl1 2 dl2 1 dl1 2 dl2] LC4 DL} def
/LT5 {PL [3 dl1 3 dl2 1 dl1 3 dl2] LC5 DL} def
/LT6 {PL [2 dl1 2 dl2 2 dl1 6 dl2] LC6 DL} def
/LT7 {PL [1 dl1 2 dl2 6 dl1 2 dl2 1 dl1 2 dl2] LC7 DL} def
/LT8 {PL [2 dl1 2 dl2 2 dl1 2 dl2 2 dl1 2 dl2 2 dl1 4 dl2] LC8 DL} def
/Pnt {stroke [] 0 setdash gsave 1 setlinecap M 0 0 V stroke grestore} def
/Dia {stroke [] 0 setdash 2 copy vpt add M
  hpt neg vpt neg V hpt vpt neg V
  hpt vpt V hpt neg vpt V closepath stroke
  Pnt} def
/Pls {stroke [] 0 setdash vpt sub M 0 vpt2 V
  currentpoint stroke M
  hpt neg vpt neg R hpt2 0 V stroke
 } def
/Box {stroke [] 0 setdash 2 copy exch hpt sub exch vpt add M
  0 vpt2 neg V hpt2 0 V 0 vpt2 V
  hpt2 neg 0 V closepath stroke
  Pnt} def
/Crs {stroke [] 0 setdash exch hpt sub exch vpt add M
  hpt2 vpt2 neg V currentpoint stroke M
  hpt2 neg 0 R hpt2 vpt2 V stroke} def
/TriU {stroke [] 0 setdash 2 copy vpt 1.12 mul add M
  hpt neg vpt -1.62 mul V
  hpt 2 mul 0 V
  hpt neg vpt 1.62 mul V closepath stroke
  Pnt} def
/Star {2 copy Pls Crs} def
/BoxF {stroke [] 0 setdash exch hpt sub exch vpt add M
  0 vpt2 neg V hpt2 0 V 0 vpt2 V
  hpt2 neg 0 V closepath fill} def
/TriUF {stroke [] 0 setdash vpt 1.12 mul add M
  hpt neg vpt -1.62 mul V
  hpt 2 mul 0 V
  hpt neg vpt 1.62 mul V closepath fill} def
/TriD {stroke [] 0 setdash 2 copy vpt 1.12 mul sub M
  hpt neg vpt 1.62 mul V
  hpt 2 mul 0 V
  hpt neg vpt -1.62 mul V closepath stroke
  Pnt} def
/TriDF {stroke [] 0 setdash vpt 1.12 mul sub M
  hpt neg vpt 1.62 mul V
  hpt 2 mul 0 V
  hpt neg vpt -1.62 mul V closepath fill} def
/DiaF {stroke [] 0 setdash vpt add M
  hpt neg vpt neg V hpt vpt neg V
  hpt vpt V hpt neg vpt V closepath fill} def
/Pent {stroke [] 0 setdash 2 copy gsave
  translate 0 hpt M 4 {72 rotate 0 hpt L} repeat
  closepath stroke grestore Pnt} def
/PentF {stroke [] 0 setdash gsave
  translate 0 hpt M 4 {72 rotate 0 hpt L} repeat
  closepath fill grestore} def
/Circle {stroke [] 0 setdash 2 copy
  hpt 0 360 arc stroke Pnt} def
/CircleF {stroke [] 0 setdash hpt 0 360 arc fill} def
/C0 {BL [] 0 setdash 2 copy moveto vpt 90 450 arc} bind def
/C1 {BL [] 0 setdash 2 copy moveto
	2 copy vpt 0 90 arc closepath fill
	vpt 0 360 arc closepath} bind def
/C2 {BL [] 0 setdash 2 copy moveto
	2 copy vpt 90 180 arc closepath fill
	vpt 0 360 arc closepath} bind def
/C3 {BL [] 0 setdash 2 copy moveto
	2 copy vpt 0 180 arc closepath fill
	vpt 0 360 arc closepath} bind def
/C4 {BL [] 0 setdash 2 copy moveto
	2 copy vpt 180 270 arc closepath fill
	vpt 0 360 arc closepath} bind def
/C5 {BL [] 0 setdash 2 copy moveto
	2 copy vpt 0 90 arc
	2 copy moveto
	2 copy vpt 180 270 arc closepath fill
	vpt 0 360 arc} bind def
/C6 {BL [] 0 setdash 2 copy moveto
	2 copy vpt 90 270 arc closepath fill
	vpt 0 360 arc closepath} bind def
/C7 {BL [] 0 setdash 2 copy moveto
	2 copy vpt 0 270 arc closepath fill
	vpt 0 360 arc closepath} bind def
/C8 {BL [] 0 setdash 2 copy moveto
	2 copy vpt 270 360 arc closepath fill
	vpt 0 360 arc closepath} bind def
/C9 {BL [] 0 setdash 2 copy moveto
	2 copy vpt 270 450 arc closepath fill
	vpt 0 360 arc closepath} bind def
/C10 {BL [] 0 setdash 2 copy 2 copy moveto vpt 270 360 arc closepath fill
	2 copy moveto
	2 copy vpt 90 180 arc closepath fill
	vpt 0 360 arc closepath} bind def
/C11 {BL [] 0 setdash 2 copy moveto
	2 copy vpt 0 180 arc closepath fill
	2 copy moveto
	2 copy vpt 270 360 arc closepath fill
	vpt 0 360 arc closepath} bind def
/C12 {BL [] 0 setdash 2 copy moveto
	2 copy vpt 180 360 arc closepath fill
	vpt 0 360 arc closepath} bind def
/C13 {BL [] 0 setdash 2 copy moveto
	2 copy vpt 0 90 arc closepath fill
	2 copy moveto
	2 copy vpt 180 360 arc closepath fill
	vpt 0 360 arc closepath} bind def
/C14 {BL [] 0 setdash 2 copy moveto
	2 copy vpt 90 360 arc closepath fill
	vpt 0 360 arc} bind def
/C15 {BL [] 0 setdash 2 copy vpt 0 360 arc closepath fill
	vpt 0 360 arc closepath} bind def
/Rec {newpath 4 2 roll moveto 1 index 0 rlineto 0 exch rlineto
	neg 0 rlineto closepath} bind def
/Square {dup Rec} bind def
/Bsquare {vpt sub exch vpt sub exch vpt2 Square} bind def
/S0 {BL [] 0 setdash 2 copy moveto 0 vpt rlineto BL Bsquare} bind def
/S1 {BL [] 0 setdash 2 copy vpt Square fill Bsquare} bind def
/S2 {BL [] 0 setdash 2 copy exch vpt sub exch vpt Square fill Bsquare} bind def
/S3 {BL [] 0 setdash 2 copy exch vpt sub exch vpt2 vpt Rec fill Bsquare} bind def
/S4 {BL [] 0 setdash 2 copy exch vpt sub exch vpt sub vpt Square fill Bsquare} bind def
/S5 {BL [] 0 setdash 2 copy 2 copy vpt Square fill
	exch vpt sub exch vpt sub vpt Square fill Bsquare} bind def
/S6 {BL [] 0 setdash 2 copy exch vpt sub exch vpt sub vpt vpt2 Rec fill Bsquare} bind def
/S7 {BL [] 0 setdash 2 copy exch vpt sub exch vpt sub vpt vpt2 Rec fill
	2 copy vpt Square fill Bsquare} bind def
/S8 {BL [] 0 setdash 2 copy vpt sub vpt Square fill Bsquare} bind def
/S9 {BL [] 0 setdash 2 copy vpt sub vpt vpt2 Rec fill Bsquare} bind def
/S10 {BL [] 0 setdash 2 copy vpt sub vpt Square fill 2 copy exch vpt sub exch vpt Square fill
	Bsquare} bind def
/S11 {BL [] 0 setdash 2 copy vpt sub vpt Square fill 2 copy exch vpt sub exch vpt2 vpt Rec fill
	Bsquare} bind def
/S12 {BL [] 0 setdash 2 copy exch vpt sub exch vpt sub vpt2 vpt Rec fill Bsquare} bind def
/S13 {BL [] 0 setdash 2 copy exch vpt sub exch vpt sub vpt2 vpt Rec fill
	2 copy vpt Square fill Bsquare} bind def
/S14 {BL [] 0 setdash 2 copy exch vpt sub exch vpt sub vpt2 vpt Rec fill
	2 copy exch vpt sub exch vpt Square fill Bsquare} bind def
/S15 {BL [] 0 setdash 2 copy Bsquare fill Bsquare} bind def
/D0 {gsave translate 45 rotate 0 0 S0 stroke grestore} bind def
/D1 {gsave translate 45 rotate 0 0 S1 stroke grestore} bind def
/D2 {gsave translate 45 rotate 0 0 S2 stroke grestore} bind def
/D3 {gsave translate 45 rotate 0 0 S3 stroke grestore} bind def
/D4 {gsave translate 45 rotate 0 0 S4 stroke grestore} bind def
/D5 {gsave translate 45 rotate 0 0 S5 stroke grestore} bind def
/D6 {gsave translate 45 rotate 0 0 S6 stroke grestore} bind def
/D7 {gsave translate 45 rotate 0 0 S7 stroke grestore} bind def
/D8 {gsave translate 45 rotate 0 0 S8 stroke grestore} bind def
/D9 {gsave translate 45 rotate 0 0 S9 stroke grestore} bind def
/D10 {gsave translate 45 rotate 0 0 S10 stroke grestore} bind def
/D11 {gsave translate 45 rotate 0 0 S11 stroke grestore} bind def
/D12 {gsave translate 45 rotate 0 0 S12 stroke grestore} bind def
/D13 {gsave translate 45 rotate 0 0 S13 stroke grestore} bind def
/D14 {gsave translate 45 rotate 0 0 S14 stroke grestore} bind def
/D15 {gsave translate 45 rotate 0 0 S15 stroke grestore} bind def
/DiaE {stroke [] 0 setdash vpt add M
  hpt neg vpt neg V hpt vpt neg V
  hpt vpt V hpt neg vpt V closepath stroke} def
/BoxE {stroke [] 0 setdash exch hpt sub exch vpt add M
  0 vpt2 neg V hpt2 0 V 0 vpt2 V
  hpt2 neg 0 V closepath stroke} def
/TriUE {stroke [] 0 setdash vpt 1.12 mul add M
  hpt neg vpt -1.62 mul V
  hpt 2 mul 0 V
  hpt neg vpt 1.62 mul V closepath stroke} def
/TriDE {stroke [] 0 setdash vpt 1.12 mul sub M
  hpt neg vpt 1.62 mul V
  hpt 2 mul 0 V
  hpt neg vpt -1.62 mul V closepath stroke} def
/PentE {stroke [] 0 setdash gsave
  translate 0 hpt M 4 {72 rotate 0 hpt L} repeat
  closepath stroke grestore} def
/CircE {stroke [] 0 setdash 
  hpt 0 360 arc stroke} def
/Opaque {gsave closepath 1 setgray fill grestore 0 setgray closepath} def
/DiaW {stroke [] 0 setdash vpt add M
  hpt neg vpt neg V hpt vpt neg V
  hpt vpt V hpt neg vpt V Opaque stroke} def
/BoxW {stroke [] 0 setdash exch hpt sub exch vpt add M
  0 vpt2 neg V hpt2 0 V 0 vpt2 V
  hpt2 neg 0 V Opaque stroke} def
/TriUW {stroke [] 0 setdash vpt 1.12 mul add M
  hpt neg vpt -1.62 mul V
  hpt 2 mul 0 V
  hpt neg vpt 1.62 mul V Opaque stroke} def
/TriDW {stroke [] 0 setdash vpt 1.12 mul sub M
  hpt neg vpt 1.62 mul V
  hpt 2 mul 0 V
  hpt neg vpt -1.62 mul V Opaque stroke} def
/PentW {stroke [] 0 setdash gsave
  translate 0 hpt M 4 {72 rotate 0 hpt L} repeat
  Opaque stroke grestore} def
/CircW {stroke [] 0 setdash 
  hpt 0 360 arc Opaque stroke} def
/BoxFill {gsave Rec 1 setgray fill grestore} def
/Density {
  /Fillden exch def
  currentrgbcolor
  /ColB exch def /ColG exch def /ColR exch def
  /ColR ColR Fillden mul Fillden sub 1 add def
  /ColG ColG Fillden mul Fillden sub 1 add def
  /ColB ColB Fillden mul Fillden sub 1 add def
  ColR ColG ColB setrgbcolor} def
/BoxColFill {gsave Rec PolyFill} def
/PolyFill {gsave Density fill grestore grestore} def
/h {rlineto rlineto rlineto gsave closepath fill grestore} bind def
%
%
/PatternFill {gsave /PFa [ 9 2 roll ] def
  PFa 0 get PFa 2 get 2 div add PFa 1 get PFa 3 get 2 div add translate
  PFa 2 get -2 div PFa 3 get -2 div PFa 2 get PFa 3 get Rec
  gsave 1 setgray fill grestore clip
  currentlinewidth 0.5 mul setlinewidth
  /PFs PFa 2 get dup mul PFa 3 get dup mul add sqrt def
  0 0 M PFa 5 get rotate PFs -2 div dup translate
  0 1 PFs PFa 4 get div 1 add floor cvi
	{PFa 4 get mul 0 M 0 PFs V} for
  0 PFa 6 get ne {
	0 1 PFs PFa 4 get div 1 add floor cvi
	{PFa 4 get mul 0 2 1 roll M PFs 0 V} for
 } if
  stroke grestore} def
/languagelevel where
 {pop languagelevel} {1} ifelse
 2 lt
	{/InterpretLevel1 true def}
	{/InterpretLevel1 Level1 def}
 ifelse
%
%
/Level2PatternFill {
/Tile8x8 {/PaintType 2 /PatternType 1 /TilingType 1 /BBox [0 0 8 8] /XStep 8 /YStep 8}
	bind def
/KeepColor {currentrgbcolor [/Pattern /DeviceRGB] setcolorspace} bind def
<< Tile8x8
 /PaintProc {0.5 setlinewidth pop 0 0 M 8 8 L 0 8 M 8 0 L stroke} 
>> matrix makepattern
/Pat1 exch def
<< Tile8x8
 /PaintProc {0.5 setlinewidth pop 0 0 M 8 8 L 0 8 M 8 0 L stroke
	0 4 M 4 8 L 8 4 L 4 0 L 0 4 L stroke}
>> matrix makepattern
/Pat2 exch def
<< Tile8x8
 /PaintProc {0.5 setlinewidth pop 0 0 M 0 8 L
	8 8 L 8 0 L 0 0 L fill}
>> matrix makepattern
/Pat3 exch def
<< Tile8x8
 /PaintProc {0.5 setlinewidth pop -4 8 M 8 -4 L
	0 12 M 12 0 L stroke}
>> matrix makepattern
/Pat4 exch def
<< Tile8x8
 /PaintProc {0.5 setlinewidth pop -4 0 M 8 12 L
	0 -4 M 12 8 L stroke}
>> matrix makepattern
/Pat5 exch def
<< Tile8x8
 /PaintProc {0.5 setlinewidth pop -2 8 M 4 -4 L
	0 12 M 8 -4 L 4 12 M 10 0 L stroke}
>> matrix makepattern
/Pat6 exch def
<< Tile8x8
 /PaintProc {0.5 setlinewidth pop -2 0 M 4 12 L
	0 -4 M 8 12 L 4 -4 M 10 8 L stroke}
>> matrix makepattern
/Pat7 exch def
<< Tile8x8
 /PaintProc {0.5 setlinewidth pop 8 -2 M -4 4 L
	12 0 M -4 8 L 12 4 M 0 10 L stroke}
>> matrix makepattern
/Pat8 exch def
<< Tile8x8
 /PaintProc {0.5 setlinewidth pop 0 -2 M 12 4 L
	-4 0 M 12 8 L -4 4 M 8 10 L stroke}
>> matrix makepattern
/Pat9 exch def
/Pattern1 {PatternBgnd KeepColor Pat1 setpattern} bind def
/Pattern2 {PatternBgnd KeepColor Pat2 setpattern} bind def
/Pattern3 {PatternBgnd KeepColor Pat3 setpattern} bind def
/Pattern4 {PatternBgnd KeepColor Landscape {Pat5} {Pat4} ifelse setpattern} bind def
/Pattern5 {PatternBgnd KeepColor Landscape {Pat4} {Pat5} ifelse setpattern} bind def
/Pattern6 {PatternBgnd KeepColor Landscape {Pat9} {Pat6} ifelse setpattern} bind def
/Pattern7 {PatternBgnd KeepColor Landscape {Pat8} {Pat7} ifelse setpattern} bind def
} def
%
%
%
/PatternBgnd {
  TransparentPatterns {} {gsave 1 setgray fill grestore} ifelse
} def
%
%
/Level1PatternFill {
/Pattern1 {0.250 Density} bind def
/Pattern2 {0.500 Density} bind def
/Pattern3 {0.750 Density} bind def
/Pattern4 {0.125 Density} bind def
/Pattern5 {0.375 Density} bind def
/Pattern6 {0.625 Density} bind def
/Pattern7 {0.875 Density} bind def
} def
%
%
Level1 {Level1PatternFill} {Level2PatternFill} ifelse
/Symbol-Oblique /Symbol findfont [1 0 .167 1 0 0] makefont
dup length dict begin {1 index /FID eq {pop pop} {def} ifelse} forall
currentdict end definefont pop
Level1 SuppressPDFMark or 
{} {
/SDict 10 dict def
systemdict /pdfmark known not {
  userdict /pdfmark systemdict /cleartomark get put
} if
SDict begin [
  /Title (paper/hqs1.tex)
  /Subject (gnuplot plot)
  /Creator (gnuplot 4.6 patchlevel 0)
  /Author (conrado)
  /CreationDate (Fri Apr 26 17:40:41 2013)
  /DOCINFO pdfmark
end
} ifelse
end
gnudict begin
gsave
doclip
0 0 translate
0.050 0.050 scale
0 setgray
newpath
BackgroundColor 0 lt 3 1 roll 0 lt exch 0 lt or or not {BackgroundColor C 1.000 0 0 7200.00 5040.00 BoxColFill} if
1.000 UL
LTb
0 640 M
63 0 V
2816 0 R
-63 0 V
0 1165 M
63 0 V
2816 0 R
-63 0 V
0 1690 M
63 0 V
2816 0 R
-63 0 V
0 2216 M
63 0 V
2816 0 R
-63 0 V
0 2741 M
63 0 V
2816 0 R
-63 0 V
0 3266 M
63 0 V
2816 0 R
-63 0 V
0 3791 M
63 0 V
2816 0 R
-63 0 V
0 640 M
0 63 V
0 3791 M
0 -63 V
960 640 M
0 63 V
0 3088 R
0 -63 V
1919 640 M
0 63 V
0 3088 R
0 -63 V
2879 640 M
0 63 V
0 3088 R
0 -63 V
stroke
LTa
0 640 M
2879 0 V
stroke
LTb
0 3791 N
0 640 L
2879 0 V
0 3151 V
0 3791 L
Z stroke
LCb setrgbcolor
LTb
1.000 UP
1.000 UL
LTb
1.000 UL
LT0
LCb setrgbcolor
LT0
777 2370 M
543 0 V
1302 15 R
-2 0 V
-3 1 V
-5 2 V
-6 2 V
-7 2 V
-8 3 V
-9 2 V
-10 3 V
-10 3 V
-10 3 V
-10 4 V
-9 3 V
-9 2 V
-9 3 V
-7 2 V
-6 2 V
-5 2 V
-3 1 V
-2 0 V
-1 0 V
-2 1 V
-3 1 V
-5 2 V
-6 2 V
-7 2 V
-9 2 V
-9 3 V
-9 3 V
-10 3 V
-10 3 V
-10 4 V
-10 3 V
-9 2 V
-8 3 V
-7 2 V
-6 2 V
-5 2 V
-3 1 V
-2 0 V
-1 1 V
-2 0 V
-3 1 V
-5 2 V
-6 2 V
-8 2 V
-8 3 V
-9 2 V
-9 4 V
-10 3 V
-10 3 V
-10 3 V
-10 3 V
-9 3 V
-8 3 V
-7 2 V
-6 2 V
-5 1 V
-3 1 V
-2 1 V
-1 0 V
-2 1 V
-4 1 V
-4 1 V
-6 2 V
-8 3 V
-8 2 V
-9 3 V
-9 3 V
-10 4 V
-10 3 V
-10 3 V
-10 3 V
-9 3 V
-8 2 V
-7 3 V
-6 2 V
-5 1 V
-4 1 V
-1 1 V
-1 0 V
-2 1 V
-4 1 V
-4 2 V
-7 2 V
-7 2 V
-8 3 V
-9 2 V
-10 4 V
-9 3 V
-11 3 V
-9 3 V
-10 3 V
-9 3 V
-8 3 V
-7 2 V
-6 2 V
-5 2 V
-4 1 V
-2 0 V
0 1 V
-2 0 V
-4 2 V
stroke 1961 2595 M
-5 1 V
-6 2 V
-7 2 V
-8 3 V
-9 3 V
-10 3 V
-10 3 V
-10 4 V
-10 3 V
-9 3 V
-9 3 V
-8 3 V
-8 2 V
-6 2 V
-4 2 V
-4 1 V
-2 0 V
-1 1 V
-2 0 V
-3 1 V
-5 2 V
-6 2 V
-7 2 V
-8 3 V
-9 3 V
-10 3 V
-10 3 V
-10 4 V
-10 3 V
-9 3 V
-9 3 V
-8 3 V
-8 2 V
-6 2 V
-5 2 V
-3 1 V
-2 1 V
-1 0 V
-2 0 V
-3 2 V
-5 1 V
-6 2 V
-7 3 V
-8 2 V
-9 3 V
-10 3 V
-10 4 V
-10 3 V
-10 3 V
-9 4 V
-9 2 V
-9 3 V
-7 3 V
-6 2 V
-5 1 V
-3 1 V
-2 1 V
-1 0 V
-2 1 V
-3 1 V
-5 2 V
-6 2 V
-7 2 V
-9 3 V
-9 3 V
-9 3 V
-10 3 V
-10 4 V
-10 3 V
-10 3 V
-8 3 V
-9 3 V
-7 2 V
-6 2 V
-5 2 V
-3 1 V
-2 1 V
-1 0 V
-2 1 V
-3 1 V
-5 1 V
-6 2 V
-7 3 V
-9 3 V
-9 3 V
-9 3 V
-10 3 V
-10 4 V
-10 3 V
-10 3 V
-9 3 V
-8 3 V
-7 2 V
-6 2 V
-5 2 V
-3 1 V
-2 1 V
-1 0 V
-2 1 V
-3 1 V
-5 1 V
-6 2 V
-8 3 V
-8 3 V
stroke 1279 2821 M
-9 3 V
-9 3 V
-10 3 V
-10 4 V
-10 3 V
-10 3 V
-9 3 V
-8 3 V
-7 2 V
-6 3 V
-5 1 V
-3 1 V
-2 1 V
-1 0 V
-2 1 V
-4 1 V
-4 2 V
-6 2 V
-8 2 V
-8 3 V
-9 3 V
-9 3 V
-10 4 V
-10 3 V
-10 3 V
-10 3 V
-9 4 V
-8 2 V
-7 3 V
-6 2 V
-5 1 V
-4 2 V
-2 0 V
0 1 V
-2 0 V
-4 1 V
-5 2 V
-6 2 V
-7 2 V
-8 3 V
-9 3 V
-10 3 V
-10 4 V
-10 3 V
-9 4 V
-10 3 V
-9 3 V
-8 3 V
-7 2 V
-7 2 V
-4 2 V
-4 1 V
-2 1 V
-1 0 V
-1 1 V
-4 1 V
-5 1 V
-6 2 V
-7 3 V
-8 3 V
-9 3 V
-10 3 V
-10 3 V
-10 4 V
-10 3 V
-9 3 V
-9 3 V
-8 3 V
-8 2 V
-6 3 V
-4 1 V
-4 1 V
-2 1 V
-1 0 V
-2 1 V
-3 1 V
-5 2 V
-6 2 V
-7 2 V
-8 3 V
-9 3 V
-10 3 V
-10 4 V
-10 3 V
-10 3 V
-9 4 V
-9 3 V
-8 2 V
-8 3 V
-6 2 V
-5 2 V
-3 1 V
-2 0 V
-1 1 V
-2 0 V
-3 1 V
-5 2 V
-6 2 V
-7 3 V
-8 2 V
-9 3 V
-10 4 V
-10 3 V
-10 3 V
stroke 585 3055 M
-10 4 V
-9 3 V
-9 3 V
-9 3 V
-7 2 V
-6 2 V
-5 2 V
-3 1 V
-2 1 V
-1 0 V
-2 0 V
-3 2 V
-5 1 V
-6 2 V
-7 3 V
-9 2 V
-9 3 V
-9 4 V
-10 3 V
-10 3 V
-10 4 V
-10 3 V
-9 3 V
-8 3 V
-7 2 V
-6 2 V
-5 2 V
-3 1 V
-2 1 V
-1 0 V
-2 0 V
-3 2 V
-5 1 V
-6 2 V
-7 3 V
-9 2 V
-9 3 V
-9 4 V
-10 3 V
-10 3 V
-10 4 V
-10 3 V
-9 3 V
-8 2 V
-7 3 V
-6 2 V
-5 1 V
-3 2 V
-2 0 V
-1 1 V
-2 0 V
-3 1 V
-5 2 V
-6 2 V
-8 2 V
-8 3 V
-9 3 V
-9 3 V
-10 3 V
-10 4 V
-10 3 V
-10 3 V
-9 3 V
-8 3 V
-7 2 V
-6 2 V
-5 2 V
-4 1 V
-1 1 V
-1 0 V
-2 0 V
-4 2 V
-4 1 V
-7 2 V
-7 3 V
-8 2 V
-9 3 V
-10 3 V
-9 4 V
-10 3 V
-10 3 V
-10 3 V
-9 3 V
-8 3 V
-7 2 V
-6 2 V
-5 2 V
-4 1 V
-2 1 V
stroke
LT1
LCb setrgbcolor
LT1
777 2170 M
543 0 V
2622 885 M
-2 1 V
-3 0 V
-5 0 V
-6 0 V
-7 1 V
-8 0 V
-9 0 V
-10 1 V
-10 0 V
-10 1 V
-10 0 V
-9 1 V
-9 0 V
-9 1 V
-7 0 V
-6 0 V
-5 0 V
-3 1 V
-2 0 V
-1 0 V
-2 0 V
-3 0 V
-5 0 V
-6 1 V
-7 0 V
-9 0 V
-9 1 V
-9 0 V
-10 1 V
-10 0 V
-10 1 V
-10 0 V
-9 1 V
-8 0 V
-7 0 V
-6 1 V
-5 0 V
-3 0 V
-2 0 V
-1 0 V
-2 0 V
-3 0 V
-5 1 V
-6 0 V
-8 0 V
-8 1 V
-9 0 V
-9 1 V
-10 0 V
-10 1 V
-10 0 V
-10 1 V
-9 0 V
-8 1 V
-7 0 V
-6 0 V
-5 1 V
-3 0 V
-2 0 V
-1 0 V
-2 0 V
-4 0 V
-4 0 V
-6 1 V
-8 0 V
-8 0 V
-9 1 V
-9 0 V
-10 1 V
-10 1 V
-10 0 V
-10 1 V
-9 0 V
-8 0 V
-7 1 V
-6 0 V
-5 0 V
-4 1 V
-1 0 V
-1 0 V
-2 0 V
-4 0 V
-4 0 V
-7 1 V
-7 0 V
-8 0 V
-9 1 V
-10 0 V
-9 1 V
-11 0 V
-9 1 V
-10 0 V
-9 1 V
-8 0 V
-7 1 V
-6 0 V
-5 0 V
-4 1 V
-2 0 V
-2 0 V
-4 0 V
-5 0 V
stroke 1956 919 M
-6 1 V
-7 0 V
-8 0 V
-9 1 V
-10 0 V
-10 1 V
-10 1 V
-10 0 V
-9 1 V
-9 0 V
-8 1 V
-8 0 V
-6 0 V
-4 1 V
-4 0 V
-2 0 V
-1 0 V
-2 0 V
-3 0 V
-5 0 V
-6 1 V
-7 0 V
-8 1 V
-9 0 V
-10 1 V
-10 0 V
-10 1 V
-10 0 V
-9 1 V
-9 0 V
-8 1 V
-8 0 V
-6 1 V
-5 0 V
-3 0 V
-2 0 V
-1 0 V
-2 0 V
-3 1 V
-5 0 V
-6 0 V
-7 1 V
-8 0 V
-9 1 V
-10 0 V
-10 1 V
-10 0 V
-10 1 V
-9 0 V
-9 1 V
-9 0 V
-7 1 V
-6 0 V
-5 1 V
-3 0 V
-2 0 V
-1 0 V
-2 0 V
-3 0 V
-5 0 V
-6 1 V
-7 0 V
-9 1 V
-9 0 V
-9 1 V
-10 0 V
-10 1 V
-10 1 V
-10 0 V
-8 1 V
-9 0 V
-7 1 V
-6 0 V
-5 0 V
-3 1 V
-2 0 V
-1 0 V
-2 0 V
-3 0 V
-5 0 V
-6 1 V
-7 0 V
-9 1 V
-9 0 V
-9 1 V
-10 0 V
-10 1 V
-10 1 V
-10 0 V
-9 1 V
-8 0 V
-7 1 V
-6 0 V
-5 0 V
-3 1 V
-2 0 V
-1 0 V
-2 0 V
-3 0 V
-5 0 V
-6 1 V
-8 0 V
-8 1 V
-9 0 V
stroke 1270 959 M
-9 1 V
-10 1 V
-10 0 V
-10 1 V
-10 0 V
-9 1 V
-8 0 V
-7 1 V
-6 0 V
-5 1 V
-3 0 V
-2 0 V
-1 0 V
-2 0 V
-4 0 V
-4 1 V
-6 0 V
-8 1 V
-8 0 V
-9 1 V
-9 0 V
-10 1 V
-10 1 V
-10 0 V
-10 1 V
-9 0 V
-8 1 V
-7 0 V
-6 1 V
-5 0 V
-4 0 V
-2 0 V
-2 1 V
-4 0 V
-5 0 V
-6 1 V
-7 0 V
-8 1 V
-9 0 V
-10 1 V
-10 0 V
-10 1 V
-9 1 V
-10 0 V
-9 1 V
-8 1 V
-7 0 V
-7 0 V
-4 1 V
-4 0 V
-2 0 V
-1 0 V
-1 0 V
-4 1 V
-5 0 V
-6 0 V
-7 1 V
-8 0 V
-9 1 V
-10 1 V
-10 0 V
-10 1 V
-10 1 V
-9 0 V
-9 1 V
-8 1 V
-8 0 V
-6 0 V
-4 1 V
-4 0 V
-2 0 V
-1 0 V
-2 0 V
-3 1 V
-5 0 V
-6 0 V
-7 1 V
-8 0 V
-9 1 V
-10 1 V
-10 0 V
-10 1 V
-10 1 V
-9 0 V
-9 1 V
-8 1 V
-8 0 V
-6 1 V
-5 0 V
-3 0 V
-2 0 V
-1 0 V
-2 1 V
-3 0 V
-5 0 V
-6 1 V
-7 0 V
-8 1 V
-9 0 V
-10 1 V
-10 1 V
-10 0 V
-10 1 V
-9 1 V
stroke 566 1007 M
-9 1 V
-9 0 V
-7 1 V
-6 0 V
-5 0 V
-3 1 V
-2 0 V
-1 0 V
-2 0 V
-3 0 V
-5 1 V
-6 0 V
-7 1 V
-9 0 V
-9 1 V
-9 1 V
-10 0 V
-10 1 V
-10 1 V
-10 1 V
-9 0 V
-8 1 V
-7 0 V
-6 1 V
-5 0 V
-3 1 V
-2 0 V
-1 0 V
-2 0 V
-3 0 V
-5 0 V
-6 1 V
-7 1 V
-9 0 V
-9 1 V
-9 1 V
-10 0 V
-10 1 V
-10 1 V
-10 1 V
-9 0 V
-8 1 V
-7 0 V
-6 1 V
-5 0 V
-3 1 V
-2 0 V
-1 0 V
-2 0 V
-3 0 V
-5 1 V
-6 0 V
-8 1 V
-8 0 V
-9 1 V
-9 1 V
-10 1 V
-10 0 V
-10 1 V
-10 1 V
-9 1 V
-8 0 V
-7 1 V
-6 0 V
-5 1 V
-4 0 V
-1 0 V
-1 0 V
-2 0 V
-4 1 V
-4 0 V
-7 0 V
-7 1 V
-8 1 V
-9 0 V
-10 1 V
-9 1 V
-10 1 V
-10 1 V
-10 0 V
-9 1 V
-8 1 V
-7 0 V
-6 1 V
-5 0 V
-4 1 V
-2 0 V
stroke
LT2
LCb setrgbcolor
LT2
777 1970 M
543 0 V
921 337 R
-1 0 V
-3 1 V
-4 2 V
-6 2 V
-6 2 V
-7 2 V
-8 3 V
-8 3 V
-8 3 V
-9 3 V
-8 3 V
-8 3 V
-8 3 V
-7 2 V
-6 3 V
-5 1 V
-5 2 V
-2 1 V
-2 0 V
-1 1 V
-2 0 V
-2 1 V
-5 2 V
-5 2 V
-6 2 V
-7 2 V
-8 3 V
-8 3 V
-8 3 V
-9 3 V
-8 3 V
-8 3 V
-8 3 V
-7 3 V
-6 2 V
-6 2 V
-4 1 V
-3 1 V
-1 1 V
-1 0 V
-2 1 V
-3 1 V
-4 1 V
-5 2 V
-6 2 V
-7 3 V
-8 3 V
-8 3 V
-8 3 V
-9 3 V
-8 3 V
-9 3 V
-7 3 V
-7 3 V
-6 2 V
-6 2 V
-4 1 V
-3 1 V
-1 1 V
-1 0 V
-2 1 V
-3 1 V
-4 1 V
-5 2 V
-6 3 V
-7 2 V
-8 3 V
-8 3 V
-8 3 V
-9 3 V
-8 3 V
-9 3 V
-7 3 V
-7 3 V
-7 2 V
-5 2 V
-4 2 V
-3 1 V
-1 0 V
-1 1 V
-2 0 V
-3 1 V
-4 2 V
-5 2 V
-6 2 V
-7 3 V
-8 3 V
-8 3 V
-8 3 V
-9 3 V
-9 3 V
-8 3 V
-7 3 V
-7 3 V
-7 2 V
-5 2 V
-4 2 V
-3 1 V
-1 0 V
-1 1 V
-2 0 V
-3 1 V
stroke 1676 2513 M
-4 2 V
-5 2 V
-6 2 V
-7 3 V
-8 3 V
-8 3 V
-9 3 V
-8 3 V
-9 4 V
-8 3 V
-7 3 V
-8 2 V
-6 3 V
-5 2 V
-4 1 V
-3 1 V
-2 1 V
-2 1 V
-3 1 V
-4 2 V
-5 2 V
-6 2 V
-7 3 V
-8 3 V
-8 3 V
-9 3 V
-8 3 V
-9 4 V
-8 3 V
-8 3 V
-7 2 V
-6 3 V
-5 2 V
-4 1 V
-3 1 V
-2 1 V
-2 1 V
-3 1 V
-4 2 V
-5 2 V
-6 2 V
-7 3 V
-8 3 V
-8 3 V
-9 3 V
-8 4 V
-9 3 V
-8 3 V
-8 3 V
-7 3 V
-6 2 V
-5 2 V
-4 2 V
-3 1 V
-2 1 V
-2 1 V
-3 1 V
-4 1 V
-5 2 V
-6 3 V
-7 2 V
-8 3 V
-8 4 V
-9 3 V
-8 3 V
-9 4 V
-8 3 V
-8 3 V
-7 3 V
-6 2 V
-5 2 V
-4 2 V
-3 1 V
-2 1 V
-1 0 V
-1 1 V
-3 1 V
-4 1 V
-5 2 V
-7 3 V
-7 3 V
-7 3 V
-8 3 V
-9 3 V
-8 4 V
-9 3 V
-8 3 V
-8 3 V
-7 3 V
-6 2 V
-5 3 V
-4 1 V
-3 1 V
-2 1 V
-1 0 V
-1 1 V
-3 1 V
-4 2 V
-5 2 V
-7 2 V
-7 3 V
-7 3 V
-8 3 V
-9 4 V
stroke 1069 2750 M
-9 3 V
-8 4 V
-8 3 V
-8 3 V
-7 3 V
-6 2 V
-5 2 V
-4 2 V
-3 1 V
-2 1 V
-1 0 V
-1 1 V
-3 1 V
-4 2 V
-6 2 V
-6 2 V
-7 3 V
-7 3 V
-9 3 V
-8 4 V
-9 3 V
-8 4 V
-8 3 V
-8 3 V
-7 3 V
-6 2 V
-5 3 V
-4 1 V
-3 1 V
-2 1 V
-1 0 V
-1 1 V
-3 1 V
-4 2 V
-6 2 V
-6 2 V
-7 3 V
-7 3 V
-9 4 V
-8 3 V
-9 4 V
-8 3 V
-8 3 V
-8 4 V
-7 2 V
-6 3 V
-5 2 V
-5 2 V
-2 1 V
-2 1 V
-1 0 V
-1 1 V
-3 1 V
-4 1 V
-6 3 V
-6 2 V
-7 3 V
-8 3 V
-8 3 V
-8 4 V
-9 3 V
-8 4 V
-8 3 V
-8 3 V
-7 3 V
-6 3 V
-6 2 V
-4 2 V
-3 1 V
-1 1 V
-1 0 V
-2 0 V
-2 2 V
-5 1 V
-5 3 V
-6 2 V
-7 3 V
-8 3 V
-8 3 V
-8 4 V
-9 3 V
-8 4 V
-9 3 V
-7 4 V
-7 2 V
-6 3 V
-6 2 V
-4 2 V
-3 1 V
-1 1 V
-1 0 V
-2 1 V
-3 1 V
-4 2 V
-5 2 V
-6 2 V
-7 3 V
-8 3 V
-8 4 V
-8 3 V
-9 4 V
-8 3 V
-9 4 V
-7 3 V
stroke 476 2992 M
-7 3 V
-6 2 V
-6 3 V
-4 1 V
-3 1 V
-1 1 V
-1 0 V
-2 1 V
-3 1 V
-4 2 V
-5 2 V
-6 3 V
-7 3 V
-8 3 V
-8 3 V
-8 4 V
-9 3 V
-8 4 V
-9 3 V
-7 4 V
-7 3 V
-7 2 V
-5 2 V
-4 2 V
-3 1 V
-1 1 V
-1 0 V
-2 1 V
-3 1 V
-4 2 V
-5 2 V
-6 3 V
-7 2 V
-8 4 V
-8 3 V
-9 4 V
-8 3 V
-9 4 V
-8 3 V
-7 3 V
-7 3 V
-7 3 V
-5 2 V
-4 2 V
-3 1 V
-1 1 V
-1 0 V
-2 1 V
-3 1 V
-4 2 V
-5 2 V
-6 2 V
-7 3 V
-8 4 V
-8 3 V
-9 4 V
-8 3 V
-9 4 V
-8 3 V
-8 3 V
-7 3 V
-6 3 V
-5 2 V
-4 2 V
-3 1 V
-2 1 V
-2 1 V
-3 1 V
-4 2 V
-5 2 V
-6 2 V
-7 3 V
-8 4 V
-8 3 V
-9 4 V
-8 3 V
-9 4 V
-8 3 V
-8 4 V
-7 2 V
-6 3 V
-5 2 V
-4 2 V
-3 1 V
-2 1 V
stroke
LT3
LCb setrgbcolor
LT3
777 1770 M
543 0 V
2241 670 M
-1 0 V
-3 0 V
-4 0 V
-6 0 V
-6 1 V
-7 0 V
-8 0 V
-8 0 V
-8 0 V
-9 0 V
-8 0 V
-8 0 V
-8 1 V
-7 0 V
-6 0 V
-5 0 V
-5 0 V
-2 0 V
-2 0 V
-1 0 V
-2 0 V
-2 0 V
-5 0 V
-5 0 V
-6 0 V
-7 0 V
-8 1 V
-8 0 V
-8 0 V
-9 0 V
-8 0 V
-8 0 V
-8 0 V
-7 1 V
-6 0 V
-6 0 V
-4 0 V
-3 0 V
-1 0 V
-1 0 V
-2 0 V
-3 0 V
-4 0 V
-5 0 V
-6 0 V
-7 0 V
-8 0 V
-8 1 V
-8 0 V
-9 0 V
-8 0 V
-9 0 V
-7 0 V
-7 0 V
-6 1 V
-6 0 V
-4 0 V
-3 0 V
-1 0 V
-1 0 V
-2 0 V
-3 0 V
-4 0 V
-5 0 V
-6 0 V
-7 0 V
-8 1 V
-8 0 V
-8 0 V
-9 0 V
-8 0 V
-9 0 V
-7 0 V
-7 1 V
-7 0 V
-5 0 V
-4 0 V
-3 0 V
-1 0 V
-1 0 V
-2 0 V
-3 0 V
-4 0 V
-5 0 V
-6 0 V
-7 1 V
-8 0 V
-8 0 V
-8 0 V
-9 0 V
-9 0 V
-8 1 V
-7 0 V
-7 0 V
-7 0 V
-5 0 V
-4 0 V
-3 0 V
-1 0 V
-1 0 V
-2 0 V
-3 0 V
stroke 1676 680 M
-4 1 V
-5 0 V
-6 0 V
-7 0 V
-8 0 V
-8 0 V
-9 0 V
-8 1 V
-9 0 V
-8 0 V
-7 0 V
-8 0 V
-6 0 V
-5 1 V
-4 0 V
-3 0 V
-2 0 V
-2 0 V
-3 0 V
-4 0 V
-5 0 V
-6 0 V
-7 0 V
-8 1 V
-8 0 V
-9 0 V
-8 0 V
-9 0 V
-8 1 V
-8 0 V
-7 0 V
-6 0 V
-5 0 V
-4 0 V
-3 0 V
-2 0 V
-2 0 V
-3 1 V
-4 0 V
-5 0 V
-6 0 V
-7 0 V
-8 0 V
-8 0 V
-9 1 V
-8 0 V
-9 0 V
-8 0 V
-8 0 V
-7 1 V
-6 0 V
-5 0 V
-4 0 V
-3 0 V
-2 0 V
-2 0 V
-3 0 V
-4 0 V
-5 1 V
-6 0 V
-7 0 V
-8 0 V
-8 0 V
-9 1 V
-8 0 V
-9 0 V
-8 0 V
-8 0 V
-7 1 V
-6 0 V
-5 0 V
-4 0 V
-3 0 V
-2 0 V
-1 0 V
-1 0 V
-3 0 V
-4 0 V
-5 1 V
-7 0 V
-7 0 V
-7 0 V
-8 0 V
-9 1 V
-8 0 V
-9 0 V
-8 0 V
-8 1 V
-7 0 V
-6 0 V
-5 0 V
-4 0 V
-3 0 V
-2 0 V
-1 0 V
-1 0 V
-3 1 V
-4 0 V
-5 0 V
-7 0 V
-7 0 V
-7 0 V
-8 1 V
-9 0 V
stroke 1069 696 M
-9 0 V
-8 0 V
-8 1 V
-8 0 V
-7 0 V
-6 0 V
-5 1 V
-4 0 V
-3 0 V
-2 0 V
-1 0 V
-1 0 V
-3 0 V
-4 0 V
-6 0 V
-6 1 V
-7 0 V
-7 0 V
-9 0 V
-8 1 V
-9 0 V
-8 0 V
-8 0 V
-8 1 V
-7 0 V
-6 0 V
-5 0 V
-4 0 V
-3 0 V
-2 1 V
-1 0 V
-1 0 V
-3 0 V
-4 0 V
-6 0 V
-6 0 V
-7 0 V
-7 1 V
-9 0 V
-8 0 V
-9 1 V
-8 0 V
-8 0 V
-8 1 V
-7 0 V
-6 0 V
-5 0 V
-5 0 V
-2 0 V
-2 0 V
-1 1 V
-1 0 V
-3 0 V
-4 0 V
-6 0 V
-6 0 V
-7 1 V
-8 0 V
-8 0 V
-8 0 V
-9 1 V
-8 0 V
-8 0 V
-8 1 V
-7 0 V
-6 0 V
-6 0 V
-4 1 V
-3 0 V
-1 0 V
-1 0 V
-2 0 V
-2 0 V
-5 0 V
-5 0 V
-6 1 V
-7 0 V
-8 0 V
-8 0 V
-8 1 V
-9 0 V
-8 0 V
-9 1 V
-7 0 V
-7 0 V
-6 1 V
-6 0 V
-4 0 V
-3 0 V
-1 0 V
-1 0 V
-2 0 V
-3 1 V
-4 0 V
-5 0 V
-6 0 V
-7 0 V
-8 1 V
-8 0 V
-8 1 V
-9 0 V
-8 0 V
-9 1 V
-7 0 V
stroke 476 718 M
-7 0 V
-6 1 V
-6 0 V
-4 0 V
-3 0 V
-1 0 V
-1 0 V
-2 0 V
-3 0 V
-4 1 V
-5 0 V
-6 0 V
-7 0 V
-8 1 V
-8 0 V
-8 1 V
-9 0 V
-8 0 V
-9 1 V
-7 0 V
-7 0 V
-7 1 V
-5 0 V
-4 0 V
-3 0 V
-1 0 V
-1 0 V
-2 1 V
-3 0 V
-4 0 V
-5 0 V
-6 0 V
-7 1 V
-8 0 V
-8 1 V
-9 0 V
-8 0 V
-9 1 V
-8 0 V
-7 1 V
-7 0 V
-7 0 V
-5 1 V
-4 0 V
-3 0 V
-1 0 V
-1 0 V
-2 0 V
-3 0 V
-4 1 V
-5 0 V
-6 0 V
-7 0 V
-8 1 V
-8 0 V
-9 1 V
-8 0 V
-9 1 V
-8 0 V
-8 1 V
-7 0 V
-6 0 V
-5 1 V
-4 0 V
-3 0 V
-2 0 V
-2 0 V
-3 0 V
-4 1 V
-5 0 V
-6 0 V
-7 1 V
-8 0 V
-8 0 V
-9 1 V
-8 0 V
-9 1 V
-8 0 V
-8 1 V
-7 0 V
-6 0 V
-5 0 V
-4 1 V
2 741 L
-2 0 V
stroke
LT4
LCb setrgbcolor
LT4
777 1570 M
543 0 V
921 829 R
-1 1 V
-3 1 V
-4 1 V
-6 2 V
-6 3 V
-7 2 V
-8 3 V
-8 3 V
-8 4 V
-9 3 V
-8 3 V
-8 3 V
-8 3 V
-7 3 V
-6 2 V
-5 2 V
-5 1 V
-2 2 V
-2 0 V
-1 0 V
-2 1 V
-2 1 V
-5 2 V
-5 2 V
-6 2 V
-7 3 V
-8 3 V
-8 3 V
-8 3 V
-9 3 V
-8 4 V
-8 3 V
-8 3 V
-7 2 V
-6 3 V
-6 2 V
-4 1 V
-3 1 V
-1 1 V
-1 0 V
-2 1 V
-3 1 V
-4 2 V
-5 2 V
-6 2 V
-7 3 V
-8 3 V
-8 3 V
-8 3 V
-9 3 V
-8 4 V
-9 3 V
-7 3 V
-7 3 V
-6 2 V
-6 2 V
-4 2 V
-3 1 V
-1 0 V
-1 1 V
-2 0 V
-3 1 V
-4 2 V
-5 2 V
-6 2 V
-7 3 V
-8 3 V
-8 3 V
-8 4 V
-9 3 V
-8 3 V
-9 4 V
-7 3 V
-7 2 V
-7 3 V
-5 2 V
-4 2 V
-3 1 V
-1 0 V
-1 1 V
-2 0 V
-3 1 V
-4 2 V
-5 2 V
-6 3 V
-7 2 V
-8 3 V
-8 4 V
-8 3 V
-9 3 V
-9 4 V
-8 3 V
-7 3 V
-7 3 V
-7 2 V
-5 2 V
-4 2 V
-3 1 V
-1 1 V
-1 0 V
-2 1 V
-3 1 V
stroke 1676 2618 M
-4 2 V
-5 2 V
-6 2 V
-7 3 V
-8 3 V
-8 3 V
-9 4 V
-8 3 V
-9 4 V
-8 3 V
-7 3 V
-8 3 V
-6 2 V
-5 2 V
-4 2 V
-3 1 V
-2 1 V
-2 1 V
-3 1 V
-4 2 V
-5 2 V
-6 2 V
-7 3 V
-8 3 V
-8 4 V
-9 3 V
-8 4 V
-9 3 V
-8 3 V
-8 4 V
-7 2 V
-6 3 V
-5 2 V
-4 2 V
-3 1 V
-2 1 V
-2 1 V
-3 1 V
-4 1 V
-5 3 V
-6 2 V
-7 3 V
-8 3 V
-8 3 V
-9 4 V
-8 4 V
-9 3 V
-8 3 V
-8 4 V
-7 2 V
-6 3 V
-5 2 V
-4 2 V
-3 1 V
-2 1 V
-2 1 V
-3 1 V
-4 2 V
-5 2 V
-6 2 V
-7 3 V
-8 3 V
-8 4 V
-9 3 V
-8 4 V
-9 3 V
-8 4 V
-8 3 V
-7 3 V
-6 3 V
-5 2 V
-4 1 V
-3 2 V
-2 0 V
-1 1 V
-1 0 V
-3 2 V
-4 1 V
-5 2 V
-7 3 V
-7 3 V
-7 3 V
-8 4 V
-9 3 V
-8 4 V
-9 3 V
-8 4 V
-8 3 V
-7 3 V
-6 3 V
-5 2 V
-4 2 V
-3 1 V
-2 0 V
-1 1 V
-1 0 V
-3 2 V
-4 1 V
-5 3 V
-7 2 V
-7 3 V
-7 3 V
-8 4 V
-9 3 V
stroke 1069 2868 M
-9 4 V
-8 4 V
-8 3 V
-8 3 V
-7 3 V
-6 3 V
-5 2 V
-4 2 V
-3 1 V
-2 1 V
-1 0 V
-1 1 V
-3 1 V
-4 2 V
-6 2 V
-6 3 V
-7 3 V
-7 3 V
-9 4 V
-8 3 V
-9 4 V
-8 4 V
-8 3 V
-8 3 V
-7 3 V
-6 3 V
-5 2 V
-4 2 V
-3 1 V
-2 1 V
-1 0 V
-1 1 V
-3 1 V
-4 2 V
-6 2 V
-6 3 V
-7 3 V
-7 3 V
-9 4 V
-8 3 V
-9 4 V
-8 4 V
-8 3 V
-8 4 V
-7 3 V
-6 2 V
-5 3 V
-5 1 V
-2 2 V
-2 0 V
-1 1 V
-1 0 V
-3 2 V
-4 1 V
-6 3 V
-6 2 V
-7 3 V
-8 4 V
-8 3 V
-8 4 V
-9 4 V
-8 3 V
-8 4 V
-8 3 V
-7 3 V
-6 3 V
-6 2 V
-4 2 V
-3 1 V
-1 1 V
-1 0 V
-2 1 V
-2 1 V
-5 2 V
-5 2 V
-6 3 V
-7 3 V
-8 3 V
-8 4 V
-8 4 V
-9 4 V
-8 3 V
-9 4 V
-7 3 V
-7 3 V
-6 3 V
-6 2 V
-4 2 V
-3 1 V
-1 1 V
-1 0 V
-2 1 V
-3 1 V
-4 2 V
-5 2 V
-6 3 V
-7 3 V
-8 4 V
-8 3 V
-8 4 V
-9 4 V
-8 3 V
-9 4 V
-7 3 V
stroke 476 3125 M
-7 4 V
-6 2 V
-6 3 V
-4 1 V
-3 2 V
-1 0 V
-1 1 V
-2 0 V
-3 2 V
-4 1 V
-5 3 V
-6 3 V
-7 3 V
-8 3 V
-8 4 V
-8 3 V
-9 4 V
-8 4 V
-9 4 V
-7 3 V
-7 3 V
-7 3 V
-5 2 V
-4 2 V
-3 1 V
-1 1 V
-1 0 V
-2 1 V
-3 1 V
-4 2 V
-5 2 V
-6 3 V
-7 3 V
-8 4 V
-8 3 V
-9 4 V
-8 4 V
-9 4 V
-8 3 V
-7 4 V
-7 3 V
-7 3 V
-5 2 V
-4 2 V
-3 1 V
-1 1 V
-1 0 V
-2 1 V
-3 1 V
-4 2 V
-5 2 V
-6 3 V
-7 3 V
-8 4 V
-8 3 V
-9 4 V
-8 4 V
-9 4 V
-8 3 V
-8 4 V
-7 3 V
-6 3 V
-5 2 V
-4 2 V
-3 1 V
-2 1 V
-2 1 V
-3 1 V
-4 2 V
-5 2 V
-6 3 V
-7 3 V
-8 4 V
-8 3 V
-9 4 V
-8 4 V
-9 4 V
-8 3 V
-8 4 V
-7 3 V
-6 3 V
-5 3 V
-4 1 V
-3 3 V
-2 1 V
stroke
LT5
LCb setrgbcolor
LT5
777 1370 M
543 0 V
921 994 R
-1 1 V
-3 1 V
-4 1 V
-6 2 V
-6 2 V
-7 3 V
-8 3 V
-8 2 V
-8 4 V
-9 3 V
-8 3 V
-8 3 V
-8 2 V
-7 3 V
-6 2 V
-5 2 V
-5 2 V
-2 1 V
-2 0 V
-1 0 V
-2 1 V
-2 1 V
-5 2 V
-5 1 V
-6 3 V
-7 2 V
-8 3 V
-8 3 V
-8 3 V
-9 3 V
-8 3 V
-8 3 V
-8 3 V
-7 3 V
-6 2 V
-6 2 V
-4 1 V
-3 1 V
-1 1 V
-1 0 V
-2 1 V
-3 1 V
-4 1 V
-5 2 V
-6 3 V
-7 2 V
-8 3 V
-8 3 V
-8 3 V
-9 3 V
-8 3 V
-9 3 V
-7 3 V
-7 3 V
-6 2 V
-6 2 V
-4 2 V
-3 1 V
-1 0 V
-1 0 V
-2 1 V
-3 1 V
-4 2 V
-5 2 V
-6 2 V
-7 3 V
-8 2 V
-8 3 V
-8 4 V
-9 3 V
-8 3 V
-9 3 V
-7 3 V
-7 2 V
-7 3 V
-5 2 V
-4 1 V
-3 1 V
-1 1 V
-1 0 V
-2 1 V
-3 1 V
-4 1 V
-5 2 V
-6 3 V
-7 2 V
-8 3 V
-8 3 V
-8 3 V
-9 4 V
-9 3 V
-8 3 V
-7 3 V
-7 2 V
-7 3 V
-5 2 V
-4 1 V
-3 1 V
-1 1 V
-1 0 V
-2 1 V
-3 1 V
stroke 1676 2572 M
-4 2 V
-5 2 V
-6 2 V
-7 3 V
-8 2 V
-8 4 V
-9 3 V
-8 3 V
-9 3 V
-8 3 V
-7 3 V
-8 3 V
-6 2 V
-5 2 V
-4 2 V
-3 1 V
-2 1 V
-2 0 V
-3 2 V
-4 1 V
-5 2 V
-6 2 V
-7 3 V
-8 3 V
-8 3 V
-9 3 V
-8 4 V
-9 3 V
-8 3 V
-8 3 V
-7 3 V
-6 2 V
-5 2 V
-4 2 V
-3 1 V
-2 0 V
0 1 V
-2 0 V
-3 1 V
-4 2 V
-5 2 V
-6 2 V
-7 3 V
-8 3 V
-8 3 V
-9 3 V
-8 4 V
-9 3 V
-8 3 V
-8 3 V
-7 3 V
-6 2 V
-5 2 V
-4 2 V
-3 1 V
-2 1 V
-2 1 V
-3 1 V
-4 1 V
-5 2 V
-6 3 V
-7 2 V
-8 3 V
-8 4 V
-9 3 V
-8 3 V
-9 3 V
-8 4 V
-8 3 V
-7 2 V
-6 3 V
-5 2 V
-4 1 V
-3 1 V
-2 1 V
-1 0 V
-1 1 V
-3 1 V
-4 2 V
-5 2 V
-7 2 V
-7 3 V
-7 3 V
-8 3 V
-9 3 V
-8 4 V
-9 3 V
-8 3 V
-8 3 V
-7 3 V
-6 2 V
-5 3 V
-4 1 V
-3 1 V
-2 1 V
-1 0 V
-1 1 V
-3 1 V
-4 2 V
-5 2 V
-7 2 V
-7 3 V
-7 3 V
-8 3 V
stroke 1078 2803 M
-9 3 V
-9 4 V
-8 3 V
-8 3 V
-8 3 V
-7 3 V
-6 2 V
-5 3 V
-4 1 V
-3 1 V
-2 1 V
-1 0 V
-1 1 V
-3 1 V
-4 2 V
-6 2 V
-6 2 V
-7 3 V
-7 3 V
-9 3 V
-8 3 V
-9 4 V
-8 3 V
-8 3 V
-8 3 V
-7 3 V
-6 3 V
-5 2 V
-4 1 V
-3 1 V
-2 1 V
-1 0 V
-1 1 V
-3 1 V
-4 2 V
-6 2 V
-6 2 V
-7 3 V
-7 3 V
-9 3 V
-8 4 V
-9 3 V
-8 3 V
-8 4 V
-8 3 V
-7 2 V
-6 3 V
-5 2 V
-5 2 V
-2 1 V
-2 0 V
-1 1 V
-1 0 V
-3 2 V
-4 1 V
-6 2 V
-6 3 V
-7 2 V
-8 3 V
-8 4 V
-8 3 V
-9 3 V
-8 4 V
-8 3 V
-8 3 V
-7 3 V
-6 2 V
-6 2 V
-4 2 V
-3 1 V
-1 1 V
-1 0 V
-2 1 V
-2 1 V
-5 2 V
-5 2 V
-6 2 V
-7 3 V
-8 3 V
-8 3 V
-8 4 V
-9 3 V
-8 3 V
-9 4 V
-7 3 V
-7 2 V
-6 3 V
-6 2 V
-4 1 V
-3 2 V
-1 0 V
-1 1 V
-2 0 V
-3 1 V
-4 2 V
-5 2 V
-6 3 V
-7 2 V
-8 3 V
-8 4 V
-8 3 V
-9 3 V
-8 4 V
-9 3 V
stroke 483 3038 M
-7 3 V
-7 3 V
-6 2 V
-6 2 V
-4 2 V
-3 1 V
-1 1 V
-1 0 V
-2 1 V
-3 1 V
-4 1 V
-5 2 V
-6 3 V
-7 3 V
-8 3 V
-8 3 V
-8 3 V
-9 4 V
-8 3 V
-9 3 V
-7 3 V
-7 3 V
-7 3 V
-5 2 V
-4 1 V
-3 1 V
-1 1 V
-1 0 V
-2 1 V
-3 1 V
-4 2 V
-5 2 V
-6 2 V
-7 3 V
-8 3 V
-8 3 V
-9 4 V
-8 3 V
-9 3 V
-8 4 V
-7 3 V
-7 2 V
-7 3 V
-5 2 V
-4 1 V
-3 2 V
-1 0 V
-1 0 V
-2 1 V
-3 1 V
-4 2 V
-5 2 V
-6 2 V
-7 3 V
-8 3 V
-8 3 V
-9 4 V
-8 3 V
-9 3 V
-8 4 V
-8 3 V
-7 2 V
-6 3 V
-5 2 V
-4 1 V
-3 1 V
-2 1 V
-2 1 V
-3 1 V
-4 2 V
-5 2 V
-6 2 V
-7 3 V
-8 3 V
-8 3 V
-9 3 V
-8 4 V
-9 3 V
-8 3 V
-8 3 V
-7 3 V
-6 2 V
-5 2 V
-4 2 V
-3 1 V
-2 1 V
stroke
LTb
0 3791 N
0 640 L
2879 0 V
0 3151 V
0 3791 L
Z stroke
1.000 UP
1.000 UL
LTb
1.000 UL
LTb
4320 640 M
63 0 V
2816 0 R
-63 0 V
4320 1090 M
63 0 V
2816 0 R
-63 0 V
4320 1540 M
63 0 V
2816 0 R
-63 0 V
4320 1990 M
63 0 V
2816 0 R
-63 0 V
4320 2441 M
63 0 V
2816 0 R
-63 0 V
4320 2891 M
63 0 V
2816 0 R
-63 0 V
4320 3341 M
63 0 V
2816 0 R
-63 0 V
4320 3791 M
63 0 V
2816 0 R
-63 0 V
4320 640 M
0 63 V
0 3088 R
0 -63 V
5280 640 M
0 63 V
0 3088 R
0 -63 V
6239 640 M
0 63 V
0 3088 R
0 -63 V
7199 640 M
0 63 V
0 3088 R
0 -63 V
stroke
4320 3791 N
0 -3151 V
2879 0 V
0 3151 V
-2879 0 V
Z stroke
LCb setrgbcolor
LTb
1.000 UP
1.000 UL
LTb
1.000 UL
LT0
LCb setrgbcolor
LT0
5960 1255 M
543 0 V
58 481 R
-1 0 V
-3 -1 V
-4 -1 V
-6 -1 V
-6 -1 V
-7 -2 V
-8 -1 V
-8 -2 V
-8 -2 V
-9 -2 V
-8 -2 V
-8 -2 V
-8 -1 V
-7 -2 V
-6 -1 V
-5 -1 V
-5 -1 V
-2 -1 V
-2 0 V
-1 -1 V
-2 0 V
-2 -1 V
-5 -1 V
-5 -1 V
-6 -1 V
-7 -2 V
-8 -1 V
-8 -2 V
-8 -2 V
-9 -2 V
-8 -2 V
-8 -2 V
-8 -2 V
-7 -2 V
-6 -1 V
-6 -1 V
-4 -1 V
-3 -1 V
-1 0 V
-1 -1 V
-2 0 V
-3 -1 V
-4 -1 V
-5 -1 V
-6 -1 V
-7 -2 V
-8 -2 V
-8 -2 V
-8 -2 V
-9 -2 V
-8 -2 V
-9 -2 V
-7 -2 V
-7 -2 V
-6 -1 V
-6 -2 V
-4 -1 V
-3 0 V
-1 -1 V
-1 0 V
-2 -1 V
-3 0 V
-4 -1 V
-5 -2 V
-6 -1 V
-7 -2 V
-8 -2 V
-8 -2 V
-8 -2 V
-9 -2 V
-8 -3 V
-9 -2 V
-7 -2 V
-7 -2 V
-7 -1 V
-5 -2 V
-4 -1 V
-3 0 V
-1 -1 V
-1 0 V
-2 0 V
-3 -1 V
-4 -1 V
-5 -2 V
-6 -1 V
-7 -2 V
-8 -2 V
-8 -3 V
-8 -2 V
-9 -2 V
-9 -3 V
-8 -2 V
-7 -2 V
-7 -2 V
-7 -2 V
-5 -1 V
-4 -1 V
-3 -1 V
-1 0 V
-1 -1 V
-2 0 V
-3 -1 V
stroke 5996 1597 M
-4 -1 V
-5 -2 V
-6 -1 V
-7 -2 V
-8 -3 V
-8 -2 V
-9 -2 V
-8 -3 V
-9 -2 V
-8 -3 V
-7 -2 V
-8 -2 V
-6 -2 V
-5 -1 V
-4 -1 V
-3 -1 V
-2 -1 V
-2 -1 V
-3 0 V
-4 -2 V
-5 -1 V
-6 -2 V
-7 -2 V
-8 -2 V
-8 -3 V
-9 -2 V
-8 -3 V
-9 -3 V
-8 -2 V
-8 -3 V
-7 -2 V
-6 -2 V
-5 -1 V
-4 -2 V
-3 -1 V
-2 0 V
-2 -1 V
-3 -1 V
-4 -1 V
-5 -2 V
-6 -2 V
-7 -2 V
-8 -2 V
-8 -3 V
-9 -3 V
-8 -2 V
-9 -3 V
-8 -3 V
-8 -2 V
-7 -3 V
-6 -2 V
-5 -2 V
-4 -1 V
-3 -1 V
-2 0 V
0 -1 V
-2 0 V
-3 -1 V
-4 -1 V
-5 -2 V
-6 -2 V
-7 -3 V
-8 -2 V
-8 -3 V
-9 -3 V
-8 -3 V
-9 -3 V
-8 -3 V
-8 -2 V
-7 -3 V
-6 -2 V
-5 -2 V
-4 -1 V
-3 -1 V
-2 -1 V
-1 0 V
-1 -1 V
-3 -1 V
-4 -1 V
-5 -2 V
-7 -2 V
-7 -3 V
-7 -2 V
-8 -3 V
-9 -3 V
-8 -3 V
-9 -4 V
-8 -3 V
-8 -2 V
-7 -3 V
-6 -2 V
-5 -2 V
-4 -2 V
-3 -1 V
-2 0 V
-1 -1 V
-1 0 V
-3 -1 V
-4 -2 V
-5 -2 V
-7 -2 V
-7 -3 V
-7 -3 V
-8 -3 V
stroke 5398 1400 M
-9 -3 V
-9 -3 V
-8 -4 V
-8 -3 V
-8 -3 V
-7 -3 V
-6 -2 V
-5 -2 V
-4 -2 V
-3 -1 V
-2 -1 V
-1 0 V
-1 -1 V
-3 -1 V
-4 -1 V
-6 -2 V
-6 -3 V
-7 -3 V
-7 -3 V
-9 -3 V
-8 -4 V
-9 -3 V
-8 -4 V
-8 -3 V
-8 -3 V
-7 -3 V
-6 -3 V
-5 -2 V
-4 -1 V
-3 -2 V
-2 0 V
-1 -1 V
-1 0 V
-3 -2 V
-4 -1 V
-6 -3 V
-6 -2 V
-7 -3 V
-7 -3 V
-9 -4 V
-8 -4 V
-9 -3 V
-8 -4 V
-8 -4 V
-8 -3 V
-7 -3 V
-6 -3 V
-5 -2 V
-5 -2 V
-2 -1 V
-2 -1 V
-1 0 V
-1 -1 V
-3 -1 V
-4 -2 V
-6 -3 V
-6 -2 V
-7 -3 V
-8 -4 V
-8 -4 V
-8 -4 V
-9 -4 V
-8 -3 V
-8 -4 V
-8 -4 V
-7 -3 V
-6 -3 V
-6 -2 V
-4 -2 V
-3 -2 V
-1 -1 V
-1 0 V
-2 -1 V
-2 -1 V
-5 -2 V
-5 -3 V
-6 -3 V
-7 -3 V
-8 -4 V
-8 -4 V
-8 -4 V
-9 -4 V
-8 -4 V
-9 -4 V
-7 -4 V
-7 -3 V
-6 -3 V
-6 -3 V
-4 -2 V
-3 -2 V
-1 0 V
-1 -1 V
-2 -1 V
-3 -1 V
-4 -2 V
-5 -3 V
-6 -3 V
-7 -3 V
-8 -5 V
-8 -4 V
-8 -4 V
-9 -5 V
-8 -4 V
-9 -4 V
stroke 4803 1133 M
-7 -4 V
-7 -4 V
-6 -3 V
-6 -3 V
-4 -2 V
-3 -2 V
-1 -1 V
-1 0 V
-2 -1 V
-3 -2 V
-4 -2 V
-5 -3 V
-6 -3 V
-7 -4 V
-8 -4 V
-8 -4 V
-8 -5 V
-9 -5 V
-8 -5 V
-9 -4 V
-7 -4 V
-7 -5 V
-7 -3 V
-5 -3 V
-4 -2 V
-3 -2 V
-1 -1 V
-1 0 V
-2 -1 V
-3 -2 V
-4 -2 V
-5 -3 V
-6 -4 V
-7 -4 V
-8 -4 V
-8 -5 V
-9 -5 V
-8 -5 V
-9 -5 V
-8 -5 V
-7 -5 V
-7 -4 V
-7 -4 V
-5 -3 V
-4 -2 V
-3 -2 V
-1 -1 V
-1 0 V
-2 -1 V
-3 -2 V
-4 -3 V
-5 -3 V
-6 -4 V
-7 -4 V
-8 -5 V
-8 -5 V
-9 -5 V
-8 -6 V
-9 -5 V
-8 -5 V
-8 -5 V
-7 -4 V
-6 -4 V
-5 -4 V
-4 -2 V
-3 -2 V
-2 -1 V
0 -1 V
-2 -1 V
-3 -2 V
-4 -2 V
-5 -4 V
-6 -4 V
-7 -4 V
-8 -6 V
-8 -5 V
-9 -6 V
-8 -5 V
-9 -6 V
-8 -6 V
-8 -5 V
-7 -5 V
-6 -4 V
-5 -3 V
-4 -4 V
-3 -1 V
-2 -2 V
stroke
LT1
LCb setrgbcolor
LT1
5960 1055 M
543 0 V
58 1950 R
-1 0 V
-3 1 V
-4 0 V
-6 1 V
-6 1 V
-7 1 V
-8 1 V
-8 1 V
-8 1 V
-9 1 V
-8 1 V
-8 1 V
-8 1 V
-7 1 V
-6 1 V
-5 0 V
-5 1 V
-2 0 V
-2 0 V
-1 0 V
-2 1 V
-2 0 V
-5 1 V
-5 0 V
-6 1 V
-7 1 V
-8 1 V
-8 1 V
-8 1 V
-9 1 V
-8 1 V
-8 1 V
-8 1 V
-7 1 V
-6 1 V
-6 1 V
-4 1 V
-3 0 V
-1 0 V
-1 0 V
-2 0 V
-3 1 V
-4 0 V
-5 1 V
-6 1 V
-7 1 V
-8 1 V
-8 1 V
-8 1 V
-9 1 V
-8 2 V
-9 1 V
-7 1 V
-7 1 V
-6 1 V
-6 0 V
-4 1 V
-3 0 V
-1 1 V
-1 0 V
-2 0 V
-3 0 V
-4 1 V
-5 1 V
-6 1 V
-7 1 V
-8 1 V
-8 1 V
-8 1 V
-9 1 V
-8 2 V
-9 1 V
-7 1 V
-7 1 V
-7 1 V
-5 1 V
-4 0 V
-3 1 V
-1 0 V
-1 0 V
-2 0 V
-3 1 V
-4 0 V
-5 1 V
-6 1 V
-7 1 V
-8 1 V
-8 1 V
-8 2 V
-9 1 V
-9 1 V
-8 2 V
-7 1 V
-7 1 V
-7 1 V
-5 1 V
-4 0 V
-3 1 V
-1 0 V
-1 0 V
-2 0 V
-3 1 V
stroke 5996 3084 M
-4 1 V
-5 0 V
-6 1 V
-7 1 V
-8 2 V
-8 1 V
-9 1 V
-8 2 V
-9 1 V
-8 1 V
-7 2 V
-8 1 V
-6 1 V
-5 0 V
-4 1 V
-3 1 V
-2 0 V
-2 0 V
-3 1 V
-4 0 V
-5 1 V
-6 1 V
-7 2 V
-8 1 V
-8 1 V
-9 2 V
-8 1 V
-9 1 V
-8 2 V
-8 1 V
-7 1 V
-6 1 V
-5 1 V
-4 1 V
-3 0 V
-2 1 V
-2 0 V
-3 1 V
-4 0 V
-5 1 V
-6 1 V
-7 2 V
-8 1 V
-8 1 V
-9 2 V
-8 1 V
-9 2 V
-8 1 V
-8 2 V
-7 1 V
-6 1 V
-5 1 V
-4 1 V
-3 0 V
-2 0 V
0 1 V
-2 0 V
-3 0 V
-4 1 V
-5 1 V
-6 1 V
-7 1 V
-8 2 V
-8 1 V
-9 2 V
-8 2 V
-9 1 V
-8 2 V
-8 1 V
-7 1 V
-6 2 V
-5 1 V
-4 0 V
-3 1 V
-2 0 V
-1 0 V
-1 1 V
-3 0 V
-4 1 V
-5 1 V
-7 1 V
-7 1 V
-7 2 V
-8 2 V
-9 1 V
-8 2 V
-9 1 V
-8 2 V
-8 2 V
-7 1 V
-6 1 V
-5 1 V
-4 1 V
-3 1 V
-2 0 V
-1 0 V
-1 0 V
-3 1 V
-4 1 V
-5 1 V
-7 1 V
-7 2 V
-7 1 V
-8 2 V
stroke 5398 3191 M
-9 1 V
-9 2 V
-8 2 V
-8 2 V
-8 1 V
-7 2 V
-6 1 V
-5 1 V
-4 1 V
-3 1 V
-2 0 V
-1 0 V
-1 0 V
-3 1 V
-4 1 V
-6 1 V
-6 1 V
-7 2 V
-7 1 V
-9 2 V
-8 2 V
-9 2 V
-8 2 V
-8 1 V
-8 2 V
-7 2 V
-6 1 V
-5 1 V
-4 1 V
-3 1 V
-2 0 V
-1 0 V
-1 1 V
-3 0 V
-4 1 V
-6 1 V
-6 2 V
-7 1 V
-7 2 V
-9 2 V
-8 2 V
-9 2 V
-8 2 V
-8 1 V
-8 2 V
-7 2 V
-6 1 V
-5 1 V
-5 1 V
-2 1 V
-2 0 V
-1 1 V
-1 0 V
-3 1 V
-4 1 V
-6 1 V
-6 1 V
-7 2 V
-8 2 V
-8 2 V
-8 2 V
-9 2 V
-8 2 V
-8 2 V
-8 2 V
-7 1 V
-6 2 V
-6 1 V
-4 1 V
-3 1 V
-1 0 V
-1 0 V
-2 1 V
-2 0 V
-5 1 V
-5 2 V
-6 1 V
-7 2 V
-8 2 V
-8 2 V
-8 2 V
-9 2 V
-8 2 V
-9 2 V
-7 2 V
-7 2 V
-6 2 V
-6 1 V
-4 1 V
-3 1 V
-1 0 V
-1 1 V
-2 0 V
-3 1 V
-4 1 V
-5 1 V
-6 2 V
-7 2 V
-8 2 V
-8 2 V
-8 2 V
-9 2 V
-8 3 V
-9 2 V
stroke 4803 3330 M
-7 2 V
-7 2 V
-6 1 V
-6 2 V
-4 1 V
-3 1 V
-1 0 V
-1 0 V
-2 1 V
-3 1 V
-4 1 V
-5 1 V
-6 2 V
-7 2 V
-8 2 V
-8 2 V
-8 3 V
-9 2 V
-8 2 V
-9 3 V
-7 2 V
-7 2 V
-7 2 V
-5 1 V
-4 1 V
-3 1 V
-1 1 V
-1 0 V
-2 0 V
-3 1 V
-4 1 V
-5 2 V
-6 2 V
-7 2 V
-8 2 V
-8 2 V
-9 3 V
-8 2 V
-9 3 V
-8 2 V
-7 3 V
-7 2 V
-7 2 V
-5 1 V
-4 1 V
-3 1 V
-1 1 V
-1 0 V
-2 1 V
-3 0 V
-4 2 V
-5 1 V
-6 2 V
-7 2 V
-8 3 V
-8 2 V
-9 3 V
-8 3 V
-9 2 V
-8 3 V
-8 3 V
-7 2 V
-6 2 V
-5 2 V
-4 1 V
-3 1 V
-2 0 V
0 1 V
-2 0 V
-3 1 V
-4 1 V
-5 2 V
-6 2 V
-7 3 V
-8 2 V
-8 3 V
-9 3 V
-8 3 V
-9 3 V
-8 3 V
-8 3 V
-7 4 V
-6 2 V
-5 4 V
-4 3 V
-3 10 V
-2 5 V
stroke
LTb
4320 3791 N
0 -3151 V
2879 0 V
0 3151 V
-2879 0 V
Z stroke
1.000 UP
1.000 UL
LTb
stroke
grestore
end
showpage
  }}%
  \put(5840,1055){\makebox(0,0)[r]{\strut{}$R_2$}}%
  \put(5840,1255){\makebox(0,0)[r]{\strut{}$R_1$}}%
  \put(5759,140){\makebox(0,0){\strut{}$\omega$}}%
  \put(7199,440){\makebox(0,0){\strut{} 1.6}}%
  \put(6239,440){\makebox(0,0){\strut{} 1.4}}%
  \put(5280,440){\makebox(0,0){\strut{} 1.2}}%
  \put(4320,440){\makebox(0,0){\strut{} 1}}%
  \put(4200,3791){\makebox(0,0)[r]{\strut{} 1.08}}%
  \put(4200,3341){\makebox(0,0)[r]{\strut{} 1.07}}%
  \put(4200,2891){\makebox(0,0)[r]{\strut{} 1.06}}%
  \put(4200,2441){\makebox(0,0)[r]{\strut{} 1.05}}%
  \put(4200,1990){\makebox(0,0)[r]{\strut{} 1.04}}%
  \put(4200,1540){\makebox(0,0)[r]{\strut{} 1.03}}%
  \put(4200,1090){\makebox(0,0)[r]{\strut{} 1.02}}%
  \put(4200,640){\makebox(0,0)[r]{\strut{} 1.01}}%
  \put(657,1370){\makebox(0,0)[r]{\strut{}$h_{V}$}}%
  \put(657,1570){\makebox(0,0)[r]{\strut{}$h_{A_3}$}}%
  \put(657,1770){\makebox(0,0)[r]{\strut{}$h_{A_2}$}}%
  \put(657,1970){\makebox(0,0)[r]{\strut{}$h_{A_1}$}}%
  \put(657,2170){\makebox(0,0)[r]{\strut{}$h_-$}}%
  \put(657,2370){\makebox(0,0)[r]{\strut{}$h_+$}}%
  \put(1439,140){\makebox(0,0){\strut{}$\omega$}}%
  \put(2879,440){\makebox(0,0){\strut{} 1.6}}%
  \put(1919,440){\makebox(0,0){\strut{} 1.4}}%
  \put(960,440){\makebox(0,0){\strut{} 1.2}}%
  \put(0,440){\makebox(0,0){\strut{} 1}}%
  \put(-120,3791){\makebox(0,0)[r]{\strut{} 1.2}}%
  \put(-120,3266){\makebox(0,0)[r]{\strut{} 1}}%
  \put(-120,2741){\makebox(0,0)[r]{\strut{} 0.8}}%
  \put(-120,2216){\makebox(0,0)[r]{\strut{} 0.6}}%
  \put(-120,1690){\makebox(0,0)[r]{\strut{} 0.4}}%
  \put(-120,1165){\makebox(0,0)[r]{\strut{} 0.2}}%
  \put(-120,640){\makebox(0,0)[r]{\strut{} 0}}%
\end{picture}%
\endgroup
 